\documentclass[superscriptaddress,10pt,amsmath,amssymb,aps,pra,twocolumn,preprintnumbers,a4paper]{revtex4-2}
\usepackage[utf8]{inputenc}
\usepackage[T1]{fontenc}
\usepackage{amsmath}
\usepackage{amsfonts}
\usepackage{amssymb}
\usepackage{amsthm}
\usepackage{dsfont}
\usepackage{graphicx}
\usepackage[dvipsnames]{xcolor}
\usepackage{enumitem}
\usepackage{hyperref}
\usepackage{physics}
\usepackage{bbm}
\usepackage{indentfirst}
\usepackage{thmtools}
\usepackage{thm-restate}
\usepackage{soul}
\usepackage{comment}
\setstcolor{red}
\usepackage{breakurl}
\usepackage{mathtools}
\usepackage{chngcntr}
\usepackage[normalem]{ulem}

\usepackage[left=0.78in, right=0.78in, bottom=1in, top=1in]{geometry}

\usepackage{silence}
\usepackage{fancyhdr}
\fancypagestyle{header}{
  \fancyhf{}%
  \fancyhead[R]{\hyperref[toc]{\thepage}}%
}

\usepackage{tikz}
\usetikzlibrary{arrows.meta,positioning,calc}

\usepackage{needspace}
\usepackage{tcolorbox}
\tcbuselibrary{breakable}

\newtcolorbox{graydefinitionbox}{
    before={\Needspace{8\baselineskip}\par\medskip},
    after={\par\medskip},
    colback=gray!4,
    colframe=gray!80,
    boxrule=1.4pt,
    arc=6pt,
    left=8pt,
    right=8pt,
    top=8pt,
    bottom=8pt,
    breakable
}

\NewDocumentEnvironment{eqsplit}{b}{%
    \begin{equation}%
    \begin{split}%
        #1
    \end{split}%
    \end{equation}%
}{}

\definecolor{blueviolet}{rgb}{0.2, 0.2, 0.6}
\definecolor{webgreen}{rgb}{0,.5,0}
\definecolor{webbrown}{rgb}{.6,0,0}

\hypersetup{
  bookmarks=true,
  colorlinks=true,
  urlcolor=webbrown,
  linkcolor=blueviolet,
  citecolor=webgreen,
  pdfstartpage=1,
  pdfstartview={XYZ},
  bookmarksopen=true,
  bookmarksopenlevel=1,
  pdfpagemode=UseOutlines,
  pdfview=XYZ
}

\usepackage{appendix}
\usepackage{cleveref}

\makeatletter

\newcommand{\apptocfile}{atoc}

\let\apptoc@orig@appendix\appendix
\renewcommand{\appendix}{%
  \apptoc@orig@appendix
  \let\apptoc@orig@addtocontents\addtocontents
  \long\def\addtocontents##1##2{%
    \def\apptoc@ext{##1}%
    \def\apptoc@toc{toc}%
    \ifx\apptoc@ext\apptoc@toc
      \apptoc@orig@addtocontents{\apptocfile}{##2}%
    \else
      \apptoc@orig@addtocontents{##1}{##2}%
    \fi
  }%
}
\newcommand{\appendixtableofcontents}{%
  \begingroup
    \raggedbottom
    \setcounter{tocdepth}{3}%
    \phantomsection
    \let\addcontentsline\@gobblethree
    \pdfbookmark[1]{Appendices}{apxcontents}%

    \renewcommand*\numberline[1]{##1.\enspace}%

    \newcommand{\apptocentry}[4]{%
      \par\addvspace{##1}%
      \noindent\hspace*{##2}%
      \parbox[t]{\dimexpr\linewidth-##2-2.4em\relax}{##3}%
      \hfill ##4\par
    }%

    \renewcommand*\l@section[2]{%
      \apptocentry{0.55em plus 0.1em}{0em}{\bfseries ##1}{##2}%
    }%

    \renewcommand*\l@subsection[2]{%
      \apptocentry{0em}{1.5em}{##1}{##2}%
    }%

    \renewcommand*\l@subsubsection[2]{%
      \apptocentry{0em}{3.5em}{##1}{##2}%
    }%

    \begin{center}
      \textbf{Contents \& Roadmap}
    \end{center}
    \vspace{0.5em}

    \@starttoc{\apptocfile}%
  \endgroup
}
\makeatother

\usepackage{bibunits}
\defaultbibliographystyle{science/aps}
\defaultbibliography{science/refs_science}

\usepackage{algpseudocode}

\makeatletter
\newcounter{algorithm}
\renewcommand{\thealgorithm}{\arabic{algorithm}}
\newcommand{\algorithmname}{Algorithm}

\newenvironment{algorithm}[1][]{%
  \par\addvspace{\intextsep}%
  \noindent\begin{minipage}{\linewidth}%
  \small
  \hrule height 0.8pt
  \vskip 2pt
  \def\@captype{algorithm}%
  \renewcommand{\caption}[2][]{%
    \refstepcounter{algorithm}%
    \vskip 1pt
    \noindent\hfil
      \textbf{\algorithmname~\thealgorithm} ##2%
    \hfil\par
    \vskip 2pt
    \hrule height 0.4pt
    \vskip 4pt
  }%
}{%
  \vskip 2pt
  \hrule height 0.8pt
  \end{minipage}%
  \par\addvspace{\intextsep}%
}
\makeatother

\algrenewcommand\algorithmicrequire{\textbf{Input:}}
\algrenewcommand\algorithmicensure{\textbf{Output:}}
\usepackage{setspace}

\newtheorem{theorem}{Theorem}

\newtheorem{fact}{Fact}
\newtheorem{lemma}[theorem]{Lemma}
\newtheorem{corollary}[theorem]{Corollary}
\newtheorem{definition}[theorem]{Definition}
\newtheorem{remark}[theorem]{Remark}

\usepackage{xcolor}

\makeatletter
\newcommand{\bigboxplus}{\mathop{\vphantom{\sum}\mathchoice
  {\vcenter{\hbox{\Large$\m@th\boxplus$}}}
  {\vcenter{\hbox{\large$\m@th\boxplus$}}}
  {\vcenter{\hbox{\normalsize$\m@th\boxplus$}}}
  {\vcenter{\hbox{\scriptsize$\m@th\boxplus$}}}
}\displaylimits}
\makeatother

\makeatletter
\def\@hangfrom@section#1#2#3{\@hangfrom{#1#2}#3}%
\def\@hangfroms@section#1#2{#1#2}%
\makeatother

\makeatletter
\renewcommand\footnoterule{%
  \kern -3pt
  \hrule width 1.6in height 0.4pt
  \kern 3pt
}
\makeatother

\begin{document}
\pagestyle{header}

\addtocontents{toc}{\protect\setcounter{tocdepth}{-1}} 

\title{Exponential quantum advantage for learning signals with a single qubit}

\author{Ishaan~Kannan}
\thanks{\raggedright Equal contribution. Contact: ishaan\_kannan@g.harvard.edu, svp36@cornell.edu}
\affiliation{Harvard Quantum Initiative, Harvard University, Cambridge, MA 02138, USA}
\affiliation{Department of Physics, Harvard University, Cambridge, MA 02138, USA}

\author{Sridhar~Prabhu}
\thanks{\raggedright Equal contribution. Contact: ishaan\_kannan@g.harvard.edu, svp36@cornell.edu}
\affiliation{School of Applied and Engineering Physics, Cornell University, Ithaca, NY 14853, USA}
\affiliation{Department of Physics, Cornell University, Ithaca, NY 14853, USA}

\author{Saeed~A.~Khan}
\affiliation{School of Applied and Engineering Physics, Cornell University, Ithaca, NY 14853, USA}

\author{Mandar~M.~Sohoni}
\affiliation{School of Applied and Engineering Physics, Cornell University, Ithaca, NY 14853, USA}

\author{Xingrui~Song}
\affiliation{School of Applied and Engineering Physics, Cornell University, Ithaca, NY 14853, USA}

\author{Saswata~Roy}
\affiliation{School of Applied and Engineering Physics, Cornell University, Ithaca, NY 14853, USA}
\affiliation{Department of Physics, Cornell University, Ithaca, NY 14853, USA}

\author{Alen~Senanian}
\thanks{\raggedright Present address: Diraq, Chicago, IL 60617, USA}
\affiliation{School of Applied and Engineering Physics, Cornell University, Ithaca, NY 14853, USA}
\affiliation{Department of Physics, Cornell University, Ithaca, NY 14853, USA}

\author{Valla~Fatemi}
\affiliation{School of Applied and Engineering Physics, Cornell University, Ithaca, NY 14853, USA}

\author{Peter~L.~McMahon}
\thanks{\raggedright These authors jointly supervised this work. Contact: jcotler@fas.harvard.edu, pmcmahon@cornell.edu}
\affiliation{School of Applied and Engineering Physics, Cornell University, Ithaca, NY 14853, USA}
\affiliation{Kavli Institute at Cornell for Nanoscale Science, Cornell University, Ithaca, NY 14853, USA}

\author{Jordan~Cotler}
\thanks{\raggedright These authors jointly supervised this work. Contact: jcotler@fas.harvard.edu, pmcmahon@cornell.edu}
\affiliation{Harvard Quantum Initiative, Harvard University, Cambridge, MA 02138, USA}
\affiliation{Department of Physics, Harvard University, Cambridge, MA 02138, USA}

\begin{abstract}
Quantum technology has the potential to transform scientific discovery, but quantum advantages often require processing capabilities well beyond the reach of experimental platforms.  We show that coupling a single controllable qubit to an otherwise conventional sensor can exponentially reduce the number of measurements required to learn classical signals. These rigorous quantum advantages apply to fundamental sensing tasks, including learning Fourier coefficients, extracting temporal correlations from time-varying signals, and estimating transformations of physical observables. Using a superconducting cavity--qubit architecture, we experimentally demonstrate $10^7$-fold reductions in the number of measurements required for Fourier-amplitude and time-varying signal learning. Our \textit{quantum feature sensing} algorithms further enable orders-of-magnitude improvements in simulations of weak-signal dark matter detection and wireless communication applications. These quantum advantages are derived from Quantum Phase-Space Inference (Q$\Psi$), a unifying theory of quantum-enhanced experiments that simultaneously converts a set of experimental objectives and constraints into tight lower bounds and optimal quantum-enhanced learning algorithms while producing a certificate of quantum advantage. Q$\Psi$ extends beyond the regimes captured by quantum Fisher information and provides a framework for systematically identifying rigorous quantum advantages in practical experimental tasks. Together, our results establish that near-term quantum technology can exponentially enhance our ability to learn from classical signals.
\end{abstract}

\maketitle

\begin{bibunit}
\setcounter{tocdepth}{-1} 

\section{Introduction}
\vspace{-2mm}
Despite decades of progress since quantum computing was first proposed, examples of quantum advantage with clear practical utility remain scarce. A growing body of work has suggested a new route to quantum speedups: using experiments enhanced by quantum information processing to learn features of the natural world~\cite{Aaronson2018ShadowTomography, cotler2020quantum, classical_shadow_2020,chen2021exponential, Aharonov_Cotler_Qi_2022, Huang_2022, nisq_2023}. These foundational results largely establish worst-case separations or existence theorems rather than near-term practical advantages, and many rely on extensive entanglement together with high fidelity, fast local control. Translating such learning advantages into experimental applications requires augmenting quantum sensors with quantum information processing, despite the more limited coherence and control available on most sensing platforms \cite{Huelga_1997, Giovannetti_2011, Degen_2017, cotler2026noisy}. The resulting experimental demands place technologically meaningful implementations of many existing protocols beyond current capabilities. 

\begin{figure*}[ht!]
    \centering
    \includegraphics[width=\linewidth]{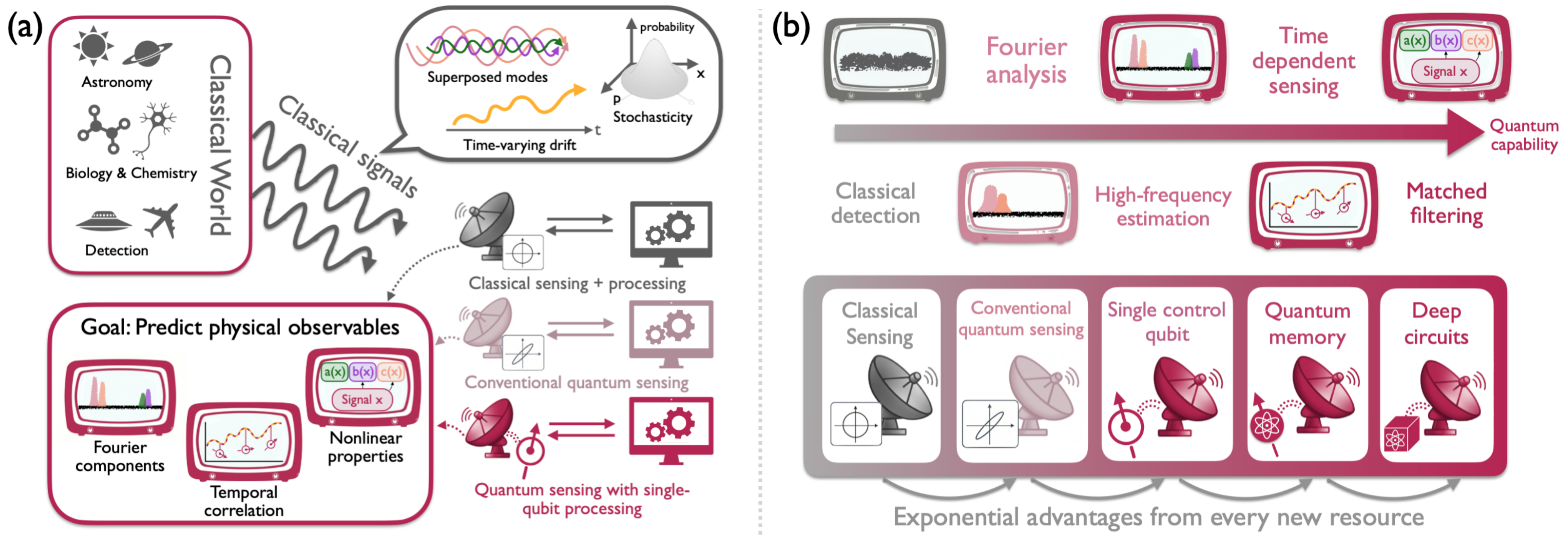}
    \caption{\textbf{Overview of exponential advantages in quantum-enhanced sensing.} (a) \textit{Architectures for sensing the classical world.} Data across experimental science and modern technology is transmitted through classical signals which may have superposed frequencies, time-varying drift, and stochasticity. In these realistic settings we consider sensing with near-term quantum architectures and predicting physical observables such as Fourier components, temporal correlations, and nonlinear transformation of the measured signals. Our $Q\Psi$ framework encapsulates all of these settings. (b) \textit{Hierarchy of quantum-enhanced sensing architectures.} As the sensing tasks become richer, sensor architectures with increasing computational capabilities enable a corresponding hierarchy of provable exponential quantum advantages.}
    \label{fig:schematic_main}
    \vspace{-3mm}
\end{figure*}

Meanwhile, quantum metrology has traditionally focused on achieving Heisenberg-limited precision, which provides a quadratic improvement over the classical scaling for estimating a well-specified parameter \cite{Giovannetti_2004, Giovannetti_2006, Demkowicz_Dobrza_ski_2012, Zhou_2018}. This framework is well suited to settings in which the signal model is known and estimation precision is the dominant asymptotic quantity. Many experimentally relevant sensing tasks, however, involve learning poorly characterized signals, resolving structured properties, or operating under platform-specific constraints that fall outside this standard parameter-estimation setting \cite{Allen_2012,Tsang_2011, Alvarez_2011,Mamin_2013, Backes_2021, Chin_2021,chen2025quantumprobetomography, cotler2026quantumadvantagesensingproperties, prabhu2026quantumcomputationaldisplacementsensing, khan2025quantum, khan2025quantum_novel, prabhu2025exponential, transversal_sensing}. A central open question is whether these broader regimes admit a rigorous framework in which near-term quantum information processing, integrated with modern quantum sensors, enables \textit{superpolynomial} improvements over classical methods for learning physical signals.

In this paper, we show that a single controllable ancilla qubit can exponentially reduce the measurement complexity of learning classical signals with a conventional single-mode sensor. We establish this advantage for several basic tasks, including estimating Fourier coefficients, extracting temporal correlations from time-varying signals, and evaluating nonlinear functionals of a signal distribution (Figure~\ref{fig:schematic_main}(a)). The required architecture consists only of the sensor and one controllable qubit, making the protocols compatible with existing experimental platforms~\cite{krantz2019quantum, bruzewicz2019trapped, aspelmeyer2014cavity}. We demonstrate these quantum advantages in proof-of-principle superconducting circuit experiments and illustrate their potential broader utility by numerically simulating applications to axionic dark matter search and wireless communication. Our protocols are designed using Quantum Phase-Space Inference (Q$\Psi$), a rigorous mathematical framework for experimentally constrained learning tasks. Q$\Psi$ identifies broad classes of sensing problems for which integrating near-term quantum information processing with modern quantum sensors yields superpolynomial advantages over both classical and conventional quantum methods.

Q$\Psi$ unifies learning, metrology, and estimation theory by mapping experimentally motivated tasks to precise complexity bounds and optimal quantum-enhanced strategies subject to the constraints of the experimental platform. Its central quantity is the \textit{accessible feature information} (AFI), a phase-space statistical overlap which determines both a tight lower bound on the resources required for a given task and an algorithm that attains this bound under the specified experimental constraints. When Q$\Psi$ is applied to learn properties of classical signals with quantum control, we refer to the resulting family of sensing strategies as \textit{quantum feature sensing} (QFS). By extending beyond the local parameter-estimation regime described by quantum Fisher information, the AFI provides a systematic way to identify quadratic, beyond-quadratic, and superpolynomial quantum advantages in realistic sensing experiments. Q$\Psi$ thereby provides an operational theory of quantum experimental advantage in which sensing architectures and learning protocols can be co-designed, encompassing sensing tasks that previously lay beyond the purview of canonical quantum metrology.

\vspace{-3mm}
\section{Exponential quantum advantage}
\vspace{-2mm}

Many sensing tasks can be formulated as learning a classical signal from the quantum channel that it induces on a quantum sensor \cite{Thomson_1982, Allen_1999, Haykin_2005, Clerk_2010, Bylander2011NoiseSpectroscopy}. Within this framework, Fourier coefficients characterize the coupling of the sensor to distinct phase-space modes of the channel, with higher-order coefficients resolving progressively finer signal structure. We show that coupling a conventional quantum sensor to a single ancilla qubit allows any phase-space Fourier coefficient to be learned exponentially more efficiently than is possible with sensing protocols that do not use quantum information processing.

To formulate this task in a model that applies to contemporary sensing platforms, we consider systems such as precision interferometers that control continuous electromagnetic degrees of freedom. The sensor is described by a quantum continuous-variable mode, and the classical signal acts on a probe state $\rho$ through a displacement channel, $\rho \mapsto \int d^2\alpha\,P(\alpha)D(\alpha)\rho D^\dagger(\alpha)$. Here $\alpha = x + ip$ labels the canonical quadrature coordinates, $D(\alpha)$ is the displacement operator, and the probability density $P(\alpha)$ completely characterizes the signal. For example, the $k$th Fourier coefficient of the signal along the $x$ quadrature is $\mathbb{E}_{x+ip\sim P}[e^{ikx}]$. 

\begin{figure*}[ht!]
    \centering
    \includegraphics[width=\linewidth]{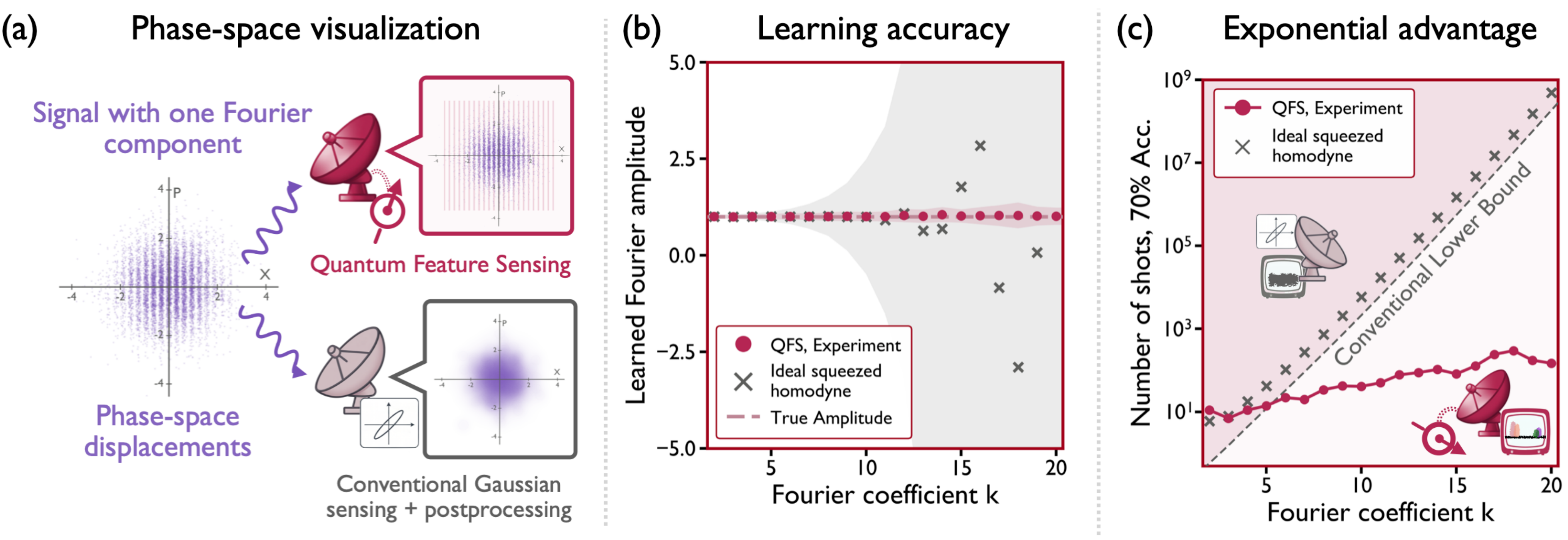}
    \caption{\textbf{Exponential quantum advantage of single-qubit control for learning Fourier amplitudes.} \textit{(a) Visualization of quantum advantage.} The signal contains a high-frequency phase-space Fourier component. The pictured distribution of displacements $P(\alpha)$, for $k=20$, is implemented in our experiment. The response of single-qubit quantum feature sensing (top) acts like a frequency comb, probing fine oscillatory structure in each measurement. The response of a conventional Gaussian protocol (bottom) sees the signal severely blurred by Gaussian convolution noise.  Plotted is the same signal as seen by $10^4$ shots from QFS vs. an entangled two-mode squeezed sensor with the same energy. \textit{(b) Exponentially improved accuracy in learning Fourier amplitudes using single-qubit QFS.} We sense a family of classical signals with $k^{\rm th}$ Fourier amplitude of $1.0$ using $10^4$ shots. Our experimental implementation of QFS accurately learns the amplitude for all frequencies $k$, whereas an idealized, noiseless simulation of squeezed homodyne sensing produces erratic estimates. Shaded regions depict one standard deviation of uncertainty in the learned amplitude (pink: QFS, experiment; gray: ideal squeezed homodyne, simulation). \textit{(c) Experimental demonstration of exponential advantage.} We experimentally realize Theorem \ref{thm:one_qubit_advantage_informal} using the protocol given in Appendix \ref{sec:quantum_control_upper_bounds} and the rigorous lower bound of Appendix \ref{sec:conventional_gaussian_lb}. For the signals depicted in (a), single-qubit QFS distinguishes the signals with prediction accuracy $70\%$ using only $\sim100$ shots, a factor of $10^7$ improvement over any conventional Gaussian approach with equal energy. We use a $70\%$ convention throughout, because success probability $\geq 1-\delta$ only incurs logarithmic shot overhead $\sim\log(\delta^{-1})$; other values of $\delta$ are shown in Appendix \ref{sec:experiment}.}
    \label{fig:single_qubit_advantage}
    \vspace{-3mm}
\end{figure*}
Because a physical signal is available for only a finite duration, the relevant resource is the number of times a sensing protocol must query the displacement channel to estimate properties of $P(\alpha)$. For conventional quantum sensors with energy $O(k)$, we prove that estimating the $k^{\rm th}$ Fourier coefficient requires a number of signal queries that grows exponentially with $k$. Gaussian resources such as squeezing, despite reducing measurement noise, leave this asymptotic scaling unchanged. We show that coupling a sensor that is otherwise capable only of classical operation to a single controllable ancilla qubit reduces the query complexity to $O(k)$:

\begin{theorem}[Exponential quantum advantage with a single qubit] \label{thm:one_qubit_advantage_informal}
A sensing protocol with probe energy $O(k)$, one ancilla qubit, and one control operation can learn the $k^{\rm th}$ Fourier coefficient of a signal using $O(k)$ signal queries. Any conventional protocol with the same energy scaling requires $\exp(\Omega(k))$ queries.
\end{theorem}

\noindent This result is directly relevant to modern quantum sensors with Gaussian probes, including squeezed states that suppress readout variance below the standard quantum limit imposed by the uncertainty principle on unsqueezed probes. Theorem~\ref{thm:one_qubit_advantage_informal} shows that coupling a single controllable qubit to sensors provides a qualitatively new resource, reducing the measurement complexity exponentially rather than further improving the precision of a Gaussian probe. A Gaussian protocol can match this query complexity only by increasing the probe energy by a polynomial factor in $k$. We demonstrate this advantage with a device comprising a transmon qubit coupled to the fundamental electromagnetic mode of a high-purity aluminum cavity for up to $k = 20$ in Figure~\ref{fig:single_qubit_advantage}. This device is used in all subsequent demonstrations. Even in the presence of decoherence, our quantum-enhanced sensor requires seven orders of magnitude fewer signal queries than a decoherence-free conventional Gaussian protocol. 

Having established that a single control operation can provide an exponential advantage for learning classical signals, we next ask whether other quantum processing resources enable comparable gains for different sensing primitives. We organize these possibilities into a hierarchy of quantum sensing architectures, with each level obtained from the previous one by adding a single quantum resource, and examine whether ascending the hierarchy makes progressively richer sensing tasks amenable to quantum advantage. Figure~\ref{fig:schematic_main}(b) summarizes this structure.

Our first result showed how a single controllable qubit can enhance the capabilities of modern quantum sensors. We next ask whether conventional quantum sensors themselves already offer exponential advantages over fully classical sensing strategies, in which the probe is restricted to classical coherent states of light. We show that they do: sensing with simple nonclassical Gaussian probe states and conventional measurements can yield exponential improvements over all classical sensing strategies.

\vspace{-0.5mm}
\begin{theorem}[Exponential advantage of quantum probes] \label{thm:conventional_vs_classical_informal}
A conventional Gaussian sensing protocol with energy $O(k)$ can learn the $k^{\rm th}$ angular Fourier coefficient of a signal using $O(1)$ queries. Any classical sensing protocol, regardless of available energy, requires $\exp(\Omega(k))$ queries to do so. 
\end{theorem}
\vspace{-0.5mm}

\begin{figure*}[ht!]
    \centering
    \includegraphics[width=\linewidth]{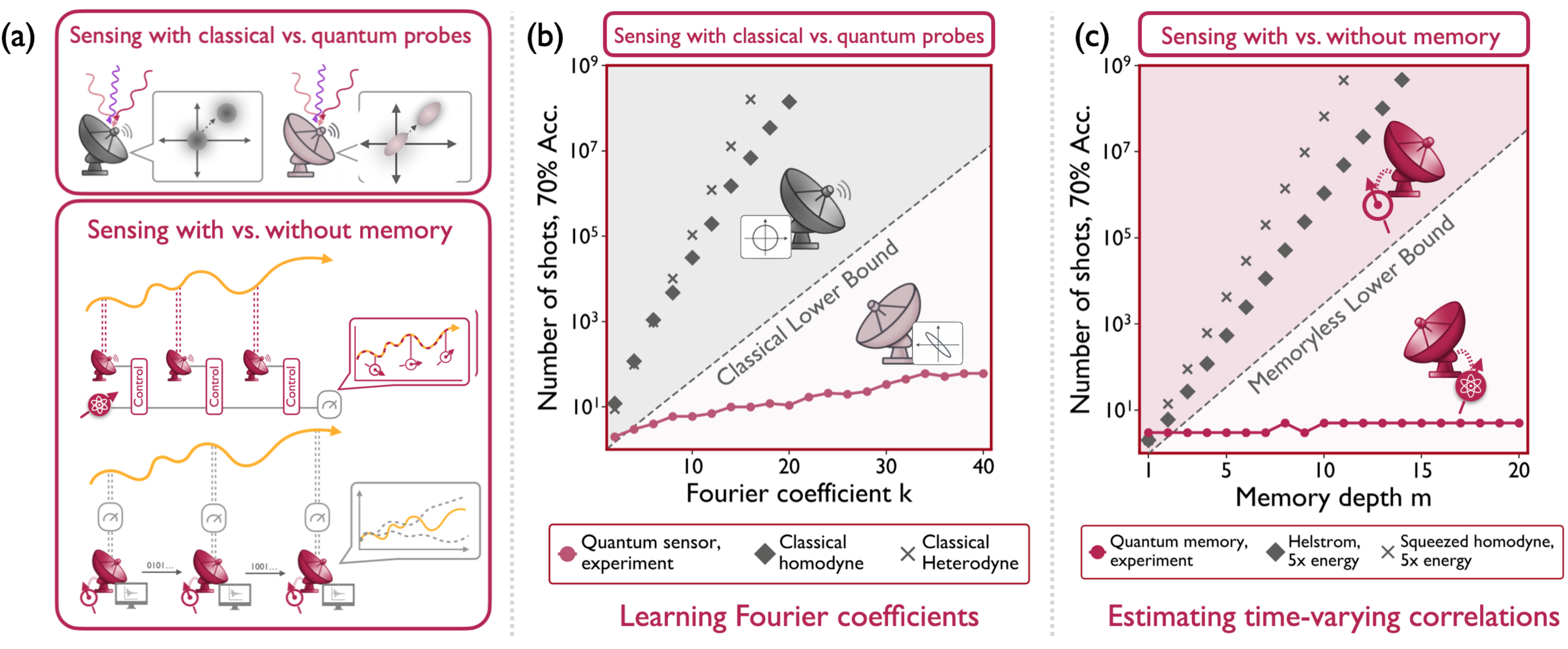}
    \caption{\textbf{Exponential quantum advantages with quantum probes and quantum memory.} (a) \textit{Top: schematic of sensing with classical vs.~conventional Gaussian quantum probes}. Classical sensors are modeled as preparing coherent probe states, whereas conventional Gaussian sensors can utilize squeezed probes and generaldyne readout. \textit{Bottom: schematic of quantum sensing with vs. without quantum memory}. A sensor equipped with quantum memory coherently accumulates temporal correlations, whereas a memoryless sensor must measure destructively after each interaction and carry forward only the resulting classical information. (b) \textit{Experimental demonstration of separation between conventional Gaussian sensor and classical probes.} A conventional Gaussian sensor estimates phase-space Fourier coefficients exponentially faster than any strategy using classical coherent-state probes and arbitrarily powerful measurements. An exponential advantage is also achieved by a protocol using a control qubit, detailed in Appendix \ref{sec:exp_proxy}, which we implement in experiment. (c) \textit{Experimental demonstration of exponential separation between quantum-enhanced, but memoryless, sensing and sensing with quantum memory.} A sensor coupled to a single memory qubit efficiently estimates temporal correlations of a signal which varies across time, while any memoryless strategy of comparable energy and arbitrary processing is exponentially costly. The simulated benchmarks include the Helstrom measurement, which is the optimal binary non-Gaussian measurement at each time step, and squeezed homodyne sensing. Both benchmarks are given five times the energy available to the memory-enabled protocol.}
    \label{fig:coherent_and_memory_experiments}
    \vspace{-3mm}
\end{figure*}

\noindent This separation is proved in Appendix \ref{sec:coherent_probe_lb} for angular Fourier coefficients, which are equivalent to \textit{moments} of the density $P$. Our experimental demonstration of this result is presented in Figure \ref{fig:coherent_and_memory_experiments}(b), where we implemented a modified quantum protocol with an ideal query complexity of $O(k^{2/3})$. 

Although we have shown that conventional quantum sensors that use quantum oscillators, such as optical or microwave cavities prepared in squeezed states, can achieve dramatic improvements beyond classical limits, they are unprotected quantum degrees of freedom generally limited to performing short-time destructive measurements due to decoherence. While the setting of Theorem \ref{thm:one_qubit_advantage_informal} considers coupling a sensor to a single physical qubit, future quantum information processing-enabled sensors may have access to encoded logical ancillas that can function both as controls and quantum memories, enabling coherent accumulation of signal over time. This capability aligns with sensing of time-dependent processes that are governed by displacement channels over correlated sequences of displacement variables $\alpha_{\Vec{t}}=(\alpha_{t_1}, \alpha_{t_2},...,\alpha_{t_m})$ for times $t_1 < t_2 <  \cdots < t_m$. The corresponding channel acts as $\rho\mapsto \int d^{2m}\alpha_{\Vec{t}} \ P_{\Vec{t}}(\alpha_{\Vec{t}}) D(\alpha_{\Vec{t}}) \rho D^\dagger(\alpha_{\Vec{t}})$, where $D(\alpha_{\Vec{t}})$ is the (time-ordered) product $\prod_{j=1}^m D(\alpha_{t_j})$. If $\chi_{t_j}$ represents the characteristic function of the signal at time $t_j$, such a time-varying signal is characterized by multi-time correlations of the form $\int d^{2m}\alpha_{\Vec{t}} \  P_{\Vec{t}}(\alpha_{\Vec{t}}) \prod_{j=1}^m \chi_{t_j}(\alpha_{t_j})$. We establish the following:

\vspace{-1mm}
\begin{theorem}[Exponential advantage from a single qubit of quantum memory] \label{thm:memory_informal}
    Let $t_1 < t_2 < \cdots < t_m$ be a list of times separated by at most $\Delta$. A sensor that decoheres on a timescale much shorter than $\Delta$, when coupled to a single qubit that remains coherent up to time $t_m$, can estimate the corresponding $m$-point temporal correlator using $O(1)$ queries to the time-varying signal. By contrast, any protocol with the same energy and no quantum degree of freedom coherent for times much longer than $\Delta$ requires $\exp(\Omega(m))$ queries, even given access to arbitrarily many qubits or modes.
\end{theorem}

\noindent This result identifies long-lived quantum memory as a resource that enables efficient characterization of time-dependent signals, with potential applications to radar, noise spectroscopy, and biomedical sensing \cite{melvin_2004,Buckley2014DiffuseCorrelation, Norris_2016}. In prominent previous works \cite{chen2021exponential, Huang_2022}, the size of the quantum advantage grows with the number of memory qubits or modes, and an exponential separation requires a memory register whose size increases with the problem. Here, the sensor itself may decohere on a much shorter timescale; repeated interactions with a single coherent qubit are sufficient to preserve the accumulated quantum information. To our knowledge, Theorem~\ref{thm:memory_informal} provides the first exponential quantum advantage obtained solely by adding one qubit of quantum memory. Figure~\ref{fig:coherent_and_memory_experiments}(c) presents a proof-of-principle implementation within the coherence times available in our experimental system.
\begin{figure*}
    \centering
    \includegraphics[width=\linewidth]{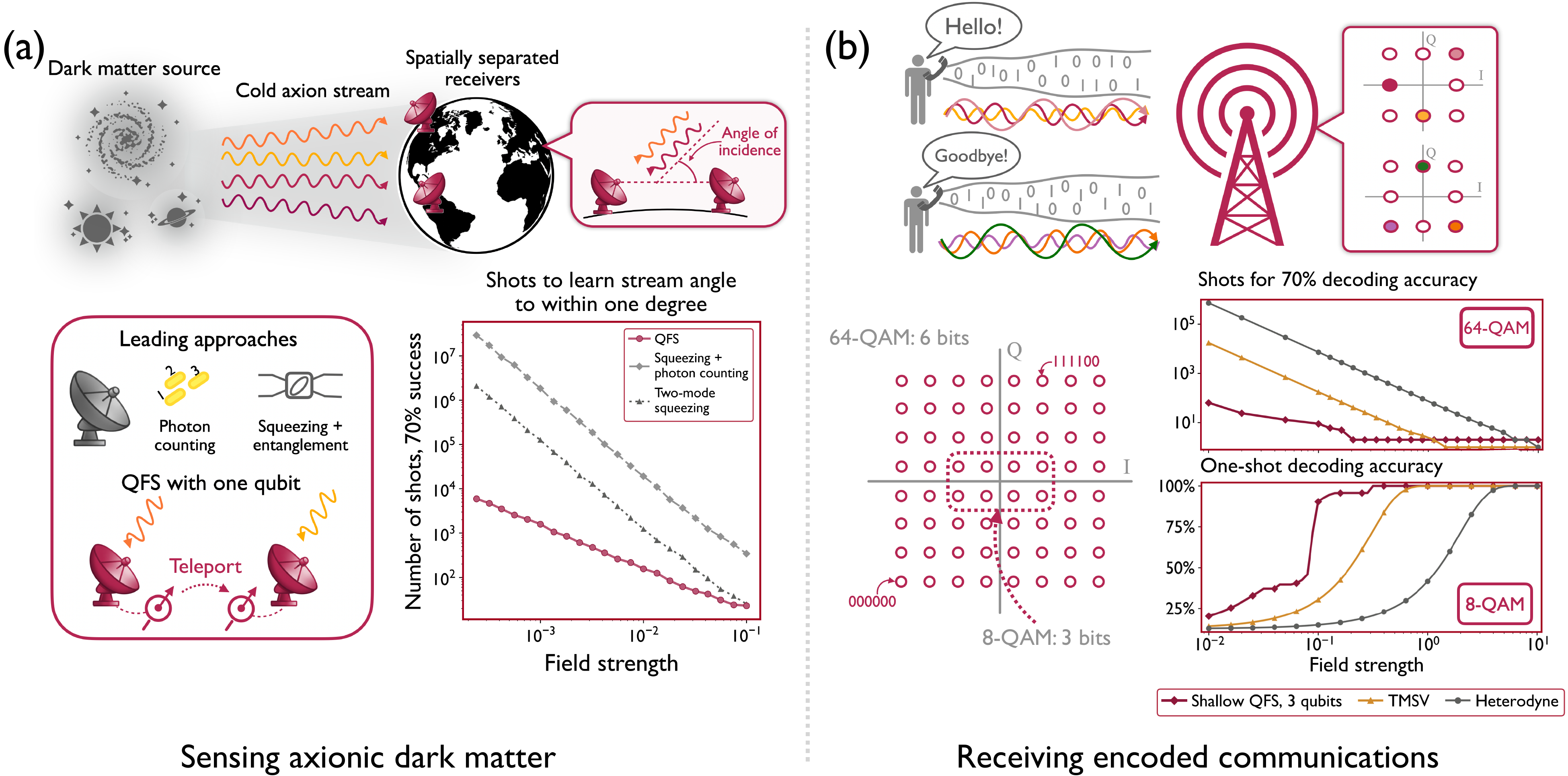}
    \caption{\textbf{Demonstrating speedups in practical applications through numerical simulations.} (a) \textit{Characterizing cold axion streams}. Daily periodicity in the angle of an incident field would constitute smoking-gun evidence of axionic dark matter. This angle manifests as Gaussian covariance between spatially separated receivers. Using estimated velocity parameters of the Sagittarius stream \cite{sagittarius}, we simulate QFS vs.~the leading practical approaches that do not use long-range, long-lived entangled sensors. The QFS protocol only requires a single qubit that is separately coupled to each detector at different times; a realization could use state teleportation of the ancilla, but no long-range-entangled probe is used for sensing. QFS accesses the covariance and resolves the stream angle to within $1^\circ$ quadratically faster than even highly non-Gaussian photon counting or entangled two-mode squeezing. (b) \textit{A quantum-enhanced receiver for wireless communication.} Wireless communications commonly encode information by modulating the amplitude and phase of a carrier wavepacket of known frequency. We simulate identification of an unknown transmitted symbol from 8-QAM and 64-QAM communication schemes, common in digital cable television, cellular networks, and satellite communications. We simulate a QFS receiver equipped with three qubits, so that each readout can perform at least 8-class classification. To identify one of 64 candidate symbols from a weak signal, QFS requires $\sim 100\times$ fewer shots than a two-mode-entangled quantum receiver, which requires $\sim 100\times$ fewer shots than a classical heterodyne receiver. For single-shot decoding, QFS achieves large success probabilities at far smaller field strengths.}
    \label{fig:applications}
    \vspace{-3mm}
\end{figure*}

We extend the hierarchy of quantum sensing architectures in the Appendices, and show that modest increases in quantum circuit depth can exponentially improve the estimation of polynomial functionals of the signal distribution $P$, the standard objective in matched filtering and estimation of signal moments. We also construct single-mode and multimode sensing problems for which control using a single qubit yields quantum advantages over all Gaussian strategies, even when those strategies have access to arbitrarily large energy and entanglement, and demonstrate that the entanglement-enabled advantages of Refs.~\cite{Oh_2024, disp_channel_learning_experiment} can be simplified to entanglement-free quantum feature sensing advantages in many cases. Whereas the demonstration of Ref.~\cite{disp_channel_learning_experiment}
focused on full tomography of a particular worst-case multimode displacement channel, the exponential speedups we demonstrate apply to elementary signal-learning primitives, such as Fourier coefficients and temporal correlations, of common single-mode displacement signals. 

In all of our results and demonstrations, the signal is a purely classical process inducing Gaussian displacements of the sensor; no exotic non-Gaussianity or nonclassical interaction is engineered. Moreover, our approach allows noise in the signal to be treated as part of the distribution, so that our distribution-agnostic protocols retain exponential advantage in the presence of Markovian, white, or thermal noise backgrounds.

Our quantum advantages and QFS protocols bear on many tasks in experimental science and technology. In Figure \ref{fig:applications}(a), we apply, in numerical simulations, QFS to the sensing of particle streams from primordial dark matter sources, for which accurate characterization of the incident angle would constitute smoking gun evidence of axionic dark matter \cite{Foster_2021}. We find that QFS, requiring only a single qubit and no long-range-entangled sensing resource, quadratically outperforms the longstanding leading approaches of squeezed photon-number-resolving measurements and two-mode entangled squeezing, at stream parameters estimated directly from astronomical measurements and calculations \cite{sagittarius, Foster_2021}. In Figure \ref{fig:applications}(b), we turn to a modern technological application in wireless communications, where the task is to decode transmitted signals. We simulate transmission of signals from standard 3-bit and 6-bit communication schemes used across digital streaming, cellular networks, and satellite communication and observe that in the weak-signal regime, a shallow-depth QFS receiver identifies a transmitted symbol using a factor of $\sim100$ fewer shots than a receiver based on a two-mode-squeezed-vacuum interferometer~\cite{braunstein2000dense} of equal energy, and $\sim 10^4$ fewer shots than a classical receiver. Details of our simulations are given in Appendix~\ref{sec:experiment}. These examples illustrate that the applicability of QFS is not limited to Theorems \ref{thm:one_qubit_advantage_informal}, \ref{thm:conventional_vs_classical_informal}, and \ref{thm:memory_informal}; the quantum advantages from our approach may become realizable in a wide range of frontier scientific and technological applications.

\vspace{-3mm}
\section{Methods}
\vspace{-3mm}
\subsection{Discovery of advantages with Quantum Phase-Space Inference}
\vspace{-2mm}

Proving an exponential quantum advantage requires both an efficient quantum-enhanced protocol and a lower bound that applies to every strategy in the specified class of conventional sensing protocols, including adaptive measurements with unrestricted classical postprocessing. Such lower bounds have so far been tractable in settings where each conventional measurement reveals only a microscopic amount of information relevant to the task, even when the protocol has access to large energy, substantial quantum depth, or arbitrarily sophisticated postprocessing \cite{chen2021exponential, Huang_2022, Oh_2024, cotler2026noisy, kannan2026exponentialspeedupsfaulttolerantprocessing}. These limitations on proving general lower bounds have confined rigorous separations with practical relevance to a narrow range of problems. A general method that derives the conventional lower bound and an efficient quantum-enhanced strategy from the same experimental objective has remained unavailable.

We introduce Quantum Phase-Space Inference (Q$\Psi$) to produce optimal lower bounds and efficient quantum-enhanced protocols within a common framework. A Q$\Psi$ instance consists of a conventional experimental family $\mathcal{M}_{\rm conv}$, defined by constraints on resources such as energy, entanglement, or non-Gaussianity; a quantum information processing-enabled family $\mathcal{M}_{\rm QIP}$ with access to additional resources; and a target physical observable $\mathcal{O}$. From these ingredients, Q$\Psi$ constructs phase-space representations of the observable and of the response functions generated by protocols in each family.

The central quantity in Q$\Psi$ is the \textit{accessible feature information} (AFI), which measures how strongly the response functions available to an experimental family overlap with the phase-space representation of the target observable. For a protocol family $\mathcal{M}$ and observable $\mathcal{O}$, the accessible feature information is denoted $\mathsf{AFI}(\mathcal{M};\mathcal{O})$; its formal definition is given in Appendix~\ref{sec:defining_qro}. We prove that any strategy restricted to $\mathcal{M}$, including arbitrary adaptive measurements and classical postprocessing, requires $\Omega\!\left(1/\mathsf{AFI}(\mathcal{M};\mathcal{O})\right)$ measurements to learn $\mathcal{O}$. The same phase-space construction identifies a protocol within $\mathcal{M}$ that attains this scaling, so the AFI characterizes the optimal measurement complexity of the experimental family.

For a sequence of observables $\mathcal{O}^{(k)}$, an exponential advantage of the quantum information processing-enabled family over the conventional family occurs precisely when $\mathsf{AFI}(\mathcal{M}_{\rm QIP};\mathcal{O}^{(k)})/\mathsf{AFI}(\mathcal{M}_{\rm conv};\mathcal{O}^{(k)}) \geq \exp(\Omega(k))$. When this condition holds, Q$\Psi$ produces both a tight lower bound for every conventional strategy and an optimal quantum information processing-enabled protocol achieving the corresponding upper bound. 
Once $\mathcal{M}_{\rm conv}$, $\mathcal{M}_{\rm QIP}$, and $\mathcal{O}$ have been specified, the quantum learning problem is thereby reduced to a classical statistical comparison of overlaps between real-valued phase-space functions.

Q$\Psi$ extends the range of settings in which quantum advantages can be identified and proved. For example, quantum learning theory has offered few general tools for characterizing the learning power of bounded-depth quantum circuits outside full tomography~\cite{Zhao_2024}; in Appendix~\ref{sec:depth_lower_bounds}, we use Q$\Psi$ to derive a fine-grained family of exponential separations between sensing architectures whose circuit depths increase incrementally. Several known separations also follow as special cases of the framework, often with substantially simpler proofs, as illustrated in Appendix~\ref{sec:qes_reproof_multimode_displacement}. Q$\Psi$ thus provides a systematic method for discovering rigorous quantum advantages from the structure of experimentally motivated learning tasks and the physical constraints imposed on the available platforms.

\vspace{-1em}
\subsection{Experimental demonstration of \\ quantum advantage}
\vspace{-2mm}

The modest control requirements of our protocols allow their experimental realization with existing superconducting-circuit technology. Our device is an ancilla qubit capacitively coupled to a continuous-variable microwave sensor. The qubit is measured via an on-chip readout resonator (see Appendix~\ref{sec:experiment} for a description of the experiment setup and system Hamiltonian, along with the experimental results). The signal displacement is implemented by applying a calibrated resonant microwave pulse to the continuous-variable mode, while entangling operations are realized by exploiting the cross-Kerr interaction between the qubit and cavity. The control unitaries are composed of single-qubit rotations and echoed conditional displacements, $\operatorname{ECD}(\beta) = D(\beta/2)\ket{g}\!\bra{e}+D(-\beta/2)\ket{e}\!\bra{g}$. Together, these operations provide programmable non-Gaussian control and readout of the qubit--cavity system, the key resource underlying the exponential quantum sensing advantages described above.

\vspace{-3mm}
\section{Discussion}
\vspace{-2mm}

We have demonstrated that a single qubit can exponentially enhance our ability to sense the classical world. Our results further show that, across a broad range of fundamental sensing tasks, modest amounts of coherent control, quantum memory, or circuit depth are sufficient to produce additional exponential advantages. Useful quantum advantages need not rely on large entangled systems or local control across many quantum degrees of freedom, and can instead arise in architectures built from a conventional oscillator and a small number of quantum information-processing resources. 

Our experimental demonstrations illustrate that strong single-photon nonlinearities, such as those readily achievable in superconducting circuits~\cite{krantz2019quantum}, enable non-Gaussian sensing protocols that offer an exponential advantage over Gaussian protocols at equivalent energy, highlighting single-photon nonlinearities as an important resource for quantum sensing alongside their established roles in quantum computing and control~\cite{yanagimoto2024mesoscopic, kimble1998strong}.

The quantum advantages presented here are produced by the Q$\Psi$ framework, which provides a unified theory for designing quantum-enhanced experiments from explicit operational constraints. Its central quantity, the AFI, connects conventional hardness with quantum achievability by simultaneously lower-bounding the measurement complexity of protocols within a specified experimental class and identifying strategies that achieve the corresponding scaling. Whereas quantum Fisher information characterizes local sensitivity within a chosen parameterization, the AFI applies to general learning objectives while incorporating restrictions on the available experimental architecture. This broader scope makes it natural to adapt tools from Bayesian experimental design \cite{Macieszczak_2014}, multiparameter optimization \cite{Humphreys_2013, Ragy_2016}, and semidefinite programming \cite{Zhou_2018, Albarelli_2019} to the systematic search for experiments with provable advantages beyond the regimes described by quantum Fisher information \cite{Demkowicz_Dobrza_ski_2020}. Since Q$\Psi$ recasts the search over sensing architectures and learning protocols as an optimization of classical statistical overlaps, it may provide the foundation for a broader program centered on the computational design of quantum-enhanced experiments with rigorous, certifiable advantages. Appendix~\ref{sec:open_questions} develops these directions in greater detail.

By identifying sensing tasks for which conventional architectures require exponentially more measurements, and by supplying experimentally accessible protocols with improved scaling, our framework opens a path toward quantum-enhanced scientific instruments with provable advantages. When these gains translate into an overall practical advantage depends on the full set of resources relevant to a given application, including achievable state preparation and control as well as experimental overhead. In particular, identifying settings in which QFS protocols outperform conventional sensors  requires understanding when the non-Gaussian control or quantum memory available on a given platform provides gains beyond those attainable with comparatively mature Gaussian architectures capable of preparing highly squeezed states. The proof-of-principle experiments reported here already demonstrate reductions of several orders of magnitude in sample complexity under equal-resource comparisons. We further identify several promising domains of application, including fundamental particle detection, noise spectroscopy, and polyspectrum learning, in Appendix~\ref{sec:open_questions}, while a broader range of scientific applications remains to be explored.

Relatively simple quantum control can therefore provide classical learners a quantum-mechanical interface to the natural world that is exponentially more expressive than any classical instrument. Whereas existing exponential separations have relied upon structured quantum features of tailored, worst-case quantum processes \cite{chen2021exponential, Huang_2022, Oh_2024}, our work gives the first examples of provable exponential quantum advantages for learning practical classical signals and stochastic processes. Through the Q$\Psi$ framework, modest quantum resources allow existing devices to realize these advantages while retaining rigorous performance guarantees. These results provide a path toward quantum-enhanced measurements in radar, astronomy, communication, chemistry and other fields in which the features of interest are encoded in classical signals.

\vspace{3mm}
\section*{Data and code availability}
\vspace{-2mm}

All code and data produced during the experiments and numerical simulations in this work are available at \href{https://doi.org/10.5281/zenodo.21896087}{https://doi.org/10.5281/zenodo.21896087}.

\vspace{-3mm}
\section*{Acknowledgements}
\vspace{-2mm}

We thank Senrui Chen and Luke Coffman for helpful comments and are grateful to the organizers of the EPFL workshop on quantum learning and sensing for stimulating discussions. IK is supported in part by the Nobile Research Initiative.  JC is supported by a fellowship from the Alfred P.~Sloan Foundation. The authors would also like to thank Bradley Cole, Clayton Larson, Britton Plourde, Eric Yelton, and Luojia Zhang for the fabrication of the transmon and on-chip resonator, Chris Wang for the design of the transmon, the on-chip resonator and the 3D superconducting cavity, and Nord Quantique for the fabrication of the 3D superconducting cavity. We gratefully acknowledge the Army Research Office for support under award number W911NF-25-1-0261; MIT Lincoln Laboratory for supplying the Josephson traveling-wave parametric amplifier (TWPA) used in our experiments; and the Air Force Office of Scientific Research for equipment purchased using a DURIP award with award number FA9550-22-1-0080.

\vspace{-2mm}
\section*{Author contributions}
\vspace{-2mm}

I.K. conceived the main theoretical ideas underlying this work and conducted the theoretical investigation. S.P. and I.K. designed the experiments, with contributions from S.A.K. and X.S.. S.P. conducted the experiments. M.M.S. suggested the protocol in Appendix~\ref{sec:gbs_alternate}. V.F. oversaw the design and fabrication of the superconducting devices by S.R. and others, which were characterized by S.R., A.S. and S.P.. I.K., S.P., J.C., P.L.M. wrote the manuscript with input from all authors. J.C. and P.L.M. supervised the work.

\putbib
\end{bibunit}
\clearpage
\onecolumngrid
\raggedbottom

\begingroup
\fontsize{10.3pt}{12.1pt}\selectfont

\renewcommand*\appendixpagename{Appendices}
\appendix
\appendixpage
\appendixtableofcontents
\label{toc}

\counterwithin{theorem}{section}

\setcounter{section}{0}
\setcounter{equation}{0}
\setcounter{figure}{0}
\setcounter{table}{0}

\allowdisplaybreaks

\begin{bibunit}
\vspace{8pt}
\begin{center}
    \textbf{Roadmap}
\end{center}

Our main takeaways can be understood from just a few short sections of these appendices. We first provide a roadmap, then recommend a guided reading path through selected sections. A high-level understanding of each section can be obtained by reading the gray boxes, and readers can return to this guide by clicking on the page number at the top-right of each page.
\vspace{2mm}

In Section \ref{sec:overview-related-works}, we place this paper in the broader context of quantum learning theory and metrology. In Section \ref{sec:experiment}, we provide experimental details. In Section \ref{sec:preliminaries}, we provide preliminaries in quantum information and statistical learning theory, then lay out the mathematical models of sensing architectures used throughout our work. In Section \ref{sec:qes}, we introduce the central Q$\Psi$ framework. In Sections \ref{sec:hardness} and \ref{sec:advantage_upper_bounds} we apply our Q$\Psi$ machinery to prove separations between architectures at increasing levels of the sensing hierarchy. Section \ref{sec:all_results} is a short, self-contained reference of all of our results. In Section \ref{sec:open_questions}, we give concrete statements of research programs that emerge from this work.

\vspace{2mm}
\noindent \textbf{Reader's guide}. The Q$\Psi$ framework introduced in this work, and its applications to sensing classical signals, are the central theoretical contributions of our work. We believe that this framework may motivate new research directions in quantum metrology and learning theory, and we recommend the following reading path for understanding the core concepts without delving into the full technical details.

\begin{enumerate}
    \item \textit{Contextualize this work:} Begin with Section \ref{sec:overview-related-works}. This section contextualizes the timeliness of our work in the history of quantum learning and metrology, elucidates how Q$\Psi$ unifies these fields, and lays out the natural research programs that arise from this unification.
    \item \textit{Understand the scope:} Skim Section \ref{sec:non_Fourier_QSL} to understand how we model general sensing tasks and observables in a common framework. Then read Section \ref{sec:sensing-hierarchy} to understand the hierarchy of sensing architectures, which frames the scope of our separations.
    \item \textit{Understand Q$\Psi$:} Read Section \ref{sec:weyl_symbols}, \textit{``Wigner functions and Weyl symbols''}, for important intuition on Q$\Psi$. Then jump to Section \ref{sec:qes}, and read the gray boxes. These are the workhorse definitions and theorems that drive Q$\Psi$, including the definition of the AFI, the upper and lower bound guarantees, the general Q$\Psi$ learning algorithm, and a template for using Q$\Psi$ to design quantum advantages. Readers interested in applying Q$\Psi$ to their own research can then read Section \ref{sec:qes} in further detail.
    \item \textit{Work through an example:} Readers interested in additional detail can skim one of our several exponential separations as a worked example of Q$\Psi$. Appendices \ref{sec:hardness} and \ref{sec:advantage_upper_bounds} contain the exponential lower bounds and efficient upper bounds respectively; each subsection is paired by number across appendices. We suggest reading the paired Sections \ref{sec:conventional_gaussian_lb} and \ref{sec:quantum_control_upper_bounds}, which together prove Theorem \ref{thm:one_qubit_advantage_informal}.
    \item \textit{Recap:} A short list of the ten main results of this work, including the Q$\Psi$ toolkit, is provided in Section~\ref{sec:all_results}. 
    \item \textit{Explore future directions:} Finally, jump to Section \ref{sec:open_questions}, where we lay out concrete statements of open questions at the intersection of quantum metrology, learning theory, and experimental physics that arise from our work. These include applications of our sensing protocols to disciplines including fundamental physics, cosmology, signal detection, and device characterization, as well as a broader program aimed at automating the discovery of practical quantum advantages.
\end{enumerate}
Readers with more specific interests may also consider the following sections. For experimental details, refer to Section \ref{sec:experiment}. Readers with background in bosonic quantum learning theory can refer to Sections \ref{sec:multimode_charfun_lower_bound} and \ref{sec:char_function_learning_gkp}, where we show that the well-known entanglement-enabled learning advantage presented in Ref.~\cite{Oh_2024} can in fact be achieved without entanglement for most instances, and instead reduces to a separation between Gaussian and non-Gaussian strategies. We give a single-qubit-enabled algorithm which realizes this separation. Moreover in Section \ref{sec:qes_reproof_multimode_displacement}, we show how Q$\Psi$ simplifies the lower bound proof of Ref.~\cite{Oh_2024}.

\newpage
\section{Overview and related works}
\label{sec:overview-related-works}
\subsection{Quantum learning theory}

A motivating thesis of modern quantum learning theory (QLT) is that \textit{quantum information processing can enable scientific discoveries that are otherwise impossible}. Early work in the field, inspired by rapid development of experimental quantum platforms, was often aimed at state-tomographic tasks that would enable efficient readout of quantum data from a quantum device or future quantum simulation \cite{vogel_89, smithey_93, hradil_quantum_1997,Chuang_1997, luis_complete_1999, cramer_efficient_2010, gross_quantum_2010, merkel_self-consistent_2013, Aaronson2018ShadowTomography}. Over the last decade, this operational goal has evolved into an organizing principle: using quantum information processing in unison with statistical learning theory to accelerate scientific discovery while obtaining provable guarantees of quantum advantage \cite{classical_shadow_2020, Huang_2022}. Thus far, however, the thesis has been established more as an epistemological claim than a practically actionable one.
 
Seminal works have shown that there \textit{exist} quantum states and processes which encode data that is provably more difficult to learn without quantum information-processing resources. The relevant data ranged from a few specific observables, such as fidelity or purity, up to tomographically complete descriptions of a state or channel, with separations most often enabled by large quantum memories and entanglement \cite{montanaro2017learningstabilizerstatesbell, Aharonov_Cotler_Qi_2022, chen2021exponential,Huang_2022, nisq_2023, Chen_2024, cotler2026noisy, kannan2026exponentialspeedupsfaulttolerantprocessing}. Intended as proofs of concept rather than practical use cases of QLT, these results often assumed worst-case classes of states or channels that seldom align with experimental reality, alongside noiseless, arbitrarily fast quantum processing and unbounded classical computation \cite{cotler2026noisy}.
 
The natural, and thus far unestablished, refinement of the core thesis is therefore the question: \textit{can quantum information processing enable large speedups in experimentally useful applications?} Recent work has clarified that answering it is not a matter of addressing minor details in the existing literature: in most cases, the idealized assumptions that enable an exponential separation lie at its core, and the advantage does not persist once they are relaxed \cite{cotler2026noisy, kannan2026exponentialspeedupsfaulttolerantprocessing}. The hard instances are frequently ensembles of highly entangled states or processes that do not arise in the experiments one wishes to enhance, and relaxing this structure causes the separation to disappear, while many separations scale only in the number of qubits and apply only to observables, such as purity, whose hardness derives from worst-case structure rather than from physically occurring signals.
 
Moreover, instantiating quantum information processing in real-world experiments requires interfacing it with Nature through a \textit{quantum sensor}: the bridge between the natural system under study and the processor. Practical quantum learning advantages must therefore operate within the constraints of sensing platforms, whose coherence and control are far more limited than the capabilities assumed in QLT \cite{Huelga_1997, Giovannetti_2011, Degen_2017, cotler2026noisy}. When the necessary capabilities include deep multi-qubit control operations between sensors and processors, any realization is placed far in the future.
 
More recent lines of work have extended QLT from the qubit setting to bosonic and fermionic systems, which describe many of the platforms and signals encountered outside of quantum simulation \cite{Wu_2024, Oh_2024, bittel2025energyindependenttomographygaussianstates, Mele_2025, Mele_2025_2, disp_channel_learning_experiment, chen2026optimaltomographybosonicfermionic}. These extensions constitute important progress: they demonstrate that learning separations are not artifacts of qubit structure, develop the phase-space and Gaussian machinery on which parts of the present work directly build, and sharpen the understanding of entanglement and quantum memory as learning resources in continuous-variable settings. Their objectives, however, have naturally mirrored those of the founding qubit results, full tomography of multimode Gaussian states \cite{chen2026optimaltomographybosonicfermionic}, for instance, or worst-case learning of displacement channels \cite{Oh_2024}, which are rarely the quantities most relevant to experimental practice and often presume control beyond what sensing platforms currently provide except for proof-of-concept demonstrations \cite{seif2024entanglementenhancedlearningquantumprocesses, disp_channel_learning_experiment}. The gap between provable learning advantage and experimental utility thus persists across qubit, bosonic, and fermionic settings, and closing it requires rethinking how the learning tasks themselves are formulated.
 
Part of why this gap has persisted is structural. A provable advantage requires two ingredients that work against one another: an efficient quantum-enhanced protocol, which favors instances with clean, exploitable structure, and a lower bound applying to every conventional strategy, including adaptive measurements with unrestricted classical postprocessing, which favors instances that conceal information from an entire measurement class. Satisfying both simultaneously has historically forced constructions organized around the available proof techniques (such as the learning tree formalism \cite{chen2021exponential} or Fano's inequality \cite{yu_assouad_1997, Zhao_2024}) rather than around any experimental setting, so that quantum learning advantages have been discovered through manual worst-case construction, with experimental relevance assessed only after the fact.
 
What is needed, then, is a single framework into which realistic experimental constraints and objectives can be inserted, and out of which tight separations --- matching upper and lower bounds --- emerge simultaneously, so that finding quantum learning advantages becomes experimentally guided discovery rather than manual worst-case construction. Such a framework must moreover interface naturally with quantum sensing, the point of contact between quantum information processing and the natural world, while accounting for a range of experimental constraints more amenably than the learning tree formalism. Q$\Psi$ accomplishes all of this at once: an instance encodes the learning objective, the admissible conventional and quantum-enhanced experiments, and their operational constraints, and its central quantity, the accessible feature information (AFI), simultaneously lower-bounds every conventional protocol and certifies a matched strategy attaining the same scaling. Realistic constraints appear as natural restrictions on the Fourier support of real-valued phase-space functions, making the obstructions they pose quantitative and transparent. The results of the present work are correspondingly different in character from the separations surveyed above: to our knowledge, they are the first exponential quantum advantages in learning that use only a single qubit and modest quantum control, and they apply to basic experimental tasks such as estimating Fourier coefficients and temporal correlations of classical signals. The separations also apply to realistic signal instances: our hard instances merely isolate particular Fourier frequencies and temporal correlations, and the lower bounds continue to apply in practice because realistic signals generically contain these hard components; their hardness relies only on standard properties such as high-frequency and multi-time structure, not on highly entangled constructions that do not appear in nature.

\subsection{Quantum sensing} \label{subsec:sensing_RW}

Modern quantum metrology developed alongside advances in quantum optics and atomic interferometry, together with a quantum information-theoretic formulation of sensing in terms of parameterized unitary dynamics~\cite{caves_quantum-mechanical_1981, yurke_su2_1986, wineland_spin_1992, holland_interferometric_1993, bollinger_optimal_1996, Giovannetti_2004}. As the study of quantum computation with qubit systems grew more prominent, these metrological concepts whose roots lay in continuous-variable systems transferred naturally to the qubit DC and AC phase-sensing settings often studied today \cite{Giovannetti_2004, leibfried_toward_2004, Giovannetti_2006}. This quantum metrological reformulation enabled a unified asymptotic theory of quantum sensing, and at its core lay the canonical quadratic separation between the standard quantum limit (SQL) achieved by entanglement-free sensors and the Heisenberg limit (HL) achieved by ideal, maximally-entangled quantum sensors. These seminal results spurred a long line of work delineating when quantum mechanics could enable a sensing speedup, largely centered around the mathematical formalism of quantum Fisher information and quantum Cramer-Rao bounds \cite{braunstein_statistical_1994, paris_quantum_2008, Giovannetti_2011, pezze_quantum_2018}. Bosonic sensing also continued to develop, often in task-specific manners motivated by major experimental objectives like gravitational-wave detection \cite{kimble_conversion_2001, schnabel_quantum_2010, Aasi_2013}.

However, an asymptotic obstruction to the canonical quadratic speedup was clear: when Markovian noise present during sensing is supported by the same generators as the signal, the Heisenberg speedup generally reduces to a constant factor \cite{Huelga_1997, escher_2011, Demkowicz_Dobrza_ski_2012, transversal_sensing}. Resolutions and even ideas for beyond-Heisenberg limited metrology were proposed, drawing upon special physical phenomena such as quantum criticality and nonlinear phase-shift interactions, but these apparent workarounds often broke down once all physical resources were counted carefully \cite{zwierz_general_2010, giovannetti_sub-heisenberg_2012, hall_does_2012, rams_at_2018}, hence why our careful accounting of mean photon number is essential in this work. Quantum error correction then emerged as a natural route through which to understand the limits of canonical quantum metrology in realistic noise environments, and enabled a study of when noise can be corrected without also erasing the signal \cite{Kessler_2014, Zhou_2018}. 

Recent work, however, has clarified two important limitations of the canonical quantum metrology framework~\cite{transversal_sensing}. Firstly, the Cramer-Rao framework cannot comprehensively be used to certify or rule out quantum metrological speedups, because it assumes it operates within a differentiable statistical model and is concerned with unbiased estimation. In machine learning and statistics, biased estimators can surpass many of the limitations of unbiased estimators; moreover, while the mathematical formalism of metrology utilizes unbiasedness to simplify analysis, statistical learning theory has developed many techniques to understand the resource cost of biased estimators. This observation suggests a natural unification between quantum sensing and quantum learning theory. The second observation of Ref.~\cite{transversal_sensing} is that for the canonical DC and AC sensing problems which have become the centerpiece of theoretical quantum metrology, asymptotic beyond-SQL sensing can be ruled out whenever signal and noise are aligned, agnostic to the assumptions of Cramer-Rao. 

Taken together, these developments show that quantum metrology has progressed through several distinct paradigms, each organized around a sharper understanding of the physical resources responsible for quantum advantage. The limitations of the canonical phase-sensing problem are now sufficiently well understood to motivate a new question: which experimentally natural sensing problems admit asymptotic quantum advantages once the learning objective and the available experimental architecture are treated as part of the problem itself? Rather than seeking further ways to recover Heisenberg scaling within one metrological model, one may ask whether different sensing objectives support advantages whose mechanisms and resource scalings are qualitatively distinct from coherent phase accumulation and which scale beyond-quadratically.

Realistic sensing experiments motivate a new, broader formulation. The signal generator may be only partially known, the interrogation time fixed by the source or sensor, and the objective a Fourier coefficient, temporal correlation, or another functional of a stochastic or multiparameter signal. Any practical advantage must also respect constraints on energy, coherence, quantum control, and accessible non-Gaussian operations. These considerations arise across bosonic, qubit, and distributed sensing platforms, but have largely been treated case by case because no encompassing theoretical framework captures both the learning task and the experimental architecture in a unified way. Quantum Fisher information remains effective for quantifying local sensitivity within a specified differentiable model, but it does not by itself formulate global learning objectives or lower-bound every protocol obeying a collection of operational constraints. Such questions generally require separate analyses of the signal model, probe family, and accessible measurements, making it difficult to compare resources such as squeezing, non-Gaussian control, quantum memory, entanglement, and coherent communication within one theory.

The Q$\Psi$ framework is designed to fill this role and unify a wide range of problems across quantum learning, metrology, and estimation theory, including their practical experimental constraints, in a single operational picture. A Q$\Psi$ instance specifies the signal family, learning objective, admissible conventional and quantum-enhanced experiments, and the physical resource to be optimized. Once the learning task and its admissible experiments can be represented mathematically, Q$\Psi$ absorbs their complexity into a purely classical inference problem. The AFI then simultaneously lower-bounds every protocol in the specified class and identifies a matched strategy attaining the same scaling, with constraints on energy, memory, control, and measurement entering directly through the optimization domain. This construction provides a common language for quantum learning and quantum metrology. The present work is one application of Q$\Psi$ to the problem of bosonic sensing of classical fields, where the QSL phase-space representation introduced in Ref.~\cite{cotler2026quantumadvantagesensingproperties} provides the mathematical interface to convert arbitrary signals, objectives, and architectural constraints into a form that can be processed by Q$\Psi$.

Several open problems in quantum sensing of classical fields admit natural formulations within our framework. In non-Gaussian noise spectroscopy, one may ask for the measurement complexity of estimating a selected higher-order correlation or polyspectral coefficient under constraints on pulse bandwidth, sensor coherence, and quantum memory~\cite{Norris_2016, sung_non-gaussian_2019, von_lupke_two-qubit_2020}. In broadband searches for weak stochastic signals such as in axion or gravitational wave detection, one may compare conventional, squeezed, entanglement-assisted, and non-Gaussian receivers while measuring performance through discovery probability, localization error, scan bandwidth, and total integration time~\cite{malnou_squeezed_2018,Backes_2021,brady_entangled_2022,Shi_2023}. Our work illustrates that even rudimentary quantum control is a qualitatively new resource that can bring new insights to these well-studied problems.

Although bosonic phase-space structure makes the present application especially direct, Q$\Psi$ is not restricted to continuous-variable systems. For a qubit sensor or sensor network, the unknown instance may specify a family of Hamiltonians or quantum channels, and the objective may be a global property rather than the full parameter vector. A natural next step is to develop a QSL formalism for qubit sensors and apply Q$\Psi$ to quantum sensing with qubits. This can enable the study of several tasks beyond phase sensing, such as identifying a weak AC signal from a bank of temporal templates, estimating a spatial Fourier component or nonlinear feature of a distributed field, learning a higher-order cross-spectrum, and directly learning a selected property of an unknown many-body Hamiltonian rather than first learning all of its coefficients~\cite{proctor_multiparameter_2018, eldredge_optimal_2018, qian_optimal_2021, Wiebe_2014}. The corresponding protocol classes may restrict entanglement, coherent control, quantum communication, memory, or accessible measurements. An AFI analysis could then determine whether these resources change the scaling with the number of candidate signals, correlation order, network size, or desired accuracy, rather than only the constant or quadratic precision improvement associated with canonical phase estimation.

\newpage
\section{Experiment and simulation details and results} 
\label{sec:experiment}

\subsection{Experimental setup, system Hamiltonian and parameters}

\begin{figure}[h!]
    \centering
    \includegraphics[width=\textwidth]{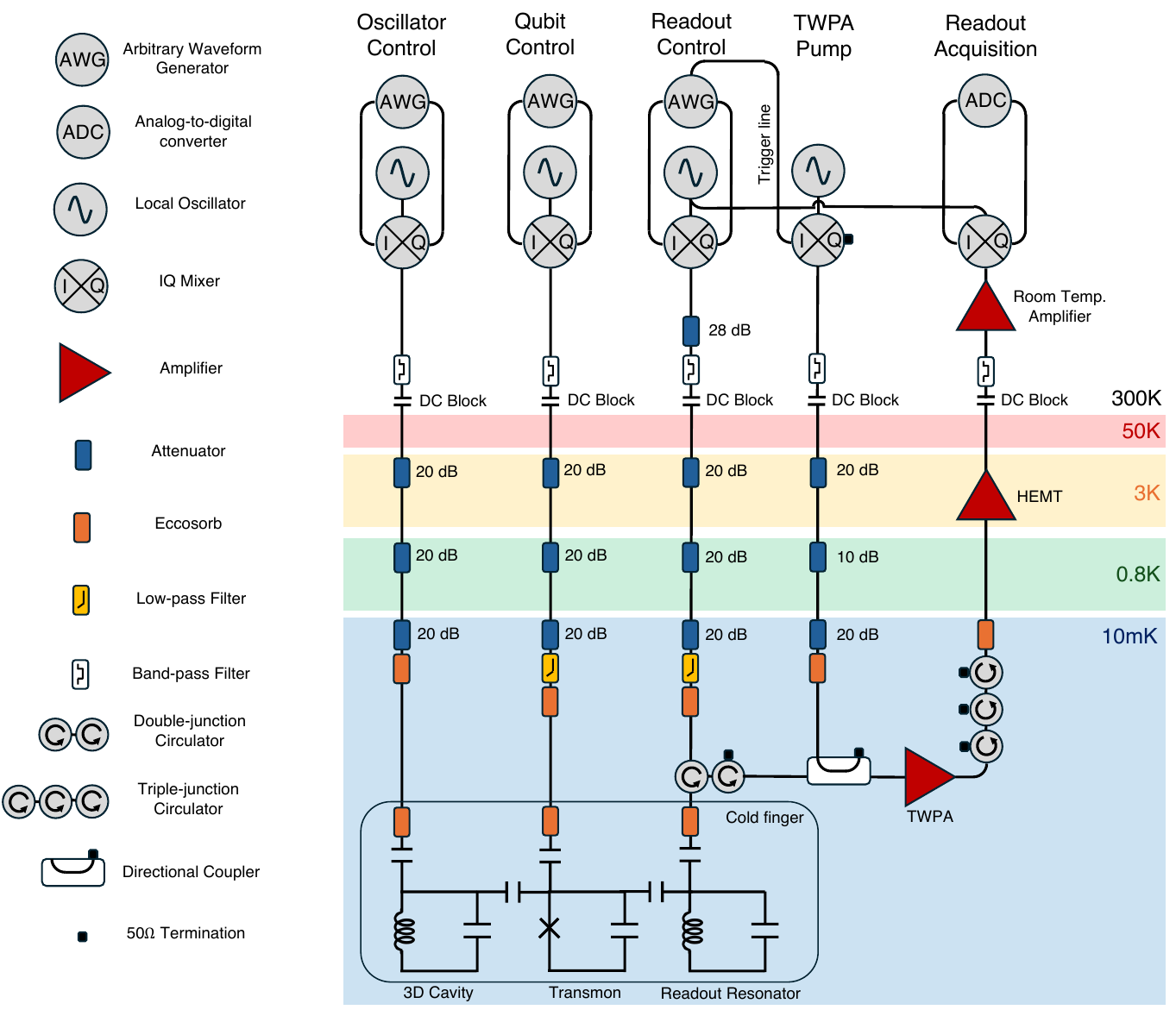}
    \caption{\textbf{Wiring Diagram of the experiment.} Experiment setup for microwave hardware and cabling for the device used in this work.}
    \label{fig:wiring}
\end{figure}

In this work, we use a transmon qubit coupled to the fundamental mode of a 3D stub-post cavity made of high-purity 4N Aluminum. The transmon qubit (the same device used in Ref.~\cite{prabhu2026quantumcomputationaldisplacementsensing}) is fabricated on resistive silicon chip, and is made of Niobium, with an Aluminum - Aluminum Oxide Josephson junction. The chip also contains a readout resonator also made out of Niobium. This chip is inserted in the Aluminum cavity and is held by a copper clamp thermalized to the cold finger made of Oxygen-free high conductivity (OFHC) copper (see Fig.~\ref{fig:image}). The cold finger is enclosed in a Berkeley black coated copper shield and an Aluminum shield. The outermost can of the dilution fridge is lined with a cryoperm shield on the inside. The device is mounted to the cold finger at the base plate of a dilution fridge and is cooled to $10~\rm mK$. Details of the cryogenic setup are shown in Fig.~\ref{fig:wiring}.

\begin{figure}[h!]
    \centering
    \includegraphics[width=\textwidth]{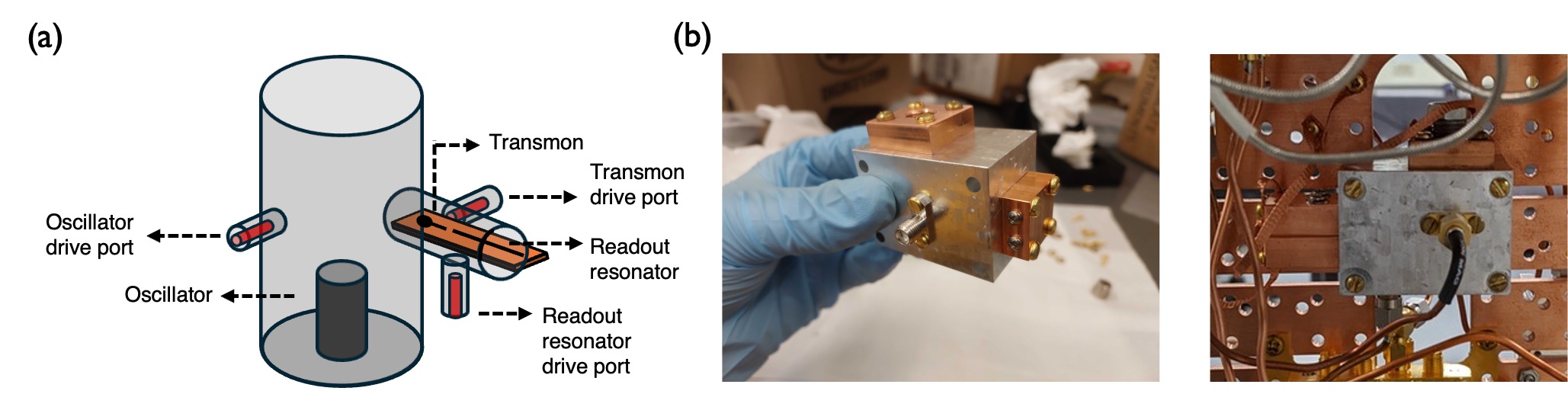}
    \caption{\textbf{Schematic illustration and photographs.} (a) Schematic illustration of the superconducting device, showing the 3D oscillator, the transmon qubit, and the transmon's readout resonator (along with their corresponding drive ports). (b) Photographs of the device and the cold finger it is mounted on.}
    \label{fig:image}
\end{figure}

Microwave pulses are generated using Zurich Instruments (ZI) HDAWG at baseband. Acquisition of microwave pulses from the devices are received by ZI UHFQA. All up and down conversions are performed using Rohde \& Schwarz SGS100A, with in-built IQ mixers. The readout pulse generation and acquisition use the same SGS100A local oscillator (using a splitter). This is done with external Marki IQ mixers (MMIQ-0416LSM-2). The pump for the traveling-wave Parametric Amplifier (TWPA) is gated with a marker line from the ZI HDAWG which is sent to the corresponding SGS100A which produces the pump signal. The readout signal, after amplification by the TWPA, is further amplified by a high-electron mobility transistor (HEMT) at the $4\rm K$ stage (LNF-LNC$0.3\_14$B from Low Noise Factory), and finally at room temperature outside the fridge with a high gain amplifier (ZVA-1W-10 3+ from Mini Circuits). The microwave lines into and out of the fridge are filtered with a series of DC blocks and Mini Circuits bandpass filters centered around the frequency of the tones. For qubit readout, a constant pulse of $0.5~\rm \mu\text{s}$ duration is sent, and the signal is measured for $1~\rm \mu\text{s}$. The TWPA is gated for $2~\rm \mu\text{s}$. We wait for $1~\rm ms$ between runs of the protocol to provide plenty of time for the qubit to thermalize to the ground state, and the oscillator to the vacuum state.

\begin{table}[h!]
    \centering  
    \setlength{\tabcolsep}{8pt}            
    \renewcommand{\arraystretch}{1.2}      

    \begin{tabular}{|l|l|}
    \hline
    \textbf{Parameter} & \textbf{Value} \\ 
    \hline\hline 

    Transmon g--e transition frequency & $\omega_q = 2\pi \times 7.332~\rm GHz$ \\ 
    \hline
    
    Transmon self-Kerr & $K = 2K_q = 2\pi \times 340~\rm MHz$ \\ 
    \hline

    Transmon relaxation time & $T_1 = 30~\rm \mu\text{s}$ \\
    \hline

    Transmon Ramsey coherence time & $T_{2R} = 30~\rm \mu\text{s}$ \\
    \hline

    Transmon Echo coherence time & $T_{2E} = 40~\rm \mu\text{s}$ \\
    \hline

    Transmon readout fidelity for $\ket{g}$ & $95\%$ \\
    \hline

    Transmon readout fidelity for $\ket{e}$ & $95\%$ \\
    \hline

    Readout frequency & $\omega_r = 2\pi \times 8.920~\rm GHz$ \\
    \hline

    Readout-Transmon dispersive shift & $\chi_{rq} \sim 2\pi \times 1 \rm~MHz$ \\
    \hline

    TWPA pump frequency & $\omega_{\rm pump} = 2\pi \times 8.550~\rm GHz$ \\
    \hline

    Oscillator frequency & $\omega_a = 2\pi \times 6.072~\rm GHz$ \\
    \hline

    Oscillator-Transmon dispersive shift  & $\chi = 2\pi \times (15.5 \pm 0.3)~\rm kHz$\\
    \hline

    Oscillator relaxation time & $T_{1c} = 250~\rm \mu\text{s}$ \\
    \hline
    
    \end{tabular}
    
    \caption{\textbf{Measured experiment parameters and readout fidelity.} These parameters are obtained using standard spectroscopic and time-resolved measurements following~\cite{chou2018teleported, eickbusch2022fast}.}
    \label{table:system_parameters}
\end{table}

For the experiments demonstrated in this work, the Hamiltonian of our device is well-described by:
\begin{equation}
    \hat{H}/\hbar = \omega_q \hat{q}^\dagger \hat{q} + \omega_a \hat{a}^\dagger \hat{a} - K_q \hat{q}^{\dagger 2} \hat{q}^2 - K_a \hat{a}^{\dagger 2} \hat{a}^2 -  \chi \hat{q}^\dagger \hat{q} \hat{a}^\dagger \hat{a} - \chi' \hat{q}^\dagger \hat{q} \hat{a}^{\dagger 2} \hat{a}^2 + \Omega(t)\hat{q}^\dagger + \varepsilon(t)\hat{a}^\dagger + \rm{h.c.},
    \label{appeq:hsysq}
\end{equation}
where $\hat{q}$ is the annihilation operator in the transmon Hilbert space, while $\hat{a}$ is the annihilation operator in the oscillator Hilbert space, describing a transmon and an oscillator mode with frequencies $\omega_q$ and $\omega_a$ respectively. The transmon anharmonicity $K_q$ is large relative to its drive bandwidth, allowing the transmon mode to be approximated as a two-level qubit system, with spin-$Z$ operator $\hat{\sigma}_z = 1 - 2\hat{q}^\dagger \hat{q}$. The interaction between the transmon and the oscillator is described by the cross-Kerr interaction $\chi$, which has strength of $O(10)~$kHz. In this situation, the higher-order terms of oscillator anharmonicity $K_a$ and second-order cross-Kerr interaction strength $\chi'$ are even smaller (of $O(10)~$Hz), and their effects can be ignored for the regime of experiments demonstrated in this work. The term $\Omega(t)$ describes a generally time-dependent pulse on the transmon, while $\varepsilon(t)$ describes a generally time-dependent pulse on the oscillator. In Table~\ref{table:system_parameters}, we list the measured values of the relevant terms of the Hamiltonian, alongside other parameters.

In this work, the sensing displacement $D(\alpha)$ is implemented by setting the amplitude and phase of a calibrated Gaussian pulse (truncated at $2\sigma$ on either side) with a total duration of $50$ns. The pulse is calibrated such that the magnitude of the displacement $|\alpha| = 11.92$ when the A.W.G. amplitude of the pulse is $1$, and the displacement is linear in amplitude (see Ref.~\cite{prabhu2026quantumcomputationaldisplacementsensing} for further details of this characterization). 

In Figure~\ref{fig:ecd_pulse}, we show how we realize the echoed conditional displacement gate $\mathrm{ECD}(\beta) = D(\beta/2)|g\rangle \langle e| + D(-\beta/2)|e\rangle \langle g|$. Our protocol, based on that introduced in Ref.~\cite{eickbusch2022fast}, involves large ``out-and-back'' displacements on the oscillator. We optimize the pulse shapes to generate large values of $\beta$ as efficiently as possible (that is, with the highest sensitivity to sensing displacements). The pulses for the oscillator are square-wave, with a duration of $t$. The out and back pulses are spaced by a delay of duration $d$. The qubit pulses are in the shape of a Gaussian, truncated at 2$\sigma$ at either side (and a total duration of $100$ns). For each experiment, we set the values of $t$ and $d$ (according to the requirements of the task and the limit of the performance of the device). This sets the largest value of $|\beta| = \beta_{\mathrm{max}}$ (achieved when $s=1$). We then can tune the value of $|\beta| \le \beta_{\mathrm{max}}$ by setting the value of $s$. The phase of $\beta$ is determined by the phase of the pulses.

\begin{figure}[h!]
    \centering
    \includegraphics[width=0.8\textwidth]{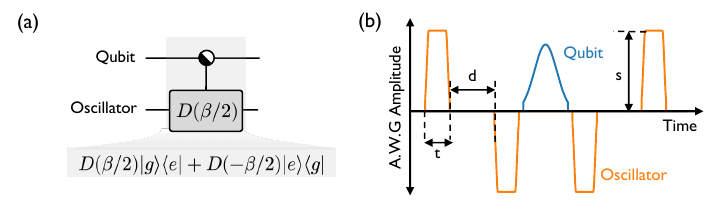}
    \caption{\textbf{Realization of the echoed conditional displacement gate.} (a) \textit{Circuit diagram of the $\mathrm{ECD}(\beta)$ gate, which acts on both the qubit with the oscillator.} (b) \textit{Pulse-level diagram.} The entire gate involves $4$ pulses on the oscillator, and a single $\pi$ pulse on the qubit. The magnitude of $\beta$ depends on the choice of the timescales $t$ and $d$, and the amplitude $s$.}
    \label{fig:ecd_pulse}
\end{figure}

\subsection{Demonstration of separation between classical and quantum probes}

We discuss the experimental implementation of the single-qubit ancilla quantum protocol for the task and results presented in Figure~\ref{fig:coherent_and_memory_experiments}(a), for which the protocol obtains an exponential advantage over classical sensors. We begin with a discussion of the characterization of the experiment, before discussing the demonstration of the task.

\begin{center}
    \textit{Characterization}
\end{center}

\begin{figure}[h!]
    \centering
    \includegraphics[width=\textwidth]{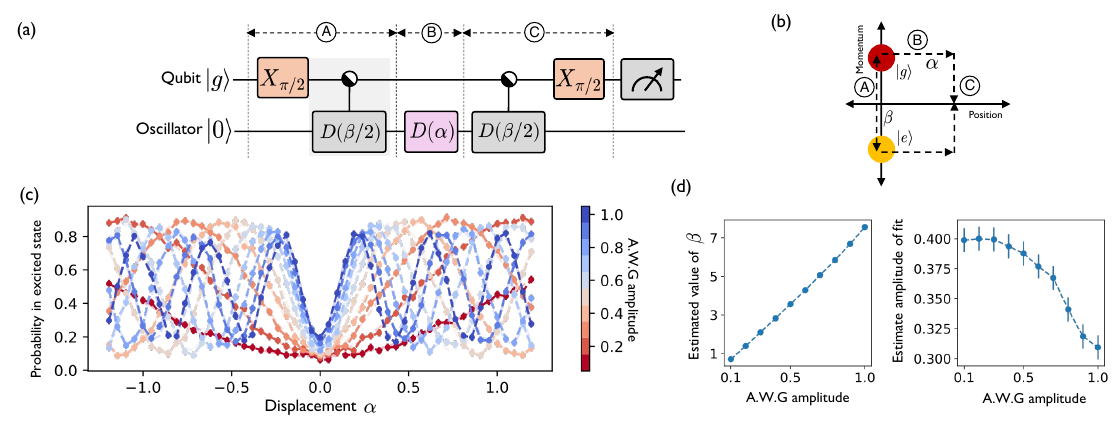}
    \caption{\textbf{Experimental characterization of the cat-state sensing protocol.} (a) \textit{Circuit diagram.} (b) \textit{Schematic illustration of the trajectories of Wigner function of the oscillator conditioned on the qubit state along the corresponding stages of the protocol.} (c) \textit{Measured qubit probability as a function of the sensed displacement $\alpha$ (orthogonal to the axis of the cat), for different values of the cat.} The cat size is varied by changing the amplitude of the pulse generated by the arbitrary waveform generator (A.W.G.). (d) \textit{Best-fit parameters as a function of the A.W.G. amplitude.}}
    \label{fig:exp1_calibration}
\end{figure}

The circuit diagram we implement for this experiment is illustrated in Figure~\ref{fig:exp1_calibration}(a). The protocol can be divided into three stages. In the first stage, a single-qubit rotation and an echoed conditional displacement prepare an entangled cat-qubit state. Then, in the second stage, the oscillator senses a displacement. Finally, in the third stage, an echoed conditional displacement with the same parameter is implemented, which disentangles the qubit from the oscillator. The information of the displacement is imparted onto the phase of the qubit, which is inferred from a qubit measurement at the end of the protocol. Figure~\ref{fig:exp1_calibration}(b) represents the Wigner function of the oscillator conditioned on the state of the qubit. The trajectories enclose an area in phase-space proportional to the size of the cat $\beta$ and the displacement orthogonal to the cat axis $\alpha$. The phase imparted onto the qubit is proportional to this area. In Figure~\ref{fig:exp1_calibration}(c), we plot the measured qubit probability $P_e$ as a function of the sensed displacement, for $t=110$ns and $d=800$ns. The frequency of the response determines the size of the cat. We vary the value of $\beta$ by changing the A.W.G. amplitude on the pulses to the oscillator (which form the conditional displacement gate)~\cite{eickbusch2022fast}. We fit the data to $P_e = A_0 + A_1 \cos{(2\beta\alpha + \phi)}$, and extract the best-fit parameters $A_0, A_1, \beta, \phi$. In Figure~\ref{fig:exp1_calibration}(d), we plot the dependence of $\beta$ (left) and $A_1$ (right) on the A.W.G. amplitude. The value of $\beta$ (which corresponds to the size of the cat) scales linearly with the amplitude. We use this relationship to set the size of the cat required for a corresponding experiment. However, due to experimental non-idealities, such as decoherence, the contrast $A_1$ reduces with increasing $\beta$. We notice $\phi$ has a quadratic dependence on $\beta$. This corresponds to a shift in the the displacement of most sensitivity. We correct this shift by applying an appropriate phase roll in the final qubit rotation. The data presented in Figure~\ref{fig:exp1_calibration} represents the results after this correction. This correction helps reduce the simulation-experiment gap, resulting in a better sample performance of the experiment.

\begin{center}
    \textit{Experiment}
\end{center}

\begin{figure}[h!]
    \centering
    \includegraphics[width=\textwidth]{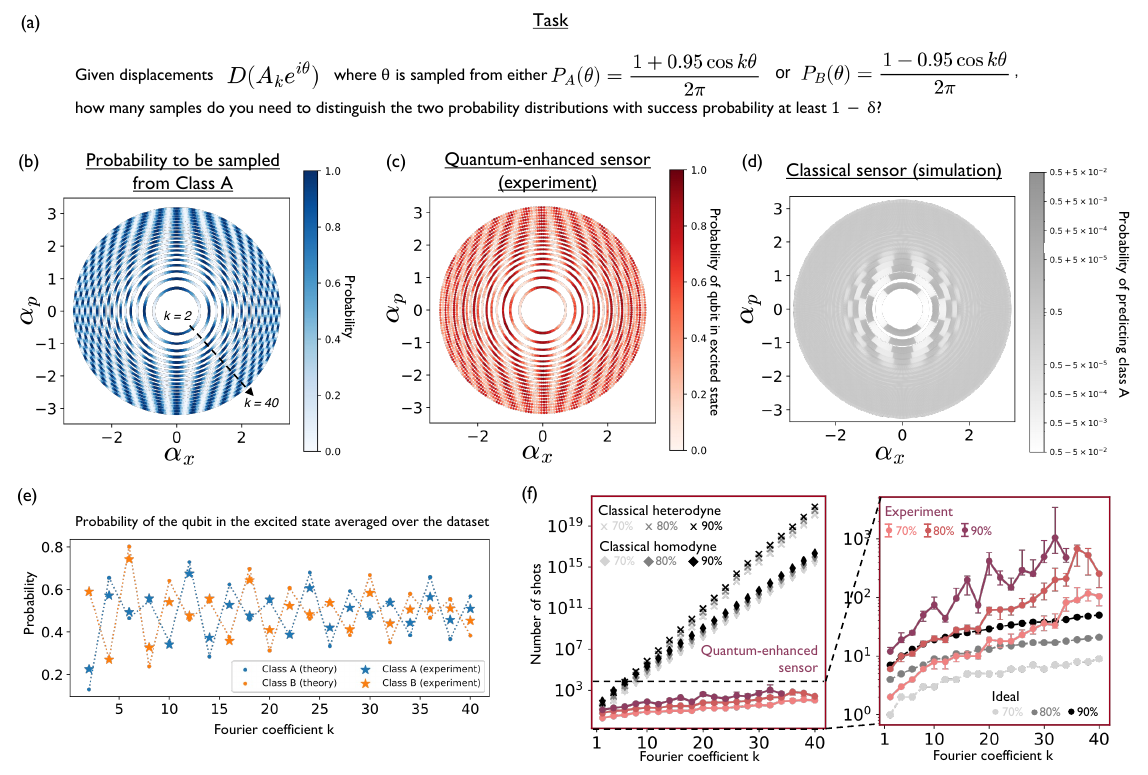}
    \caption{\textbf{Experimental demonstration of separation between classical and quantum probes.} (a) \textit{We consider the sample requirement for correctly classifying the distribution with a success probability within $\delta$ of $1$.} (b) \textit{Illustration of the probability distribution of a signal given to originate from Class A.} If the probability is $1$, then if a displacement with the corresponding value is sensed, it can only originate from Class A. (c) \textit{Response of the quantum-enhanced sensor in experiment.} (d) \textit{Response of the classical sensor in simulation.} The response of the quantum-enhanced sensor visually shows how it can efficiently distinguish the two classes. (e) \textit{Qubit excitation probability as a function of the Fourier coefficient $k$ in theory and experiment for the two classes.} (f) (Left) \textit{Corresponding sample complexity of the quantum-enhanced sensor in experiment and the classical baseline in simulation to achieve classification accuracies $70\%$ , $80\%$ and $90\%$.} (Right) \textit{Zoom in of the sample complexity of the quantum-enhanced sensor in experiment, along with the expected performance in the ideal limit.}}
    \label{fig:exp1_results}
\end{figure}

We consider a binary classification task, where the goal is to distinguish between two distributions of displacement (see Figure~\ref{fig:exp1_results}(a)). The magnitude of the displacements is fixed to be $A_k = \sqrt{k}/2$, where $k$ is the integer-valued Fourier coefficient. All the information which distinguishes the two distributions is in the phase of the displacement. We illustrate this in Figure~\ref{fig:exp1_results}(b) where we plot the probability of the displacement to originate from Class A (where the different rings correspond to different values of $k$). When the probability is close to $1$, that particular displacement is more likely to originate from Class A (and vice versa). We plot the distributions for even values of $k$ ranging between $2$ and $40$. The experiment proceeds as follows:

\begin{enumerate}
    \item For each Fourier coefficient $k$, construct a dataset of phases $\theta$ size $D$ (in this experiment $D=10^4$) for Class A and B via rejection sampling.

    \item Stream the dataset to the experiment, where we run the experiment once per dataset (therefore generating $D$ measurements in total for each class). The sensed displacement $D(\alpha)$ has a value $\alpha = A_k e^{i\theta}$ for the particular value $\theta$ of the dataset.
    
    \item The optimal value of the conditional displacement is $\beta_k = z_k/(2A_k)$, where $z_k$ is the first positive zero of the derivative $J_k(z)$ (which is the $k$th Bessel function of the first kind). We realize this value of $\beta_k$ by setting the A.W.G. amplitude for the $\mathrm{ECD}(\beta_k)$ gate to be $s_k = \beta/\beta_{\mathrm{max}}$, where $\beta_{\mathrm{max}} = 7.5$ is the maximum value obtained when $s=1$, obtained from the calibration results presented in Figure~\ref{fig:exp1_calibration}(d).

    \item Run the protocol cat-state protocol with this parameter (illustrated in Figure~\ref{fig:exp1_calibration}(a)), and record the qubit measurement outcome (that is, whether it was measured to be in the ground or excited state).
\end{enumerate}

\begin{center}
    \textit{Results}
\end{center}

A protocol which has a response which matches this distribution will efficiently distinguish the two distributions. This is indeed the case with the quantum-enhanced sensor. In Figure~\ref{fig:exp1_results}(c), we plot the measured qubit probability for the same range of displacements as the dataset. The probability matches the periodicity of the dataset for a significant range of phase. On the other hand, as illustrated in Figure~\ref{fig:exp1_results}(d), the contrast of the classical sensor's response exponentially decays with $k$ towards $0.5$. In Figure~\ref{fig:exp1_results}(e), we plot the theoretical and experimental probability of the qubit in the excited state for each of the class. The experimental probability is obtained by simply averaging over the entire dataset of the corresponding class (the errorbars are too small to be visible on the plot). For a large range of $k$, the experimental probability is close to the theoretical value. The significant source of discrepancy is the readout fidelity, whose effect is to push the experimental value closer to $0.5$. From this dataset, we can then estimate how many samples would be required to correctly identify the class label $70\%$, $80\%$ and $90\%$ of the time (see Figure~\ref{fig:exp1_results}(f)). Due to the lower separation of probabilities in experiment compared to the ideal value, the corresponding sample requirement is larger. However, even for $k = 40$, the sample requirement for achieving an accuracy of at least $70\%$ is around $100$, several orders of magnitude below that of a classical heterodyne or homodyne protocols.

\subsection{Demonstration of separation between conventional quantum experiments and quantum control}

We discuss the experimental implementation for the task and results presented in Figure~\ref{fig:single_qubit_advantage}. In this scenario, the experiment (single-qubit ancilla entangled with the sensing oscillator) achieves an exponential advantage over conventional quantum sensing protocols with access to only Gaussian resources with the same energy.

\begin{center}
    \textit{Characterization}
\end{center}

\begin{figure}[h!]
    \centering
    \includegraphics[width=\textwidth]{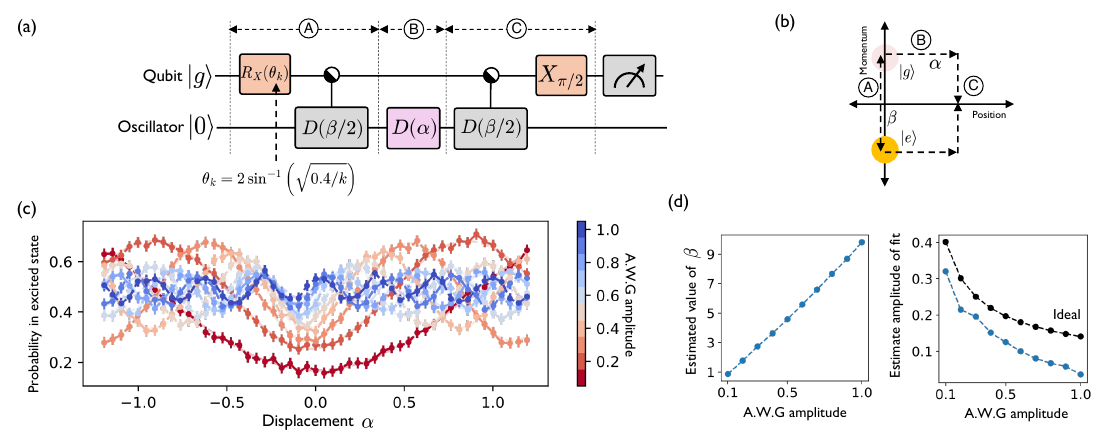}
    \caption{\textbf{Experimental characterization of the unbalanced cat-state sensing protocol.} (a) \textit{Circuit diagram.} (b) \textit{Schematic illustration of the trajectories of Wigner function of the oscillator conditioned on the qubit state along the corresponding stage of the protocol.} (c) \textit{Measured qubit probability as a function of the sensed displacement $\alpha$ (orthogonal to the axis of the cat), for different values of the cat.}. (d) \textit{Best-fit parameters as a function of the A.W.G.~amplitude, along the theory-expected value under ideal conditions of no decoherence.}}
    \label{fig:exp2_calibration}
\end{figure}

The protocol, illustrated in Figure~\ref{fig:exp2_calibration}(a), is essentially identical to that in Figure~\ref{fig:exp1_calibration}(a). The main difference is the amplitude of the initial qubit rotation, which depends on the Fourier coefficient $k$. We choose this value such that the number of photons in the displaced frame scales linearly with $k$ (see Appendix~\ref{sec:sensing-hierarchy} for an explanation for the choice of this resource metric). This in general therefore realizes an ``unbalanced'' cat, where the coefficient of wavefunction conditioned on the ground and excited states of the qubit are different. This is schematically illustrated in Figure~\ref{fig:exp2_calibration}(b), where the faded color scale of the Wigner function conditioned on $\ket{g}$ is due to the smaller weight of the wavefunction overlap with this state. This protocol results in a Ramsey response for the qubit excitation probability, but with a smaller amplitude, as shown in Figure~\ref{fig:exp2_calibration}(c), where we set $t=110$ns and $d=1100$ns. Like before, we plot the best-fit parameters to a sinusoidal response, shown in Figure~\ref{fig:exp2_calibration}(d). We also plot the value of the amplitude expected as a function of $k$ in the idealized scenario without experimental imperfections.

\begin{center}
    \textit{Experiment}
\end{center}

\begin{figure}[h!]
    \centering
    \includegraphics[width=\textwidth]{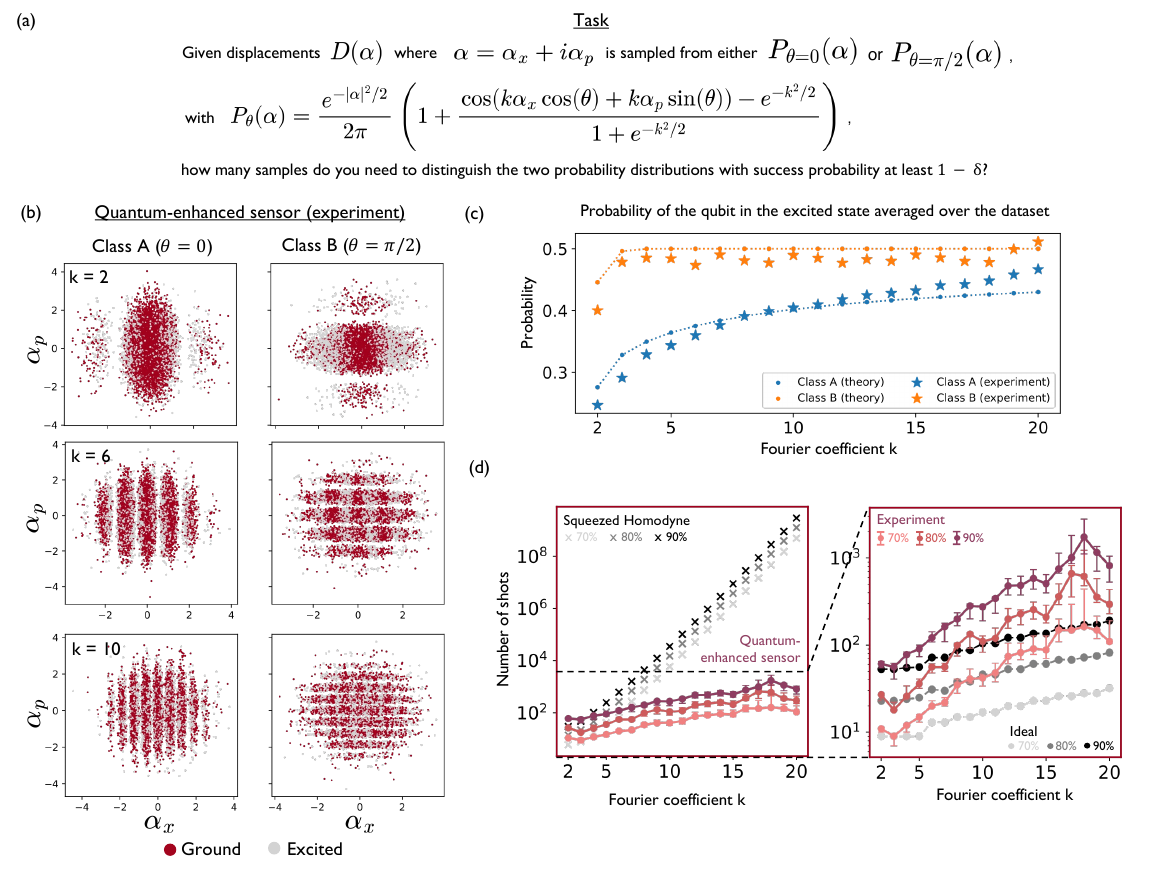}
    \caption{\textbf{Experimental demonstration of separation between conventional quantum experiments and quantum control.} (a) \textit{We consider the sample requirement for correctly classifying the distribution with a success probability within $\delta$ of $1$.} (b) \textit{Illustration of the datasets of each class for $k = 2,6,10$.} The data points are colored based on the qubit measurement outcomes of that particular experiment. (c) \textit{Qubit excitation probability as a function of the Fourier coefficient $k$ in theory and experiment for the two classes.} (d) (Left) \textit{Corresponding sample complexity of the quantum-enhanced sensor in experiment and the Gaussian baseline in simulation to achieve classification accuracies $70\%$, $80\%$, and $90\%$.} (Right) \textit{Zoom in of the sample complexity of the quantum-enhanced sensor in experiment, along with the expected performance in the ideal limit.}}
    \label{fig:exp2_results}
\end{figure}

For this task, the two classes are described by the distribution written in Figure~\ref{fig:exp2_results}(a) at the parameters $\theta = 0$ and $\theta = \pi/2$. In this case, the distributions are spread out over both the position and momentum components of the displacement. Figure~\ref{fig:exp2_results}(b) are scatter plots of the dataset of each task, for select values of the Fourier coefficient $k$. Here $k$ is a proxy for the number of ``fringes" in the dataset. Our protocol can efficiently distinguish the two distributions but matching the periodicity of these fringes. The experiment proceeds as follows:

\begin{enumerate}
    \item For each coefficient $k$, construct a dataset for Class A and B of size $D$ (in this experiment $D = 10^4$) via rejection sampling.

    \item Stream the dataset to the experiment, where we run the experiment once per dataset (therefore generating $D$ measurements in total for each class).

    \item The value of the conditional displacement is $\beta_k = k/2$, while the qubit-rotation amplitude is chosen such that the corresponding rotation operator $R_X(\theta_k)$ has a value $\theta_k = 2\sin^{-1}(\sqrt{0.4/k})$. We realize $\beta_k$ by setting the A.W.G. amplitude for the $\mathrm{ECD}(\beta_k)$ gate to be $s_k = \beta/\beta_{\mathrm{max}}$, where $\beta_{\mathrm{max}} = 10$ is the maximum value obtained when $s=1$, obtained from the calibration results presented in Figure~\ref{fig:exp2_calibration}(d).

    \item Run the protocol imbalanced cat-state protocol with this parameter (illustrated in Figure~\ref{fig:exp2_calibration}(a)), and record the qubit measurement outcome (that is, whether is was measured to be in the ground or excited state).
\end{enumerate}

\begin{center}
    \textit{Results}
\end{center}

Figure~\ref{fig:exp2_results}(b) visually shows how the two classes can be distinguished. The orientation and size of the cat is such that the red dots (corresponding to the qubit being measured in the ground state) are more frequent for Class A. On the other hand, for Class B, there are similar frequency of both outcomes. This results in the qubit excitation probability in experiment (see Figure~\ref{fig:exp2_results}(c)) to be close to $0.5$. On the other hand, the qubit excitation probability for Class A is smaller. Similar to before, the errorbars are too small to be visible on the plot. From this dataset, we estimate the sample requirement for a classification accuracy of $70\%$, $80\%$ and $90\%$. In Figure~\ref{fig:exp2_results}(d) we plot this as a function of $k \in [2,20]$ for both the experimental and the theoretically ideal results. Similar to before, we observe a sample advantage of several orders of magnitude, persisting for $k = 20$.

\subsection{Demonstration of quantum memory advantage}

We discuss the experiment results of the task presented in Figure~\ref{fig:coherent_and_memory_experiments}(c), for the scenario where the ancilla qubit has coherent quantum memory across the duration of the sensing events. Our protocol achieves an exponential advantage over protocols without access to quantum memory during sensing, with the same photon number.

\begin{center}
    \textit{Experiment}
\end{center}

\begin{figure}[h!]
    \centering
    \includegraphics[width=\textwidth]{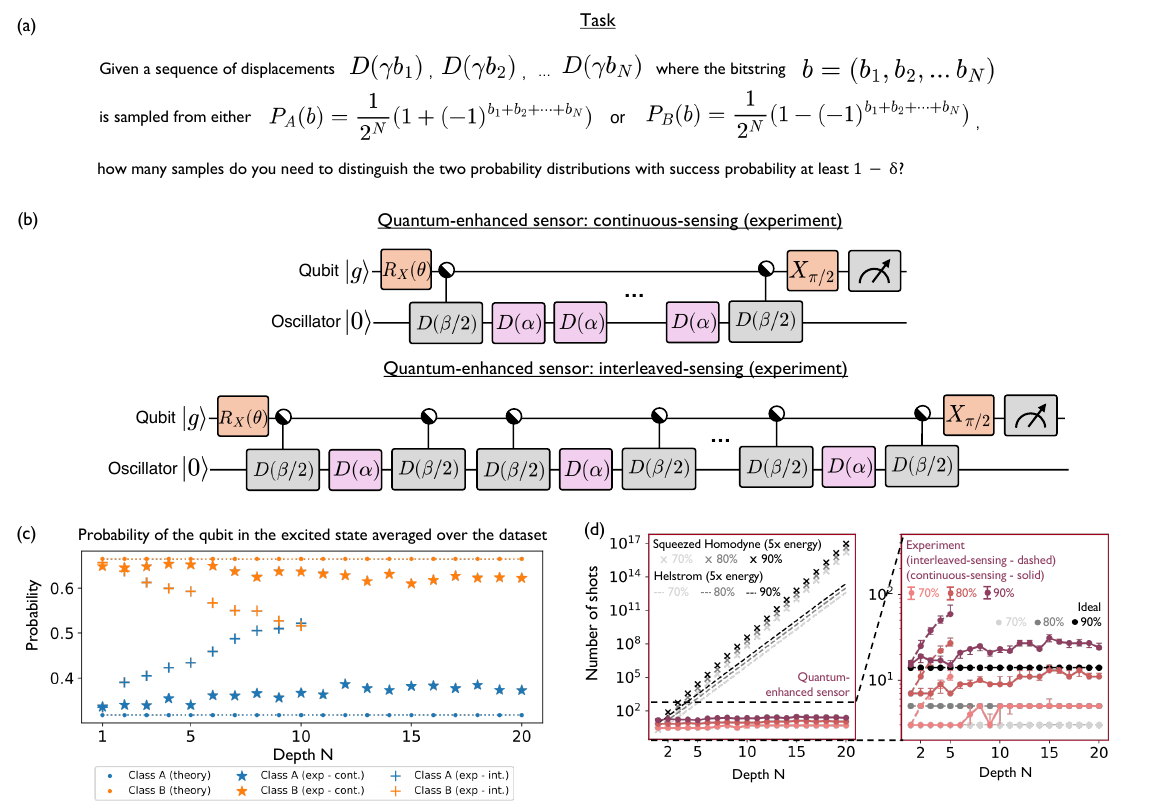}
    \caption{\textbf{Experimental demonstration of quantum memory advantage.} (a) \textit{We consider the sample requirement for correctly classifying the distribution with a success probability within $\delta$ of $1$.} (b) \textit{Circuit diagrams of the two protocols we consider.} The two protocols consider the scenarios where the sensor is or is not coherent across multiple sensing rounds. (c) \textit{Qubit excitation probability as a function of the depth $N$ in theory and experiment (for both choices of protocols) for the two classes.} (d) (Left) \textit{Corresponding sample complexity of the quantum-enhanced sensor in experiment and the baseline without quantum memory in simulation to achieve classification accuracies $70\%$, $80\%$, and $90\%$.} (Right) \textit{Zoom in of the sample complexity of the quantum-enhanced sensor in experiment, along with the expected performance in the ideal limit.}}
    \label{fig:exp3_results}
\end{figure}

We consider a binary classification task where the two distributions are distinguished in terms of what sequence of displacements are allowed. $b = (b_1, b_2, \dots, b_N)$ represents a bitstring which denotes the sequence of displacements on the oscillator. The displacement on the oscillator is $D(\gamma b_i)$ for bit $b_i$. Therefore, the two possibilities are the oscillator senses a displacement or not. We consider a binary classification task in distinguishing two distributions, represented by Class A and Class B. Each class is a probability distribution over all possible $2^N$ bitstrings for depth $N$. As defined in Figure~\ref{fig:exp3_results}(a), Class A consists of all bitstrings with even parity, while Class B consists of all bitstrings with odd parity. We consider two variations of quantum-enhanced protocols with memory: continuous-time and interleaved-time. In the first version, we allow the quantum sensor to continuously accrue displacements over time. Such a protocol relies on both the oscillator and qubit to remain coherent on the time scale of the sensing. On the other hand, one can consider the situation where the oscillator cannot stay coherent over the entire sensing duration (only the ancilla qubit can). This corresponds to interleaved-sensing, which is designed to disentangle the qubit from the oscillator after a sensing round. Therefore, the sensor can be reset without destroying the information stored in the qubit. For both protocols, we set the amplitude $s=1$, time scales $t=100$ ns and $d=400$ ns for the parameters of the ECD gate. This corresponds to $\beta = 4.15$. The experiment proceed as follows:

\begin{enumerate}
    \item For each depth $N$, construct a dataset for Class A and B of size $D$ (in this experiment $D = 10^4$) via rejection sampling.

    \item Stream the dataset to the experiment, where we run the experiment once per dataset (therefore generating $D$ measurements in total for each class).

    \item The value of the qubit rotation angle is $\theta = 2\sin^{-1}\!\big( \sqrt{E/\beta^2} \big)$, with $E = ((2\nu\beta/\pi)^2 - 1)/2$ and $\nu = 0.41$. This is not unique. Instead these were chosen to maximize the quantum advantage over the baseline.

    \item Run each of the two protocols (continuous-sensing and interleaved-sensing). For each protocol, record the qubit measurement outcome (that is, whether is was measured to be in the ground or excited state).
\end{enumerate}

\begin{center}
    \textit{Results}
\end{center}

In Figure~\ref{fig:exp3_results}, we plot the qubit excitation probability, averaged over the dataset, for each class, as a function of the depth $N$. The theoretical value for both protocols are identical. As the depth $N$ increases, the probabilities in experiment tend towards $0.5$, with the separation reducing. This is due to decoherence, since the lifetime of the experiment is finite. This is more pronounced for the interleaved-sensing experiment, since the duration of the protocol is much longer due to the inclusion of additional ECD gates during the sensing stage. For our experimental regime, the oscillator has a longer coherence time than the qubit -- therefore, the device does not benefit from the motivation of this protocol to disentangle the ancilla qubit from the sensor. However, the results of this protocol provide a proof-of-principle demonstration for devices which operate in the regime of the lifetime of the ancilla mode much longer than the sensing mode. The effect of decoherence is marginal for the continuous-sensing protocol. Since each sensing duration takes place in $50$ns, there is only a slight enhancement in the overall decoherence for the range of depths considered. This is reflected in Figure~\ref{fig:exp3_results}, where we plot the sample requirement of the experiment, and theoretical baseline, for achieving $70\%$, $80\%$, and $90\%$ accuracy. The sample requirement of both the continuous-sensing and theoretical baseline are $O(1)$ for achieving an accuracy at least $70\%$. On the other hand, a sensor without access to quantum memory will have a sample complexity scaling exponential in the depth $N$.

\subsection{Simulation of cold axion stream characterization} \label{sec:axion_stream}

Here we describe the numerical experiment depicted in Figure \ref{fig:applications}(a). We will provide a high-level description of the axion-physics models used in our numerics, which are derived from the results of Ref.~\cite{Foster_2021}. We refer readers to this work for a detailed understanding of how the formulae and physical parameters are obtained.

\begin{center}
    \textit{Derivation of task from axion physics}
\end{center}

A nonrelativistic axion propagating through free space is a real scalar field of mass $m_a$ with plane-wave solutions
\begin{equation}
    a_{\mathbf{v}}(\mathbf{x},t) = a_{\bf v}(\cos (\omega_{\bf v}t -{\bf k}_{\bf v}\cdot {\bf x} - \phi_{\bf v}) )\ ,
\end{equation}
where $\bf v$ is the 3-velocity, $\bf k_v = m_a \bf v$, and $\omega_{\bf v} = \sqrt{m_a^2 + |{\bf k_v}|^2}$. The oscillation induced in a detector, then, is of the form
\begin{equation} \label{eq:axion_detector_form}
    \Phi({\bf x}, t) = m_a\cdot \kappa \cdot a_{\bf v}({\bf x}, t) \ ,
\end{equation}
where $\kappa$ captures the physics of the coupling of the detector to the field. A realistic axion field is not simply one plane wave, but a mixture of many incoming waves at different velocities (frequencies) and phases. Ref.~\cite{Foster_2021} shows that under the assumption of a uniform phase distribution, the field is well-described by
\begin{equation} \label{eq:axion_field}
    a_{\bf v}({\bf x}, t) = \frac{\sqrt{\rho_{\rm DM}}}{m_a}\sum_d \alpha_d \sqrt{f({\bf v}_d) (\Delta v)^3}(\cos (\omega_{d}t -{\bf k}_{d}\cdot {\bf x} - \phi_{d})) \ ,
\end{equation}
where velocity space is discretized into intervals of $\Delta v$ in each direction and each speed is labeled by $d$. Moreover, $\rho_{\rm DM}$ is the local dark matter energy density, $f(\bf v)$ is the normalized three-dimensional lab-frame velocity density, $\alpha_d$ is Rayleigh distributed as $p(\alpha_d) = \alpha_d e^{-\alpha_d^2/2}$, every $\phi_d$ is uniformly distributed and we assume that different velocity increments have independent $\alpha_d,\phi_d$. 

This model gives quadrature amplitudes which are exactly Gaussian, which is evident upon setting $X_d = \alpha_d \cos \phi_d, Y_d = \alpha_d\sin \phi_d$ and computing the joint distribution. Upon combining equations \eqref{eq:axion_detector_form} and \eqref{eq:axion_field}, the response of a detector $i$ is
\begin{equation}
    \Phi_i({\bf x}, t) = \sqrt{\rho \kappa_i^2}\sum_d \alpha_d \sqrt{f({\bf v}_d) (\Delta v)^3}(\cos \omega_{d}t -{\bf k}_{d}\cdot {\bf x} - \phi_{d}) \ .
\end{equation}
One can then discretize interrogation time into intervals of $\Delta t$ such that $t_n = n\Delta t$ for $n = 0, 1, ..., M-1$, and write $\Phi_{i,n} = \Phi_i(t_n)$, where we omit the position coordinate and take the detector's position to be fixed in the laboratory frame. Then one can compute a discrete Fourier transform of $\Phi_{i, n}$:  
\begin{equation}
    \widetilde{\Phi}_{i,\ell} = \sum_n \Phi_{i, n}e^{-i2\pi \ell n/M} \ ,
\end{equation}
and define the frequency-domain quadrature coordinates
\begin{equation}
    R_{i, \ell} = \frac{\Delta t}{\sqrt{M\Delta t}} {\rm Re}(\Tilde{\Phi}_{i, \ell}), \qquad I_{i, \ell} = \frac{\Delta t}{\sqrt{M\Delta t}} {\rm Im}(\Tilde{\Phi}_{i, \ell})  \ .
\end{equation}
Note that both expressions are functions of the angular frequency $\omega_\ell=2\pi\ell/(M\Delta t)$. The normalization is chosen precisely so that $R_{i, \ell}^2 + I_{i, \ell} = \frac{(\Delta t)^2}{T} |\widetilde{\Phi}_{i, \ell}|^2$, which is exactly the power spectral density in frequency bin $\ell$. Ref.\cite{Foster_2021} then shows that the random variables $R_{i, \ell}, I_{i, \ell}$ have exactly zero mean, and moreover that
\begin{equation}
    {\rm Var}(R_{i, \ell}) = {\rm Var}(I_{i, \ell}) = \frac{\pi A_i}{2 m_a v_\omega}f_{\rm sp}(v_\omega) \ ,
\end{equation}
where $A_i = \rho_{\rm DM}\kappa_i^2$, $v_\omega$ is the velocity which satisfies $\omega = m_a(1+ v_{\omega}^2/2)$, and
\begin{equation}
    f_{\rm sp}(v) = \int d^3 {\bf u} \ f(u) \delta(|{\bf u}| - v) \ .
\end{equation}
Moreover, the covariance is zero. Any additive background does not correlate the quadratures, and only contributes a possibly frequency-dependent variance $\lambda_{i}(\omega)/2$. Therefore, the detector response is a mean-0 random variable with covariance matrix, in these quadrature coordinates, of
\begin{equation}
    \Sigma_i(\omega) = \frac{1}{2}\left(\frac{\pi A_i}{m_a v_\omega}f_{\rm sp}(v_\omega) + \lambda_i(\omega)\right)I_2 \ .
\end{equation}
Now, we can describe this physics using bosonic phase space. We introduce the complex detector amplitude coordinate $z_i(\omega) = R_i(\omega) + i I_i(\omega)$. Then, it immediately follows that
\begin{equation}
    z_i(\omega) \sim \mathcal{N}(0, P_i(\omega) + \lambda_i(\omega))
\end{equation}
with $P_i(\omega) = \frac{\pi A_i}{m_a v_\omega}f_{\rm sp}(v_\omega)$. Therefore, the detector maps a classical field amplitude to a displacement of a single bosonic mode in the appropriate quadrature coordinates. Characterizing the axion field then amounts to learning properties of the Gaussian displacement channel that describes these physics. 

With this context, we now understand the main motivating question of Ref.~\cite{Foster_2021}. Notice that $\Sigma_i(\omega)$ depends on the velocity distribution $v_\omega$ only through $f_{\rm sp}$, but this distribution integrates over all velocities with fixed magnitude $|\mathbf{v}|$ and therefore contains no directional information. As such, a single detector cannot characterize the direction of an anisotropic axion stream. This is essential, because as described in Ref.~\cite{Foster_2021}, the daily modulation that would be observed in the incident angle of a detected field would imply the existence of a fixed-orientation stream and would constitute smoking-gun evidence of an axion field. 

The key observation of Ref.~\cite{Foster_2021} is that while a single detector is insensitive to the stream's direction, \textit{two spatially-separated detectors} can ascertain the angle of incidence. We now omit the derivation and state the conclusion: letting $z_i, z_j$ denote the complex displacement coordinate of two spatially separated detectors located at ${\bf x}_i, {\bf x}_j$ and separated by vector ${\bf d} = {\bf x}_i -  {\bf x}_j$, the covariance matrix of the two coordinates has the form
\begin{equation}
    \Sigma_\theta (v) = P(v) \begin{pmatrix}
        1+\beta(v) & \gamma_\theta(v) \\ \gamma_\theta(v)^* & 1+\beta(v)
    \end{pmatrix} \ ,
\end{equation}
where $P(v), \beta(v)$ are independent of the stream direction, but $\gamma_\theta(v)$ is sensitive to the angle of incidence $\theta$ between the stream and the detector plane:
\begin{align}
\gamma_{\theta}(v)
&=
\frac{
F_{12}^{c}(v;\theta)
-
iF_{12}^{s}(v;\theta)
}{
f_{\mathrm{sp}}(v)
}, \quad \textnormal{ where } \\ 
F_{12}^{c}(v)
&=
\int d^{3}\mathbf{v}'\,
f(\mathbf{v}')\,
\cos\!\left(m_{a}\mathbf{v}'\cdot\mathbf{d}\right)
\delta\!\left(\lvert\mathbf{v}'\rvert-v\right), \\ 
F_{12}^{s}(v) &= \int d^{3}\mathbf{v}'\, f(\mathbf{v}')\, \sin\!\left(m_{a}\mathbf{v}'\cdot\mathbf{d}\right) \delta\!\left(\lvert\mathbf{v}'\rvert-v\right).
\end{align}
Here the functions $F_{12}$ depend on $\theta$ because of the term $\mathbf{v'} \cdot \mathbf{d}$. The entire task of learning the angle of incidence therefore reduces to sensing the off-diagonal covariance in a displacement channel's distribution.

These expressions can be simplified to numerical values given real physical parameters. Ref.~\cite{Foster_2021} models the Sagittarius stream \cite{sagittarius} using a Maxwellian speed distribution with boost vector $(0, 93.2, -388)$ km/s, dispersion $v_0 = 10$ km/s, and velocity $u = 400$ km/s. Substituting a Maxwellian speed distribution and simplifying gives us
\begin{align}
    \gamma_{\theta}(v) &=\frac{s(v)}{\chi_\theta(v)}\frac{\sinh \chi_\theta(v)}{\sinh s(v)} \\ 
    s(v) &= \frac{2vu}{v_0^2} \ , \\
    q(v) &= m_a d v\\
    \chi_\theta(v) &= \sqrt{a(v)^2 - q(v)^2 - 2ia(v)q(v)\cos \theta} \ .
\end{align}
Using an axion mass estimate of $25.2  \ \mu$eV alongside the estimated Sagittarius velocity parameters, we can compute numerical values for $\gamma_\theta$. In line with Ref.~\cite{Foster_2021}, we consider learning the incident angle $\theta = 45^\circ$. Substituting $\theta = 45^\circ$ and $46^\circ$ into our covariance matrix expression and evaluating numerically, we are left with the task of distinguishing two two-mode Gaussian displacement channels with equal mean and differing covariance matrices
\begin{align}
    &\Sigma_{45} = \Sigma_0 + \epsilon\begin{pmatrix}
        1 & 0.6065648 - 0.0066916 i \\ 0.6065648 + 0.0066916 i & 1
    \end{pmatrix} \ , \\
    &\Sigma_{46} = \Sigma_0 + \epsilon\begin{pmatrix}
        1 & 0.3298756 + 0.4965210 i \\ 0.3298756 - 0.4965210 i & 1 
    \end{pmatrix} \ .
\end{align}
Here, $\epsilon$ is the normalized value corresponding to the mapping of the physical scale parameters onto the bosonic phase space. $\epsilon$ therefore captures the total strength of interaction, which depends broadly on the axion field's power and its coupling constant to the detector. This constant cannot be evaluated without a particular detector specification, but in phase-space coordinates it is generally $\ll 1$ because the off-diagonal covariance is much weaker than the diagonal variance components due to background and power broadening. $\epsilon$ controls the hardness of the sensing task, since weaker fields are naturally more difficult to characterize.  

This task, derived directly from the physical model of axion detection of Ref.~\cite{Foster_2021} and evaluated with estimated parameters of the Sagittarius stream, therefore captures the challenge of resolving the angle of the incident axion stream to within one degree. 

\begin{center}
    \textit{Numerical implementation}
\end{center}

We compare QFS against two receivers that act locally at the detector sites, fixing the total mean photon number in the two sensing modes to $E=1$. The first prepares an independent single-mode squeezed vacuum at each detector, applies the inverse local squeezing after sensing, and performs local photon-number-resolving measurements. This combines phase-sensitive Gaussian probes with a non-Gaussian readout. The second prepares an entangled two-mode squeezed vacuum (TMSV) state between each sensing mode and a local idler, and performs a local continuous-variable Bell measurement after sensing. This converts each displacement into a noisy complex-amplitude sample and is the canonical entanglement-assisted Gaussian receiver. The two signal arms again contain total energy $E=1$, while the stored idlers contain one additional photon on average in total. Together, these protocols use the principal Gaussian and non-Gaussian resources available locally, and therefore provide stringent benchmarks for strategies which do not perform coherent operations between detector sites. Squeezing, photon-counting receivers, and TMSV-based entanglement assistance are among the principal quantum-enhanced architectures studied for stochastic sensing and axion detection and are established as near-optimal in many cases~\cite{Gardner_2025,Shi_2023,Sushkov_2023,Backes_2021,brady_entangled_2022}.

One could instead distribute entanglement between the probe states used at two detector sites or perform a joint quantum measurement across them. Such strategies may access the cross-covariance more directly, but require the generation, distribution, and stabilization of entangled bosonic modes over the detector separation, which is far more demanding than the significant challenges already posed by the above benchmarks. QFS does not require an entangled bosonic probe shared across this baseline. Instead, a single ancilla qubit coherently controls conditional displacements of the two sensing modes, mapping the selected collective quadrature onto a single-qubit measurement after one layer of control. While ancillary preshared entanglement could facilitate execution by quantum teleportation of the control qubit (rather than physically shuttling the qubit between detector sites), these resources are never exposed to the signal and can therefore be protected by e.g. quantum error correction or error mitigation, unlike entangled sensor probes that must remain sensitive to their environment. 

In our simulation, one shot consists of one independent draw of the two-mode Gaussian displacement under either angular hypothesis. For QFS, we prepare the depth-one conditional coherent-state probe, apply the displacement, invert the conditional displacement, and measure the ancilla in the $X$ basis. A shot of the squeezed photon-counting receiver returns the pair of local photon numbers obtained after inverse squeezing, while a shot of the TMSV receiver returns the pair of complex outcomes from the two local Bell measurements, which corresponds to sampling a Gaussian convolution of the displacement channel. For an episode of length $N$, we repeat the corresponding shot independently $N$ times and distinguish the $45^\circ$ and $46^\circ$ hypotheses using the exact likelihood ratio of the complete measurement record. At each field strength and candidate value of $N$, we estimate the classification accuracy using $5{,}000$ balanced Monte Carlo episodes and report the smallest tested value for which the empirical accuracy is at least $70\%$.

\begin{center}
    \textit{Origin of quadratic separation}
\end{center}
The origin of the quadratic separation in Fig.~\ref{fig:applications} is most transparent in the weak-field limit. In our numerical experiment, we write the stream-induced covariance under hypothesis $j$ as $\Gamma_j(\epsilon)=\epsilon S_j$. For a QFS filter with conditional displacement $\boldsymbol{\beta}$, let $c_j=\boldsymbol{\beta}^{\dagger}S_j\boldsymbol{\beta}$. The probability of the negative qubit outcome is
\begin{equation}
    1-p_j=\frac{1-\exp(-\epsilon \,c_j)}{2}=\frac{\epsilon \,c_j}{2}+O(\epsilon^2) \ .
\end{equation}
Provided that $c_0\neq c_1$, the hypotheses therefore produce different rare-event rates at first order in $\epsilon$. After $N$ shots, both the mean number and variance of these events scale as $N\epsilon$, so constant discrimination accuracy requires $N\epsilon=\Theta(1)$.

The baseline receivers behave differently. For the TMSV receiver, the measured covariance has the form $V_j=\nu I_2+\epsilon S_j$ with $\nu>0$. The two outcome distributions coincide at $\epsilon=0$ in the presence of this fixed measurement noise, so their statistical divergence begins at second order in the covariance perturbation. For squeezed photon counting, the local photon-number marginals are identical under the two hypotheses, and the first hypothesis-dependent information appears in joint photon-number events whose probabilities scale as $\epsilon^2$. Consequently, the one-shot Chernoff informations $C$ and the corresponding sample requirements satisfy
\begin{align}
    C_{\rm QFS}(\epsilon) &= \Theta(\epsilon), & N_{\rm QFS} &= \Theta(\epsilon^{-1}), \\
    C_{\rm SMSV}(\epsilon)=C_{\rm TMSV}(\epsilon) &= \Theta(\epsilon^2), & N_{\rm SMSV},N_{\rm TMSV} &= \Theta(\epsilon^{-2})=\Theta(N_{\rm QFS}^2) \ .
\end{align}
Thus QFS converts the covariance difference into a first-order rare-event rate and achieves Heisenberg scaling in the field strength, while the local receivers, despite using different resources even beyond QFS (more non-Gaussinity for the photon-counting receiver, and an entangled idler mode for the TMSV receiver), they only see the signal through second-order changes in their measurement statistics. This produces the quadratic separation observed in Figure \ref{fig:applications}(a). The core message is that programmable single-qubit control, even if only a depth-1 quantum circuit, can engineer filter functions which are asymptotically more sensitive to a signal of interest than even significantly sophisticated conventional strategies that may still use non-Gaussianity or entanglement. For axion detection in particular, this suggests that superconducting-circuit receivers, enabled by QFS design principles, may enable significant improvements to the detection of axionic dark matter, even when compared to longstanding and well-established quantum sensing strategies.

\subsection{Simulation of wireless communication receiver}

\begin{center}
    \textit{Task motivation: quadrature amplitude modulation}
\end{center}

Quadrature amplitude modulation (QAM) is a standard method for encoding digital information in wireless signals. A passband waveform associated with symbol $m$ may be written as
\begin{equation}
    s_m(t)=I_m\cos(\omega_c t)-Q_m\sin(\omega_c t),
\end{equation}
where $I_m$ and $Q_m$ are its in-phase and quadrature amplitudes. In a frame rotating at the carrier frequency, these amplitudes are precisely the two quadratures of a single electromagnetic mode, and the complex envelope $I_m+iQ_m$ may be identified with the bosonic phase-space coordinate $\alpha_m$. An $M$-QAM scheme encodes each symbol as one of $M$ predetermined points in the I-Q plane. In square 64-QAM, each quadrature takes one of eight possible values, giving $64$ symbols and therefore six encoded bits per symbol. Channel noise ordinarily broadens the received location, and the receiver must determine which constellation point was transmitted. We model this task directly as discrimination among the displacement channels associated with the QAM constellation.

\begin{center}
    \textit{Numerical implementation}
\end{center}

We consider the square 64-QAM constellation and a rectangular 8-QAM constellation,
\begin{align}
    \mathcal A_{64}(A)&=\left\{\alpha_{ab}=\frac{A}{\sqrt{42}}(a+ib)\mathrel{\Big|}a,b\in\{-7,-5,-3,-1,1,3,5,7\}\right\},\\
    \mathcal A_8(A)&=\left\{\alpha_{ab}=\frac{A}{\sqrt{6}}(a+ib)\mathrel{\Big|}a\in\{-3,-1,1,3\},\ b\in\{-1,1\}\right\}.
\end{align}
Both constellations are normalized such that
\begin{equation}
    \frac{1}{|\mathcal A_M|}\sum_{\alpha\in\mathcal A_M}|\alpha|^2=A^2.
\end{equation}
The 8-QAM constellation is therefore a $4\times2$ rectangular constellation containing three encoded bits. Its unnormalized coordinates form a subset of the 64-QAM lattice, but the two constellations are normalized independently. Consequently, $\mathcal A_8(A)$ coincides with the corresponding physical subset of $\mathcal A_{64}(\sqrt{7}A)$ rather than $\mathcal A_{64}(A)$.

Symbol $m$ implements the deterministic displacement channel
\begin{equation}
    \mathcal E_m(\rho) = D(\alpha_m)\,\rho\, D(\alpha_m)^\dagger.
\end{equation}
In each episode, the transmitted symbol is sampled uniformly from the relevant constellation and held fixed over all channel interactions. The receiver is given the constellation but not the transmitted symbol. The random-guessing probabilities are therefore $1/64$ and $1/8$ for the two tasks.

The shallow QFS receiver uses three ancilla qubits coupled to the received bosonic mode. Each qubit undergoes a depth-ten sequence of equatorial rotations and signal-interleaved controlled displacements before all three qubits are measured. Let $\mathbf q_{js}$ be the phase-space wavevector applied to qubit $j$ in layer $s$, and let $R(\theta_{js},\phi_{js})$ denote the corresponding equatorial qubit rotation. After the unknown displacement, the effective action of one layer on candidate $m$ can be represented as
\begin{equation}
    \ket{\psi_{mj}^{(s)}}=Z(\mathbf q_{js}\cdot\boldsymbol{\alpha}_m)R(\theta_{js},\phi_{js})\ket{\psi_{mj}^{(s-1)}},
\end{equation}
where $\boldsymbol{\alpha}_m=(I_m,Q_m)$ and $Z(\varphi)=\ket{0}\bra{0}+e^{i\varphi}\ket{1}\bra{1}$. In our displacement convention, a wavevector $\mathbf q=(q_I,q_Q)$ corresponds to the controlled-displacement amplitude $\beta=(q_Q-iq_I)/2$. We constrain the mean oscillator energy immediately before every signal interaction according to
\begin{equation}
    E_{mjs}=P_{e,mjs}\frac{\|\mathbf q_{js}\|^2}{4}\leq E_{\mathrm{cap}},
\end{equation}
where $P_{e,mjs}$ is the excited-state population of qubit $j$ before layer $s$. We take $E_{\mathrm{cap}}=10$ throughout.

If $p_{mj}$ denotes the terminal probability of outcome one on qubit $j$, one terminal measurement produces a three-bit outcome $\mathbf b\in\{0,1\}^3$ with conditional probability
\begin{equation}
    P(\mathbf b|m)=\prod_{j=1}^3p_{mj}^{b_j}(1-p_{mj})^{1-b_j}.
\end{equation}
We refer to one joint terminal measurement of the three qubits as one shot. A depth-ten QFS shot contains at most ten signal interactions per qubit and therefore thirty signal interactions in total.

For the 64-QAM sample-complexity calculation, we optimize a symmetry-restricted but explicitly realizable subclass of these circuits. Each qubit is prepared in an equal superposition, the same phase gradient $\mathbf q_j$ is applied in all ten layers, and the intermediate rotations are set to the identity. Choosing the final analysis phase appropriately gives
\begin{equation}
    p_{mj}=\frac{1}{2}\left[1+\sin\!\left(10\,\mathbf q_j\cdot\boldsymbol{\alpha}_m\right)\right],
\end{equation}
with $\|\mathbf q_j\|^2/8\leq E_{\mathrm{cap}}$. We optimize the three accumulated phase gradients over four orientation families,
\begin{equation}
    \left\{\left(0,\frac{\pi}{2},\frac{\pi}{4}\right),\left(0,\frac{\pi}{3},\frac{2\pi}{3}\right),\left(\frac{\pi}{6},\frac{\pi}{2},\frac{5\pi}{6}\right),\left(\frac{\pi}{4},\frac{3\pi}{4},0\right)\right\}.
\end{equation}
For each family, the three gradient magnitudes are chosen to maximize the worst-case pairwise Bhattacharyya information,
\begin{align}
    \max_{\{\mathbf q_j\}}\min_{m<n}B_{mn}\,,\qquad
    B_{mn}&=-\log\sum_{\mathbf b}\sqrt{P(\mathbf b|m)P(\mathbf b|n)}\,.
\end{align}

For a total of $N$ terminal shots, let $n_{\mathbf b}$ denote the number of occurrences of joint outcome $\mathbf b$. Candidate $m$ is assigned the exact multinomial log likelihood
\begin{equation}
    \mathcal L_m^{\mathrm{QFS}}=\sum_{\mathbf b}n_{\mathbf b}\log P(\mathbf b|m)\,,
\end{equation}
and the transmitted symbol is decoded as the candidate with the largest likelihood. We estimate the 64-QAM decoding probability using $100{,}000$ uniformly sampled truth and measurement episodes at each queried value of $N$, and report the smallest integer shot count attaining $70\%$ success. At larger field amplitudes, we also include the achievable response obtained by attenuating the signal to a previously optimized amplitude. This preserves the conditional response of the earlier circuit and prevents an artificial degradation in the strong-signal regime.

For the one-shot calculation, we use the rectangular 8-QAM constellation. Its three Gray-coded binary partitions specify the sign of the I quadrature, whether the I quadrature is on an inner or outer level, and the sign of the Q quadrature. We train one depth-ten circuit for each binary partition at seven logarithmically spaced amplitudes between $10^{-2}$ and $10$. At every plotted amplitude, we evaluate the resulting response bank and choose one circuit for each qubit to maximize the exact one-shot decoding probability,
\begin{equation}
    P_{\mathrm{QFS}}^{(1)}=\frac{1}{8}\sum_{\mathbf b}\max_m P(\mathbf b|m)\,.
\end{equation}
The plotted curve again includes the response-preserving attenuation envelope. Since all eight possible three-qubit outcomes are retained, a single terminal measurement can encode the three bits required to distinguish the eight candidate symbols.

We compare QFS against ideal heterodyne detection and an energy-matched two-mode squeezed vacuum receiver. A Gaussian receiver observation is modeled as
\begin{equation}
    \mathbf y_t=\boldsymbol{\alpha}_m+\mathbf z_t\,,\qquad \mathbf z_t\sim\mathcal N(0,\nu I_2).
\end{equation}
For TMSV, one mode probes the displacement while the second mode is retained as an idler and a joint Gaussian measurement is performed at the receiver. We take $\sinh^2r = E_{\mathrm{cap}}=10$ in the signal arm, with the retained idler energy not included in the quoted constraint. Its effective noise variance is $\nu_{\mathrm{TMSV}}=e^{-2r}\nu_{\mathrm{het}}$.

The benchmark Gaussian success probabilities for TMSV and heterodyne are evaluated exactly from the nearest-neighbor decision regions. Defining
\begin{equation}
    S_L(u)=\frac{2(L-1)\Phi(u)-(L-2)}{L}\,,
\end{equation}
the success probability for square 64-QAM after $N$ observations is
\begin{equation}
    P_{64}(N;\nu) = S_8\!\left(A\sqrt{\frac{N}{42\nu}}\right)^2.
\end{equation}
For rectangular 8-QAM, the one-shot probability is
\begin{equation}
    P_8(1;\nu)=S_4\!\left(\frac{A}{\sqrt{6\nu}}\right)S_2\!\left(\frac{A}{\sqrt{6\nu}}\right).
\end{equation}
We use the conventional complex-amplitude normalization $\nu_{\mathrm{het}}=1/2$. The TMSV variance is reduced by the same factor $e^{-2r}$ relative to heterodyne in each calculation.

The upper panel of Figure~\ref{fig:applications} reports the number of terminal shots required to identify a uniformly sampled 64-QAM symbol with $70\%$ probability. At $A=0.1$, QFS requires $12$ terminal shots, compared with $175$ TMSV shots and $7324$ heterodyne shots. At $A=0.01$, the respective requirements are $64$, $17{,}447$, and $732{,}343$. The lower panel reports the exact one-shot decoding probability for rectangular 8-QAM. At $A=0.1$, QFS identifies the transmitted symbol with probability $0.903$, compared with $0.303$ for TMSV and $0.149$ for heterodyne.

\begin{center}
    \textit{Origin of speedup}
\end{center}

For two symbols separated by distance $d$, Gaussian observations with covariance $\nu I_2$ have one-shot Bhattacharyya information
\begin{equation}
    B_{\mathrm{het}}(d)=\frac{d^2}{8\nu}\,.
\end{equation}
Heterodyne detection must resolve this displacement through a noisy phase-space estimate whose variance decreases as $1/N$. QFS instead converts the displacement difference into an accumulated relative phase,
\begin{equation}
    \Delta\varphi_{mn}=\sum_s\mathbf q_s\cdot\left(\boldsymbol{\alpha}_m-\boldsymbol{\alpha}_n\right).
\end{equation}
Quantum control allows this phase to accumulate coherently over several interactions before the ancilla is measured. The three-qubit receiver additionally produces eight possible terminal outcomes rather than one binary outcome. For 8-QAM, the three qubits can be matched directly to the three binary partitions of the constellation, allowing one joint outcome to resolve all three encoded bits. For 64-QAM, repeated three-bit outcomes provide a likelihood signature that distinguishes the complete set of $64$ symbols. Therefore, the high-level intuition for the observed advantage is that QFS can trade locally-resolved measurement of any point in phase space for amplification of signal in demarcated, noncontiguous regions of tunable geometry, at the cost of a loss of resolution within the regions themselves. Thus, whereas Gaussian strategies must resolve every point equally well, a QFS receiver can combine various settings to amplify the detected response on each QAM symbol and achieve a trainable task-specific advantage at equal energy. 

\newpage
\section{Preliminaries and definitions}
\label{sec:preliminaries}

In this section we begin by reviewing concepts in quantum information theory useful for understanding our work. We next prove a number of results in Gaussian statistics relevant to Theorem \ref{thm:one_qubit_advantage_informal}. Then we lay out the core operational definitions that underlie our separations: we recall the Quantum Signal Learning framework from Ref.~\cite{cotler2026quantumadvantagesensingproperties}, understand how QSL allows us to mathematically study the sensing of classical observables, and finally explain the mathematical formulation of the sensing hierarchy in Figure~\ref{fig:schematic_main}.

\subsection{Continuous-variable quantum information}
\label{subsec:cv_prelims}
Here we collect the standard facts in continuous-variable quantum mechanics used throughout this work. We set $\hbar = 1$ and work with a single bosonic mode unless stated otherwise; all statements extend to $n$ modes by promoting phase-space coordinates to $\mathbb{R}^{2n}$ and taking direct sums.
 
\begin{center}
    \textit{Quantum measurements}
\end{center}
The most general measurement permitted by quantum mechanics is described by a positive operator-valued measure.
 
\begin{definition}[POVM]
A positive operator-valued measure (POVM) on a Hilbert space $\mathcal{H}$ with outcome set $\mathcal{Y}$ is a collection of positive semidefinite operators $\{M_y\}_{y\in\mathcal{Y}}$ satisfying $\sum_y M_y = I$; for continuous outcomes one works with an operator-valued measure $M(dy)$ with $\int_{\mathcal{Y}} M(dy) = I$. Measuring a state $\rho$ yields outcome $y$ with probability $\Pr[y] = \Tr[M_y \rho]$.
\end{definition}
 
While a POVM specifies the statistics a measurement produces, it does not specify how the measurement is physically enacted. That information is carried by its dilation.
 
\begin{fact}[Naimark dilation]
\label{fact:naimark}
For every POVM $\{M_y\}$ on $\mathcal{H}$ there exist an ancilla Hilbert space $\mathcal{H}_A$ with a pure state $\sigma_A$, a unitary $U$ on $\mathcal{H}\otimes\mathcal{H}_A$, and a projective measurement $\{\Pi_y\}$ such that
\begin{equation}
    \Tr[M_y \rho] = \Tr\!\left[\Pi_y\, U (\rho \otimes \sigma_A) U^\dagger\right]
\end{equation}
for every state $\rho$ and outcome $y$.
\end{fact}
 
Fact~\ref{fact:naimark} gives POVMs an operational reading that will recur throughout this work: the dilation is an accounting of the ancillary resources used to make the measurement. Constraining a family of sensing protocols is then, mathematically, the act of constraining the allowed POVM family through its dilations: which ancilla states may be prepared, which joint unitaries may be applied, and which projective readouts are available. As a concrete example, the class of Gaussian (generaldyne) measurements reviewed below consists, in full generality, of exactly those POVMs admitting a dilation in which the ancilla state is Gaussian, the joint unitary is Gaussian, and the final readout is homodyne, followed by classical postprocessing; the energy constraint of our conventional class then amounts to bounding the total mean photon number of probe and ancilla in the dilation. Even non-continuous-variable resources fit into this picture: a protocol that couples the sensor mode to an ancilla qubit and reads out the qubit is, from the perspective of the mode alone, simply another POVM on the mode --- one obtained from a dilation with a two-dimensional ancilla, and, as we will see in Appendix~E, one that no Gaussian dilation can reproduce. The sensing hierarchy of Figure~\ref{fig:schematic_main}(b) is in this sense a nested family of dilation constraints, and quantum-enhanced sensing is the enlargement of the accessible POVM family by new dilation resources. Finally, a multi-round protocol is a sequence of such measurements in which the POVM applied in each round may be chosen adaptively from the classical transcript of earlier outcomes; this is the form in which adaptivity enters all of our lower bounds.
 
\begin{center}
    \textit{Displacement, squeezing, and Gaussian states}
\end{center}
A bosonic mode is described by canonical quadrature operators $\hat{x},\hat{p}$ obeying $[\hat{x},\hat{p}]=i$, related to the annihilation operator by $\hat{x}=(\hat{a}^\dagger+\hat{a})/\sqrt{2}$ and $\hat{p}=i(\hat{a}^\dagger-\hat{a})/\sqrt{2}$, and collected into the quadrature vector $\hat{z}=(\hat{x},\hat{p})$. This convention fixes the vacuum noise floor: the vacuum state $\ket{0}$ satisfies $\Delta x^2 = \Delta p^2 = 1/2$.
 
We use the single-mode phase-space convention $\alpha=(q_\alpha,p_\alpha)\in\mathbb R^2$. The standard symplectic form is
\begin{equation}
    \Omega(\zeta,\alpha) = q_\zeta p_\alpha - p_\zeta q_\alpha\,.
\end{equation}
If $J$ is the standard symplectic matrix, then $\Omega(\zeta,\alpha)=\zeta^T J\alpha$. For each $\zeta\in\mathbb R^2$, define the \textit{Weyl character}
\begin{equation}
\chi_\zeta(\alpha)=e^{i\Omega(\zeta,\alpha)}.
\end{equation}
For each phase-space point $\alpha$, the displacement operator is
\begin{equation}
    D(\alpha) = e^{-i\Omega(\alpha,\hat z)} = e^{i(p_\alpha \hat{x} - q_\alpha \hat{p})} ,
\end{equation}
which shifts the quadratures as $D(\alpha)^\dagger \hat{z} D(\alpha) = \hat{z} + \alpha$ and generates the coherent states $\ket{\alpha} = D(\alpha)\ket{0}$ from the vacuum. These characters are the natural Fourier modes of phase-space displacement sensing. The reason is the Weyl relation
\begin{equation} \label{eq:weyl_relation}
    D(a)D(b)=e^{-i\Omega(a,b)/2}D(a+b),
\end{equation}
which says that the noncommutativity of phase-space displacements is exactly measured by the symplectic phase $\Omega(a,b)$. Thus whenever a known control displacement is coherently compared with an unknown signal displacement, the qubit response is built from phases of the form $e^{i\Omega(\zeta,\alpha)}$.
 
The second elementary Gaussian resource is squeezing. The single-mode squeezer $S(r)$ is the unitary defined by its action on the quadratures,
\begin{equation}
    S(r)^\dagger \hat{x} S(r) = e^{-r}\hat{x}, \qquad S(r)^\dagger \hat{p} S(r) = e^{r}\hat{p},
\end{equation}
so that the squeezed vacuum $S(r)\ket{0}$ has quadrature variances $\Delta x^2 = e^{-2r}/2$ and $\Delta p^2 = e^{2r}/2$, purchasing sub-vacuum resolution along one axis at a mean photon cost of $\sinh^2 r$.
 
A Gaussian state is one whose Wigner function, reviewed below, is a Gaussian density on phase space. Such states are fully specified by their mean vector and covariance matrix,
\begin{equation}
    m_j = \Tr\!\left[\rho\, \hat{z}_j\right], \qquad V_{jk} = \tfrac{1}{2}\Tr\!\left[\rho\,\{\hat{z}_j - m_j, \hat{z}_k - m_k\}\right],
\end{equation}
with the vacuum corresponding to $m = 0$ and $V = I_2/2$. Gaussian states are exactly those generated from the vacuum by Gaussian unitaries --- unitaries generated by Hamiltonians at most quadratic in the quadratures --- which for a single mode are compositions of displacements, phase-space rotations, and squeezers, supplemented by beamsplitters in the multimode setting. The following elementary fact governs all of the energy accounting in this work.
 
\begin{fact}[Energy of a Gaussian state]
\label{fact:energy}
A single-mode Gaussian state with mean vector $m$ and covariance matrix $V$ has mean photon number
\begin{equation}
    \bar{n} = \Tr[\rho\, \hat{a}^\dagger \hat{a}] = \tfrac{1}{2}\left(\Tr V + \|m\|^2 - 1\right),
\end{equation}
which follows directly from $\hat{a}^\dagger\hat{a} = (\hat{x}^2 + \hat{p}^2 - 1)/2$. In particular, an energy budget $\bar{n} \leq E$ enforces $\Tr V + \|m\|^2 \leq 2E + 1$.
\end{fact}
 
Fact~\ref{fact:energy} cleanly separates the two ways a Gaussian protocol can spend energy. The term $\frac{1}{2}\|m\|^2$ is the classical share of the budget, spent on coherent displacement, while $\frac{1}{2}(\Tr V - 1)\geq 0$ counts the photons invested in squeezing (or absorbed as thermal noise); the uncertainty relation $\det V \geq 1/4$ forces $\Tr V \geq 1$, with equality precisely for coherent states. It is the latter, nonclassical share of the budget that powers conventional quantum sensing, and holding the total budget fixed across protocol classes is what makes our separations statements about quantum information processing rather than about energy. We remark that a few proofs in the appendices rescale phase-space coordinates so that a convenient reference Gaussian has identity covariance; wherever this is done, the convention is stated locally.
 
\begin{center}
    \textit{Homodyne, heterodyne, and generaldyne measurements}    
\end{center}
The two ubiquitous measurement primitives of quantum optics sit at opposite extremes of a single Gaussian family. Homodyne measurement is the projective measurement of a single rotated quadrature $\hat{x}_\varphi = \cos\varphi\, \hat{x} + \sin\varphi\, \hat{p}$, with POVM elements $\ket{x_\varphi}\bra{x_\varphi}$ and outcome density $P^\varphi_\rho(x) = \langle x_\varphi | \rho | x_\varphi\rangle$; equivalently, as recalled below, $P^\varphi_\rho$ is the marginal of the Wigner function of $\rho$ along the direction conjugate to $\hat{x}_\varphi$. Because the outcome variance is exactly the intrinsic quadrature variance of the state, homodyne detection resolves squeezed fluctuations at the level $e^{-2r}/2$ when the squeezing axis is aligned with the measured quadrature; but it reveals only a one-dimensional marginal per shot. Physically, it is implemented by interfering the signal with a strong local oscillator at phase $\varphi$ on a balanced beamsplitter and subtracting photocurrents; in microwave platforms, IQ mixing and phase-sensitive amplification play the same role.
 
Heterodyne measurement is the coherent-state POVM
\begin{equation}
    M(d^2\alpha) = \frac{1}{2\pi}\,|\alpha\rangle \langle \alpha|\, d^2\alpha \,,
\end{equation}
which measures both quadratures at once. Its outcome density is the Wigner function of $\rho$ convolved with the vacuum Gaussian, so every heterodyne record carries an irreducible unit of vacuum noise that is independent of the probe state and cannot be removed by squeezing. In exchange, the record is informationally complete. Physically, heterodyne is realized as double homodyne: the signal is split with vacuum on a balanced beamsplitter and the two outputs are homodyned along conjugate axes.
 
The full Gaussian measurement family interpolating between these extremes is the generaldyne class, which is most usefully defined through its dilation.
 
\begin{definition}[Generaldyne measurement]
\label{def:generaldyne}
A generaldyne measurement is any POVM realized by appending ancilla modes prepared in Gaussian states, applying a Gaussian unitary to the joint system, and performing homodyne measurements on a commuting set of output quadratures.
\end{definition}
 
Definition~\ref{def:generaldyne} makes the resource content of conventional bosonic sensing explicit, and it has a structural consequence that drives our conventional lower bounds: because Gaussian dilations act affinely on quadratures, the outcome of any generaldyne experiment performed on a Gaussian probe subject to a deterministic displacement $z$ is itself a Gaussian random variable, $Y \sim \mathcal{N}(Az + b,\, C)$, with $(A, b, C)$ determined by the probe, dilation, and readout (this is restated and proven in Appendix~E, where it is first used). Together with classical adaptivity between shots and the energy constraint of Fact~\ref{fact:energy}, generaldyne measurements on Gaussian probes constitute the class of conventional finite-energy protocols against which our quantum-enhanced separations are proven.
 
\begin{center}
    \textit{Wigner functions and Weyl symbols}
\end{center} \label{sec:weyl_symbols}
Phase space furnishes a faithful representation of bosonic states and observables, and it is the representation in which all of our arguments take place. For states, the standard object is the Wigner function.
 
\begin{definition}[Wigner function]
The Wigner function of a single-mode state $\rho$ is
\begin{equation}
    W_\rho(\alpha) = \frac{1}{2\pi}\int_{\mathbb{R}} dy\; e^{i p_\alpha y} \langle q_\alpha - y/2 |\, \rho \,| q_\alpha + y/2\rangle .
\end{equation}
\end{definition}
 
\begin{fact}[Properties of the Wigner function]
\label{fact:wigner}
The Wigner function is real and normalized, $\int d^2\alpha\, W_\rho(\alpha) = 1$, though it may take negative values; by Hudson's theorem, a pure state has a nonnegative Wigner function if and only if it is Gaussian. Its marginals reproduce homodyne statistics: integrating $W_\rho$ along the direction conjugate to $\hat{x}_\varphi$ yields $P^\varphi_\rho$. It is covariant under displacements, $W_{D(\beta)\rho D(\beta)^\dagger}(\alpha) = W_\rho(\alpha - \beta)$, and a Gaussian state with moments $(m, V)$ has Wigner function $\mathcal{N}(\alpha;\, m, V)$. Finally, overlaps are computed by the rule $\Tr[\rho\sigma] = 2\pi \int d^2\alpha\, W_\rho(\alpha) W_\sigma(\alpha)$.
\end{fact}
 
The Wigner function extends from states to arbitrary operators. This extension, the Weyl symbol, provides the phase-space intuition underlying the Q$\Psi$ framework of Appendix~D.
 
\begin{definition}[Weyl symbol]
The Weyl symbol of an operator $O$ is
\begin{equation}
    W_O(\alpha) = \int_{\mathbb{R}} dy\; e^{i p_\alpha y} \langle q_\alpha - y/2 |\, O \,| q_\alpha + y/2\rangle ,
\end{equation}
so that the Wigner function is $1/2\pi$ times the Weyl symbol of the density operator. The mismatched normalization is chosen so that states integrate to one while expectation values obey the clean trace rule
\begin{equation}
\label{eq:trace_rule}
    \Tr[O\rho] = \int d^2\alpha\; W_O(\alpha)\, W_\rho(\alpha) .
\end{equation}
\end{definition}
 
\begin{fact}[Properties of Weyl symbols]
\label{fact:weyl_symbols}
The Weyl symbol is linear in $O$, real for Hermitian $O$, and satisfies $W_I = 1$. Consequently, the symbols of a POVM form a partition of unity on phase space, $\sum_y W_{M_y}(\alpha) = 1$ pointwise --- but positivity of $M_y$ does \emph{not} imply positivity of its symbol, and this signed structure is precisely the non-Gaussian resource exploited in this work. The displacement operators are the operator representation of phase-space Fourier modes:
\begin{equation}
    W_{D(\zeta)}(\alpha) = e^{-i\Omega(\zeta,\alpha)} = \overline{\chi_\zeta(\alpha)} \,,
\end{equation}
so that expanding an operator in the displacement basis, $O = \frac{1}{2\pi}\int d^2\zeta\, \Tr[O D(\zeta)^\dagger]\, D(\zeta)$, is the same as symplectically Fourier-expanding its Weyl symbol. For a state, this Fourier data is the characteristic function $\chi_\rho(\zeta) = \Tr[\rho D(\zeta)] = \int d^2\alpha\, W_\rho(\alpha)\, \overline{\chi_\zeta(\alpha)}$, the symplectic Fourier transform of the Wigner function.
\end{fact}
 
Gaussian states and generaldyne POVM elements have Gaussian symbols; by contrast, the photon-number parity operator $\hat\Pi$ satisfies $\Tr[\hat\Pi D(\zeta)] = 1/2$ for every $\zeta$, so its Weyl symbol is the distribution $\pi\delta^2(\alpha)$.  It therefore has uniform weight across all symplectic frequencies, in sharp contrast to a Gaussian symbol. This is why photon-counting measurements can solve qualitatively different problems than conventional Gaussian sensing.

We say that a measurement is Wigner-positive if its Weyl symbol is nonnegative everywhere in phase space; all Gaussian measurements are Wigner-positive. Moreover, Wigner-positivity has an important learning-theoretic consequence.
\begin{fact}
    Given $N$ copies of any state $\rho$, the outcome of any strategy making Wigner-positive measurements of $\rho$, interleaved with classical adaptivity, is exactly equivalent to classical postprocessing of the Wigner function of $\rho^{\otimes N}$.
\end{fact}
This fact arises because positive Weyl symbols exactly form a Markov kernel, so any distribution $P^{(N)}$ over measurement transcripts can be written as
\begin{equation} \label{eq:wigner_positive_transcript}
    P^{(N)}(y_{1...N}) = \int \kappa^{(N)}(y_{1...N}|z_{1...N}) W_\rho^{\otimes N}(z_{1...N}) \,dz_{1...N} \ .
\end{equation}
 
Weyl symbols and the measurement kernels at the heart of Q$\Psi$ go hand in hand. Consider any protocol that prepares a probe $\rho$, exposes it to a deterministic signal displacement $D(\alpha)$, and measures a POVM element $M_y$. Its response function is, by the trace rule~\eqref{eq:trace_rule} and displacement covariance,
\begin{equation}
\label{eq:response_convolution}
    f_{M,y}(\alpha) = \Tr\!\left[M_y\, D(\alpha)\rho D(\alpha)^\dagger\right] = \int d^2 z\; W_{M_y}(z)\, W_\rho(z - \alpha) \ .
\end{equation}
The classical statistical response of any experiment is the Weyl symbol of its measurement, smoothed by the Wigner function of its probe. Taking symplectic Fourier transforms, the convolution theorem factorizes the response frequency by frequency,
\begin{equation}
\label{eq:response_factorization}
    \hat{f}_{M,y}(\zeta) = 2\pi\, \Tr[M_y D(\zeta)]\; \overline{\chi_\rho(\zeta)} \,, \qquad \hat{f}(\zeta) := \int d^2\alpha\, f(\alpha)\, e^{-i\Omega(\zeta,\alpha)} .
\end{equation}
When the probe is entangled with ancillas, or the measurement is qubit-assisted, the product on the right is replaced by a joint coefficient, but the conclusion is unchanged: every response function is a superposition of Weyl characters, and its amplitude at frequency $\zeta$ must be supplied jointly by the probe and the measurement. This identity is the intuition underlying the entire Q$\Psi$ formalism, and it holds on any phase space -- not only on bosonic ones. The functions $f_{M,y}$ are exactly the response kernels $K_M(dy|\alpha)$ of Appendix~D, and the accessible feature information measures their overlap with the phase-space representation of the target observable. 

Q$\Psi$ is, at heart, harmonic analysis of Weyl symbols under the dilation constraints of a physical experiment -- or in this paper, constraints of the sensing hierarchy. It also renders our separations intuitive before any calculation. A Gaussian state with covariance $V$ has $|\chi_\rho(\zeta)| = \exp(-\tfrac{1}{2}\zeta^T (J V J^T)\zeta)$, and Fact~\ref{fact:energy} bounds $\Tr V \leq 2E+1$, so an energy-$E$ Gaussian probe (and, by Definition~\ref{def:generaldyne}, an energy-$E$ generaldyne measurement) carries appreciable Fourier weight only within an ellipse whose longest semi-axis is $O(\sqrt{E})$; its response to a frequency-$k$ character is consequently suppressed as $\exp(-\Omega(k^2/E))$ even when the squeezing is optimally aligned, which is the mechanism behind the lower bound of Theorem \ref{thm:one_qubit_advantage_informal}. A single echoed conditional displacement with gate parameter $\beta$, by contrast, imprints the character $\chi_\beta$ on the qubit coherence directly, so that frequency $k$ is reached with response amplitude of order $\sqrt{E}/k$, which is polynomially, rather than exponentially, small. The exponential separations of this work are precisely the formalization of this gap in a variety of settings.

\subsection{Gaussian learning and estimation theory}
Theorem \ref{thm:one_qubit_advantage_informal} establishes an exponential lower bound against conventional bosonic quantum sensors using Gaussian operations. To prove this bound, we will prove a number of preliminary lemmas in Gaussian statistical learning theory and linear algebra. The role of these results is to control the measurement distribution of energy-constrained Gaussian protocols.

A general conventional Gaussian sensing experiment allows the preparation of an arbitrary Gaussian probe state, possibly entangled with many Gaussian ancillas, followed by interrogation of the signal channel and arbitrary Gaussian measurement. For full generality, we allow the final measurement to be any generaldyne measurement using ancilla systems, and we allow the collected classical data to be used to adaptively inform subsequent signal queries. The formal definition is given in Definition \ref{def:gaussian_bosonic_sensing_formal}. The following lemma allows us to characterize the output of any such quantum experiment using classical Gaussian statistics.

\begin{lemma} \label{lemma:Gaussian_experiment_form}
Let $Y$ be the random variable associated with the outcome of any single-mode Gaussian experiment in which the signal acts as a displacement $D_S(z)$, where the state before measurement is
\begin{equation}
    D_S(z) \rho_{SA} D_S^\dagger(z) \ .
\end{equation}
Then there exists a real matrix $A$, real vector $b$, and covariance matrix $C$ such that $Y$ is distributed as $Y\sim \mathcal{N}(Az + b, C)$. 
\end{lemma}
\begin{proof}
    Fix a single experiment, including the choice made from any previous transcript. A generaldyne measurement, by definition, admits a Gaussian dilation: one appends Gaussian ancillas, applies a Gaussian unitary, and homodynes a collection of commuting quadratures. Let $R$ be the quadrature vector of all modes in this dilation before the final Gaussian unitary. Since the probe and measurement ancillas are Gaussian, the joint state is Gaussian with some mean vector $m$ and covariance matrix $V$. The signal displacement translates only the signal quadratures, so there is a fixed real matrix $L$ such that after applying $D_S(z)$ the mean is
    \begin{equation}
        m(z)=m+Lz
    \end{equation}
    and the covariance remains $V$.
    The Gaussian unitary in the measurement dilation acts on quadratures as
    \begin{equation}
        R \longmapsto SR+d
    \end{equation}
    for a real symplectic matrix $S$ and a real vector $d$. The homodyne readout then selects a fixed real linear projection of these quadratures, so for some real matrix $M$, the measurement outcome is distributed as the linear marginal
    \begin{equation}
        Y=M(SR+d) \ .
    \end{equation}
    Linear marginals of Gaussian states are Gaussian. Hence, conditional on the displacement $z$, the outcome has mean and covariance
    \begin{align}
        \mathbb{E}[Y\mid z] &= MS(m+Lz)+Md \ ,\\
        \operatorname{Cov}(Y\mid z) &= MSVS^TM^T \ .
    \end{align}
    Setting
    \begin{equation}
        A = MSL\,, \qquad b = MSm + Md\,, \qquad C = MSVS^TM^T\,,
    \end{equation}
    gives $Y\sim\mathcal{N}(Az+b,C)$.
\end{proof}
This lemma tells us that regardless of the adaptivity we allow between measurements, any single experiment in a conventional Gaussian protocol with a deterministic displacement has an outcome distribution that follows some Gaussian distribution; we can then use linearity to compute the output distribution of a displacement channel. This fact heavily constrains the kinds of integrals we need to bound when controlling the total variation. However, we need a way to quantify the effect of an experimental energy constraint on this measurement distribution. This is done with the following lemmas.

First, we recall the definition of the classical and quantum Fisher information, which are used in the next lemma. 
\begin{definition}[Classical and quantum Fisher information]
Let $\{\rho_z\}_{z\in\mathbb R^d}$ be a differentiable family of quantum states, and let a fixed measurement produce an outcome distribution with density $p_z(y)$. The classical Fisher information matrix of the outcome distribution is
\begin{equation}
    [F_c(z)]_{jk}
    =
    \mathbb E_{Y\sim p_z}\left[
    \frac{\partial\log p_z(Y)}{\partial z_j}
    \frac{\partial\log p_z(Y)}{\partial z_k}
    \right].
\end{equation}
For each $j$, define the symmetric logarithmic derivative (SLD) $L_j$ to be a Hermitian operator satisfying
\begin{equation}
\frac{\partial \rho_z}{\partial z_j} = \frac{1}{2}\left(L_j \rho_z + \rho_z L_j\right).
\end{equation}
The quantum Fisher information matrix is then
\begin{equation}
[F_Q(z)]_{jk} = \frac{1}{2}\Tr\left(\rho_z \{L_j,L_k\}\right).
\end{equation}
\end{definition}
\noindent For any family of quantum states and any fixed measurement, the classical Fisher information is bounded by the quantum Fisher information, $F_c(z) \preceq F_Q(z)$. Moreover, the following specialization to single-parameter unitary families is a well-known fact in quantum metrology.
\begin{lemma}[Quantum Fisher information of a unitary family] \label{lemma:QFI_variance}
Let $\rho_\theta=e^{-i\theta G}\rho e^{i\theta G}$.
Then the quantum Fisher information of $\{\rho_\theta\}_\theta$ satisfies
\begin{equation}
    F_Q(\rho,G)\leq 4\operatorname{Var}_\rho(G).
\end{equation}
\end{lemma}

\begin{proof}
Write $\rho=\sum_j p_j\ketbra{j}{j}$. The quantum Fisher information of the unitary family is
\begin{equation}
    F_Q(\rho,G)
    =
    2\sum_{j,k:p_j+p_k>0}
    \frac{(p_j-p_k)^2}{p_j+p_k}
    |\langle j|G|k\rangle|^2.
\end{equation}
Since $(p_j-p_k)^2\leq(p_j+p_k)^2$,
\begin{equation}
    F_Q(\rho,G)
    \leq
    2\sum_{j,k}(p_j+p_k)|\langle j|G|k\rangle|^2
    =
    4\Tr(\rho G^2).
\end{equation}
Adding a scalar multiple of the identity to $G$ does not change $\rho_\theta$. Applying the preceding bound to $\widetilde G=G-\Tr(\rho G)I$ therefore gives
\begin{equation}
    F_Q(\rho,G) = F_Q(\rho,\widetilde G) \leq 4\Tr(\rho\,\widetilde G^2) = 4\operatorname{Var}_\rho(G).
\end{equation}
\end{proof}
Finally, we record a special feature of classical Fisher information of Gaussian random variables.
\begin{lemma} \label{lemma:classical_FI_cov}
Let $Y\sim \mathcal{N}(Az +b, C)$ be a Gaussian random variable where $C$ is full rank and where $z \in \mathbb{R}^m$ for any $m\in\mathbb Z_+$. Then the classical Fisher information of $Y$ with respect to $z$ is
\begin{equation}
    F_c(z) = A^T C^+ A \ ,
\end{equation}
where $C^+$ is the Moore-Penrose pseudoinverse of the covariance matrix $C$.
\end{lemma}
\begin{proof}
    Let $X = Y - b$; the Fisher information of $Y$ is the same as the Fisher information of $X$ since they exactly determine each other. Then:
    \begin{align}
        F_c(z) &= \mathbb{E}_{x\sim p_z(x)}\left[\nabla \log p_z(x) (\nabla \log p_z(x))^T\right] \ .
    \end{align}
    Directly evaluating the derivative of the log-likelihood using the PDF of a multivariate Gaussian,
    \begin{align}
        \nabla \log p_z(x) &= \nabla \left[-\frac{1}{2}(x-Az)^T C^{+} (x-Az) + \textnormal{const.} \right] \\
        &= \frac{1}{2}\left[A^T C^{+} (x-Az) +  (x-Az)^T C^{+}A\right] \\ 
        &= A^T C^+ (x-Az)
    \end{align}
    where in the final equality we use that $C^+$ is symmetric to rearrange terms. Substituting back into the Fisher information,
    \begin{align}
        F_c(z) &= \mathbb{E}_{x\sim p_z(x)}\left[A^T C^+ (x-Az)(x-Az)^T C^+ A\right] \\ 
        &= A^T C^+ C C^+ A = A^TC^+ A
    \end{align}
    where the second equality uses the definition of the covariance matrix. This completes the proof.
\end{proof}
We require a lemma which allows us to relate the obtained classical distribution to the energy utilized in the quantum experiment. This is the conceptual core of the separation.
\begin{lemma} \label{lemma:covariance_normbound}
Let $Y\sim \mathcal{N}(Az+b, C)$ be the random variable distributed according to the outcome of any conventional single-mode protocol limited to energy $E$, for a single displacement. Then the covariance matrix $C$ necessarily satisfies
\begin{equation}
    C\succeq (8E+4)^{-1}AA^T \ .
\end{equation}
Equivalently,
\begin{equation}
    A^T C^+ A \preceq 8E+4 I_2 \ ,
\end{equation}
where $C^+$ is the Moore-Penrose pseudoinverse of $C$. 
\end{lemma}
\begin{proof}
    In this proof, we use the quantum Fisher information and some of its known properties as an intermediate step to obtain the bound. Recall that the state before measurement is
    \begin{equation}
        \rho_z =  D_S(z) \rho_{SA} D_S^\dagger(z) \ .
    \end{equation}
    Let the direction of $z$ be an unknown parameter. That is, let $u = (u_x, u_p)= z / \|z\|_2$. The generator of this normalized displacement is the operator
    \begin{equation}
        G_u = u_x \hat{p}_S - u_p \hat{x}_S \ .
    \end{equation}
   By Lemma \ref{lemma:QFI_variance}, $F_Q(u) \leq 4 \textnormal{Var}_{\rho_{SA}}(G_u)$. Now notice that $\textnormal{Var}_{\rho_{SA}}(G_u) \leq \langle \hat{x}_S^2 + \hat{p}_S^2\rangle = \Tr(V)$, where $V$ is the covariance matrix of $\rho_S$. By Fact \ref{fact:covariance_energy}, we have $\textnormal{Var}_{\rho_{SA}}(G_u) \leq 2E+1$, and therefore,
    \begin{equation}
        F_Q \preceq (8E+4)I_2 \ .
    \end{equation}
     Since $C$ results from a Gaussian measurement, it is full-rank, so $\textnormal{im}(A)\subseteq \textnormal{im}(C)$, so by Lemma \ref{lemma:classical_FI_cov}, $F_c$ of the outcome $Y$ satisfies $F_c = A^T C^+ A$. Because $F_c\preceq F_Q$, it follows that
    \begin{equation}
        A^TC^+ A \preceq (8E+4) I_2 \ .
    \end{equation}
\end{proof}
\begin{lemma} \label{fact:covariance_energy}
A one-mode Gaussian state with mean vector $m$ and covariance $V$ has mean photon number \begin{equation}
    \bar{n} = \frac{1}{2}(\Tr V + \|m\|^2 - 1) \ .
\end{equation}
For an energy constraint $\bar{n} \leq E$ (working in $\hbar = 1$ units), this implies
\begin{equation}
    \Tr V \leq 2E+1
\end{equation}
\end{lemma}
\begin{proof}
    Let $R=(\hat{x},\hat{p})^T$, with mean vector $m=\langle R\rangle$ and covariance matrix
    \begin{equation}
        V_{ij}=\frac{1}{2}\langle \{R_i-m_i,R_j-m_j\}\rangle \ .
    \end{equation}
    In the convention $a=(\hat{x}+i\hat{p})/\sqrt{2}$, the number operator is
    \begin{equation}
        \hat{n}=a^\dagger a=\frac{1}{2}(\hat{x}^2+\hat{p}^2-1) \ .
    \end{equation}
    Therefore
    \begin{align}
        \bar{n}
        &= \frac{1}{2}\left(\langle \hat{x}^2\rangle+\langle \hat{p}^2\rangle-1\right) \\
        &= \frac{1}{2}\left(V_{11}+V_{22}+m_1^2+m_2^2-1\right) \\
        &= \frac{1}{2}\left(\Tr V+\|m\|^2-1\right) \ .
    \end{align}
    If $\bar{n}\leq E$, then
    \begin{equation}
        \Tr V+\|m\|^2\leq 2E+1 \ .
    \end{equation}
    Since $\|m\|^2\geq 0$, this implies $\Tr V\leq 2E+1$.
\end{proof}
Our bounds will also require the following results in classical Gaussian estimation theory and linear algebra.

\begin{lemma}[Gaussian conditioning formula]\label{lemma:conditional_gaussian_estimation}
     Let $(A, B)$ be random variables that are jointly Gaussian (each may itself be multivariate). Let $(\bar{A}, \bar{B})$ denote the mean of the joint distribution, and let
     \begin{equation}
         \Sigma = \begin{pmatrix} \Sigma_{AA} & \Sigma_{AB} \\ \Sigma_{BA} & \Sigma_{BB} 
         \end{pmatrix}
     \end{equation}
    be the joint covariance matrix in block-matrix form. Then the conditional variable $A|B=b$ is distributed as $\mathcal{N}(\bar{\mu}, \bar{\Sigma})$, where
    \begin{align}
        &\bar{\mu} = \mu_A + \Sigma_{AB}\Sigma_{BB}^{-1}(b-\mu_B)  \ ,\\
        & \bar{\Sigma} = \Sigma_{AA} - \Sigma_{AB}\Sigma_{BB}^{-1}\Sigma_{BA} \ .
    \end{align}
\end{lemma}
\begin{proof}
    Assume first that $\Sigma_{BB}$ is invertible. Define
    \begin{equation}
        K=\Sigma_{AB}\Sigma_{BB}^{-1}
    \end{equation}
    and write
    \begin{equation}
        A=\mu_A+K(B-\mu_B)+R
    \end{equation}
    where
    \begin{equation}
        R=A-\mu_A-K(B-\mu_B) \ .
    \end{equation}
    Since $(A,B)$ is jointly Gaussian, the pair $(R,B)$ is jointly Gaussian. We now compute the mean and covariance of $R$. First,
    \begin{align}
        \mathbb{E}[R]
        &= \mathbb{E}[A]-\mu_A-K(\mathbb{E}[B]-\mu_B) \\
        &= 0 \ .
    \end{align}
    Next,
    \begin{align}
        \operatorname{Cov}(R,B)
        &= \operatorname{Cov}(A-\mu_A-K(B-\mu_B),B) \\
        &= \Sigma_{AB}-K\Sigma_{BB} \\
        &= \Sigma_{AB}-\Sigma_{AB}\Sigma_{BB}^{-1}\Sigma_{BB} \\
        &= 0 \ .
    \end{align}
    Thus $R$ and $B$ are uncorrelated jointly Gaussian random variables, so they are independent. Conditioning on $B=b$ therefore makes the term $\mu_A+K(B-\mu_B)$ deterministic, but leaves the distribution over $R$ unchanged, allowing us to reduce the randomness in $A|B$ to randomness in $R$. Hence
    \begin{equation}
        A\mid (B=b) =\mu_A+K(b-\mu_B)+R
    \end{equation}
    in distribution. The conditional mean is therefore
    \begin{align}
        \mathbb{E}[A\mid B=b]
        &= \mu_A+K(b-\mu_B)+\mathbb{E}[R] \\
        &= \mu_A+\Sigma_{AB}\Sigma_{BB}^{-1}(b-\mu_B) \ .
    \end{align}
    The conditional covariance is
    \begin{align}
        \operatorname{Cov}(A\mid B=b)
        &= \operatorname{Cov}(R) \\
        &= \operatorname{Cov}(A-KB) \\
        &= \Sigma_{AA}-K\Sigma_{BA}-\Sigma_{AB}K^T+K\Sigma_{BB}K^T \\
        &= \Sigma_{AA}-\Sigma_{AB}\Sigma_{BB}^{-1}\Sigma_{BA} \ .
    \end{align}
    Since an affine transformation of a jointly Gaussian random variable is Gaussian, this proves
    \begin{equation}
        A\mid B=b\sim \mathcal{N}\left(\mu_A+\Sigma_{AB}\Sigma_{BB}^{-1}(b-\mu_B),\Sigma_{AA}-\Sigma_{AB}\Sigma_{BB}^{-1}\Sigma_{BA}\right) \ .
    \end{equation}
\end{proof}

\begin{lemma}[Woodbury identity]\label{lemma:woodbury}
    Let $C\succ 0$ be a real covariance matrix and let $A$ be a real matrix with compatible dimensions. Then
    \begin{equation}
        I_2-A^T(AA^T+C)^{-1}A=(I_2+A^TC^{-1}A)^{-1} \ .
    \end{equation}
\end{lemma}
\begin{proof}
    Define
    \begin{equation}
        M=I_2+A^TC^{-1}A \ .
    \end{equation}
    We first claim that
    \begin{equation}
        (AA^T+C)^{-1}A=C^{-1}AM^{-1} \ .
    \end{equation}
    Indeed,
    \begin{align}
        (AA^T+C)C^{-1}AM^{-1}
        &= (A+AA^TC^{-1}A)M^{-1} \\
        &= A(I_2+A^TC^{-1}A)M^{-1} \\
        &= AMM^{-1} \\
        &= A \ .
    \end{align}
    Since $AA^T+C$ is invertible, the claim follows. Therefore,
    \begin{align}
        A^T(AA^T+C)^{-1}A
        &= A^TC^{-1}AM^{-1} \\
        &= (M-I_2)M^{-1} \\
        &= I_2-M^{-1} \ .
    \end{align}
    Rearranging gives
    \begin{equation}
        I_2-A^T(AA^T+C)^{-1}A=M^{-1}=(I_2+A^TC^{-1}A)^{-1} \ .
    \end{equation}
\end{proof}
The above lemmas will be used in Section \ref{sec:conventional_gaussian_lb}. For Section \ref{sec:gaussian_testing_lower_bound}, we further require the following results.

\begin{lemma} \label{lemma:Gaussian_KL}
    Let $P = \mathcal{N}(0, \Sigma_0), Q = \mathcal{N}(0, \Sigma_1)$ be two $d$-dimensional centered Gaussians with positive definite covariance matrices. Then,
    \begin{equation}
        D_{\rm KL}(P\| Q) = \frac{1}{2}\left[\Tr(\Sigma_1^{-1}\Sigma_0) - \log \det(\Sigma_1^{-1} \Sigma_0) -d\right] \ .
    \end{equation}
\end{lemma}
\begin{proof}
    Evaluating the log likelihood ratio:
    \begin{align}
        \log \frac{p(x)}{q(x)}& = \frac{1}{2}\left[ - \log \det \Sigma_0 - x^T\Sigma_0^{-1} x +  \log \det \Sigma_1 + x^T\Sigma_1^{-1}  x\right] \\
        &= \frac{1}{2}\left(\log \frac{\det \Sigma_1}{\det \Sigma_0} + x^T(\Sigma_1^{-1} - \Sigma_0^{-1})x\right) \ .
    \end{align}
    The KL divergence is
    \begin{equation}
        D_{\rm KL}(P\| Q) = \mathbb E_{X\sim P}\left[\log \frac{p(X)}{q(X)}\right] = \frac{1}{2}\left(\log \frac{\det \Sigma_1}{\det \Sigma_0} + \mathbb E_{X\sim P}[X^T \Sigma_1^{-1} X - X^T \Sigma_0^{-1} X]\right) \ .
    \end{equation}
    Using $x^T A x = \Tr(A xx^T)$, $\log \frac{\det \Sigma_1}{\det \Sigma_0} = -\log \det(\Sigma_1^{-1}\Sigma_0)$, and $\mathbb E[XX^T] = \Sigma_0$ for centered Gaussians gives us the desired conclusion.
\end{proof}

\subsection{Formalizing sensing of classical signals through Quantum Signal Learning} \label{sec:non_Fourier_QSL}

This work considers the setting of learning from classical signals using continuous-variable quantum modes. Such quantum systems are utilized in ubiquitous sensing platforms because they naturally arise as quantized electromagnetic, mechanical, and vibrational degrees of freedom that coherently accumulate weak signals. Consequently, they underpin a wide range of quantum sensors. Optical cavity modes enable displacement and gravitational-wave detection in interferometers such as LIGO~\cite{Aasi_2013}. Microwave cavity modes form the basis of superconducting-circuit sensors and quantum-limited microwave measurements~\cite{Clerk_2010, krantz2019quantum} and mechanical resonators are used for force and mass sensing \cite{Degen_2017}. These sensing protocols can be realized across a variety of hardware platforms, including superconducting circuits, optical and optomechanical cavities, trapped ions and nanomechanical resonators~\cite{Degen_2017, aspelmeyer2014cavity, bruzewicz2019trapped}.

In all such instances, the interaction between the probe, which is a single bosonic mode, and a classical signal is governed by a Hamiltonian of the following form:
\begin{equation} \label{eq:onemode_Hamiltonian}
    H(t) = f_x(t)\hat{x} + f_p(t)\hat{p} \ .
\end{equation}
where the c-number coefficients $f_x(t), f_p(t)$ may be time-dependent and are usually stochastic, due to noise and uncertain priors. Importantly, Hamiltonians for classical-field sensing are \textit{linear} in the canonical quadrature operators, or equivalently in raising-lowering operators. This interaction persists up to some timescale $T$ after which point the signal has passed. Ref.~\cite{cotler2026quantumadvantagesensingproperties} shows that this composite evolution is always described by a \textit{displacement channel}.

\begin{fact}[Classical-field sensing is displacement channel learning] \label{fact:classical_sensing_equals_displacement}
    Let $\rho$ be the initial state of a bosonic mode, and let it evolve under any Hamiltonian of the form in \eqref{eq:onemode_Hamiltonian} for time $T$. Then the final state can always be written in the form
    \begin{equation}
        \mathcal{E}_H^{(T)}(\rho) = \int d^{2}\alpha \ P_H^{(T)}(\alpha) D(\alpha)\rho D(\alpha)^\dagger
    \end{equation}
    where $P_H^{(T)}$ is a positive, real probability density over $\mathbb{C}$ or $\mathbb{R}^2$, depending on the phase-space convention.
\end{fact}

A proof of this standard fact and its multimode generalization is shown in \cite{cotler2026quantumadvantagesensingproperties}. In this work, we largely study the single-mode ($N = 1$) setting (aside from Sections \ref{sec:multimode_charfun_lower_bound} and \ref{sec:char_function_learning_gkp} which address a canonical multimode learning task), but a future direction is to extend the speedups demonstrated in this work to the multimode setting and understand whether further advantages in mode number can be discovered.

Importantly, Fact \ref{fact:classical_sensing_equals_displacement} tells us that classical sensing with continuous-variable quantum sensors equates to learning observables from query access to a displacement channel, and that all information about the signal accessible at interrogation time $T$ is contained in $P_H^{(T)}$. This is even the case classically: a classical sensing protocol can be described as probing the channel $\mathcal{E}_H^{(T)}$ with a coherent state, a classical state of light without any quantum nonlinearity. 

To formulate the most general possible sensing task, Ref.~\cite{cotler2026quantumadvantagesensingproperties} then introduces the following definition of a signal property.

\begin{definition} [Bell smoothing operator]\label{def:Bell_smoothing}
    Let $\varphi_r$ denote the density $\mathcal{N}_{\mathbb{C}}(0, e^{-2r}\mathds{1}_n)$ over $\mathbb{C}^n$, and let $p * q$ denote the convolution between two densities over $\mathbb{C}^n$,
    \begin{equation}
        (p * q)(\xi) = \int d^{2n}\alpha \ p(\alpha) q(\xi - \alpha) \ .
    \end{equation}
    Then the Bell smoothing operator $T_r$ is the Gaussian convolution acting on functions $f:\mathbb{C}^n\rightarrow \mathbb{C}$ as 
    \begin{equation}
        (T_rf)(\alpha) = (f*\varphi_r)(\alpha) \ .
    \end{equation}
\end{definition}
\begin{definition}[Property of a classical signal] \label{def:property_formal}
Fix an $N$-mode linear Hamiltonian $H$ generated by an unknown classical signal and an evolution time $T$. Moreover, fix a model class $\mathcal{P}$ of distributions, such that any $P_H^{(T)}\in\mathcal{P}$. A property of the signal is the linear functional
\begin{equation}
    \Psi\!\left(P_H^{(T)}\right) = \mathbb{E}_{\alpha \sim P_H^{(T)}}\left[\psi(\alpha) \right]
    \label{eq:property_defn}
\end{equation}
where we refer to the function $\psi(\alpha)$ as the property kernel. For any valid property kernel, there must exist some $r\geq 0$ such that
\begin{enumerate}
    \item \textnormal{($\psi$ can be recovered from the measurement outcomes)} There exists a measurable function $g_\psi$ such that \begin{equation}
        T_rg_\psi = \psi
    \end{equation}
    \item \textnormal{(The recovery map has bounded variance for all relevant signals)} For every $P\in \mathcal{P}$, 
    \begin{equation}
    \mathbb{E}_{\xi\sim(P*\varphi_r)}\left[|g_\psi(\xi)|^2\right] < \infty \ .
    \end{equation}
\end{enumerate}
\end{definition}

The operationally meaningful part of Definition \ref{def:property_formal} is simply that a physical observable corresponds to a linear functional of the displacement distributions. The two qualifiers are basic regularity conditions to ensure that the learning task is physically well defined. That is, one introduces the Bell smoothing operator in Definition \ref{def:Bell_smoothing} to model the measurement response of the continuous-variable Bell sampling protocol used in Ref.~\cite{cotler2026quantumadvantagesensingproperties} assuming that input states have been squeezed with a squeezing strength $e^{-2r}$, because this measurement is informationally complete. As a result, if a particular property is to be estimable at all, there should exist some bounded, measurable function which recovers the property's kernel under smoothing by the Bell distribution. This is only a technical detail that will not appear in any of our subsequent discussion as all physically well-defined properties will satisfy these regularity conditions.

To gain some basic intuition for this definition, we look at some special cases. If one chooses the kernel $\psi_{\rm Fourier}^{(k)}(x + ip) = \exp(ikx)$, note that the corresponding property is exactly the $k$-th Fourier coefficient of the signal along the $x$ quadrature. Moments, polyspectra, and correlations all appear as kernels which are polynomials of the components of $\alpha$. Alternatively, by choosing the kernel to be an appropriate Dirac delta function, one can extract quantities such as the amplitude or phase of the underlying signal. Upon imposing additional structure on $H$ as is done in conventional AC or DC metrology, a delta-function kernel recovers the canonical metrology problem. More generally, matched filtering tasks considered in e.g. gravitational wave sensing or fundamental particle detection correspond to more complex choices of $\psi$ which may be polynomials or Lorentzian functions of $\alpha$.

We can also extend the notion of a property to \textit{nonlinear} functionals of a signal. This is done by allowing properties of the form
\begin{equation}
    \Psi\!\left(P\right) = \int \psi(\alpha_1, \alpha_2, ..., \alpha_m) \prod_{j=1}^m P(d^{2n}\alpha_j) \ .
\end{equation}
Such a definition includes properties that cannot be extracted from any experiment making only a single query to the signal, including features like the quantum channel purity of the signal. More practically, properties of time-varying signals such as multi-time correlations fall under this more general notion via joint distributions $P(d\alpha_1 \cdots d\alpha_m)$, which we explore in this work. Now, to package this definition into a learning task, we introduce the full QSL problem.

\begin{graydefinitionbox}
\begin{definition}[Quantum Signal Learning] \label{def:QSL_formal}
Fix a linear Hamiltonian $H$ and an evolution time $T$, and a property $\Psi$. The Quantum Signal Learning problem \textnormal{\textsc{QSL}}$(H, T, \Psi, \epsilon, \delta)$ is to estimate $\Psi(P_H^{(T)})$ to within absolute error $\epsilon$, with success probability at least $1-\delta$, given query access to $\mathcal{E}_{H}^{(T)}$. 
\end{definition}    
\end{graydefinitionbox}

Ref.~\cite{cotler2026quantumadvantagesensingproperties} studied QSL in the \textit{post-hoc} setting, where one only receives the list of desired properties after making all measurements. However, the more experimentally motivated setting considered in this work considers learning a few predetermined properties that are set before deciding the sensing procedure. This also opens up the potential for exponential quantum advantages that do not rely on entangled measurements. The main results of this work are concerned with several practically meaningful instances of QSL -- namely, the properties of interest include directional and angular Fourier coefficients, temporal correlation functions, and other polynomial transformations of the displacement channel. 

We will study the resource complexity of learning these fundamental observables when sensors possess practically-motivated, near-term quantum processing resources, vs. when they do not. Our separations parameterize the properties of interest by a positive number $k$, and show that any protocol, regardless of adaptive control, that lacks a particular quantum resource requires $S\geq \exp(\Omega(k))$ signal shots to solve QSL. Meanwhile, appending a single, modest resource enables a protocol to solve the same QSL task using $S = \textnormal{poly}(k)$ shots. This hierarchy of resources is depicted in Figure \ref{fig:schematic_main}(b) and detailed in the next section.

To obtain these quantum advantages, we introduce Quantum Phase-Space Inference (Q$\Psi$) in Section \ref{sec:qes}. However, we remark that $Q\Psi$ applies more broadly beyond the results of this paper: it is a flexible framework for discovering quantum speedups across a broad range of learning tasks. To apply Q$\Psi$ to a particular experimentally-grounded setting, such as sensing classical fields with single-mode sensors and additional QIP resources, one requires a mathematical model for the tasks of interest; upon defining the appropriate model, Q$\Psi$ handles the rest of the analysis. QSL, with property kernels chosen to produce the basic sensing primitives above, is the appropriate $Q\Psi$ model for this work.

\subsection{A hierarchy of conventional and quantum-enhanced sensing}
\label{sec:sensing-hierarchy}

This work rigorously demonstrates that even rudimentary quantum control, coupled to conventional quantum sensors, can confer dramatic advantages in our ability to learn a wide array of signals. To formalize this claim, it will be important to clearly understand the distinction between conventional and quantum-enhanced sensing. Moreover, as quantum-enhanced sensing technology continues to develop, more sophisticated experiments will have access to more powerful forms of quantum information processing, such as increased control depth or long-lived quantum memory. This motivates us to define the \textit{hierarchy} of quantum-enhanced sensing in Figure \ref{fig:schematic_main}(b).

To prove an exponential separation between each level in Figure \ref{fig:schematic_main}(b), our strategy is to define classes of sensing frameworks which include each lower level as a strict subset, while adding a single quantum processing capability.

\begin{center}
\textit{Classical vs. conventional quantum sensing.} 
\end{center}
The first level of the sensing hierarchy contains all classical sensing protocols, which can only perform classical probe state preparation and measurement. For our lower bound, however, the following definition is convenient.

\begin{definition}[Coherent-probe protocol.]\label{def:coherent_probe_protocol} A coherent-probe protocol for quantum sensing of a channel $\mathcal{E}_P$ is one in which any query to $\mathcal{E}_P$ can only be enacted upon a coherent state $\rho = D(\alpha)\ketbra{0}{0}D(\alpha)^\dagger$. Arbitrary quantum information processing and measurement (i.e. any POVM) is allowed on the state $\mathcal{E}_P(\rho)$.
\end{definition}

Coherent-probe protocols simply require that the state used to interrogate the signal be classical. Beyond this requirement, all forms of adaptivity and quantum measurement are allowed. Of course, this circumscribes all purely classical sensing protocols in which not only the probe state, but the measurements, must be made classically. 

When we discuss quantum sensing with conventional quantum probes, we are concerned with the class of \textit{Gaussian} measurements, defined as follows.

\begin{definition}[Gaussian sensing]
    A Gaussian sensing protocol is one in which $\rho$ may be any Gaussian state, and any generaldyne measurement can be made on the state $\mathcal{E}_P(\rho)$. Moreover, classical adaptivity is permitted between measurements.
\end{definition}

Experimentally, these operations can be implemented using highly developed technologies in the optical domain such as optical parametric oscillators for generating squeezed light, beam splitters and phase shifters for Gaussian interferometry, and balanced homodyne or heterodyne detection for readout. Fast electronic processing further enables measurement outcomes to be used in real time to modify subsequent operations, making adaptive Gaussian protocols experimentally practical. Gaussian sensing therefore encompasses the realistic protocols which can be implemented in mature technologies (such as optics) but which cannot currently access the single-photon nonlinear regime~\cite{gerry2005introductory}. 

To establish the first separation, we show that a simple protocol using squeezed input states and homodyne measurements requires exponentially fewer measurements than any coherent-probe protocol to estimate any angular Fourier coefficient. We demonstrate that other simple quantum-enhanced probes also suffice to instantiate this separation by proposing a simple protocol using a cat probe state. We realize this quantum advantage experimentally, depicted in Fig.~\ref{fig:coherent_and_memory_experiments}(b); due to the limitations on performing homodyne measurements on our superconducting-circuit platform, we instead implement the protocol given in Section \ref{sec:exp_proxy}. 

\begin{center}
\textit{Conventional quantum sensing vs. constant-depth single-qubit control.} 
\end{center}

Up to this point we have discussed classes of sensing strategies which can be readily realized in e.g. conventional optical cavity platforms. Before we introduce our first quantum processing-enhanced class, it is important to understand resource accounting. Namely, the performance of conventional bosonic sensing platforms is usually constrained by the amount of non-classicality they can generate. In the Gaussian setting, this is encapsulated by the amount of squeezing that can be prepared in probe and ancilla states, since passive linear optics and displacements are classical operations that, in conjunction with squeezing, realize the full class of Gaussian protocols. 

The quantity that operationally constrains conventional protocols is therefore the maximum accessible photon number $\Bar{n}$, which is equivalently the maximum energy $E$ used by the protocol upon multiplying by Planck's constant $\hbar$. To discuss a quantum enhancement, it will be important to ensure that $E$ is held constant in all comparisons so that the advantage is truly due to a quantum information processing capability and not an energy increase. 

The constraint of constant $E$ is also apparent when considering the experimental implementation of this protocol with superconducting circuits. Here, the dominant source of decoherence of the bosonic mode is photon loss, whose effect scales with $E$. Therefore, the coherence timescale of the mode sets an effective maximum value of $E$ which can be obtained. For our experimental implementation, the relevant definition of $E$ depends on the displaced frame where this number is minimized. With photon loss rate $\kappa$, the Lindblad master equation describing the evolution of the density matrix is:
\begin{equation}
    \partial_t \rho = - i[H, \rho] + \kappa \,\mathcal{D}[a]\rho,
\end{equation}
where $\mathcal{D}[L] = L \rho L^\dagger - (1/2)\{L^\dagger L, \rho \}$ is the disspator and $H$ is the Hamiltonian. Here we have set $\hbar = 1$. When moving into a time-dependent displaced frame $\tilde{\rho} = D^\dagger(\alpha) \rho D(\alpha)$, the the dissipator term transforms as:
\begin{equation}
    \kappa \mathcal{D}[a + \alpha] \tilde{\rho} = \kappa\mathcal{D}[a]\tilde{\rho} - i \left[ i \frac{\kappa}{2} (\alpha^* a - \alpha a^\dagger), \tilde{\rho} \right],
\end{equation}
which results in two contributions: a deterministic re-centering force at a rate $\kappa |\alpha|/2$ (which can be corrected for), and a dissipator whose effect is strength is unaffected. Therefore, changing the displaced frame (or equivalently, displacing the oscillator state) does not enhance the effective decoherence rate. Therefore, the true resource, for this description of decoherence, is the value of $E$ in the displaced frame. 

Now, keeping all sensing resources constant, our next level in the hierarchy is to add a single qubit supported in the two-dimensional linear subspace $\{|g\rangle,\,|e\rangle\}$. For concreteness, we allow two unitary operations of the following form: 
\begin{equation} \label{eq:gate_defns}
    \textnormal{Any } U(\theta, \phi) \in \textsf{SU}(2), \qquad \textnormal{JC}(\beta_1, \beta_2) = D(\beta_1)\otimes \ketbra{e}{g} + D(\beta_2) \otimes \ketbra{g}{e} \ .
\end{equation}
The first is simply any rotation of the qubit, which can be easily realized in most qubit platforms. The second is a generalized Jaynes-Cummings-like interaction which can be realized in constant depth by combining a symmetric Echoed Conditional Displacement (ECD) gate 
\begin{equation}
    \textnormal{ECD}(\beta) = D(\beta)\otimes \ketbra{e}{g} + D(-\beta) \otimes \ketbra{g}{e}
\end{equation}
with a displacement of the bosonic mode. Our results are not unique to this choice of gateset, as will become clear in the following sections; we simply choose to work with this model theoretically for consistency, as it is the gate model implemented in our experiments (see Figure \ref{fig:ecd_pulse}). At the first quantum-enhanced level of the hierarchy, we use only a single application of each of the two gates to realize an exponential advantage in learning Fourier amplitudes.

\begin{center}
    \textit{A single memory qubit vs.~quantum processing-enhanced, memoryless sensing}
\end{center}

Another limiting feature of conventional experiments is that they perform destructive measurements on the joint system of qubit and mode. Beyond basic control capabilities, adding a single qubit also gives us the ability to retain a small quantum \textit{memory} that stores a quantum state as the mode interacts with the signal multiple times. Mathematically, the distinction is as follows. In all previous levels of the hierarchy, a single signal query is made, followed by POVMs defined by the hierarchy class; the resulting classical outcome is stored and can be used to inform the next POVM. However when memory is allowed, we are permitted to perform measurements of (or trace out) \textit{only} the bosonic mode, while the qubit retains its state across queries to the signal and can be measured at any time. We show that a simple protocol which only requires the qubit to remain coherent over time even while the sensor mode decoheres quickly can achieve an exponential advantage over any protocol which measures conventional resources destructively, even given arbitrarily powerful but short-lived quantum processing. We demonstrate this experimentally in Fig. \ref{fig:coherent_and_memory_experiments}(c).

\begin{center}
\textit{Beyond shallow depth} 
\end{center}

The separations at previous levels of the hierarchy establish that with rudimentary quantum information processing already accessible with modern technology, exponential advantages can be realized for learning classical signals. However, future sensing platforms are likely to have more sophisticated quantum processing capabilities, which can enable deep, coherent quantum circuits. 

Ref.~\cite{prabhu2026quantumcomputationaldisplacementsensing} experimentally demonstrated that trainable, coherent ECD circuits could realize displacement-sensing protocols that outperform conventional baselines by performing sophisticated QIP before readout. For concreteness, we work within our current circuit model, and consider sensing protocols which interleave control operations of the form 
\begin{equation}
U_r = \prod_{j=1}^r \textnormal{ECD}(\beta_j)(I\otimes U(\theta_j, \phi_j)) \ ,
\end{equation}
with queries to a signal and measure the qubit in any basis. This constitutes a powerful class of non-Gaussian sensing protocols that include, for instance, any strategy in the Quantum Computational Displacement Sensing (QCDS) framework \cite{prabhu2026quantumcomputationaldisplacementsensing} that builds on Quantum Signal Processing (QSP) \cite{Low_2017}. Here, we establish that protocols with access to coherent depth $r$ can estimate properties of classical signals using $\exp(\Omega(\gamma r))$ fewer signal queries than any any protocol with depth $\leq \gamma r$, for a constant $\gamma$ less than 1. 

\newpage

\section{Quantum Phase-Space Inference (Q$\Psi$)} \label{sec:qes}
In this section, we introduce Quantum Phase-Space Inference (Q$\Psi$), a framework for analyzing and designing quantum learning experiments under explicit physical constraints. Q$\Psi$ starts from three ingredients: a family of possible physical systems, a class of experiments allowed by the available architecture, and a property to be learned. It represents these ingredients on a common phase space and quantifies how strongly the allowed experiments can respond to the property of interest. This produces lower bounds on the resource complexity of any experiment in the specified architecture and identifies quantum-enhanced experiments that saturate the optimal scaling. Comparing the resulting bounds across architectures then certifies the advantage provided by additional quantum resources. This perspective allows us to derive exponential quantum advantages for basic sensing tasks using only a single qubit coupled to a conventional sensor.

We develop Q$\Psi$ here beyond its specific application to sensing with a single-mode oscillator. First, we build a phase-space formalism for understanding the outcomes of general quantum experiments and define the \textit{accessible feature information}  (AFI), a flexible information quantity which controls downstream Q$\Psi$ bounds. We then review information-theoretic learning lower bound techniques, which allow us to prove the first of three main tools in the Q$\Psi$ toolkit: the \textit{semiclassical measurement lemma}. This lemma turns a sensing or learning task into a phase-space description and outputs a tight query-complexity lower bound based on the AFI. We next build the upper-bound machinery, and show the \textit{Q$\Psi$ saturability guarantee}; this theorem proves that our Q$\Psi$ lower bounds are saturable and produces a hypothesis-testing algorithm which saturates them. Finally, we extend from hypothesis testing to global learning tasks, and show that Q$\Psi$ systematically translates quantum experimental design into classical statistical optimization of Fourier-domain overlaps between phase-space functions. We conclude with a template for Q$\Psi$ quantum advantage design that combines these tools.

\subsection{Building the language of Q$\Psi$}

The basic idea of Q$\Psi$ can be stated before introducing the formalism. A point $x$ in phase space labels a possible realization of the physical system. An experiment converts that realization into a classical outcome $Y$, described by a conditional distribution $K_{\mathsf E}(dy|x)$, while the property to be learned is represented by a function of the same realization $x$. The learning power of the experiment is therefore controlled by how much information about the target function survives the map from $x$ to the observed outcome $Y$. The purpose of this subsection is to formalize these three objects: the physical realization, the experimental response, and the property of interest.

\subsubsection{Experiments, observables, and uncharacterized systems} \label{sec:qpsi_setup}

To formalize how general quantum experiments, observables, and uncharacterized systems interact on a single phase-space, we introduce tools from probability theory and harmonic analysis. Note that these definitions let us work with generic phase-spaces, not only single-mode bosonic ones. 

\begin{definition}[Probability space]
A probability space is a triple $(\mathcal{X}, \mathcal{F}, P_0)$, where $\mathcal{X}$ is the sample space, $\mathcal{F}$ is the $\sigma$-algebra of measurable subsets of $\mathcal{X}$, and $P_0$ is a probability measure on $(\mathcal{X}, \mathcal{F})$.
\end{definition}
The reference measure $P_0$ will play two roles in Q$\Psi$. It defines the inner product used to compare functions on phase space, and it provides the reference law around which we formulate local learning problems.

In Q$\Psi$, we work with phase-space equipped with a normalized probability measure, thus treating phase-space as a probability space. In the single-mode bosonic setting, $\mathcal{X}$ is $\mathbb{R}^2$, and $P_0$ can, for instance, be a Gaussian distribution over points in phase-space corresponding to the vacuum noise distribution. When we represent bosonic density matrices as functions on phase-space, we are generally working with representations on the Hilbert space
\begin{equation}
L^2(\mathbb{C}) = \left\{f:\mathbb{C} \rightarrow \mathbb{C} : \int |f(\alpha)|^2 d^2\alpha < \infty\right\} \ ,
\end{equation}
i.e. normalized functions relative to the Lebesgue measure. However, in Q$\Psi$ it will be important to work on the Hilbert space
\begin{equation}
L^2(P_0) = \left\{f:\mathbb{C} \rightarrow \mathbb{C} : \int |f(\alpha)|^2 dP_0 < \infty\right\}
\end{equation}
equipped with the inner product
\begin{equation}
    \langle f, g\rangle_{P_0} = \int \overline{f(x)} g(x) dP_0(x) \ .
\end{equation}
Here we identify the sample space $\mathcal{X}$ with $\mathbb{R}^2$ or $\mathbb{C}$ and the $\sigma$-algebra with the corresponding Borel $\sigma$-algebra. When working with other quantum systems such as qubits, the sample space may be a finite field like $\mathbb{F}_2^{2n}$, or even a nonabelian group like $\textsf{SU}(2)$. Next we define a Fourier character family, which generalizes the notion of a complete basis of functions supporting a Fourier transform on an arbitrary sample space.

\begin{definition}[Fourier character family]
A Fourier character family on a sample space $\mathcal{X}$ is a family
\begin{equation}
    \mathcal{D} = \{\chi_\omega: \mathcal{X} \rightarrow \mathbb{C}\}_{\omega \in \Omega}
\end{equation}
where $\Omega$ is an abelian label group such that $\chi_0(x)=1$, $|\chi_\omega(x)|=1$, $\chi_{\omega+\nu}(x)=\chi_\omega(x)\chi_\nu(x)$, and $\chi_{-\omega}(x)=\overline{\chi_\omega(x)}$. Given a measure $P_0$ on $\mathcal{X}$, the mean of a character is
\begin{equation}
    \mu_\omega = \int \chi_\omega(x) dP_0(x)
\end{equation}
and the centered character is $\widetilde\chi_\omega(x) = \chi_\omega(x) - \mu_\omega$. The centered character family $\widetilde{\mathcal{D}}_{P_0}$ is the set of all centered characters on $\mathcal{X}$ with respect to measure $P_0$.
\end{definition}

For the single-mode bosonic phase-space, a canonical Fourier character family is the \textit{Weyl family} $\chi_\zeta(\alpha) = e^{i\Omega_{\rm sp}(\zeta,\alpha)}$, where $\Omega_{\rm sp}$ is the symplectic form. For qubits, one often uses $\chi_s(z) = (-1)^{s\cdot z}$. Operationally, the character family gives Q$\Psi$ a way to decompose different quantum operators, such as states, channels, or measurements, which live on the same phase-space into a common basis. This makes it much easier to bound overlaps.

\begin{definition}[Gram matrix of centered character family]\label{def:character_gram}
Given a probability space $\mathcal{X}$ equipped with measure $P_0$ and a centered character dictionary
\begin{equation}
    \widetilde{\mathcal{D}}_{P_0} = \{\widetilde\chi_\omega: \omega \in \Omega, \widetilde\chi_\omega \neq 0\} \ ,
\end{equation}
the Gram matrix of the family is
\begin{equation}
    G(\omega, \nu) = \langle \widetilde\chi_\omega, \widetilde\chi_\nu\rangle_{P_0} \ .
\end{equation}
\end{definition}

The Gram matrix controls overlaps between Fourier basis elements. For an orthonormal centered character family, its off-diagonal entries vanish and its diagonal entries are one.

\begin{lemma}[Orthogonality of character family from compactness] \label{lemma:compact_orthog}
Let $G$ be a compact abelian group with normalized Haar measure $P_0$, and let $\{\chi_\omega\}_{\omega\in\widehat G}$ be its character group. Then $\mu_\omega=0$ for every nontrivial character, and
\begin{equation}
    \langle \chi_\omega,\chi_\nu\rangle_{P_0}=\mathds{1}[\omega=\nu] \ .
\end{equation}
\end{lemma}

\begin{proof}
If $\omega\neq 0$, choose $a\in G$ such that $\chi_\omega(a)\neq 1$. Using Haar's theorem since we assume compactness, we have
\begin{equation}
    \int \chi_\omega(x)dP_0(x)=\int \chi_\omega(x+a)dP_0(x)=\chi_\omega(a)\int \chi_\omega(x)dP_0(x) \ ,
\end{equation}
and hence $\mu_\omega=0$. Applying this to $\chi_{\nu-\omega}$ gives the stated orthogonality relation.
\end{proof}

\noindent The lemma provides a useful intuition: if a physical process admits a representation on a compact abelian group in phase space, it also admits an orthonormal Fourier decomposition. This is convenient because quantum measurements and the uncharacterized system of interest can then be related in the same orthonormal basis. Unfortunately, the natural Weyl basis for bosonic phase space is not orthonormal on $L^2(\mathbb{R}^{2n})$. However, it satisfies an approximate orthonormality condition relative to a Gaussian measure, allowing us to retain many of the same advantages by equipping phase space with a Gaussian reference measure equipped with a Gaussian reference.

\begin{lemma}[Approximate orthonormality of Weyl basis relative to Gaussian reference]
Let $P_0=\mathcal N(0,\Sigma)$ on $\mathbb R^{2n}$, and let $\chi_\zeta(x)=e^{i\zeta^T\Omega x}$, where $\Omega$ is the symplectic form. Then
\begin{equation}
    \mu_\zeta=\exp\left(-\frac{1}{2}\zeta^T\Omega\Sigma\Omega^T\zeta\right)
\end{equation}
and
\begin{equation}
    G(\zeta,\nu)=\exp\left(-\frac{1}{2}(\nu-\zeta)^T\Omega\Sigma\Omega^T(\nu-\zeta)\right)-\overline{\mu_\zeta}\mu_\nu \ .
\end{equation}
\end{lemma}

\begin{proof}
The key point is that the Gram matrix of bosonic Weyl characters relative to a distribution $P_0$ is exactly the Fourier transform of $P_0$. Indeed,
\begin{equation}
    \langle \chi_\zeta,\chi_\nu\rangle_{P_0}=\int e^{i(\nu-\zeta)^T\Omega x}dP_0(x) \ .
\end{equation}
For a centered Gaussian, this characteristic function is the first term in the claimed expression. Centering the characters subtracts $\overline{\mu_\zeta}\mu_\nu$.
\end{proof}

The above simple fact will allow us to argue that when measurements and channels are both decomposed in the Weyl basis, the amount of information gained from a measurement is directly quantified by a Fourier overlap.

We now connect the abstract phase space to the physical system being learned. The useful picture is that $x\in\mathcal X$ is a hidden classical label specifying one possible realization of the system, while $\mathcal N_x$ describes the quantum operation associated with that realization. The family $x\mapsto\mathcal N_x$ is known; what is unknown is the probability law $P$ governing which realizations occur. For example, in displacement sensing, $x=\alpha$ is a displacement amplitude and $\mathcal N_\alpha$ is the corresponding deterministic displacement channel. This motivates the following definition.

\begin{definition}[Phase-space oracle model]
Let $(\mathcal X,\mathcal F)$ be a measurable space. A phase-space oracle model is a measurable family of quantum channels
\begin{equation}
    x\longmapsto\mathcal N_x
\end{equation}
from an input system $A$ to an output system $B$. 
\end{definition}
Through this definition, our choice of phase space parameterizes specific realizations of our system of interest. In an experiment, we have oracle access to a particular probability law supported the phase-space, which specifies the uncharacterized system itself.

\begin{definition}[Q$\Psi$ quantum oracle]
Given a phase-space oracle model $x\mapsto\mathcal N_x$ and a probability law $P$ on $(\mathcal X,\mathcal F)$, the corresponding quantum oracle is
\begin{equation}
    \mathcal N_P(\rho)
    =
    \int_{\mathcal X}\mathcal N_x(\rho)dP(x) \ .
\end{equation}
Query access to the uncharacterized physical system means query access to $\mathcal N_P$. A quantum state is included as the special case in which $A$ is one-dimensional and $\mathcal N_x(1)=\rho_x$.
\end{definition}

All ``quantumness'' is therefore baked into the definition of the phase space itself. This representation is useful because the phase-space $\mathcal X$ can be chosen to match the natural degrees of freedom of the physical problem. For single-mode displacement sensing, we take $\mathcal X=\mathbb R^2\simeq\mathbb C$ and associate $\alpha\in\mathbb C$ with
\begin{equation}
    \mathcal N_\alpha(\rho)=D(\alpha)\rho D(\alpha)^\dagger \ ,
\end{equation}
so that each displacement coordinate maps to its corresponding deterministic displacement channel. An unknown displacement distribution $P$ then produces the usual random displacement channel. A deterministic displacement is contained in the same model by taking $P=\delta_\alpha$.

For intuition we can also consider an example with qubit systems, for which one can instead use a discrete phase-space. For example, $\mathcal X=\mathbb F_2^{2n}$ naturally labels the $n$-qubit Pauli operators $P_x$, and the map
\begin{equation}
    \mathcal N_x(\rho)=P_x\rho P_x^\dagger
\end{equation}
gives the usual representation of a Pauli channel as a probability law over discrete phase-space. More general state-learning problems are obtained by treating states as channels with trivial input: one chooses $\mathcal X$ to parameterize the relevant family of states and sets $\mathcal N_x(1)=\rho_x$. For instance, in a Pauli-tomography lower bound, $x$ may label the Pauli-structured family of candidate states being distinguished, while the Pauli characters provide the natural Fourier dictionary on that family \cite{chen2021exponential, cotler2026noisy}.

We next formalize the operations available to the experimentalist. A single use of a channel differs from a single copy of a state because the experimentalist may choose which probe is supplied to the channel. More generally, if coherent quantum memory is available, the output of one query may be retained and processed coherently before a later query. Both cases are described by the same object.

\begin{definition}[Q$\Psi$ quantum experiment]
An $r$-query quantum experiment $\mathsf E$ with query access to an oracle $\mathcal{N}_P$ is a tuple $(M, \{C_j\}_{j=1}^{r-1}, \sigma_{R_0 A})$ consisting of an initial state $\sigma_{R_0A_1}$, quantum channels
\begin{equation}
    \mathcal C_j:R_{j-1}B_j \longrightarrow R_jA_{j+1},
    \qquad
    j=1,\ldots,r-1,
\end{equation}
and a final POVM $M(dy)$ on $R_{r-1}B_r$. The systems $A_j$ and $B_j$ are respectively the input and output of the $j$-th oracle query, while $R_j$ contains any quantum memory retained by the experiment. The experiment proceeds by alternately applying $\mathcal{N}_P$ to $A_j$ and the processing channel $\mathcal C_j$, followed after the $r$-th query by the POVM $M$.
\end{definition}

This definition includes classical feedforward without additional notation, since intermediate measurement outcomes may be stored in classical registers contained in $R_j$. It also includes arbitrary reference systems and entangled probes. 

In many near-term physical experiments, including the first three levels of the sensing hierarchy, coherent quantum memory interleaving quantum control is not available. These cases are $r=1$ quantum experiments, which are fully specified by the input state $\sigma_{RA}$ and a POVM $M$. If the oracle is a quantum state, $\sigma_{RA}$ is no longer a degree of freedom, so a memoryless experiment is entirely specified by a choice of POVM. 

A physical experiment therefore specifies how each realization of the uncharacterized system is converted into classical data. This conversion is the central object in the Q$\Psi$ representation.

\begin{definition}[Experimental response kernel]
Let $\mathsf E$ be an $r$-query quantum experiment. For fixed realizations $x_1,\ldots,x_r\in\mathcal X$, let $\rho_{\mathsf E}(x_1,\ldots,x_r)$ denote the final quantum state obtained by inserting $\mathcal N_{x_j}$ into the $j$-th oracle slot of $\mathsf E$. The experimental response kernel is
\begin{equation}
    K_{\mathsf E}^{(r)}(dy|x_1,\ldots,x_r)
    =
    \Tr\left[M(dy)\rho_{\mathsf E}(x_1,\ldots,x_r)\right] \ .
\end{equation}
For $r=1$, if $\mathsf E$ consists of the probe $\sigma_{RA}$ and POVM $M(dy)$, this becomes
\begin{equation}
    K_{\mathsf E}(dy|x)
    =
    \Tr\left[
        M(dy)
        (\operatorname{id}_R\otimes\mathcal N_x)(\sigma_{RA})
    \right] \ .
\end{equation}
\end{definition}

For $r=1$, the experimental response kernel has a simple interpretation: $K_{\mathsf E}(dy|x)$ is the conditional distribution of the classical outcome $Y$ given the physical realization $X=x$. Thus, if the unknown realization is drawn according to $X \sim P$, the experimentally observed distribution is simply the marginal law
\begin{equation}
    Q_{\mathsf E,P}(dy) = \int_{\mathcal X}K_{\mathsf E}(dy|x)dP(x) \ .
\end{equation}
For general $r$, the same interpretation applies to the tuple of realizations $(x_1,\ldots,x_r)$. The response kernel is therefore the object that converts the underlying physical realization into experimentally accessible classical data.

This is the basic Q$\Psi$ representation of an experiment. The physical system is encoded by its law $P$, the quantum architecture determines which kernels $K_{\mathsf E}$ can be realized, and all information available to the experimentalist is contained in the resulting classical outcome law $Q_{\mathsf E,P}$. 

The final ingredient is the quantity the experimenter wishes to learn. At the broadest level, this is a functional of the unknown physical law. The properties most useful for Q$\Psi$ admit a particularly simple representation on the same realization space.

\begin{definition}[Property and property kernel]
A property of an uncharacterized physical system is a functional $\Psi$ of its law $P$. We say that $\Psi$ admits an $m$-replica property kernel if there is a measurable function $\psi:\mathcal X^m\rightarrow\mathbb C
$ such that
\begin{equation}
    \Psi_\psi(P)
    =
    \int_{\mathcal X^m}
    \psi(x_1,\ldots,x_m)
    dP(x_1)\cdots dP(x_m) \ .
\end{equation}
For $m=1$, we simply write
\begin{equation}
    \Psi_\psi(P)=\mathbb E_{X\sim P}[\psi(X)]
\end{equation}
and call $\psi$ the property kernel.
\end{definition}
\noindent The important point is that the experiment and the learning objective are now represented on the same realization space: $K_{\mathsf E}(dy|x)$ describes how the experiment responds to $x$, while $\psi(x)$ describes how strongly the property of interest depends on $x$. This common representation is what will allow Q$\Psi$ to compare experiments and properties directly.

The above definition generalizes quantum signal learning (QSL), as introduced in Section \ref{sec:non_Fourier_QSL}, to arbitrary kinds of physical experiments through an appropriate redefinition of the phase space. 

The case $m=1$ is the most physically common setting of properties that representable as linear functionals of the quantum system; we consider some examples here for intuition. For state learning, an observable $O$ induces
\begin{equation}
    \psi_O(x)=\Tr(O\rho_x),
\end{equation}
so that
\begin{equation}
    \Psi_O(P)=\Tr(O\rho_P),
    \qquad
    \rho_P=\int\rho_xdP(x) \ .
\end{equation}
Pauli shadow tomography is obtained by taking $O$ from a collection of Pauli operators. For channel learning, a linear functional of the channel similarly induces a property kernel $x\mapsto\psi(x)$; in finite dimensions one may equivalently view these as linear functionals of the Choi operator. Full state or channel tomography corresponds to learning a sufficiently rich family of such linear properties that separates the possible physical objects.

Properties nonlinear in the averaged state or channel naturally require $m>1$. For example, if $\rho_P=\int\rho_xdP(x)$ and $O^{(m)}$ is an observable on $m$ replicas, then
\begin{equation}
    \Tr\!\left[O^{(m)}\rho_P^{\otimes m}\right]
    =
    \int
    \Tr\!\left[
        O^{(m)}
        \rho_{x_1}\otimes\cdots\otimes\rho_{x_m}
    \right]
    dP(x_1)\cdots dP(x_m) \ ,
\end{equation}
which is precisely an $m$-replica property kernel. Purity, moments of a state, and other nonlinear quantum observables fit naturally into this form.

For the semiclassical learning problems studied below, we will primarily take $r=m=1$. An allowed quantum experiment $\mathsf E$ then produces a Markov kernel $K_{\mathsf E}(dy|x)$ on the realization space, while the learning objective is specified by a property or perturbation $h(x)$ on the same space. Q$\Psi$ quantifies how much of this target feature survives the map from the physical realization $X$ to the classical measurement outcome $Y$. The information quantity which makes this comparison precise is the accessible feature information.

\begin{graydefinitionbox}
\begin{remark}
\textnormal{
        To summarize, Q$\Psi$ identifies all ingredients of a quantum experimental task with objects on the same phase space $\chi$, whose definition is tailored to the task at hand. The uncharacterized system is an oracle $\mathcal{N}_P = \int dx \ P(x) \mathcal{N}_x$, where $\mathcal{N}_x$ is a known quantum state or channel for each point $x\in \chi$, while $P$ is unknown. An experiment is specified by an initial state, a POVM, and given coherent quantum memory, additional control operations. Then, the outcomes of an experiment $\mathsf E$ are cleanly described by a classical Markov kernel $K_{\mathsf E}$, which we call the experimental response kernel. Experimental outcomes are obtained by integrating the Markov kernel against $P$, while physical properties of the oracle are obtained by integrating the \textit{observable}'s property kernel against $P$.}
    \end{remark}
\end{graydefinitionbox}

We now have an experiment kernel and a property kernel. The intuitive, but crucial step is to realize that their overlap, weighted by the density of $P$, quantifies how well the experiment learns the property. The appropriately-defined overlap quantity is the \textit{accessible feature information}.

\subsubsection{Defining the Accessible Feature Information (AFI)} \label{sec:defining_qro}

We now have a way to relate experiments, uncharacterized physical processes, and physical features through kernels in the same phase-space picture. The following object will be the quantity central to our lower bounds. For a one-query experiment $\mathsf E$, let $Q_{\mathsf E,0}=Q_{\mathsf E,P_0}$ denote its outcome distribution under the reference law:
\begin{equation}
    Q_{\mathsf E,0}(dy)=\int K_{\mathsf E}(dy|x)dP_0(x) \ .
\end{equation}
Moreover, for a property kernel $h$, we define
\begin{equation}
    A_{\mathsf E,h}(dy)=\int h(x)K_{\mathsf E}(dy|x)dP_0(x) \ .
\end{equation}
Because $K_{\mathsf E}$ is a Markov kernel, $A_{\mathsf E,h}$ is absolutely continuous with respect to $Q_{\mathsf E,0}$, so $dA_{\mathsf E,h}/dQ_{\mathsf E,0}$ is well defined on the support of $Q_{\mathsf E,0}$.

\begin{graydefinitionbox}
\begin{definition}[Accessible Feature Information] \label{def:AFI}
For any function $h\in L^2_0(P_0)$, define the accessible feature information of an experiment $\mathsf E$ about $h$ by
\begin{equation}
    \mathsf A_{\mathsf E}(h;P_0)=\int \left|\frac{dA_{\mathsf E,\bar{h}}}{dQ_{\mathsf E,0}}(y)\right|^2 dQ_{\mathsf E,0}(y) \ ,
\end{equation}
where $\bar{h} = h - \int h \ dP_0$. For an entire class of experiments $\mathcal M$, define
\begin{equation}
    \mathsf A_{\mathcal M}(h;P_0)=\sup_{\mathsf E\in\mathcal M}\mathsf A_{\mathsf E}(h;P_0) \ .
\end{equation}
If $\mathsf E$ has finite or countable outcomes, write $f_{\mathsf E,y}(x)=K_{\mathsf E}(y|x)$ and $q_{\mathsf E,0}(y)=\int f_{\mathsf E,y}(x)dP_0(x)$. Then
\begin{equation}
    \mathsf A_{\mathsf E}(h;P_0)=\sum_{y:q_{\mathsf E,0}(y)>0}\frac{\left|\int f_{\mathsf E,y}(x)h(x)dP_0(x)\right|^2}{q_{\mathsf E,0}(y)} \ .
\end{equation}
\end{definition}
\end{graydefinitionbox}

The definition has a direct statistical interpretation. Draw $X\sim P_0$ and then run the experiment conditioned on $X$, producing an outcome $Y$. For centered $h$, the Radon--Nikodym derivative appearing above is
\begin{equation}
    \frac{dA_{\mathsf E,h}}{dQ_{\mathsf E,0}}(Y)
    =
    \mathbb E[h(X)\mid Y],
\end{equation}
and therefore
\begin{equation}
    \mathsf A_{\mathsf E}(h;P_0)
    =
    \mathbb E\left[\left|\mathbb E[h(X)\mid Y]\right|^2\right].
\end{equation}
Since $\mathbb E[h(X)\mid Y]$ is the optimal mean-square predictor of $h(X)$ from the measurement outcome, the AFI quantifies how much of the feature $h(X)$ remains accessible after the physical realization $X$ has been converted into the classical outcome $Y$. In particular, $\mathsf A_{\mathsf E}(h;P_0)=0$ when the outcome contains no information about $h(X)$, while the contraction property of conditional expectation gives
\begin{equation}
    \mathsf A_{\mathsf E}(h;P_0) \leq \|h\|_{L^2(P_0)}^2.
\end{equation}
In the next subsection, we show that this statistical notion of accessibility directly controls the query complexity of local learning problems. Tight bounds on the AFI therefore translate directly into tight bounds on the required resources. Fourier analysis provides a useful tool for obtaining such bounds, since the overlap between response functions $f_{E,y}$ and a property kernel $h$ is often easiest to control in a common Fourier basis.

To this end, we introduce a concept relevant to Fourier analysis of the AFI, which we use later in our control-depth separation, Theorem \ref{thm:interleaved_control_lb_formal}.

\begin{definition}[Valid dictionary and sparse response class]
Let $\mathcal{M}$ be a class of finite or countable-outcome experiments associated with probability space $(\mathcal{X},\mathcal F,P_0)$, and let $\mathcal{D}$ be a Fourier character family. Then $\mathcal{D}$ is a valid dictionary for $\mathcal{M}$ if for every experiment $M\in \mathcal{M}$ and every outcome $y$ with $q_{M,0}(y)>0$, the centered response function
\begin{equation}
    f_{M,y}-q_{M,0}(y)
\end{equation}
lies in the closed linear span of the centered characters $\{\widetilde\chi_\omega\}_{\omega\in \Omega}$. We say that $\mathcal M$ is $L$-sparse in the Fourier dictionary $\mathcal D$ under $P_0$ if, for every $M\in\mathcal M$, there exists $S_M\subset\Omega$ with $|S_M|\le L$ such that every centered response function satisfies
\begin{equation}
    f_{M,y}-q_{M,0}(y)\in\operatorname{span}\{\widetilde\chi_\omega:\omega\in S_M\}.
\end{equation}
\end{definition}

\begin{definition}[Captured mass]
For a centered function $h\in L^2_0(P_0)$, define
\begin{equation}
    \mathsf C_L(h;\mathcal D,P_0)
    =
    \sup_{\substack{S\subset\Omega\\ |S|\le L}}
    \left\|\Pi_S h\right\|_{L^2(P_0)}^2,
\end{equation}
where $\Pi_S$ is the orthogonal projector onto $\operatorname{span}\{\widetilde\chi_\omega:\omega\in S\}$.
\end{definition}

These definitions have a simple geometric interpretation. An $L$-sparse experimental class can access at most $L$ Fourier directions at a time. The captured mass $\mathsf C_L(h;\mathcal D,P_0)$ measures the largest fraction of the $L^2(P_0)$ weight of the target $h$ that can lie in any such $L$-dimensional Fourier subspace. Since only the component of $h$ lying in the response subspace of a measurement can contribute to the AFI, $\mathsf C_L$ provides a direct upper bound on the information accessible to an $L$-sparse architecture.

For an exactly orthonormal dictionary, this becomes especially transparent. If
\begin{equation}
h = \sum_{\omega\in S_h}c_\omega\chi_\omega,
\end{equation}
then $\mathsf C_L$ is simply the squared Fourier weight contained in the best choice of at most $L$ frequencies:
\begin{equation}
\mathsf C_L(h;\mathcal D,P_0) = \sup_{\substack{S\subseteq S_h\\ |S|\le L}} \sum_{\omega \in S} |c_\omega|^2 \, .
\end{equation}
Thus, if the target is spread approximately uniformly over $H\gg L$ Fourier components, an $L$-sparse measurement can access only an $O(L/H)$ fraction of its total Fourier weight. In particular, $\mathsf C_L$ is bounded by $L$ times the squared magnitude of the largest Fourier coefficient of $h$.

These intuitions form the core of Q$\Psi$: the expressivity of a constrained experimental architecture appears via restrictions on the response kernels that it can realize, and these restrictions often appear naturally in an appropriate Fourier domain.

\subsubsection{Information-theoretic lower bound tools}
\label{sec:info-lower-bounds-sensing}

We now understand how to represent quantum-mechanical experiments in the phase-space picture, and how the power of an experiment is controlled by its AFI. To connect these definitions to resource complexity and prove tight lower bounds, we draw on techniques from statistical learning theory.

To establish a rigorous query-complexity lower bound, we must be able to account for any arbitrary, possibly adaptive, sequence of experiments chosen from a family. The first key step in obtaining such bounds is a reduction of the full learning task to a hypothesis test, using the following observation. We state this observation for sensing for maximum operational clarity, but it trivially applies to any property-estimation task.

\begin{fact}[Learning reduces to hypothesis testing] \label{fact:learning_to_testing_reduction} Consider two families of signals with distributions $P_0^{(k)}, P_1^{(k)}$, parameterized by integer $k > 0$, and a property of interest $\psi_k(\alpha)$, such that for all $k$, 
\begin{equation}
    \left|\mathbb{E}_{\alpha \sim P_0^{(k)}}\left[\psi_k(\alpha)\right]- \mathbb{E}_{\alpha \sim P_1^{(k)}}\left[\psi_k(\alpha)\right]\right| \geq 2\epsilon ,
\end{equation}
where $\epsilon$ is a positive constant independent of $k$. Suppose there is a learning strategy $\mathcal{L}$ which, given access to any valid signal $P$, can estimate $\mathbb E_{\alpha \sim P}[\psi_k(\alpha])$ to accuracy $\epsilon$ with success probability at least $2/3$, and that $\mathcal{L}$ uses at most $f(k)>0$ samples to do so. Then, it is possible to determine with success probability at least $2/3$ which signal was provided with $\leq f(k)$ queries.
\end{fact}
This follows because a hypothesis tester can simply apply $\mathcal{L}$ to the provided signal, and due to the guarantee that the two distributions differ in the property $\psi_k$ by at least $2\epsilon$, conclude based on the estimated value which signal was provided. By contrapositive, it follows that if one proves that any learning protocol in a family of experiments requires $\geq \Omega(f(k))$ queries to solve such a hypothesis testing problem, then any such protocol also requires $\geq \Omega(f(k))$ queries to learn the property $\psi_k$ to constant precision. When $f(k) = \exp(\Omega(k))$, this implies that exponentially many samples are required for learning. 

Having established that our separations will rely on lower-bounding the sample complexity of a hypothesis-testing task, we now record the mathematical tools we use to do so. First, we formally understand the most general description of a quantum experiment. We again provide the statement in the single-mode sensing context, but the extension to other experimental settings is immediate.

\begin{definition}[Classical transcript for an adaptive hypothesis testing protocol]\label{def:sensing_tree}
Consider any quantum sensing protocol for hypothesis testing that makes $N$ total queries to a single-mode signal, which is promised to either be $\mathcal{E}_{P_0}$ or $\mathcal{E}_{P_1}$. In the first round, the learner can prepare a probe state (possibly joint across modes and qubits), expose one mode to the signal, and perform a POVM. The outcome $q_1$ is sampled from a distribution $Q_1^{(a)}$ over all possible outcomes from this POVM, where $a = 0$ or $1$ based on whether the true signal distribution is $P_0$ or $P_1$ respectively. In each round $j= 2,...,N$ thereafter, the learner prepares a state and chooses a POVM given access to all outcomes $q_1,...,q_{j-1}$. Finally, based on the classical transcript $q_1,q_2,...,q_N$, the learner outputs either $P_0$ or $P_1$. We say that the entire transcript is sampled from the distribution
\begin{equation}
    Q^{(a)}(\Vec{q}) \coloneqq Q^{(a)}(q_1,q_2,...,q_N) = Q^{(a)}_1(q_1)Q^{(a)}_2(q_2|q_1)Q^{(a)}_3(q_3|q_2,q_1) \cdots Q^{(a)}_N(q_N|q_{N-1},...,q_1)
\end{equation}
\end{definition}

Particular classes of sensing protocols in the hierarchy refine this definition by constraining the probe states and allowed measurements. In other experimental settings, the hypothesis test may provide query access to a quantum state, multimode or multiqubit channel, Hamiltonian evolution, etc. The allowable family of learning protocols can be tuned accordingly. The main workhorse in obtaining query-complexity lower bounds is then Le Cam's two-point method. Before we state the theorem, we require definitions of some standard divergences between probability distributions.

\begin{definition}[Total variation distance]
    Let $Q^{(0)}$ and $Q^{(1)}$ be two probability laws on a common measurable
    space $(\Omega,\mathcal{F})$. Their total variation distance is
    \begin{equation}
        d_{\rm TV}\!\left(Q^{(0)},Q^{(1)}\right)
        \coloneqq
        \sup_{A\in\mathcal{F}}
        \left|Q^{(0)}(A)-Q^{(1)}(A)\right| .
    \end{equation}
    If the transcript of an $N$-round sensing protocol is
    $\Vec{q}=(q_1,\ldots,q_N)\in\mathbb{C}^N$, as is the case for generaldyne measurements, then
    \begin{equation}
        d_{\rm TV}\!\left(Q^{(0)},Q^{(1)}\right)
        =
        \frac{1}{2}
        \int_{\mathbb{C}^N}
        \left|
            Q^{(0)}(\Vec{q})-Q^{(1)}(\Vec{q})
        \right|
        \,d^{2N}\Vec{q} \ .
    \end{equation}
    For homodyne or any other single-quadrature readout, one simply replaces $\mathbb{C}^N$ with $\mathbb{R}^N$.
\end{definition}
The TV distance has a useful operational meaning in the context of quantum state hypothesis testing.
\begin{lemma}[Total Variation trace distance bound]\label{lemma:state_testing_TV_trace_distance_bound}
    Consider two hypothesis states $\rho_0, \rho_1$, and any arbitrary POVM $\{F_s\}$ which induces measurement transcripts $q_0(s) = \Tr(F_s\rho_0)$ and $q_1 =  \Tr(F_s\rho_1)$. Then
    \begin{equation}
        d_{\rm TV}(q_0, q_1) \leq \frac{1}{2}\|\rho_0 - \rho_1\| \ .
    \end{equation}
\end{lemma}
\noindent Closely related is the Kullback-Leibler (KL) divergence, which we define only on the relevant domains:
\begin{definition}[Kullback--Leibler divergence]
    Let $Q^{(0)}$ and $Q^{(1)}$ be two probability laws on $\mathbb{C}^N$. If the transcript of an $N$-round sensing protocol is
    $\Vec{q}=(q_1,\ldots,q_N)\in\mathbb{C}^N$ then
    \begin{equation}
        D_{\rm KL}\!\left(Q^{(0)}\middle\|Q^{(1)}\right)
        =
        \int_{\mathbb{C}^N}
        Q^{(0)}(\Vec{q})
        \log\!\left(
            \frac{Q^{(0)}(\Vec{q})}{Q^{(1)}(\Vec{q})}
        \right)
        \,d^{2N}\Vec{q} \ .
    \end{equation}
    For homodyne or any other single-quadrature readout, one simply replaces
    $\mathbb{C}^N$ with $\mathbb{R}^N$.
\end{definition}
\noindent These metrics are related by the following well-known inequality:
\begin{lemma}[Pinsker's inequality]\label{lemma:pinsker}
    Let $Q^{(0)}$ and $Q^{(1)}$ be two probability laws on a common measurable
    space. Then
    \begin{equation}
        d_{\rm TV}\!\left(Q^{(0)},Q^{(1)}\right)
        \leq
        \sqrt{
            \frac{1}{2}
            D_{\rm KL}\!\left(Q^{(0)}\middle\|Q^{(1)}\right)
        } \ .
    \end{equation}
\end{lemma}
Recall that our lower bounds must account for arbitrary classical adaptivity based on the collected transcript. Total variation has an important property that we will leverage to decouple the transcript distributions, so that we only need to consider bounding the TVD from a single experiment.
\begin{lemma}[Adaptive TV hybrid bound]\label{lemma:adaptive_tv_hybrid}
    Let $Q^{(0)}$ and $Q^{(1)}$ be transcript distributions on $\mathbb{C}^N$. If for every $j$ and every previous transcript $\Vec{q}_{<j}$,
    \begin{equation}
        d_{\rm TV}\left(Q^{(0)}_j(\cdot \mid \Vec{q}_{<j}), Q^{(1)}_j(\cdot \mid \Vec{q}_{<j})\right) \leq \epsilon \ ,
    \end{equation}
    then
    \begin{equation}
        d_{\rm TV}\left(Q^{(0)}, Q^{(1)}\right) \leq N\epsilon \ .
    \end{equation}
\end{lemma}

\begin{proof}
    For each $j\in\{0,\ldots,N\}$, define the hybrid transcript distribution
    \begin{equation}
        R_j(\Vec{q}) =
        \left(\prod_{\ell=1}^{j} Q^{(0)}_\ell(q_\ell \mid \Vec{q}_{<\ell})\right)
        \left(\prod_{\ell=j+1}^{N} Q^{(1)}_\ell(q_\ell \mid \Vec{q}_{<\ell})\right) \ .
    \end{equation}
    Thus $R_0=Q^{(1)}$ and $R_N=Q^{(0)}$. By the triangle inequality,
    \begin{equation}
        d_{\rm TV}\left(Q^{(0)}, Q^{(1)}\right) \leq \sum_{j=1}^{N} d_{\rm TV}\left(R_j, R_{j-1}\right) \ .
    \end{equation}
    It remains to bound a single hybrid step. The distributions $R_j$ and $R_{j-1}$ agree on the first $j-1$ rounds and differ only in the conditional law used for the $j$-th outcome. Therefore,
    \begin{align}
        d_{\rm TV}\left(R_j, R_{j-1}\right)
        &= \frac{1}{2}\int R_{j-1}(\Vec{q}_{<j})\left(\left|Q^{(0)}_j(q_j \mid \Vec{q}_{<j})-Q^{(1)}_j(q_j \mid \Vec{q}_{<j})\right|\right)d^2q_j\,d^{2(j-1)}\Vec{q}_{<j} \\
        &= \int R_{j-1}(\Vec{q}_{<j})d_{\rm TV}\left(Q^{(0)}_j(\cdot \mid \Vec{q}_{<j}), Q^{(1)}_j(\cdot \mid \Vec{q}_{<j})\right)d^{2(j-1)}\Vec{q}_{<j} \\
        &\leq \epsilon \ .
    \end{align}
    Substituting this into the telescoping bound gives
    \begin{equation}
        d_{\rm TV}\!\left(Q^{(0)}, Q^{(1)}\right) \leq N\epsilon \ ,
    \end{equation}
    as claimed.
\end{proof}

\noindent The KL divergence satisfies the same chain-rule inequality, which we state without a nearly identical proof.

\begin{lemma}[Adaptive KL hybrid bound]\label{lemma:adaptive_kl_hybrid}
    Let $Q^{(0)}$ and $Q^{(1)}$ be transcript distributions on $\mathbb{C}^N$. If for every $j$ and every previous transcript $\Vec{q}_{<j}$,
    \begin{equation}
        D_{\rm KL}\!\left(Q^{(0)}_j(\cdot \mid \Vec{q}_{<j}), Q^{(1)}_j(\cdot \mid \Vec{q}_{<j})\right) \leq \epsilon \ ,
    \end{equation}
    then
    \begin{equation}
        D_{\rm KL}\!\left(Q^{(0)}, Q^{(1)}\right) \leq N\epsilon \ .
    \end{equation}
\end{lemma}

With these results, we are prepared to state Le Cam's two-point method and understand how it lets us control query complexity in hypothesis testing.

\begin{lemma}[Le Cam's Two-Point Method]\label{lemma:le_cam}
Consider a hypothesis-testing formulation of a sensing task with signals induced by distribution $P_0$ or $P_1$, and an adaptive hypothesis testing protocol whose transcript is distributed according to $Q^{(0)}$ or $Q^{(1)}$. Then, the probability that the protocol selects the correct hypothesis is upper bounded by
\begin{equation}
    \frac{1}{2} + \frac{1}{2}d_{\rm TV}\left(Q^{(0)}, Q^{(1)}\right)
\end{equation}
\end{lemma}
\noindent Combining the TVD hybrid bound with Le Cam's method, we obtain the core lemma used in many information-theoretic learning lower bounds.
\begin{lemma}[Single-experiment TV lower bound]\label{lemma:single_experiment_tv_lower_bound}
    Consider a hypothesis-testing formulation of a sensing task with signals induced by $P_0$ or $P_1$. Suppose that for any allowed single-query experiment, and for every previous transcript $\Vec{q}_{<j}$, the resulting outcome distributions satisfy
    \begin{equation}
        d_{\rm TV}\!\left(Q^{(0)}_j(\cdot \mid \Vec{q}_{<j}),Q^{(1)}_j(\cdot \mid \Vec{q}_{<j})\right) \leq \Delta \ .
    \end{equation}
    Then any adaptive sensing protocol which distinguishes $P_0$ from $P_1$ with success probability at least $1/2+\epsilon$ must use at least
    \begin{equation}
        N \geq \frac{2\epsilon}{\Delta}
    \end{equation}
    queries to the signal. In particular, any protocol which succeeds with constant success bias $\epsilon$ requires $N \geq \Omega(1/\Delta)$ queries.
\end{lemma}

\begin{proof}
    Let $Q^{(0)}$ and $Q^{(1)}$ be the full transcript distributions induced by an arbitrary adaptive protocol using $N$ queries. By Lemma \ref{lemma:adaptive_tv_hybrid},
    \begin{equation}
        d_{\rm TV}\!\left(Q^{(0)},Q^{(1)}\right) \leq N\Delta \ .
    \end{equation}
    Lemma \ref{lemma:le_cam} then implies that the success probability of the protocol is at most
    \begin{equation}
        \frac{1}{2}+\frac{1}{2}\,d_{\rm TV}\!\left(Q^{(0)},Q^{(1)}\right) \leq \frac{1}{2}+\frac{N\Delta}{2} \ .
    \end{equation}
    Therefore, if the protocol succeeds with probability at least $1/2+\epsilon$, it must hold that
    \begin{equation}
        \frac{1}{2}+\epsilon \leq \frac{1}{2}+\frac{N\Delta}{2} \ .
    \end{equation}
    Rearranging gives
    \begin{equation}
        N \geq \frac{2\epsilon}{\Delta} \ .
    \end{equation}
    Setting $\epsilon=1/6$ gives the stated lower bound for success probability at least $2/3$.
\end{proof}

\noindent Lemma \ref{lemma:single_experiment_tv_lower_bound} is the core result that we will use to interface the AFI with information-theoretic lower bound techniques.

\subsection{Q$\Psi$ lower bounds}\label{sec:semiclassical_measurement_lemma_simple}

We now show that the AFI directly controls the difficulty of local learning. The mechanism is simple. Starting from a reference law $P_0$, we perturb it in a direction $h$ to obtain two nearby hypotheses $P_\pm=(1\pm\eta h)P_0$. For any fixed experiment, the corresponding change in its outcome distribution is governed by the conditional expectation $\mathbb E[h(X)\mid Y]$, whose squared $L^2$ norm is exactly the AFI. A small AFI therefore implies that each query produces only a small statistical separation between the two hypotheses. The chain rule for adaptive protocols then accumulates this separation over queries, and Le Cam's method converts it into a query-complexity lower bound. The following lemma formalizes this argument.

\begin{graydefinitionbox}
\begin{lemma}[Semiclassical measurement lemma]
\label{lemma:semiclassical_quantum_measurement}
Consider a probability space $(\mathcal{X}, \mathcal{F}, P_0)$. Let $h:\mathcal X\to\mathbb R$ satisfy $\int h(x)dP_0(x)=0$ and $\|h\|_\infty\le 1$. For $0<\eta< 1$, define
\begin{equation}
    P_\pm(dx)=\left(1\pm \eta h(x)\right)P_0(dx).
\end{equation}
Let $\mathcal M$ be a class of measurements with response kernels as above, and suppose
\begin{equation}
    \mathsf A_{\mathcal M}(h;P_0)\le \Lambda \ .
\end{equation}
Then any adaptive protocol using measurements from $\mathcal M$ requires
\begin{equation}
    N\ge \Omega\left(\eta^{-2}\Lambda^{-1}\right)
\end{equation}
queries to distinguish $P_+$ from $P_-$ with success probability at least $2/3$, with the convention that $\Lambda^{-1}=\infty$ if $\Lambda=0$.
\end{lemma}
\end{graydefinitionbox}

\begin{proof}
Fix a single experiment $M\in\mathcal M$. Let $Q_0$ denote the outcome distribution under $P_0$, and let $A_h$ denote the signed outcome measure
\begin{equation}
    A_h(dy)=\int h(x)K_M(dy|x)dP_0(x) \ .
\end{equation}
The outcome distributions under $P_+$ and $P_-$ are
\begin{equation}
    Q_\pm(dy)=Q_0(dy)\pm \eta A_h(dy).
\end{equation}
Let $R_h=dA_h/dQ_0$. Equivalently, $R_h(y)=\mathbb E[h(X)|Y=y]$ under $X\sim P_0$ and the measurement response $K_M(dy|X)$. Since $\|h\|_\infty\le 1$, we have $|R_h(y)|\le 1$ for $Q_0$-almost every $y$. Thus
\begin{equation}
    Q_\pm(dy)=\left(1\pm \eta R_h(y)\right)Q_0(dy)
\end{equation}
and $|\eta R_h(y)|\le 1/2$. Moreover,
\begin{equation}
    \int R_h(y)dQ_0(y)=\int h(x)dP_0(x)=0.
\end{equation}
Using the elementary bound $(1+u)\log (1+u)-u \leq u^2$ for all $u\in [-1, 1]$, we have
\begin{align}
    D_{\rm KL}(Q_+\|Q_0) &\leq \int \left(1+\eta R_h(y)\right)\log\left(1+\eta R_h(y)\right)dQ_0(y) \\
    &= \int \left[\left(1+\eta R_h(y)\right)\log\left(1+\eta R_h(y)\right)-\eta R_h\right]dQ_0(y) \\ 
    &\le \eta^2\int |R_h(y)|^2dQ_0(y) \\
    &= \eta^2\mathsf A_M(h;P_0) \\
    &\le \eta^2\Lambda \ ,
\end{align}
Now consider an adaptive $N$-query protocol. At each step of the transcript, the learner may choose a different measurement from $\mathcal M$, but the same one-query KL bound applies uniformly. As such, applying the KL chain rule for adaptive protocols gives
\begin{equation}
    D_{\rm KL}(Q_+^{(N)}\|Q_0^{(N)})\le N\eta^2\Lambda,
\end{equation}
where $Q_+^{(N)}$ is the transcript distributions under $P_+$. The same inequality holds when $Q_+$ is replaced by $Q_-$ via the above argument. If the protocol distinguishes the two hypotheses $P_\pm$ with success probability at least $2/3$, then Lemma \ref{lemma:le_cam} implies
\begin{equation}
    d_{\rm TV}(Q_+^{(N)},Q_-^{(N)})\ge \frac{1}{3},
\end{equation}
where we move from KL divergence to TVD using Pinsker's inequality (Lemma \ref{lemma:pinsker}) and by applying the triangle inequality. Applying Lemma \ref{lemma:pinsker} in the opposite direction then gives $D_{\rm KL}(Q_+^{(N)}\|Q_-^{(N)})\ge c$ for a universal constant $c>0$. Hence
\begin{equation}
    N\ge \Omega\left(\eta^{-2}\Lambda^{-1}\right).
\end{equation}
\end{proof}
The following corollary is the form of the lemma we will use when a constrained measurement class has a sparse Fourier response.

\begin{corollary}[Sparse-response semiclassical measurement lemma]
\label{cor:sparse_semiclassical_quantum_measurement}
Consider the setting of Lemma \ref{lemma:semiclassical_quantum_measurement}. Let $\mathcal D$ be a valid dictionary for $\mathcal M$, and suppose $\mathcal M$ is $L$-sparse in $\mathcal D$ under $P_0$. Then
\begin{equation}
    \mathsf A_{\mathcal M}(h;P_0)\le \mathsf C_L(h;\mathcal D,P_0).
\end{equation}
Consequently, any adaptive protocol using measurements from $\mathcal M$ requires
\begin{equation}
    N\ge \Omega\left(\eta^{-2}\mathsf C_L(h;\mathcal D,P_0)^{-1}\right)
\end{equation}
queries to distinguish $P_+$ from $P_-$ with success probability at least $2/3$.

If $\mathcal D$ is orthonormal and $h=\sum_{\omega\in S_h}c_\omega\chi_\omega$, with $|S_h|=H$ and $|c_\omega|^2\le M/H$ for every $\omega\in S_h$, then
\begin{equation}
    N\ge \Omega\left(\eta^{-2}\frac{H}{ML}\right).
\end{equation}
\end{corollary}

\begin{proof}
Fix $M\in\mathcal M$. By $L$-sparsity, there is a set $S_M\subset\Omega$ with $|S_M|\le L$ such that every centered response function lies in
\begin{equation}
    V_M=\operatorname{span}\{\widetilde\chi_\omega:\omega\in S_M\}.
\end{equation}
Let $\Pi_M$ denote the orthogonal projector onto $V_M$. For a finite or countable outcome measurement, write $q_{M,0}(y)=\int f_{M,y}(x)dP_0(x)$. Since $h$ is centered,
\begin{equation}
    \int f_{M,y}(x)h(x)dP_0(x)=\left\langle f_{M,y}-q_{M,0}(y),h\right\rangle_{P_0}.
\end{equation}
Since $f_{M,y}-q_{M,0}(y)\in V_M$, we can move the projection onto $h$:
\begin{equation}
    \left\langle f_{M,y}-q_{M,0}(y),h\right\rangle_{P_0}
    =
    \left\langle f_{M,y}-q_{M,0}(y),\Pi_M h\right\rangle_{P_0}.
\end{equation}
Therefore
\begin{equation}
    \mathsf A_M(h;P_0)=\mathsf A_M(\Pi_M h;P_0).
\end{equation}
By the contraction property of accessible feature information,
\begin{equation}
    \mathsf A_M(\Pi_M h;P_0)\le \|\Pi_M h\|_{L^2(P_0)}^2.
\end{equation}
Thus
\begin{equation}
    \mathsf A_M(h;P_0)\le \mathsf C_L(h;\mathcal D,P_0).
\end{equation}
Taking the supremum over $M\in\mathcal M$ gives $\mathsf A_{\mathcal M}(h;P_0)\le \mathsf C_L(h;\mathcal D,P_0)$, and Lemma \ref{lemma:semiclassical_quantum_measurement} gives the query lower bound. The exactly orthonormal case is an illustrative example. If $\mathcal D$ is orthonormal and $h=\sum_{\omega\in S_h}c_\omega\chi_\omega$, then for any set $S$,
\begin{equation}
    \left\|\Pi_S h\right\|_{L^2(P_0)}^2=\sum_{\omega\in S\cap S_h}|c_\omega|^2.
\end{equation}
Therefore
\begin{equation}
    \mathsf C_L(h;\mathcal D,P_0)=\max_{\substack{S\subseteq S_h\\ |S|\le L}}\sum_{\omega\in S}|c_\omega|^2.
\end{equation}
If $|c_\omega|^2\le M/H$ for all $\omega\in S_h$ and $H=|S_h|$, then
\begin{equation}
    \mathsf C_L(h;\mathcal D,P_0)\le \frac{ML}{H},
\end{equation}
which gives the final lower bound.
\end{proof}
By Fact \ref{fact:learning_to_testing_reduction}, the hypothesis-testing bound produced by the semiclassical measurement lemma applies directly to any protocol for learning the value of $h$ to absolute precision $\eta$. We therefore see that through Q$\Psi$, a bound on the AFI directly yields a lower bound on query complexity. Next, we show that this lower bound is in fact saturable: the AFI can also certify an optimal hypothesis-testing protocol. 

\subsection{Q$\Psi$ hypothesis testing guarantee}

The lower bound above is tight because the same conditional expectation that defines the AFI also provides the optimal statistic for detecting the perturbation. For a fixed experiment $M$, define
\begin{equation}
    R_{M,h}(y) = \frac{dA_{M,h}}{dQ_{M,0}}(y) = \mathbb E[h(X)\mid Y=y].
\end{equation}
Under the perturbed law $P_\eta=(1+\eta h)P_0$, the mean of $R_{M,h}(Y)$ shifts by exactly $\eta\mathsf A_M(h;P_0)$. Repeatedly measuring $M$ and averaging this statistic therefore produces an estimator whose sample complexity is proportional to the inverse AFI. We now formalize this matching upper bound.

\begin{lemma}[Variational form of accessible feature information]\label{lemma:learnable_fourier_variational}
Let $M$ be an experiment with response kernel $K_M(dy|x)$, and let $h\in L^2(P_0)$. Then
\begin{equation}
    \mathsf A_M(h;P_0)
    =
    \sup_{\substack{g:\mathcal Y_M\to\mathbb C\\ \mathbb E_{Y\sim Q_{M,0}}[|g(Y)|^2]\le 1}}
    \left|
    \int h(x)\left(\int g(y)K_M(dy|x)\right)dP_0(x)
    \right|^2 \ .
\end{equation}
If $\mathsf A_M(h;P_0)>0$, the supremum is achieved by $g(y)=R_{M,h}(y)/\sqrt{\mathsf A_M(h;P_0)}$.
\end{lemma}

\begin{proof}
By definition of $A_{M,h}$,
\begin{equation}
    \int h(x)\left(\int g(y)K_M(dy|x)\right)dP_0(x)=\int g(y)dA_{M,h}(y) \ .
\end{equation}
Since $dA_{M,h}=R_{M,h}dQ_{M,0}$, the right-hand side becomes
\begin{equation}
    \int g(y)R_{M,h}(y)dQ_{M,0}(y) \ .
\end{equation}
By Cauchy--Schwarz,
\begin{equation}
    \left|\int g(y)R_{M,h}(y)dQ_{M,0}(y)\right|^2
    \le
    \mathbb E_{Y\sim Q_{M,0}}[|g(Y)|^2]\mathbb E_{Y\sim Q_{M,0}}[|R_{M,h}(Y)|^2] \ .
\end{equation}
The first factor is at most one by assumption, while the second factor is exactly $\mathsf A_M(h;P_0)$. This proves the upper bound. If $\mathsf A_M(h;P_0)>0$, choosing $g=R_{M,h}/\sqrt{\mathsf A_M(h;P_0)}$ saturates Cauchy--Schwarz.
\end{proof}

This variational form tells us that the AFI is exactly the maximum squared overlap with $h$ that can be extracted from the measurement outcome using any classical postprocessing of that outcome. We now convert this operational statement into an explicit sample-complexity guarantee.

\begin{graydefinitionbox}
\begin{theorem}[Q$\Psi$ sample complexity guarantee]\label{thm:qes_matched_score_guarantee}
Consider a probability space $(\mathcal X,\mathcal F,P_0)$ and let $h:\mathcal X\to\mathbb R$ satisfy $\int h(x)dP_0(x)=0$ and $\|h\|_\infty\le 1$. For $|\eta|\le 1$, define
\begin{equation}
    P_\eta(dx)=\left(1+\eta h(x)\right)P_0(dx) \ .
\end{equation}
Fix an experiment $M$ such that $\mathsf A_M(h;P_0)>0$. Run $M$ independently $N$ times, obtaining outcomes $Y_1,\ldots,Y_N$, and define
\begin{equation}
    \widehat\eta
    =
    \frac{1}{N\mathsf A_M(h;P_0)}
    \sum_{j=1}^N R_{M,h}(Y_j) \ .
\end{equation}
Then $\widehat\eta$ is an unbiased estimator of $\eta$. Moreover, for every $0<\epsilon\le 1$ and $0<\delta<1$, 
\begin{equation}
    N=
    O\left(\frac{1}{\epsilon^2\mathsf A_M(h;P_0)}\log\!\left(\frac{1}{\delta}\right)\right)
\end{equation}
samples suffice to learn $\eta$ to within absolute error $\epsilon$ with probability at least $1-\delta$, and thereby distinguish the hypotheses $P_\eta$ and $P_{-\eta}$ with success probability at least $1-\delta$.
\end{theorem}
\end{graydefinitionbox}

\begin{proof}
Let $Q_\eta$ denote the outcome distribution obtained by running experiment $M$ when the quantum oracle is distributed according to $P_\eta$. By the definition of $R_{M,h}$,
\begin{equation}
    Q_\eta(dy)=\left(1+\eta R_{M,h}(y)\right)Q_{M,0}(dy) \ .
\end{equation}
Since $h$ is centered, $\int R_{M,h}(y)dQ_{M,0}(y)=0$. Therefore
\begin{align}
    \mathbb E_{Y\sim Q_\eta}[R_{M,h}(Y)]
    &=
    \int R_{M,h}(y)\left(1+\eta R_{M,h}(y)\right)dQ_{M,0}(y) \\
    &=
    \eta\,\mathsf A_M(h;P_0) \ .
\end{align}
This proves that $\widehat\eta$ is unbiased. Next we obtain the sample complexity guarantee. Since $R_{M,h}(y)=\mathbb E[h(X)|Y=y]$ and $\|h\|_\infty\le 1$, we have $|R_{M,h}(y)|\le 1$ for $Q_{M,0}$-almost every $y$. Furthermore,
\begin{align}
    \mathbb E_{Y\sim Q_\eta}[R_{M,h}(Y)^2]
    &=
    \int R_{M,h}(y)^2\left(1+\eta R_{M,h}(y)\right)dQ_{M,0}(y) \\
    &\le
    (1+|\eta|)\mathsf A_M(h;P_0) \\
    &\le
    \frac{3}{2}\mathsf A_M(h;P_0) \ .
\end{align}
Thus each summand has variance at most $\frac{3}{2}\mathsf A_M(h;P_0)$ and is bounded in absolute value by one. Bernstein's inequality gives
\begin{equation}
    \Pr\left[
    \left|
    \frac{1}{N}\sum_{j=1}^N R_{M,h}(Y_j)-\eta\mathsf A_M(h;P_0)
    \right|
    >
    \epsilon\,\mathsf A_M(h;P_0)
    \right]
    \le
    2\exp\left(-cN\epsilon^2\mathsf A_M(h;P_0)\right)
\end{equation}
for a universal constant $c>0$, where we used $0<\epsilon\le 1$ and $\mathsf A_M(h;P_0)\le \|h\|_{L^2(P_0)}^2\le 1$. Rearranging proves the stated sample complexity. The hypothesis testing guarantee follows estimating $\eta$ and deciding between the two hypotheses by its sign.
\end{proof}

Theorem \ref{thm:qes_matched_score_guarantee} shows that the accessible feature information appearing in the lower-bound lemma is exactly achievable. This is intuitive: the AFI exactly quantifies the maximum phase-space overlap achievable between a property of interest and a particular experiment. By running the experiment which maximizes the AFI within a class of interest, one saturates the lower bound. 

\subsection{AFI as certificate of advantage and global learning}
Putting together Theorems \ref{lemma:semiclassical_quantum_measurement} and \ref{thm:qes_matched_score_guarantee} leads us to the following fact.

\begin{graydefinitionbox}
\begin{corollary} \label{cor:AFI_tight}
    Let $\mathcal{M}$ be any class of quantum learning protocols and consider a function $h$ on a probability space with some measure $P_0$ as in Theorem \ref{lemma:semiclassical_quantum_measurement}. Then,
    \begin{equation}
        N = \Theta\!\left(\frac{1}{\mathsf A_{\mathcal M}(h; P_0)\epsilon^2}\right)
    \end{equation}
    samples from $P = P_\epsilon$ or $P_{-\epsilon}$ as in Theorem \ref{thm:qes_matched_score_guarantee} are necessary and sufficient to distinguish the hypotheses with constant success probability.  
\end{corollary}    
\end{graydefinitionbox}

There are several important points of interpretation. First, because of Corollary \ref{cor:AFI_tight}, the AFI functions operationally similarly to the QFI in estimation theory, despite having a different formulation and set of use cases. In particular, the AFI is an information quantity which directly controls the sample complexity of a learning protocol via our semiclassical measurement lemmas, in the same way that the QFI controls sensing time in metrology via the Cramer-Rao bounds. Like the QFI, the AFI gives a \textit{saturable} lower bound. This suggests an operational design principle: upon fixing a reference distribution $P_0$ and an observable perturbation around it $h$, one can design a provably optimal learning procedure by optimizing the AFI; this has been the subject of extensive work in the metrological setting, where many semidefinite programming-based approaches for maximizing the QFI have been studied across applications. Because the AFI concerns only real-valued functions, it may even lead to more classically tractable methods for quantum experimental design; these ideas are developed further in Section \ref{sec:open_questions}. The AFI has yet another operational similarity to the QFI: it is a quantity that controls \textit{local estimation problems}. In particular, it controls estimation of perturbations around a known reference distribution $P_0$, which is precisely the setting considered in quantum learning lower bounds. 

However, the AFI has several important characteristics that greatly differ from the QFI and make it a more flexible tool for learning-theoretic analysis. On the one hand, AFI lower bounds are agnostic to estimator bias, an important blindspot for QFI-based lower bounds \cite{transversal_sensing}. Moreover, the QFI can fail to produce meaningful results if the measurement response is not a continuous function of the underlying parameter, so discrete learning strategies can have arbitrarily small QFI despite succeeding with few samples or sensing time. The AFI does not face this issue and will succeed with arbitrary quantum POVMs and classical postprocessing, as it only relies on the phase-space representation of an experiment. Most importantly, the AFI is a flexible quantity for experimentally meaningful tasks. The QFI is most commonly used to study quadratic sensing-time separations between structured sensing protocols in the asymptotic setting, and the scaling that results from Cramer-Rao bounds is quite sensitive to small changes in the sensing task (for instance, the asymptotic scaling for Heisenberg-limited DC metrology tends to revert to the standard quantum limit under any noise that is more than inverse-polynomially small, losing track of important finite-size gains \cite{transversal_sensing}). While QFI may be able to produce exponential lower bounds in certain engineered settings, it has not found broad use as a tool for proving superpolynomial quantum advantages. AFI, on the other hand, does not encounter these issues as evidenced by the exponential separations in the remainder of this work, and excels at producing tight learning-theoretic analyses of realistic experimental tasks.

The AFI gives a sharp characterization of \textit{local} learning around a reference law $P_0$: one specifies a perturbation direction $h$ and asks how efficiently that perturbation can be detected or estimated. Many experimental objectives are instead genuinely global, with no distinguished reference distribution or local perturbation. For these problems, Q$\Psi$ uses the same phase-space idea in the Fourier domain.

The basic question is then frequency by frequency: how strongly can the experimental architecture respond to each Fourier component of the property being learned? The answer is quantified by the \textit{Fourier gain}. Once the property and the experimentally achievable responses are expanded in the same Fourier dictionary, global experimental design reduces to matching the Fourier response of the experiment to the Fourier support of the target.

\begin{definition}[Fourier gain of a sensing protocol family]
Let $\mathcal A$ be a class of experimental protocols, including classical postprocessing, and for a protocol in $\mathcal A$, let $Z$ denote the scalar output of one shot. Let
\begin{equation}
    f_Z(x)=\mathbb E[Z|X=x]
\end{equation} 
denote the protocol's response function, and let $V_Z=\sup_{x}\mathbb E[|Z|^2|X=x]$ be its maximum second moment. Let $\{\chi_\omega\}$ be a valid Fourier dictionary on the relevant sample space, such that $f_Z=\sum_{\omega}\widehat f_Z(\omega)\chi_\omega$. Then, the Fourier gain of $\mathcal A$ at frequency $\omega$ is defined as
\begin{equation}
    G_{\mathcal A}(\omega)=\sup_{Z\in\mathcal A}\frac{|\widehat f_Z(\omega)|^2}{V_Z} \ .
\end{equation}
\end{definition}
Intuitively, $G_{\mathcal A}(\omega)$ is the largest squared Fourier coefficient that the sensing protocol can place at frequency $\omega$ per unit shot noise. By decomposing the property in the same Fourier dictionary, sample complexity will be controlled by the ratio between the coefficient of the property and the Fourier gain at that frequency. This is precisely how the phase-space-overlap picture directly controls sample complexity in true global learning problems. These observations are formalized in the following theorem.

\begin{graydefinitionbox}
\begin{theorem}[Q$\Psi$ global learning guarantee]
\label{thm:qes_fourier_gain_global_observable}
Let $\psi$ be a property kernel with finite Fourier expansion
\begin{equation}
    \psi(x)=\sum_{\omega\in S}\widehat\psi(\omega)\chi_\omega(x)
\end{equation}
for some finite set $S$. Let $\mathcal P$ be a  class of signal distributions for which $\Psi_\psi(P)=\mathbb E_{X\sim P}[\psi(X)]$ is well defined. Let $\mathcal A$ be a class of sensing protocols with Fourier gain $G_{\mathcal A}(\omega)$ on each $\omega\in S$, and suppose $G_{\mathcal A}(\omega)>0$ whenever $\widehat\psi(\omega)\neq 0$. Suppose moreover that there is an integer $m\in\{1,\ldots,|S|\}$ such that, for every subset $T\subseteq S$ with $|T|\le m$ and every choice of coefficients $\{a_\omega\}_{\omega\in T}$, the protocol family $\mathcal A$ can synthesize the response
\begin{equation}
    f_T(x)=\sum_{\omega\in T}a_\omega\chi_\omega(x)
\end{equation}
with one-shot output $Z_T$ satisfying
\begin{equation}
    \mathbb E[Z_T|X=x]=f_T(x)
\end{equation}
and
\begin{equation} \label{eq:additive_envelope}
    \sup_x\mathbb E[|Z_T|^2|X=x]\le C_{\mathcal A}\sum_{\omega\in T}\frac{|a_\omega|^2}{G_{\mathcal A}(\omega)}
\end{equation}
for a constant $C_{\mathcal A}$ independent of $S$ and $\psi$. Then there is a protocol in $\mathcal A$ which estimates $\Psi_\psi(P)$ to additive error $\epsilon$ with probability at least $1-\delta$ for every $P\in\mathcal P$ using
\begin{equation}
    N=
    O\left(
        \frac{C_{\mathcal A}}{\epsilon^2}
        \min\left\{
            \frac{|S|}{m}\sum_{\omega\in S}\frac{|\widehat\psi(\omega)|^2}{G_{\mathcal A}(\omega)},
            \left(\sum_{\omega\in S}\frac{|\widehat\psi(\omega)|}{\sqrt{G_{\mathcal A}(\omega)}}\right)^2
        \right\}
        \log\!\left(\frac{1}{\delta}\right)
    \right)
\end{equation}
queries.
\end{theorem}
\end{graydefinitionbox}

Before we give the proof, let us break down the interpretation of this result intuitively. The first important object is $S$, which is the Fourier support of the property of interest. As long as the chosen Fourier basis satisfies approximate orthogonality, there is no value in synthesizing a measurement whose response kernel has support outside of $S$, since probability amplitude spent there is uninformative and can at worst bias the outcome. Then, when we say that $\mathcal{A}$ has Fourier gain $G_{\mathcal A}(\omega)$ on each $\omega\in S$, this is a statement that the \textit{best} protocol in $\mathcal{A}$ for that \textit{single frequency} $\omega$ achieves a response amplitude (per unit variance) of $G_{\mathcal A}(\omega)$. However, the property kernel can be supported on many, possibly unknown frequencies, so we do not, \textit{a priori}, know which protocol in $\mathcal{A}$ to choose. Ideally, we would like to synthesize a response which consists of many frequencies, with amplitudes tuned to maximize Fourier overlap with $\psi$. However, constraints on $\mathcal{A}$ such as energy, memory, or control depth can limit the extent to which a synthesized response can attain a large overlap. To capture this, we introduce $m$ and say that we allow any response in $\mathcal{A}$ to have at most $m$ Fourier terms. In practice, $m$ is a byproduct of constraints. 

With these definitions, we say that $Z_T$ is simply the output of a chosen sensing strategy, with expectation $f_T(x)$ being the response function. However, we still need Equation \eqref{eq:additive_envelope} for the following reason. When a protocol synthesizes a multi-frequency response, it must choose to distribute response mass in a concrete way. This can include, for instance, classical randomization of the probe state or a non-Gaussian state preparation that creates a superposition state with many branches of differing amplitude. After querying the quantum oracle and performing a measurement, it could be the case that a poorly chosen protocol, despite having desirable Fourier overlap with the target property, has a large readout variance. A toy example is as follows.

Imagine a class of experiments $\mathcal{A}$ with exactly two allowed protocols which have response functions $f_1(x) = \chi_{\omega_1}(x), f_2(x) = \chi_{\omega_2}(x)$ with unit variance, and suppose we wish to learn the property $\psi(x) = (\chi_{\omega_1} + \chi_{\omega_2})(x)/\sqrt{2}$. If $f_1 + f_2$ were included in $\mathcal{A}$, the property could be measured directly with unit variance. However, because the responses are only separately synthesizable, the optimal protocol is to randomly measure $f_1$ and $f_2$ each with probability $1/2$ and combine with classical postprocessing, which yields a variance of $2$. If one now considered a property consisting of $n$ modes, readout variance could be amplified as many as $n$ times. Therefore, when one places the constraint that the architecture can synthesize at most $m$ responses simultaneously, this is only meaningful if it can do so while still retaining a bounded variance. This is the precise meaning behind Equation \eqref{eq:additive_envelope}.

With this understood, the final sample complexity bound becomes illustrative. In short, the sample complexity is controlled by the ratio

\begin{equation}
    \sum_{\omega\in S}\frac{|\widehat\psi(\omega)|^2}{G_{\mathcal A}(\omega)}
\end{equation}
which is exactly the overlap between the property's Fourier amplitude at each frequency and the Fourier gain of the architecture at that frequency. This is adjusted for the number of frequencies that are simultaneously synthesizable with bounded variance. The second term in the minimum is the extreme case in which the protocol can only synthesize a single frequency at a time, such that randomizing over frequencies is the best option; this term is a tighter bound than simply sending $m\rightarrow 1$. With this, we now see that Fourier overlap is exactly the quantity which determines sample complexity in global learning problems, giving us a meaningful optimization design principle.

\begin{proof}
We prove two upper bounds and then take the better of the two. First, we use the fact that the architecture can synthesize $m=c|S|$ Fourier components at once (letting $c=m/|S|$). Choose a subset $T\subseteq S$ uniformly at random among all subsets of size $m$. Then every $\omega\in S$ is included in $T$ with probability $c$. Conditioned on the choice of $T$, use the assumed synthesis property with coefficients
\begin{equation}
    a_\omega=\frac{\widehat\psi(\omega)}{c}
\end{equation}
for $\omega\in T$. The resulting one-shot output $Z_T$ has conditional response
\begin{equation}
    \mathbb E[Z_T|X=x,T]=\sum_{\omega\in T}\frac{\widehat\psi(\omega)}{c}\chi_\omega(x) \ .
\end{equation}
Averaging also over the random choice of $T$ gives
\begin{align}
    \mathbb E[Z_T|X=x]
    &=
    \mathbb E_T\left[\sum_{\omega\in T}\frac{\widehat\psi(\omega)}{c}\chi_\omega(x)\right] \\
    &=
    \sum_{\omega\in S}\Pr[\omega\in T]\frac{\widehat\psi(\omega)}{c}\chi_\omega(x) \\
    &=
    \sum_{\omega\in S}\widehat\psi(\omega)\chi_\omega(x) \\
    &=
    \psi(x) \ .
\end{align}
Thus this randomized one-shot protocol is an unbiased estimator of the kernel $\psi$ pointwise, and hence an unbiased estimator of $\Psi_\psi(P)$ for every $P\in\mathcal P$. Its second moment is bounded by
\begin{align}
    \sup_x\mathbb E[|Z_T|^2|X=x]
    &\le
    \mathbb E_T\left[
        C_{\mathcal A}\sum_{\omega\in T}\frac{|\widehat\psi(\omega)|^2}{c^2G_{\mathcal A}(\omega)}
    \right] \\
    &=
    \frac{C_{\mathcal A}}{c}\sum_{\omega\in S}\frac{|\widehat\psi(\omega)|^2}{G_{\mathcal A}(\omega)} \ .
\end{align}
Averaging independent repetitions and applying the standard median-of-means estimator to the real and imaginary parts gives
\begin{equation}
    N=
    O\left(
        \frac{C_{\mathcal A}}{c\epsilon^2}
        \left(
            \sum_{\omega\in S}\frac{|\widehat\psi(\omega)|^2}{G_{\mathcal A}(\omega)}
        \right)
        \log\left(\frac{1}{\delta}\right)
    \right)
\end{equation}
queries. Second, we give an estimator which estimates every frequency independently, which corresponds to the regime where one does not use a protocol that can synthesize response functions containing multiple Fourier frequencies. For each $\omega\in S$, define
\begin{equation}
    A_\omega=\frac{|\widehat\psi(\omega)|}{\sqrt{G_{\mathcal A}(\omega)}}
\end{equation}
and let $A=\sum_{\omega\in S}A_\omega$. If $A=0$, then $\psi=0$ and the claim is trivial, so assume $A>0$. Choose frequency $\omega$ with probability $q_\omega=A_\omega/A$. Conditioned on choosing $\omega$, use the assumed synthesis property on the one-element set $\{\omega\}$ with coefficient
\begin{equation}
    a_\omega=\frac{\widehat\psi(\omega)}{q_\omega} \ .
\end{equation}
The resulting one-shot output $Z_\omega$ has averaged response
\begin{align}
    \mathbb E[Z_\omega|X=x]
    &=
    \sum_{\omega\in S}q_\omega\frac{\widehat\psi(\omega)}{q_\omega}\chi_\omega(x) \\
    &=
    \psi(x) \ .
\end{align}
Its second moment is bounded by
\begin{align}
    \sup_x\mathbb E[|Z_\omega|^2|X=x]
    &\le
    \sum_{\omega\in S}q_\omega C_{\mathcal A}\frac{|\widehat\psi(\omega)|^2}{q_\omega^2G_{\mathcal A}(\omega)} \\
    &=
    C_{\mathcal A}\sum_{\omega\in S}\frac{|\widehat\psi(\omega)|^2}{q_\omega G_{\mathcal A}(\omega)} \\
    &=
    C_{\mathcal A}\left(\sum_{\omega\in S}\frac{|\widehat\psi(\omega)|}{\sqrt{G_{\mathcal A}(\omega)}}\right)^2 .
\end{align}
Again applying median-of-means to the empirical average gives
\begin{equation}
    N=
    O\left(
        \frac{C_{\mathcal A}}{\epsilon^2}
        \left(\sum_{\omega\in S}\frac{|\widehat\psi(\omega)|}{\sqrt{G_{\mathcal A}(\omega)}}\right)^2
        \log\left(\frac{1}{\delta}\right)
    \right)
\end{equation}
queries. Taking the better of the $m$-frequency synthesis protocol and the independent-frequency protocol proves the theorem.
\end{proof}
We will see in the remainder of this work that the global learning algorithm produced by Theorem \ref{thm:qes_fourier_gain_global_observable} matches the sample complexity of the optimal hypothesis test produced by Theorem \ref{thm:qes_matched_score_guarantee} in many cases, enabling provably optimal global learning. Moreover, the Fourier-overlap picture suggests a natural numerical optimization route: given a continuous family of experiments, one can either find a low-dimensional parameterization or a finite-size subset of experiments, and then numerically optimize the achievable Fourier gain for a desired property. Optimal experimental design then becomes a classical statistical optimization problem, with sample complexity certified by Theorem \ref{thm:qes_fourier_gain_global_observable}. This idea, which suggests automated design of quantum-enhanced experiments with certifiable advantage, is further developed in Section \ref{sec:open_questions}.

To conclude, we next understand how to put the pieces of Q$\Psi$ together to design and discover experimentally-motivated quantum advantages.

\subsection{Using Q$\Psi$ to discover quantum advantages}

We now have all the tools to use Q$\Psi$ to design quantum experiments while certifying an advantage. We have seen that the AFI tightly certifies the optimal query complexity of a hypothesis-testing task based on an observable of interest, under the constraints of an experiment class. To prove a quantum speedup, we will apply our Q$\Psi$ hypothesis-testing tools to two classes of experimental protocols, with and without a particular quantum processing resource. We then prove that the added resource amplifies the AFI, establishing a hypothesis-testing speedup. We then apply our global-learning guarantee to the task of interest to produce a general-purpose learning algorithm achieving a practical quantum advantage. This procedure is detailed in the following reusable proof template. Here we assume that a phase-space $\chi$ has been fixed based on the learning task.

\begin{graydefinitionbox} \label{box:proof_template}
    \noindent \textbf{Template for Q$\Psi$ quantum advantage.} Provided an experimental task with the following inputs:
    \begin{itemize}
        \item The objective is represented as a family of functions $f^{(k)}(x)$, for $x\in \chi$.
        \item Let $P_0$ be a probability distribution supporting approximate orthogonality of some valid Fourier dictionary of $\chi$ (for continuous-variable systems, $P_0$ will often be a Gaussian, or if working on a compact phase space, a uniform distribution).  Let $h^{(k)} = f^{(k)} - \int f^{(k)} \ dP_0$.
        \item Let $\mathcal{M}_{\rm conv}$ be a class of ``conventional'' learning protocols, and let $\mathcal{M}_{\rm QIP}$ be a class of stronger protocols in which conventional experimentation is enhanced by access to a quantum processing resource. 
    \end{itemize} 
    The Q$\Psi$ procedure is as follows.
    \begin{enumerate}   
        \item Certify the optimal hypothesis-testing advantage.
        \begin{enumerate}
            \item Obtain an upper bound  $\mathsf A_{\mathcal{M}_{\rm conv}}(h^{(k)}; P_0) \leq c(k)$.
            \item Obtain a lower bound $A_{\mathcal{M}_{\rm QIP}}(h^{(k)}; P_0) \geq q(k)$.
            \item Conclude by Theorem \ref{cor:AFI_tight} that there exists a protocol in $\mathcal{M}_{\rm QIP}$ achieving a quantum advantage of at least $\Omega(q(k)/c(k))$ over every, possibly adaptive sequence of experiments from $\mathcal{M}_{\rm conv}$.
        \end{enumerate}
        \item Produce an algorithm for global learning. 
        \begin{enumerate}
            \item Choose a parameterization or discretization of $\mathcal{M}_{\rm QIP}$, and (analytically or numerically) optimize the Fourier gain relative to $h^{(k)}$. 
            \item Apply Theorem \ref{thm:qes_fourier_gain_global_observable} to design an optimal sequence of quantum-enhanced experiments. Note that the hypothesis-testing upper bound, Theorem \ref{thm:qes_matched_score_guarantee} may often already produce a valid global learning algorithm.
            \item Use Fact \ref{fact:learning_to_testing_reduction} to lift the hypothesis-testing lower bound to global learning, and conclude optimal quantum-enhanced experimental design.
        \end{enumerate}
    \end{enumerate}
\end{graydefinitionbox}

Q$\Psi$ takes as input experimentally-grounded learning tasks, architectures, and proposed quantum resources, and interfaces them using the phase-space picture we have introduced. By controlling the AFI of conventional and quantum-enhanced classes, Q$\Psi$ outputs tight bounds on the achievable quantum advantage, while producing the optimal hypothesis-testing algorithm certifying this advantage under the chosen constraints. It also provides a route to systematically design optimal global-learning algorithms. Because realistic experimental constraints on resources like energy, coherence, magic, and non-Gaussianity appear as natural convex constraints on phase-space response functions, the phase-space structure leveraged by Q$\Psi$ presents a uniquely tractable method for rigorous, certifiably optimal quantum-enhanced experimental design beyond the scope of existing methods in both quantum learning theory and quantum metrology. 

Applied to single-mode sensing, Q$\Psi$ enables the proofs of quantum advantage in the subsequent appendices, as well as the design of the experiments behind Figures \ref{fig:single_qubit_advantage} and \ref{fig:coherent_and_memory_experiments}. This new approach allows us to prove the first exponential quantum advantages in learning that use only a single qubit, as well as the first exponential separations between Gaussian and non-Gaussian strategies that are enabled by single-qubit quantum control. We also show in Appendix \ref{sec:qes_reproof_multimode_displacement} that existing quantum learning separations can emerge as special cases of the Q$\Psi$ approach with substantially simplified proofs. We believe that this framework may enable the discovery of new, practical quantum advantages in experimental tasks compatible with near-term quantum hardware. These perspectives are discussed in Appendix \ref{sec:open_questions}.

\newpage
\section{Hardness of learning from constrained experiments} \label{sec:hardness}
In this section we will establish the exponential lower bounds discussed in the main text. Our bounds reveal operational constraints on the capabilities of sensing architectures at each level of the sensing hierarchy discussed in Section \ref{sec:sensing-hierarchy} and depicted in Figure \ref{fig:schematic_main}.

First, we show that any classical sensor, irrespective of total available energy, requires $\exp(\Omega(k))$ rounds of interrogation of a signal to estimate its $k$-th angular Fourier coefficient or moment. Next, we move into the first level of quantum-enhanced sensing, and consider the class of conventional Gaussian sensing protocols implemented in modern optical sensing platforms. Here, the amount of energy which can be coherently utilized to create squeezed Gaussian probe states and reduce quadrature variance below the vacuum floor is the deciding resource: we show that any Gaussian protocol with energy $k$ per experiment and arbitrary allowed adaptivity requires $\exp(\Omega(k))$ interrogations to estimate a signal's $k$-th Fourier coefficient. Moving slightly deeper into the regime of quantum-enhanced sensing, we consider protocols which now couple ancilla qubits to the quantum sensor, enabling computationally universal control. Here, we prove that characterizing \textit{time-dependent} Fourier spectra remains hard when the sensing architecture does not support long-lived quantum memory: estimating a Fourier coefficient across $m$ points in time requires $\exp(\Omega(m))$ interrogations when all resources decohere much more quickly than the lifetime of the signal. Finally, we characterize the advantages gained by future quantum-enhanced sensing architectures which may support deep quantum control. We show that there exist degree-$k$ polynomial QSL properties of the signal such that for a constant $\gamma \in (1/2, 1)$, any sensing architecture which can create probe states using circuit depth no more than $\gamma \,k$ requires $\exp(\Omega(k))$ interrogations.

Our lower bounds make use of the novel techniques introduced in Section \ref{sec:qes}, elucidating the utility of our approach for obtaining lower bounds in various constrained learning settings. Then in Section \ref{sec:advantage_upper_bounds}, we will revisit the same tasks considered here and provide efficient sensing protocols which live at one higher level of the sensing hierarchy than each lower bound given. Together, these results rigorously establish a hierarchy of exponential advantages for classical sensing, using different quantum information processing capabilities of only a single qubit.

\subsection{Hardness of learning with classical probes} \label{sec:coherent_probe_lb}
First, we establish hardness for classical sensing protocols by proving the following theorem. As discussed in Section \ref{sec:sensing-hierarchy}, we prove a bound against the class of coherent-probe protocols from Definition \ref{def:coherent_probe_protocol}, which circumscribe all classical sensing protocols. In fact, the following bound even allows a protocol to make arbitrarily high-energy joint measurements on many probe states at once; the only constraint is that the probes themselves must be classical.

\begin{graydefinitionbox}
\begin{theorem} \label{thm:classical_hardness_formal}
    Any adaptive coherent probe protocol that can estimate the $k^{\rm th}$ angular Fourier coefficient of a signal requires least $S \geq \exp(\Omega(k))$ signal queries. More specifically, there are constants $C_1, C_2$ such that
    \begin{equation}
        S \geq C_1 \exp\left(\frac{k}{2}\operatorname{arsinh}\left(C_2k\right) - \frac{C_2}{2}\left(\sqrt{1+C_2^2k^2} - 1\right)\right)
    \end{equation}
    and such that the exponent is $\Omega(k)$ for all $k>0$.
\end{theorem}
\end{graydefinitionbox}
\begin{proof}
    Our proof will be based around use of the semiclassical quantum measurement lemma \ref{lemma:semiclassical_quantum_measurement}, beginning with the instantiation of two hypothesis families that are difficult for coherent-probe protocols to distinguish. Consider the sample space $\mathcal{X} = S^1$ to be the circle, and let $P_0 = d\theta / 2\pi$ be the uniform distribution on the circle. Then, following the proof template \ref{box:proof_template}, consider the family of functions 
    \begin{equation}
        h_k(\alpha) = 0.9\cos(k\theta) \ .
    \end{equation}
    It follows that $\int_{S^1} h_k(\theta) dP_0(\theta) = 0$ and $\|h_k\|_\infty \leq 1$. This choice of $h_k$ instantiates physically valid displacement distributions. Let us parameterize the phase-space $\mathbb{C}$ in polar coordinates $(a, \theta)$. Consider the two families of angular probabilty distributions
    \begin{equation}
    T_{\pm}^{(k)}(\theta) =\frac{1}{2\pi}(1\pm 0.9 \cos(k\theta)) \ .
    \end{equation}
    Then, consider the phase-space displacement distributions
    \begin{equation}
        P_{\pm}^{(k)}(\theta) = (Ae^{i\theta})_{\#}(T_{\pm}^{(k)}(\theta)) \ .
    \end{equation}
    These displacement distributions are the pushforward of $T_{\pm}^{(k)}(\theta)$ under the map $\theta\rightarrow Ae^{i\theta}$ for a fixed $A > 0$. We further impose that for a constant $C>0$, $A^2 \leq Ck$; in particular any constant $A$ independent of $k$ will already suffice for this separation. We first check that these are valid probability distributions. Since $|\cos(k\theta)|\leq 1$, we have
    \begin{equation}
        T_{\pm}^{(k)}(\theta)=\frac{1}{2\pi}(1\pm 0.9\cos(k\theta))\geq 0 \ .
    \end{equation}
    Moreover, because $k\in \mathbb{Z}^{+}$,
    \begin{align}
        \int_0^{2\pi}T_{\pm}^{(k)}(\theta)d\theta
        &= \frac{1}{2\pi}\left[\theta \pm \frac{0.9}{k}\sin(k\theta)\right]_0^{2\pi} \\
        &= \frac{1}{2\pi}\left(2\pi \pm \frac{0.9}{k}\sin(2\pi k)\right) \\
        &= 1 \ ,
    \end{align}
    Thus $T_{\pm}^{(k)}$ is a valid probability density on the circle, and its pushforward under $\theta\mapsto Ae^{i\theta}$ is a valid probability law $P_{\pm}^{(k)}$ on phase-space. Let $P_{+}^{(k)}(\theta), P_{-}^{(k)}(\theta)$ induce the displacement channels $\mathcal{E}_+^{(k)}, \mathcal{E}^{(k)}_-$. Physically, these channels correspond to fixed-amplitude signals whose phase distribution contains an angular harmonic of frequency $k$.
    
    Now we can proceed with applying the semiclassical measurement lemma. Recall that the AFI is defined as
    \begin{equation}
    \mathsf A_M(h;P_0)=\int \left|\frac{dA_{M,h}}{dQ_{M,0}}(y)\right|^2 dQ_{M,0}(y)        \ ,
    \end{equation}
    where for a coherent-probe protocol with POVM $M(dy)$, 
    \begin{equation}
        Q_{M, 0}(dy) = \frac{1}{2\pi}\int K_M(dy|\theta)d\theta, \qquad A_{M, h} = \frac{1}{2\pi}\int \cos(k\theta)K_M(dy|\theta)d\theta \ ,
    \end{equation}
    and where $K_M(dy|\theta)$ is the response kernel of $M(dy)$. Noting that $\frac{dA_{M,h}}{dQ_{M,0}}(y) = \mathbb{E}[\cos(k\theta) | Y= y] \leq 1$, it follows that
    \begin{equation}
        \mathsf A_M(h;P_0) \leq \int \left|\frac{dA_{M,h}}{dQ_{M,0}}(y)\right| dQ_{M,0}(y) = \int |A_{M, h}(dy)| \ .
    \end{equation}
    Now consider a coherent-probe protocol in which the coherent state $\ket{\beta}, \beta\in \mathbb{C}$ is exposed to the signal. Then under displacement $Ae^{i\theta}$, the measurement kernel is $K_M(dy|\theta) = \Tr(M(dy)\ketbra{\beta + Ae^{i\theta}})$. Using the definition of $A_{M, h}$, we have
    \begin{equation}
         \mathsf A_M(h;P_0) \leq \left\|\int \frac{d\theta}{2\pi}\  0.9\cos (k\theta) \ketbra{\beta + Ae^{i\theta}}\right\|_1
    \end{equation}
    Note that the displacements $D(\beta), D(\beta)^\dagger$ factor out to the right and left respectively, and by unitarity do not affect the magnitude of the expression. Rewrite the expression, then as $ \mathsf A_M(h;P_0) \leq \|\Delta\|_1/2$. Using the Fock basis expansion of coherent states, we have
    \begin{align}
        \Delta &= \frac{0.9}{\pi}\int d\theta \ \cos(k\theta)) \ketbra{ Ae^{i\theta}} \\ 
        &=
        \frac{0.9e^{-A^2}}{\pi} \sum_{n,m \geq 0}\frac{A^{n+m}}{\sqrt{n!m!}}\left[\int d\theta \ \cos(k\theta)e^{i(n-m)\theta}\right]|n\rangle \langle m| \\
        &= 0.9e^{-A^2}\sum_{m=0}^\infty \frac{A^{2m+k}}{\sqrt{m!(m+k)!}}(|m+k\rangle \langle m| + |m\rangle \langle m + k|)
    \end{align}
    where in the secondline we use the sifting property
    \begin{align}
        \int d\theta \ \cos(k\theta)e^{i(n-m)\theta} &=\frac{1}{2}\int d\theta \left(e^{i(n-m-k)} + e^{i(n-m+k)}\right) \\&=\pi(\delta(n-m-k)  + \delta(n-m+k))\ .
    \end{align}
        It then follows that
    \begin{equation}
        \|\Delta\|_1/2 \leq 0.9 e^{-A^2} \sum_m\frac{A^{2m+k}}{\sqrt{m!(m+k)!}} 
    \end{equation}
    by triangle inequality. Let $p_m = e^{-A^2}A^{2m}/m!$ so that 
    \begin{equation}
        \|\Delta\|_1/2 = 0.9 \sum_m \sqrt{p_mp_{m+k}} \ .
    \end{equation}
    Now note that for any constant $s>0$ we can rewrite
    \begin{equation}
        \sqrt{p_m p_{m+k}} = e^{-sk/2}\sqrt{e^{-sm}p_m e^{s(m+k)}p_{m+k}}.
    \end{equation}
    Then
    \begin{align}
        \sum_m \sqrt{p_mp_{m+k}} &= e^{-sk/2} \sum_m\sqrt{e^{-sm}p_m e^{s(m+k)}p_{m+k}} \\
        &\leq e^{-sk/2}\left(\sum_me^{-sm}p_m\right)^{1/2}\left(\sum_me^{s(m+k)}p_{m+k}\right)^{1/2}
    \end{align}
    where the last step is by Cauchy-Schwarz. At this point note that upon setting $\lambda \coloneqq A^2$, $p_m = e^{-\lambda}\lambda^m/m!$ is precisely the probability of $m$ detections from a Poisson process of expectation $\lambda$. We can then rewrite the two multiplicative factors as:
    \begin{align}
        &\sum_me^{-sm}p_m = \mathbb{E}_{m\sim \textnormal{Pois}(\lambda)}[e^{-sm}] \\
        &\sum_me^{s(m+k)}p_{m+k}\leq \sum_me^{sm}p_m= \mathbb{E}_{m\sim \textnormal{Pois}(\lambda)}[e^{sm}]
    \end{align}
    For a Poisson distribution,
    \begin{align}
        \mathbb{E}_{m\sim \textnormal{Pois}(\lambda)}[e^{tm}] &= \sum_{j=0}^\infty \frac{\exp(tj)\lambda^j}{j!}\exp(-\lambda) \\
        &= \exp(-\lambda)\sum_j\frac{(e^t\lambda)^j}{j!} \\
        &= \exp(-\lambda)\exp(e^t\lambda) \\
        &= \exp(\lambda(e^t-1)) \ .
    \end{align}
    We therefore have that
    \begin{equation}
        \sum_m \sqrt{p_mp_{m+k}} \leq \exp\left(\lambda(\cosh(s) - 1) - \frac{sk}{2}\right) \ .
    \end{equation}
    which gives us
    \begin{equation}
        \|\Delta\|_1/2 \leq 0.9\exp\left(\lambda(\cosh(s) - 1) - \frac{sk}{2}\right) \ .
    \end{equation}
    Now we choose $s$ to minimize the upper bound. Let the exponent be $-F(s)$; then
    \begin{equation}
        F'(s) = \frac{k}{2}-\lambda\sinh(s)
    \end{equation}
    At this value,
    \begin{align}
        F(s_*)
        &=
        \frac{k}{2}\operatorname{arsinh}\left(\frac{k}{2\lambda}\right)
        -\lambda\left(\sqrt{1+\frac{k^2}{4\lambda^2}}-1\right).
    \end{align}
    Letting $x \coloneqq \frac{k}{2\lambda}$, we can rewrite this as
    \begin{equation}
        F(s_*)=\lambda\left(x\operatorname{arsinh}x-\sqrt{1+x^2}+1\right).
    \end{equation}
    Define $h(x)\coloneqq x\operatorname{arsinh}x-\sqrt{1+x^2}+1$. Then \(h(0)=0\), and
    \begin{align}
        h'(x)
        &= \operatorname{arsinh}x +\frac{x}{\sqrt{1+x^2}}-\frac{x}{\sqrt{1+x^2}} \\
        &=
        \operatorname{arsinh}x.
    \end{align}
    
    Hence $h'(x)>0$ for all $x>0$, so $h(x)>0$ for all $x>0$. Therefore $F(s_*) > 0$ for every $k>0$, since $x=k/(2\lambda)>0$. Consequently,
    \begin{equation}
        \|\Delta\|/2 \leq 0.9
        \exp(-F(s_*)),
    \end{equation}
    with a strictly positive $F(s_*)$. With this AFI upper bound in hand, we can now apply Lemma \ref{lemma:semiclassical_quantum_measurement} to obtain that
    any algorithm which successfully distinguishes the two hypotheses with probability at least $p > 1/2$ requires
    \begin{equation}
        T \geq \frac{(2p-1)}{0.9}\exp(F(s_*)) \ .
    \end{equation}
    This is the rigorous lower bound plotted in Figure \ref{fig:coherent_and_memory_experiments}(b). We can see that this is exponential in $k$ as follows. Since $A^2=\lambda\leq Ck$, if $x=k/(2\lambda)$, then $x\geq 1/(2C)$, and
    \begin{equation}
        F(s_*) = k\frac{h(x)}{2x} \geq k\frac{h(1/(2C))}{2/(2C)} \eqqcolon c' k,
    \end{equation}
    where $c'$ is a constant depending only on C. Thus, for any constant success bias, so that \(2p-1=\Omega(1)\), the preceding bound implies
    \begin{equation}
        T \geq \frac{2p-1}{0.9}\exp(F(s_*)) \geq \exp(\Omega(k)).
    \end{equation}
    Using Fact \ref{fact:learning_to_testing_reduction} yields the desired hardness result for learning the $k$-th Fourier coefficient with coherent probes.
\end{proof}

In Section \ref{sec:advantage_upper_bounds} we provide quantum sensing protocols enabled by conventional squeezed probes that can learn angular Fourier coefficients using $\textnormal{poly}(k)$ queries, this lower bound demonstrates that quantum probes generically enable an exponential expansion of the bandwidth of angular Fourier coefficients and moments accessible to a sensing protocol.

\subsection{Hardness of learning with conventional quantum sensing} \label{sec:conventional_gaussian_lb}

In this section, we consider the fundamental limitations of conventional quantum sensing protocols using bosonic probes. As discussed in Section \ref{sec:preliminaries}, conventional cavity and interferometric systems use Gaussian probe states and measurements. The precision limits of these sensing platforms are controlled by the maximum mean photon number that can be reliably created in non-classical states. As such, we consider here the class of Gaussian protocols with a total energy budget $E$; in units of $\hbar = 1$, this is equivalent to the mean photon number. The class is defined as follows.

\begin{definition}[Conventional finite-energy protocol for bosonic quantum sensing] \label{def:gaussian_bosonic_sensing_formal}
    A conventional protocol for bosonic quantum sensing with access to a signal channel $\mathcal{E}$ and with energy constraint $E$ allows:
    \begin{enumerate}
        \item Any Gaussian probe state $\rho_S$,
        \item Any generaldyne measurement, performed using Gaussian ancilla state $\rho_A$,
        \item Classical communication (adaptivity) across experiments.
        \item In each experiment, it must hold that $\Tr(\rho_{SA}(\hat{n}_S + \hat{n}_A)) \leq E$, where $\hat{n} = a^\dagger a$ is the number operator.
    \end{enumerate}
\end{definition}

The amount of accessible energy is a crucial resource that must be accounted for when studying separations between conventional and quantum-enhanced sensing protocols, so our separations will parameterize the total energy as the instance size $k$, holding this constant across our upper and lower bounds. We remark that the above class can be substantially more powerful than the kinds of single-mode sensing usually available in practice, because it permits generaldyne measurements that could require a large number of ancilla modes not present in conventional platforms. We nevertheless establish an exponential lower bound against this stronger class, thereby establishing a bound against any realistic Gaussian experimental protocol. 

\begin{graydefinitionbox}
\begin{theorem}\label{thm:conventional_hardness_formal}
    Any conventional bosonic sensing protocol as in Definition \ref{def:gaussian_bosonic_sensing_formal} with per-experiment energy $k$ $S \geq \exp(\Omega(k))$ signal queries to estimate the $k^{\rm th}$ directional Fourier coefficient of a single-mode signal to constant accuracy. Specifically, for $\eta < 1$, at least
    \begin{equation}
         S\geq \frac{2p-1}{\eta\left(\exp\left(-\frac{k^2}{8k + 5}\right) + \exp\left(-k^2\right)\right)}
    \end{equation}
    samples are necessary to estimate the $k$-th Fourier coefficient $\mathbb{E}[e^{ikx}]$ to absolute accuracy $\eta/2$ with success probability at least $p>1/2$.
\end{theorem}
\end{graydefinitionbox}
\begin{proof}[Proof of Theorem \ref{thm:conventional_hardness_formal}]
    As before, our proof will begin with a definition of an appropriate family of functions to use the semiclassical measurement lemma (Lemma \ref{lemma:semiclassical_quantum_measurement}). Here, for a more intuitive proof, we will consider a task in which the hypothesis signals differ in their $k$-th Fourier coefficient along a known phase-space axis, which will be sufficient to obtain an exponential lower bound. However, we remark in practice this information is not known in the harder learning setting, and one can obtain exponential lower bounds with larger exponents in that setting. 

    To begin, let us parameterize our phase-space in real quadrature coordinates, $z = (x, p) \in \mathbb{R}^2$. In this proof, phase space will be our sample space, equipped with the Gaussian measure $P_0 = \mathcal{N}(0, I_2)$. In these coordinates, the standard vacuum-noise Gaussian distribution is
    \begin{equation}
            g(x, p) = \frac{1}{2\pi}\exp\left(-\frac{x^2+p^2}{2}\right) \ .
    \end{equation}
    For any $k\in\mathbb{Z}^+$ (where we work with integers for clarity, but positive reals would suffice), define the quantity
    \begin{align}
    c_k &= \frac{1}{2\pi}\int_{\mathbb{R}^2}e^{-(x^2+p^2)/2}\cos(kx)dxdp \\
        &= \frac{1}{\sqrt{2\pi}}\int_{\mathbb{R}}e^{-x^2/2}\cos(kx)dx \\
        &= \operatorname{Re}\left[\frac{1}{\sqrt{2\pi}}\int_{\mathbb{R}}e^{-x^2/2+ikx}dx\right] \\
        &= \operatorname{Re}\left[e^{-k^2/2}\frac{1}{\sqrt{2\pi}}\int_{\mathbb{R}}e^{-(x-ik)^2/2}dx\right] \\
        &= e^{-k^2/2} \ .
    \end{align}
    Then define the Q$\Psi$ choice of functions
    \begin{equation}
        h_k(x, p) = \frac{\cos(kx) - c_k}{1+c_k} \ .
    \end{equation}
    It follows that $\mathbb{E}_{P_0}[h_k] = 0$ and $-1\leq h_k \leq 1$, as required for the semiclassical measurement lemma. With these, we define the displacement distributions
    \begin{equation}
        P_\pm^{(k)}(x, p) = g(x, p)(1\pm \eta h_k(x, p)) \ .
    \end{equation}
    Up to the small constant $c_k$ which is present for later convenience, these are highly natural distributions for a Fourier analysis problem: they correspond to raw vacuum noise superposed with a frequency-$k$ modulation. These are also physically valid since the conditions on $h_k$ are satisfied.

    Next, let $\varphi_C(z) = \textnormal{PDF}(\mathcal{N}(0, C))$ denote the density of a bivariate mean-0 Gaussian with covariance matrix $C$. By Lemma \ref{lemma:Gaussian_experiment_form} it follows that conditioned on the displacement $z=(x,p)$, the outcome $Y$ of any conventional Gaussian protocol, where the probe state and measurement may be chosen based on classical advice, has a response kernel of the form $K_M(dy|z) = \varphi_C(y-Az-b)dy$. Here, $x$ is the first coordinate of $z$, $A$ is some real 2 by 2 matrix, and $b$ is a real 2-dimensional vector. The allowed $A,b$ will later be constrained by the accessible energy, but any conventional Gaussian protocol with access to $\mathcal{E}_\pm$ must have this general form. Then
    \begin{equation}
        Q_{M, 0}(dy) = \left(\int \varphi_C(y-Az-b)g(z) dz\right)dy \coloneqq q_0(y)dy \ ,
    \end{equation}
    and
    \begin{equation}
        A_{M,h}(dy) = \left(\int \frac{\cos(kx) - c_k}{1+c_k}\varphi_C(y-Az-b)g(z) dz\right) dy \ .
    \end{equation}
    Defining 
    \begin{equation}
    q_c(w) = \int dz \ \varphi_C(w - Az - b)g(z) \cos(kx) \ ,
    \end{equation}
    we have
    \begin{equation}
        A_{M, h}(dy) = \frac{q_c(y) - c_k q_0(y)}{1+c_k}dy \Longrightarrow \frac{dA_{M, h}}{dQ_{M, 0}}(y) = \frac{1}{1+c_k}\left(\frac{q_c(y)}{q_0(y)}-c_k\right)
    \end{equation}
    
    To control this expression, we proceed by obtaining a Bayesian interpretation of $q_c$. Let $Z$ denote the random variable corresponding to draws from $g(z)$. Let $W$ denote the random variable distributed according to $q_c(w)$. It follows that
    \begin{equation}
        W|(Z = z) \sim \mathcal{N}(Az+b, C) \ . 
    \end{equation}
    Hence the conditional density of $W = w$ is
    \begin{equation}
        p_{W|Z}(w|z) = \varphi_C(w - Az - b) \ .
     \end{equation}
     By the definition of conditional probability,
     \begin{equation}
         p_{W, Z}(w, z) = p_Z(z) p_{W|Z}(w|z) = g(z) \varphi_C(w - Az - b) \ .
     \end{equation}
     By Bayes' theorem,
     \begin{equation}
         p_{Z|W=w}(z) = \frac{
         p_{W, Z}(w, z)}{p_W(w)}
     \end{equation}
     The denominator is simply obtained by integrating over $z$ in the joint density:
     \begin{equation}
         p_W(w) = \int g(z) \varphi_C(w - Az - b) \ dz = q_0(w) \ .
     \end{equation}
     We thus have a clean expression for the density $p_{Z|W=w}(z)$. Plugging this in to compute a conditional expectation, we have the expression
     \begin{equation}
         \mathbb{E}_Z[\cos(kx)|W = w] = \frac{1}{q_0(w)}\int dz \ \varphi_C(w - Az - b)g(z) \cos(kx) = \frac{q_c(w)}{q_0(w)} \ .
     \end{equation}
     Rearranging, we find 
     \begin{equation}\label{eq:qc_decomp}
         q_c(w) = q_0(w) \mathbb{E}_Z[\cos(kx)|W = w] \ .
     \end{equation}
     This tells us that the AFI is precisely
     \begin{equation}
     \label{eq:TV_exp_two_comp}
         \mathsf A_{M}(h; P_0) = \mathbb{E}_{Y\sim Q_0}\left[\left(\frac{\mathbb{E}[\cos(kX)|Y] - c_k}{1+c_k}\right)^2\right] \leq  \mathbb{E}_{Y\sim Q_0}\left[\mathbb{E}[\cos(kX)|Y]^2\right] +  c_k^2
     \end{equation}
     where the inequality uses that $1+c_k \geq 1$. Thus to obtain our AFI bound, we need to control the expectation of $\cos(kx)$ conditioned on the value of $W$ (where $X$ is simply the first coordinate of the random variable $Z$). To do this, we must characterize the distribution $Z|W = w$. Lemma \ref{lemma:conditional_gaussian_estimation} enables precisely this computation; we use its block-matrix notation here.

     Note that $(Z, W)$ are jointly Gaussian. We know all the moments:
     \begin{equation}
         \mathbb{E}[Z] = 0, \qquad \mathbb{E}[W] = b 
     \end{equation}
     and
     \begin{equation}
         \Sigma_{ZZ} = I_2, \qquad \Sigma_{WW} = AA^T + C
     \end{equation}
     Then for $\zeta\sim \mathcal{N}(0, C)$, we have $\Sigma_{ZW} = \textnormal{Cov}(Z , AZ + b + \zeta) = \textnormal{Cov}(Z, Z)A^T  = A^T$. Using Lemma \ref{lemma:conditional_gaussian_estimation}, we then find that the covariance of $Z|W = w$ is
    \begin{equation}
        \operatorname{Cov}(Z\mid W=w)=I_2-A^T(AA^T+C)^{-1}A \ .
    \end{equation}
    Applying the Woodbury identity,
     \begin{equation}
        \Sigma_{Z|W} = (I_2 + A^T C^{-1} A)^{-1} \ .
     \end{equation}
     We're interested characterizing the first coordinate, which has variance $s^2$ given by
     \begin{equation}
         s^2 = \begin{pmatrix}
             1 & 0 
         \end{pmatrix}
         \Sigma_{Z|W} \begin{pmatrix}
             1 \\ 0
         \end{pmatrix}
     \end{equation}
     Crucially, $s^2$ is entirely independent of $w$. Now note that
     \begin{equation}
          |\mathbb{E}_Z[\cos(kx)|W = w]| \leq | \mathbb{E}_Z[e^{ikX}|W = w]| = \left|\int_{\mathbb{R}} \frac{e^{ikx}}{\sqrt{2\pi s^2}}\exp\left(-\frac{(x-\mu(w))^2}{2s^2}\right) \right|\ .
     \end{equation}
     Shift and rescale the integral to a standard normal with the change of variables $x = \mu(w) + st$. Then 
     \begin{equation}
         \mathbb{E}_Z[e^{ikX}|W = w] = e^{ik\mu(w)}\mathbb{E}_{t\sim\mathcal{N}(0, 1)}[e^{ikst}] = e^{ik\mu(w)} e^{-k^2s^2/2}
     \end{equation}
     Taking absolute magnitudes, 
     \begin{equation}
         |\mathbb{E}_Z[\cos(kx)|W = w]| \leq e^{-k^2s^2/2}
     \end{equation}
     Returning to the inequality \eqref{eq:TV_exp_two_comp}, we thus have
     \begin{equation}
         \mathsf A_{M}(h; P_0) \leq e^{-k^2 s^2} + e^{-k^2} \ .
     \end{equation}
     Now recall Lemma \ref{lemma:covariance_normbound}, which tells us that
     \begin{equation}
         (I_2 + A^T C^{-1} A)^{-1} \preceq (8k+5)I_2 \ .
     \end{equation}
     Therefore, $\Sigma_{Z|W} \succeq (8k+5)^{-1}I_2$, so $s^2 \geq (8k+5)^{-1}$. For $k \geq 1$, this makes the first term in the TV bound dominate. Together, we have
     \begin{equation}
         \mathsf A_{M}(h; P_0) \leq\exp\left(-\frac{k^2}{8k + 5}\right) + \exp\left(-k^2\right) = \exp(-O(k)) \ .
     \end{equation}
     Finally applying Lemma \ref{lemma:semiclassical_quantum_measurement}, any strategy using
     \begin{equation}
         S \leq \exp(O(k))
     \end{equation}
     experiments has success probability $1/2 + o(1)$. This establishes that at least 
     \begin{equation}
         S\geq \frac{2p-1}{\eta\left(\exp\left(-\frac{k^2}{8k + 5}\right) + \exp\left(-k^2\right)\right)} = \exp(\Omega(k))
     \end{equation}
    samples are necessary to solve the distinguishing task with constant success probability $p>1/2$. Applying Fact \ref{fact:learning_to_testing_reduction}, we lift to the desired sample lower bound for Fourier learning.
\end{proof}

\subsection{Hardness of learning time-dependent signals without memory} \label{sec:memoryless_lower_bounds}

The previous section established an exponential lower bound for static Fourier coefficients with conventional finite-energy sensing. As we will see in Section \ref{sec:advantage_upper_bounds}, the single-qubit-enhanced protocol which estimates these coefficients efficiently given the same energy uses the \textit{control} capability of the qubit to generate non-Gaussian probe states. However, coupling a qubit to the sensor mode adds another meaningful resource: quantum memory. A qubit can be retained as a memory state while the sensor mode undergoes multiple evolutions under a classical signal. Here, we focus on a separation which explicitly uses the ability of the qubit to act as a memory. The core point is that this capability can add exponential sensing power to a conventional bosonic sensing device when signals are time-dependent. 

A time-dependent signal is obtained when the generating Hamiltonian has time-varying coefficients $f_x(t), f_p(t)$.  Physically, these are stochastic processes that are continuous functions of time. However, any physical experiment that performs destructive quantum measurements accesses the induced displacement channel at a discrete set of times, which we call $\mathcal{T}$. As such, a time-dependent signal is operationally characterized by a probability distribution $P$ over the stochastic process $\{Z_t\}_{t\in \mathcal{T}}$, where $Z_t = (X_t, P_t)\in\mathbb{R}^2$ is the random variable corresponding to the induced displacement at time $t$. In the static setting, each $Z_t$ is independent and identical, whereas the time-varying case allows for different, temporally-correlated displacement distributions. Keeping with the Fourier analysis setting, we work with the following object in the time-varying case; $\Omega$ denotes the usual symplectic form.

\begin{definition}[Temporal Fourier coefficient]
    For times \(\mathbf{t}=(t_1,\ldots,t_m)\) and phase-space frequencies \(\boldsymbol{\zeta}=(\zeta_1,\ldots,\zeta_m)\), define the $(\mathbf{t}, \boldsymbol{\zeta})$-temporal Fourier coefficient is
    \begin{equation}
        \widehat P(\mathbf{t},\boldsymbol{\zeta})=
        \mathbb{E}_{P}\left[
            \exp\left(i\sum_{j=1}^m \Omega(\zeta_j,Z_{t_j})\right)
        \right] \ .
    \end{equation}
\end{definition}
This definition reduces to the usual static Fourier coefficient when \(m=1\). Moreover, as we detail in Section \ref{sec:quantum_memory_upper_bounds}, a wide array of time-dependent properties can be extracted from temporal Fourier coefficients by taking derivatives at the origin. First, we will consider estimation of the two-point temporal Fourier correlation. In particular, we restrict to coefficients of the form
\begin{equation}
    \Theta_k(t_0,t_1)=
    \mathbb{E}_{P}\left[e^{ik(X_{t_0}-X_{t_1})}\right] \ .
\end{equation}
In symplectic notation this is \(\widehat P((t_0,t_1),(\zeta_k,-\zeta_k))\), where \(\zeta_k\) is chosen so that \(\Omega(\zeta_k,Z_t)=kX_t\). 

\subsubsection{Time-varying hypothesis testing for frequency-$k$, two-point correlation}
We now describe the hypothesis-testing instance which realizes a memory-based advantage in sensing with a single qubit. Once again let
\begin{equation}
    g(x,p)=\frac{1}{2\pi}\exp\left(-\frac{x^2+p^2}{2}\right)
\end{equation}
and define $c_k=e^{-k^2/2}$. For any $\theta \in [0, 2\pi)$, define
\begin{equation}
    h_{k,\theta}(x,p)=\cos(kx+\theta)-c_k\cos(\theta) \ .
\end{equation}
The subtraction term simply ensures that \(\mathbb{E}_{g}[h_{k,\theta}(X,P)]=0\). Fix a constant \(\eta\leq 1/4\). In each realization of the time-dependent signal, draw a latent phase \(\Phi\sim{\rm Unif}[0,2\pi)\). Let $\sigma =\pm 1$ be a fixed sign, and let the time-$t$ displacement distribution conditioned on $\Phi$ correspond to
\begin{equation}
    P_{\sigma,\Phi,t}^{(k)}(x,p)=
    g(x,p)\left(1+\eta h_{k,\Phi+\sigma\Omega_0 t}(x,p)\right) \ .
\end{equation}
Here \(\Omega_0\) is a known drift rate. The two hypotheses correspond to $\sigma = \pm 1$. Physically, this is a simple model: a $k$-th order Fourier feature drifts in phase-space over time, and the direction of the drift is unknown. Upon querying the signal, one does not know the phase-space direction along which the feature resides, akin to how the signal phase is generally unknown in AC sensing, so it is assumed to be drawn from a uniform prior, after which point subsequent queries see a constant-rate drift from the initial point.

Since $|h_{k,\theta}|\leq 2$, these are valid densities for $\eta\leq 1/4$. Moreover, for every fixed time $t$,
\begin{equation}
    \int_0^{2\pi}\frac{d\Phi}{2\pi}P_{\sigma,\Phi,t}^{(k)}(x,p)=g(x,p) \ .
\end{equation}
Thus, every one-time displacement channel is identical under $\sigma=+1$ and $\sigma=-1$. The class label is only present in multi-time Fourier coefficients.

Choose two times \(t_0,t_1\) such that \(\Omega_0(t_1-t_0)=\pi/2\). We now compute the separation in the two-point coefficient. For any \(\theta\), define
\begin{equation}
    m_k(\theta)=
    \mathbb{E}_{(X,P)\sim g(1+\eta h_{k,\theta})}
    \left[e^{ikX}\right] \ .
\end{equation}
A direct Gaussian integral gives
\begin{equation}
    m_k(\theta)=c_k+\eta\left(a_k e^{-i\theta}+b_k e^{i\theta}\right)
\end{equation}
where
\begin{equation}
    a_k=\frac{1-e^{-k^2}}{2}, \qquad b_k=\frac{e^{-2k^2}-e^{-k^2}}{2} \ .
\end{equation}
Therefore
\begin{align}
    \Theta_k^{(\sigma)}(t_0,t_1)
    &=\mathbb{E}_{\Phi}\left[
        m_k(\Phi+\sigma\Omega_0 t_0)
        \overline{m_k(\Phi+\sigma\Omega_0 t_1)}
    \right] \\
    &=c_k^2+\eta^2\left(a_k^2 e^{i\sigma\Omega_0(t_1-t_0)}+b_k^2 e^{-i\sigma\Omega_0(t_1-t_0)}\right) \ .
\end{align}
Taking imaginary parts and using \(\Omega_0(t_1-t_0)=\pi/2\), we obtain
\begin{equation}
    \Im\Theta_k^{(\sigma)}(t_0,t_1)=\sigma\eta^2(a_k^2-b_k^2) \ .
\end{equation}
Finally,
\begin{equation}
    a_k^2-b_k^2=
    \frac{(1-e^{-2k^2})(1-e^{-k^2})^2}{4} \ .
\end{equation}
Thus, for $k\geq 1$, $a_k^2-b_k^2\geq \frac{1}{16}$. The two hypotheses therefore differ by a constant amount in the temporal Fourier coefficient \(\Im\Theta_k(t_0,t_1)\). As such, any protocol which can learn $\Theta_k(t_0, t_1)$ for a time-varying signal to acccuracy at least $\epsilon = \eta^2/32$ can be used to distinguish the two hypotheses, so the temporal Fourier analysis truly reduces.

\begin{theorem}[Hardness of temporal Fourier learning without memory]\label{thm:temporal_fourier_no_memory_lower}
    For any integer $k \geq 1$, any conventional Gaussian protocol with energy $k$ per experiment requires at least
    \begin{equation}
        S\geq
        \frac{c_p}{\eta\left(\exp\left(-\frac{k^2}{16k+10}\right)+\exp\left(-\frac{k^2}{2}\right)\right)} = \exp(\Omega(k))
    \end{equation}
     signal queries to successfully distinguish the two drift directions with success probability at least $p > 1/2$, where \(c_p>0\) depends only on $p$. 
\end{theorem}

\begin{proof}
    We first recall the one-experiment response bound from the proof of Theorem \ref{thm:conventional_hardness_formal}. Fix any destructive Gaussian experiment with energy at most \(k\), chosen using arbitrary previous classical advice. By Lemma \ref{lemma:Gaussian_experiment_form}, the outcome distribution conditioned on a deterministic displacement \(z=(x,p)\) has the form \(W\sim\mathcal{N}(Az+b,C)\). Define
    \begin{equation}
        q_0(w)=\int dz \ \varphi_C(w-Az-b)g(z)
    \end{equation}
    and
    \begin{equation}
        r_{\theta}(w)=\int dz \ \varphi_C(w-Az-b)g(z)h_{k,\theta}(z) \ .
    \end{equation}
    The same conditional-Gaussian calculation used in Theorem \ref{thm:conventional_hardness_formal}, with \(\cos(kx+\theta)\) in place of \(\cos(kx)\), gives the uniform bound $\|r_{\theta}\|_1\leq \varepsilon_k$ for every $\theta$, where
    \begin{equation}
        \varepsilon_k=
        \exp\left(-\frac{k^2}{16k+10}\right)+\exp\left(-\frac{k^2}{2}\right) \ .
    \end{equation}
    Now consider an arbitrary adaptive conventional protocol making $S$ signal queries to the time-dependent signal. Before query $\ell$, the protocol has a classical transcript $w_{<\ell}$, chooses a time $t_\ell$, and chooses a Gaussian experiment with energy at most $k$. Conditional on the phase $\Phi$ and hypothesis $\sigma$, the conditional density of the next outcome $w_\ell$ can be written as
    \begin{equation}
        q_{\ell,0}(w_\ell)+\eta r_{\ell,\theta_\sigma(t)}(w_\ell) \ ,
    \end{equation}
    where $\theta_\sigma(t) = \Phi+\sigma\Omega_0 t_\ell$. Hence, the full transcript density under hypothesis \(\sigma\) is
    \begin{equation}
        Q_{\sigma}(w_1,\ldots,w_S)=
        \int_0^{2\pi}\frac{d\Phi}{2\pi}
        \prod_{\ell=1}^S
        \left(q_{\ell,0}(w_\ell)+\eta r_{\ell,\theta_\sigma(t)}(w_\ell)(w_\ell)\right) \ .
    \end{equation}
    Expanding this product, the $q_{\ell, 0}^S$ term is independent of $\sigma$. All terms with exactly one factor of $r_{\ell,\theta_\sigma(t)}$ vanish after averaging over $\Phi$, since \(\int_0^{2\pi}h_{k,\Phi+\theta}(x,p)d\Phi=0\) pointwise. Thus the difference \(Q_+-Q_-\) only contains terms with at least two response factors.

    Taking the $L_1$ norm and integrating sequentially over the transcript variables $w_1...w_S$, each response factor contributes at most \(\varepsilon_k\), while each $q_{\ell, 0}$ density integrates to one. Therefore, whenever $S\eta\varepsilon_k\leq 1$,
    \begin{align}
        \|Q_+-Q_-\|_1
        &\leq 2\sum_{r=2}^S {S\choose r}(\eta\varepsilon_k)^r \\
        &\leq 4S^2\eta^2\varepsilon_k^2 \ .
    \end{align}
    As such we obtain $d_{\rm TV}(Q_+,Q_-)\leq 2S^2\eta^2\varepsilon_k^2$. If \(S\eta\varepsilon_k\) is not bounded by a sufficiently small constant, then the desired exponential lower bound is already true. Thus, Le Cam's two-point method implies that any protocol succeeding with probability at least $p>1/2$ must satisfy $S\geq \frac{c_p}{\eta\varepsilon_k}$ for a constant $c_p>0$ depending only on $p$. Substituting the expression for \(\varepsilon_k\) gives the stated bound. 
\end{proof}

This result establishes an exponential lower bound for two-point temporal Fourier coefficients, where the bound is exponential in the Fourier frequency. One can also instantiate a separation that is exponential in the temporal order of the correlation itself; we consider this next.

\subsubsection{Constant-frequency, $m$-point correlation estimation}

We now give a second time-dependent instance which realizes a different memory-based separation. The previous example was exponential in the Fourier frequency of a two-point temporal coefficient. Here, the local Fourier frequency is fixed once and for all, and the exponential hardness comes from the order of the temporal correlation itself. To avoid overloading notation, let \(m\) denote the number of time points in this construction.

The lower bound below applies to a broader memoryless class than the Gaussian class used above. We allow arbitrary probe states, arbitrary ancillas, and arbitrary destructive measurements, but we impose a constant energy constraint on the signal-coupled mode in each destructive experiment. The only forbidden operation is the retention of a quantum system from one time point to the next. Thus the protocol may use arbitrary classical adaptivity, but it has no coherent memory. 

Before proceeding with the bound, we make an important conceptual remark. Our proof here will utilize a parity-based construction inspired by Ref.~\cite{cotler2026quantumadvantagesensingproperties}; however, the context in which we do so is substantially different. In Ref.~\cite{cotler2026quantumadvantagesensingproperties}, the parity-based construction was presented as an example that without careful accounting of energetic budgets, one could prove quantum separations that are not physically meaningful to real experiments, in part because the parity-based hypothesis test hid an unphysical correlation in a high moment of a classical signal that could be accessed with higher-energy conventional probes. Our coherent-probe bounds should be viewed as a more physically meaningful form of the idea that quantum probes can outperform classical ones. Moreover, our next lower bound will use the parity construction to realize a  lower bound in a much more physically meaningful context: by bounding against protocols that can even use arbitrary quantum control resources in each measurement while carefully tracking energy budgets, this result isolates that quantum memory can be a truly indispensable sensing resource for time-varying signals. As such, in this case, realizations of an exponential lower bound that are more physically well-motivated than hidden parity correlation are likely possible, but our bound is a proof-of-concept of the operating principle that constant-sized quantum memory can yield exponential quantum advantages in learning classical signals. 

Let $\mathcal{M}$ denote any protocol for temporal Fourier analysis. We use the convention that a single query corresponds to a pass of the signal for all times in $\mathbf{t}$.

\begin{graydefinitionbox}
\begin{theorem}[Hardness of constant-frequency \(m\)-point temporal correlations without memory]\label{thm:constant_frequency_parity_memory_lower}
    Fix constants $0<\epsilon, \delta<1/2$ and $E_0 > 0$. For every $m \geq 1$ and any $E_0$ there is a constant $C(E_0)$ such that any memoryless protocol $\mathcal{M}$ with energy $E_0$ per experiment requires at least
    \begin{equation}
        S\geq 2^m\frac{1-2\delta}{\epsilon}
    \end{equation}
    queries to the signal to learn an $m$-point temporal Fourier coefficient of frequency $C$ at each point in time, to absolute precision $\epsilon$ and with success probability at least $1-\delta$. In particular, this is $\exp(\Omega(m))$ for any constant $\epsilon, \delta$.
\end{theorem}
\end{graydefinitionbox}
\begin{proof}
    We begin by defining the appropriate hypothesis testing instance. Fix a constant energy bound $E_0$ and let $\nu = 1/2$, and define
    \begin{equation}
    x_\star=\frac{\nu}{\sqrt{2E_0+1}} \ , \qquad 
    \lambda_\star=\frac{\pi}{x_\star}\,.
    \end{equation}
    Then let $\zeta_*$ be the constant phase-space frequency such that $\Omega(\zeta_\star,(x,p))=\lambda_\star x$, independent of $m$. Let $t_1,\ldots,t_m$ be $m$ designated sensing times. The $m$-point temporal Fourier coefficient we will consider is
    \begin{equation}
        \Xi_m(P)=\widehat P((t_1,\ldots,t_m),(\zeta_\star,\ldots,\zeta_\star))=
        \mathbb{E}_{P}\left[\exp\left(i\lambda_\star\sum_{j=1}^m X_{t_j}\right)\right] \ .
    \end{equation}
    The hypothesis testing problem is as follows. In each realization of the time-dependent signal, draw a bit string \(B=(B_1,\ldots,B_m)\in\{0,1\}^m\). Under the hypothesis \(\sigma\in\{+1,-1\}\), the bit string has distribution
    \begin{equation}
        \Pr_{\sigma}(B=b)=2^{-m}\left(1+\sigma\epsilon(-1)^{b_1+\cdots+b_m}\right)
    \end{equation}
    where \(\gamma\in(0,1]\) is a fixed constant. Conditioned on the bit string, the signal displacement at time \(t_j\) is $Z_{t_j}=B_j(x_\star,0)$. The distributions above are valid since
    \begin{equation}
        \sum_{b\in\{0,1\}^m}(-1)^{b_1+\cdots+b_m}=0
    \end{equation}
    and $0\leq \epsilon \leq 1$. For this instance,
    \begin{align}
        \Xi_m(P_\sigma) &=
        \mathbb{E}_{\sigma}\left[\exp\left(i\lambda_\star x_\star\sum_{j=1}^m B_j\right)\right] \\
        &=
        \mathbb{E}_{\sigma}\left[(-1)^{B_1+\cdots+B_m}\right] \\
        &= \sigma\epsilon \ .
    \end{align}
    Therefore the two hypotheses differ by $2\epsilon$ in an $m$-point temporal Fourier coefficient; any protocol which can learn an $m$-point temporal Fourier coefficient (with precision parameters $\epsilon, \delta$) can solve the hypothesis testing problem with success probability at least $1-\delta$. As such a lower bound on the query complexity of this hypothesis testing instance immediately lifts to the main claim. We say that under $P_+$, the above $m$-point temporal Fourier coefficient  has expectation $+\epsilon$, while under $P_-$, it has expectation $-\epsilon$. We now prove that any memoryless finite-energy protocol obtains only exponentially small total variation distance from one realization of the $m$-time process. Consider one destructive signal query at a fixed time \(t_j\), and condition on an arbitrary previous classical transcript. The protocol may choose any joint probe-ancilla state \(\rho\), with mean photon number at most $E_0$ in the sensor mode, and may perform any POVM after the signal displacement. If $B_j=0$, the post-signal state is $\rho$. If $B_j=1$, the post-signal state is
    \begin{equation}
        \rho_1=D((x_\star,0))\rho D((x_\star,0))^\dagger \ .
    \end{equation}
    Let \(K_0\) and \(K_1\) denote the resulting classical outcome distributions under \(B_j=0\) and \(B_j=1\). By Lemma \ref{lemma:state_testing_TV_trace_distance_bound},
    \begin{equation}
        d_{\rm TV}(K_0,K_1)\leq \frac{1}{2}\|\rho-D((x_\star,0))\rho D((x_\star,0))^\dagger\|_1 \ .
    \end{equation}
    We now bound the right-hand side using the energy constraint. Since \(D((x_\star,0))=\exp(-ix_\star \hat p)\), for any pure state \(\ket{\psi}\),
    \begin{align}
        \frac{1}{2}\left\|\ketbra{\psi}{\psi}-e^{-ix_\star \hat p}\ketbra{\psi}{\psi}e^{ix_\star \hat p}\right\|_1
        &=\sqrt{1-\left|\bra{\psi}e^{-ix_\star \hat p}\ket{\psi}\right|^2} \\
        &\leq \left\|(I-e^{-ix_\star \hat p})\ket{\psi}\right\| \\
        &\leq x_\star\sqrt{\bra{\psi}\hat p^2\ket{\psi}} \ .
    \end{align}
    The last inequality follows from the spectral theorem applied to \(\hat p\). For a mixed state, we can apply this bound to any pure-state decomposition of \(\rho\) and use convexity of the trace norm. Thus
    \begin{equation}
        \frac{1}{2}\|\rho-D((x_\star,0))\rho D((x_\star,0))^\dagger\|_1
        \leq x_\star\sqrt{\Tr(\rho \,\hat p^2)} \ .
    \end{equation}
    The energy constraint is $\Tr(\rho \hat{n}) \leq E_0$, and 
    \begin{equation}
        \Tr(\rho \hat{n}) = \frac{1}{2}\Tr(\rho(\hat{x}^2 + \hat{p}^2 - 1)) \ .
    \end{equation}
    As such we have
    \begin{equation}
        \Tr(\rho\hat{p}^2) \leq 2\Tr(\rho \hat{n}) + 1 \leq 2E_0 + 1 \ .
    \end{equation}
    Therefore
    \begin{equation}
        d_{\rm TV}(K_0,K_1)\leq x_\star\sqrt{2E_0+1}=\nu \ .
    \end{equation}
    This bound holds for every possible previous transcript and every adaptively chosen state and measurement. Now consider one full realization of the $m$-time signal. Let $Y_j$ denote the classical outcome obtained at time $t_j$, and let $Y_{<j}=(Y_1,\ldots,Y_{j-1})$. Conditioned on a history \(y_{<j}\), the protocol chooses a particular measurement protocol which results in outcome $Y_j$. Since the only input parameter of measurement $j$ is the sign $B_j$, an adaptive experiment is a classical map from $B_j$ to the distribution of $Y_j$ conditioned on $y_{<j}$:
    \begin{equation}
        K_{b}^{y_{<j}}(dy_j) = \Pr(Y_j\in dy_j| B_j = b, \textnormal{previous outcomes }y_{<j})
    \end{equation}
    Said another way, $K_{b}^{y_{<j}}$ is the density of outcome $Y_j$ conditioned on the last $j-1$ classical outcomes and the unknown bit $B_j$ being $b$. Next, define the average and difference signed kernels
    \begin{equation}
        A_j^{y_{<j}}=\frac{1}{2}\left(K_{0}^{y_{<j}}+K_{1}^{y_{<j}}\right), \qquad \Delta_j^{y_{<j}}=\frac{1}{2}\left(K_{0}^{y_{<j}}-K_{1}^{y_{<j}}\right) \ .
    \end{equation}
    These allow us to rewrite  \(K_{b}^{y_{<j}}=A_j^{y_{<j}}+(-1)^b\Delta_j^{y_{<j}}\), \(A_j^{y_{<j}}\), so that the difference in outcome densities from any step in an adaptive experiment under $B_j = +1$ vs. $-1$ is entirely encoded in $\Delta_j^{y_{<j}}$. Then we already have the following $L_1$-bound, because the $L_1$ norm of $\Delta_j^{y_{<j}}$ is exactly the total variation between $K_0$ and $K_1$ conditioned on a history:
    \begin{equation}
        \|\Delta_j^{y_{<j}}\|_1=d_{\rm TV}(K_{0}^{y_{<j}},K_{1}^{y_{<j}})\leq \nu
    \end{equation}
    for every \(j\) and every history \(y_{<j}\). Now let \(Q_\sigma\) be the distribution of the full transcript \(Y=(Y_1,\ldots,Y_m)\) obtained from one realization of the signal under hypothesis \(\sigma\). For a fixed bitstring $b_1b_2...b_m$, the outcome density is $\prod_{j=1}^m K_{b_j}^{y_{<j}}$. Averaging over randomness in the bitstring by the law of total probability,
    \begin{equation}
        Q_\sigma(dy_1,\ldots,dy_m)=2^{-m}\sum_{b\in\{0,1\}^m}\left(1+\sigma\epsilon(-1)^{b_1+\cdots+b_m}\right)\prod_{j=1}^m K_{j,b_j}^{y_{<j}}(dy_j) \ .
    \end{equation}
    Now we consider the composite densities under hypothesis $\sigma = +1$ vs. $-1$. Then we have
    \begin{equation}
        Q_+-Q_-=2\epsilon \cdot 2^{-m}\sum_{b\in\{0,1\}^m}(-1)^{b_1+\cdots+b_m}\prod_{j=1}^m K_{b_j}^{y_{<j}} \ .
    \end{equation}
    Substituting \(K_{b_j}^{y_{<j}}=A_j^{y_{<j}}+(-1)^{b_j}\Delta_j^{y_{<j}}\),
    \begin{align}
        \sum_{b\in\{0,1\}^m}(-1)^{b_1+\cdots+b_m}\prod_{j=1}^m K_{b_j}^{y_{<j}} &= \sum_{b\in\{0,1\}^{m-1}}(-1)^{b_1+\cdots + b_{m-1}}\left[(A_m^{y_{<m}}+\Delta_m^{y_{<m}})-(A_m^{y_{<m}}-\Delta_m^{y_{<m}})\right]\prod_{j=1}^{m-1} K_{b_j}^{y_{<j}}  \\
        &= 2\Delta_m^{<y_m}\sum_{b\in\{0,1\}^{m-1}}(-1)^{b_1+\cdots + b_{m-1}}\prod_{j=1}^{m-1} K_{b_j}^{y_{<j}} \\
        &\textnormal{repeating, }= 2^m \Delta_m^{<y_m}\Delta_{m-1}^{<y_{m=1}} \cdots \Delta_1
    \end{align}
    This leaves us with
    \begin{equation}
        Q_+-Q_-=2\epsilon \Delta_1(dy_1)\Delta_2^{y_1}(dy_2)\cdots \Delta_m^{y_{<m}}(dy_m) \ .
    \end{equation}
    Taking total variation and integrating sequentially,
    \begin{align}
        \|Q_+-Q_-\|_1
        &\leq 2\epsilon\int |\Delta_1|(dy_1)\int |\Delta_2^{y_1}|(dy_2)\cdots \int |\Delta_m^{y_{<m}}|(dy_m) \\
        &\leq 2\epsilon\nu^m \ .
    \end{align}
    Therefore one realization of the \(m\)-time signal yields $d_{\rm TV}(Q_+,Q_-)\leq \epsilon\,\nu^m$.

    The same bound holds conditionally on any previous realization transcripts, since the one-time response bound was uniform over all classical advice. Applying the adaptive hybrid argument of Lemma \ref{lemma:adaptive_tv_hybrid} over \(S\) independent realizations gives
    \begin{equation}
        d_{\rm TV}(Q_+^{(S)},Q_-^{(S)})\leq S\epsilon\,\nu^m \ .
    \end{equation}
    By Lemma \ref{lemma:le_cam}, any memoryless protocol using \(S\) realizations succeeds in distinguishing the two hypotheses with probability at most $\frac{1}{2}+\frac{S\gamma\nu^m}{2}$. Therefore any protocol with success probability at least $p>1/2$ must satisfy
    \begin{equation}
        S\geq \frac{2p-1}{\epsilon\,\nu^m} \ .
    \end{equation}
    Substituting $\nu = 1/2$ and $\delta = 1-p$ and lifting to a learning lower bound proves the claim.
\end{proof}

This theorem emphasizes that coherent memory-enabled quantum processing of quantum data can exponentially outperform measurement and classical postprocessing when the target is estimating temporal correlations, even if the frequency of those correlations is small. The corresponding upper bound using single-qubit memory is simple and will be given in Section \ref{sec:advantage_upper_bounds}. 

\begin{remark}
    The two mechanisms for memory-based separations are independent. Conventional protocols can struggle to estimate high-frequency temporal Fourier coefficients for the same reason that they struggle to estimate static Fourier coefficients: high-frequency response is exponentially small with Gaussian protocols, whether it lives in a single or multiple time-slices. Memoryless protocols, on the other hand, struggle to measure multi-point temporal correlations because classical postprocessing can exponentially underperform coherent quantum processing before destructive readout, independent of the frequency. In principle, estimating a frequency-$k$, $m$-point temporal Fourier coefficient results in the above bounds stacking, yielding an asymptotic lower bound of $\exp(\Omega(mk))$ for conventional sensing protocols.
\end{remark}

\subsection{Lower bound for infinite-energy conventional quantum sensing} \label{sec:gaussian_testing_lower_bound}
The previous sections established exponential energy-aware lower bounds for static and time-varying Fourier analysis tasks. A natural question is whether provable separations for single-mode sensing persist even if we remove all energy constraints and consider arbitrarily powerful Gaussian measurements. Conceptually, this would illustrate that the non-Gaussian sensing capabilities that become cheap upon obtaining single-qubit control can confer quantum advantages that cannot be replicated by any, arbitrarily high-quality conventional experiment.

In this section, we describe a task for which such a separation is possible. In particular, this task is concerned with estimating the variance of a single-mode quantum state. This is physically natural, as the broader problem of learning observables of Gaussian states is precisely what one is faced with upon sensing classical fields with Gaussian probes. We prove a quadratic lower bound for variance estimation with unconstrained Gaussian measurements. In Section \ref{sec:Gaussian_testing_upper_bound} we will then demonstrate a non-Gaussian strategy using a constant-depth qubit-enabled Ramsey protocol to estimate the same variance using linear samples, exhibiting a quadratic advantage.

This lower bound applies directly to the task of estimating the phase-space variance of a quantum state.

\begin{definition}[Variance estimation]
Let $\hat{R}=(\hat{x},\hat{p})^T$ denote the quadrature vector of a single-mode quantum state $\rho$, and let
\begin{equation}
V_{ij}(\rho)=\frac{1}{2}\Tr\!\left(\rho\{\hat{R}_i-\langle \hat{R}_i\rangle_\rho,\hat{R}_j-\langle \hat{R}_j\rangle_\rho\}\right)
\end{equation}
denote its covariance matrix. The variance of $\rho$ is
\begin{equation}
v(\rho)=\frac{1}{2}\Tr V(\rho)=\frac{1}{2}\left(\operatorname{Var}_\rho(\hat{x})+\operatorname{Var}_\rho(\hat{p})\right) \ .
\end{equation}
The $k,\delta$-variance estimation task is to estimate $v(\rho)$ for any single-mode state to within absolute error $1/(2k)$ with success probability at least $1-\delta$.
\end{definition}
Variance estimation is often a core task in sensing stochastic signals; for instance, an important task for the detection of axionic dark matter is to estimate small, narrowband power increases in a broad frequency band, and the signal encountered in each frequency bin is very well-modeled by a Gaussian displacement channel \cite{Gardner_2025}. A signal detection amounts to accurate certification of a slight broadening in one particular bin. As such, the quadratic separation we present here applies directly to this setting and provides rigorous evidence for the large empirical speedups observed in \cite{lamoreaux2013analysis, dixit2021searching} using non-Gaussian photon-counting measurements.

We prove the following.
\begin{theorem} \label{thm:Gaussian_testing_lower_bound}
Any adaptive Gaussian protocol which solves the variance estimation task for constant $0<\delta<1/2$ requires $\Omega(k^2)$ samples.
\end{theorem}
\begin{proof}
We again proceed by reduction to hypothesis testing. Let $\tau_t$ be the single-mode thermal state with mean photon number $t$ given by
\begin{equation}
\tau_t = \sum_{n=0}^\infty \frac{t^n}{(1+t)^{n+1}}\ketbra{n}{n} \ .
\end{equation}
Recall that this has covariance matrix $(1+2t)I_2$. We will consider the following two hypotheses:
\begin{equation}
\rho_0^{(k)} = \ketbra{0}{0}, \qquad \rho_1^{(k)} = \tau_{1/k} \ .
\end{equation}
Thus one hypothesis is simply the vacuum state, while the other is a thermal state whose mean photon number is $1/k$. Both states are Gaussian and have nonnegative Wigner functions. In the convention where the vacuum covariance is $I_2$, their Wigner functions are
\begin{equation}
W_0(z)=\frac{1}{2\pi}\exp\!\left(-\frac{|z|^2}{2}\right)
\end{equation}
and
\begin{equation}
W_1(z) = \frac{1}{2\pi(1+2/k)}\exp\!\left(-\frac{|z|^2}{2(1+2/k)}\right) \ .
\end{equation}
The corresponding phase-space variances are
\begin{equation}
v(\rho_0^{(k)})=1, \qquad v(\rho_1^{(k)})=1+\frac{2}{k} \ .
\end{equation}
As such, any algorithm which can estimate the variance to within accuracy $1/(2k)$ can solve this distinguishing task by thresholding its estimate at $1+1/k$.

We now proceed with the sample lower bound on the hypothesis testing problem. We will prove the stronger statement that the lower bound holds for any adaptive protocol whose measurements are Wigner-positive. This includes adaptive Gaussian measurement protocols. The reason is that a Wigner-positive measurement maps the Wigner function of the measured state to a classical outcome distribution by a stochastic map on phase-space (see Equation \eqref{eq:wigner_positive_transcript}). Thus, even if the measurement at time $t$ is chosen adaptively from all previous outcomes, the entire transcript distribution obtained from $N$ copies is a classical stochastic postprocessing of the product Wigner distribution $W_i^{\otimes N}$, where $i\in\{0,1\}$ labels the hypothesis.

Let $P_0^{(N)}$ and $P_1^{(N)}$ denote the transcript distributions induced by an arbitrary adaptive Wigner-positive protocol using $N$ copies. By data processing for total variation distance, 
\begin{equation}
    d_{\rm TV}\left(P_0^{(N)},P_1^{(N)}\right) \leq \frac{1}{2}\left\|W_0^{\otimes N}-W_1^{\otimes N}\right\|_1 \ .
\end{equation}
We now bound this $L^1$ distance using the usual KL step. Pinsker's inequality and additivity of KL divergence give
\begin{align}
    d_{\rm TV}\left(P_0^{(N)},P_1^{(N)}\right) &\leq \sqrt{\frac{1}{2}D_{KL}\left(W_0^{\otimes N}\|W_1^{\otimes N}\right)} \\
    &= \sqrt{\frac{N}{2}D_{KL}\left(W_0\|W_1\right)} \ .
\end{align}
Let $a=1+\frac{2}{k}$. Since $W_0$ and $W_1$ are centered two-dimensional Gaussians with covariance matrices $I_2$ and $aI_2$, respectively, we have by Lemma \ref{lemma:Gaussian_KL} that
\begin{equation}
    D_{KL}\left(W_0\|W_1\right) = \log a + \frac{1}{a}-1 \ .
\end{equation}
Setting $x=2/k$, this is
\begin{equation}
    D_{KL}\left(W_0\|W_1\right) = \log(1+x)+\frac{1}{1+x}-1 \ .
\end{equation}
For $k\geq 2$, $x\leq 1$, and we can use
\begin{equation}
    \log(1+x)+\frac{1}{1+x}-1 \leq x^2 \ .
\end{equation}
Therefore
\begin{equation}
    D_{KL}\left(W_0\|W_1\right) \leq \frac{4}{k^2} \ .
\end{equation}
It follows that
\begin{equation}
    d_{\rm TV}\left(P_0^{(N)},P_1^{(N)}\right) \leq O\!\left(\frac{\sqrt{N}}{k}\right) \ .
\end{equation}
Hence if $N=o(k^2)$, the total variation distance between the two transcript distributions is $o(1)$; by Lemma \ref{lemma:le_cam}, no such protocol can distinguish the two hypotheses with nonvanishing success bias. Thus any adaptive Gaussian protocol requires
\begin{equation}
    N \geq \Omega(k^2) \ .
\end{equation}
\end{proof}

\subsection{Multimode hardness of conventional characteristic function learning} \label{sec:multimode_charfun_lower_bound}
We have just established a single-mode learning task for which arbitrarily high-energy Gaussian protocols face a quadratic lower bound, while we later show a non-Gaussian strategy, enabled by single-qubit control, which succeeds using linear samples. We next discuss how such energy-independent separations between Gaussian and non-Gaussian can appear in \textit{multimode} learning tasks in more dramatic ways. 

In particular, our discussion here concerns results obtained by Ref.~\cite{Oh_2024}, which studied an entanglement-enabled quantum advantage for learning the characteristic function of a displacement channel. This task is rather different (and more difficult) than our Fourier analysis setting despite its cosmetic similarity, because it requires collecting a dataset that allows the learner to post-hoc evaluate the channel's characteristic function at any point within a phase-space radius. Rather than the physically-motivated property estimation tasks we have considered thus far, characteristic function learning is equivalent to learning a complete description of the channel itself, as one might do in full quantum tomography of states or processes. The core observation of Ref.~\cite{Oh_2024} is that due to the special structure of displacement channels as the continuous-variable analog of qubit Pauli channels, measurement with entangled EPR pairs enables an efficient learning protocol. Ref.~\cite{Oh_2024} then demonstrates that all entanglement-free strategies (i.e. those without ancillary quantum memory) must, in the worst case, utilize $\exp(\Omega(n))$ queries to learn an $n$-mode characteristic function of a displacement channel.

Here, we establish that an exponential separation enabled by single-qubit control extends to multimode sensing problems by showing that, surprisingly, the characteristic function learning problem can be efficiently solved by a polynomial-energy, entanglement-free strategy that utilizes single-qubit control of an $n$-mode sensor. The protocol that enables this advantage will be presented in Section \ref{sec:char_function_learning_gkp} by realizing another class of non-Gaussian measurements; in this section, we will restate the exponential lower bounds proven in Ref.~\cite{Oh_2024} and understand why, in light of their entanglement-free lower bounds, an exponential advantage is even possible without entangled bosonic probes.

We now restate the relevant setting and lower bounds in a form that will be useful for comparison with our single-qubit non-Gaussian protocol in Section \ref{sec:char_function_learning_gkp}. For an $n$-mode displacement channel
\begin{equation}
    \mathcal{E}_P(\rho)=\int d^{2n}\alpha \ P(\alpha)D(\alpha)\rho D(\alpha)^\dagger \ ,
\end{equation}
we write its characteristic function as
\begin{equation}
    \lambda_P(\beta)=\mathbb{E}_{\alpha\sim P}\left[e^{i\Omega(\beta,\alpha)}\right] \ .
\end{equation}
The post-hoc characteristic-function learning task is the following.

\begin{graydefinitionbox}
\begin{definition}[Post-hoc characteristic-function learning]
Fix $\kappa,\epsilon,\delta>0$. A protocol learns the $n$-mode displacement channel $\mathcal{E}_P$ on the radius-$\kappa n$ ball if, after making $N$ channel queries and storing only its classical transcript, it receives an arbitrary query $\beta\in\mathbb{R}^{2n}$ satisfying $|\beta|^2\leq \kappa n$ and outputs an estimate $\tilde{\lambda}(\beta)$ such that
\begin{equation}
    \Pr\left[\left|\tilde{\lambda}(\beta)-\lambda_P(\beta)\right|\leq \epsilon\right]\geq 1-\delta \ .
\end{equation}
\end{definition}
\end{graydefinitionbox}

The hard family used by Ref.~\cite{Oh_2024} to instantiate a hypothesis-testing lower bound is a three-peak family in characteristic-function space. In our phase-space convention, a representative form is
\begin{equation}
    \lambda_{s,\gamma}(\beta)
    =
    e^{-|\beta|^2/(2\sigma^2)}
    +2is\eta_0e^{-|\beta-\gamma|^2/(2\sigma^2)}
    -2is\eta_0e^{-|\beta+\gamma|^2/(2\sigma^2)} \ ,
\end{equation}
where $s\in\{\pm1\}$, $\gamma\in\mathbb{R}^{2n}$, and $\eta_0$ is a sufficiently small constant. The corresponding displacement density (i.e. the symplectic Fourier transform of the above characteristic function) has the form
\begin{equation}
    P_{s,\gamma}(\alpha)
    =
    \left(\frac{2\sigma^2}{\pi}\right)^n e^{-2\sigma^2|\alpha|^2}
    \left[1+4s\eta_0\sin\left(2\Omega(\gamma,\alpha)\right)\right] \ .
\end{equation}
The real parameter $\sigma$ controls the width of the Gaussian envelope in the displacement distribution. Smaller $\sigma$ means broader displacement noise, while larger $\sigma$ means the channel usually applies smaller displacements.

The main lower bound of Ref.~\cite{Oh_2024} applies to all entanglement-free strategies, including arbitrary non-Gaussian input states and arbitrary non-Gaussian destructive measurements after each channel use. However, for the finite-energy three-peak family, this lower bound requires an additional small-$\sigma$ condition. Defining
\begin{equation}
    M_\kappa
    =
    \max\left\{
        1-1.98\kappa,
        0.99\kappa\left(\sqrt{1+(0.99\kappa)^{-2}}-1\right)
    \right\} \ ,
\end{equation}
Ref.~\cite{Oh_2024} proves that if $n\geq 8$, $\epsilon\leq 0.24$, and importantly, $2\sigma^2\leq M_\kappa$, then any entanglement-free protocol that learns the 3-peak hard family (with unknown $s, \gamma$) on the ball $|\beta|^2\leq \kappa n$ must use
\begin{equation}
    N
    \geq
    0.01\epsilon^{-2}
    \left(
        1+\frac{1.98\kappa}{1+2\sigma^2}
    \right)^n
\end{equation}
queries. This should be contrasted with the lower bound Ref.~\cite{Oh_2024} obtains for Gaussian entanglement-free strategies. If the input states and destructive measurements are Gaussian, then for the same three-peak family the following lower bound holds for every $\sigma>0$, with no small-$\sigma$ assumption:
\begin{equation}
    N
    \geq
    0.01\epsilon^{-2}
    \min\left\{
        \left(1+\frac{0.99\kappa}{\sigma^2}\right)^{n/2},
        \left(1+\frac{1.98\kappa}{1+2\sigma^2}\right)^n
    \right\} \ .
\end{equation}
In particular, for any fixed $\kappa>0$ and any constant $\sigma^2=O(1)$, Gaussian entanglement-free characteristic-function learning remains exponentially hard in $n$, even allowing arbitrarily high-energy Gaussian states and measurements. Substituting $\sigma^2=C\log n$ for a fixed constant $C$ into the Gaussian lower bound, we get
\begin{equation} \label{eq:characteristic_function_lb_Gaussian}
    N_{\rm Gauss}
    \geq
    \epsilon^{-2}\exp\left(\Omega_\kappa\left(\frac{n}{\log n}\right)\right) \ .
\end{equation}
Namely, for $\sigma$ of order $\log n$, a superpolynomial lower bound still persists in the Gaussian setting, whereas no entanglement-free lower bound is known to exist. It is not a priori clear whether this missing condition is merely an artifact of the proof method or whether it marks a real transition in the power of non-Gaussian entanglement-free measurements. Surprisingly, we show in Section \ref{sec:char_function_learning_gkp} that the latter is the correct interpretation: once the displacement distribution is narrow enough (i.e. $\sigma$ is of order $\log n$), a non-Gaussian probe prepared and measured using only single-qubit control of an $n$-mode probe state can convert each channel use into an approximate classical sample of the displacement. This gives an efficient entanglement-free learner in a regime where the general non-Gaussian lower bound no longer applies, while the Gaussian lower bound still gives a superpolynomial lower bound. This result marks a superpolynomial Gaussian vs. non-Gaussian learning separation in the multimode setting, enabled by only single-qubit control.

\subsubsection{Simplified entanglement-free lower bound on multimode channel learning through Q$\Psi$}
\label{sec:qes_reproof_multimode_displacement}
At the conclusion of Appendix \ref{sec:qes}, we noted that Q$\Psi$ is a broad framework that goes beyond existing techniques in quantum learning theory in its ability to incorporate realistic constraints and yield simultaneous upper and lower bounds. In fact, many of the existing separations in quantum learning theory can be shown to arise as special cases of Q$\Psi$, and often in a far simpler manner. Here, we provide a reproof of one such canonical result, both as a worked example of Q$\Psi$-style proofs and a demonstration of how Q$\Psi$ can substantially simplify otherwise difficult separations. 

We consider the lower bound of Ref.~\cite{Oh_2024}, which proved that any entanglement-free strategy suffers an exponential query complexity to learn the full characteristic function of a worst-case instance of a multimode displacement channel under post-hoc queries. The original proof is complex and involves two overarching steps: first, relating the success probability of an arbitrary adaptive $N$-query protocol to a problem-specific information quantity via involved bosonic analysis calculations, and second, controlling the information quantity itself. The Q$\Psi$ approach greatly simplifies the former half of the proof, and it follows that the particular information quantity arises naturally as a bound on the AFI. 

We recall some notation from Ref.~\cite{Oh_2024}. Let $\gamma\in\mathbb R^{2n}$ be drawn from
\begin{equation}
    q_{\sigma_\gamma}(\gamma)=\left(\frac{1}{2\pi\sigma_\gamma^2}\right)^n\exp\left(-\frac{|\gamma|^2}{2\sigma_\gamma^2}\right).
\end{equation}
For a normalized $n$-mode state $\ket{A}$ and a nonzero vector $\ket{B}$, define
\begin{align}
    G^{A,B}(\beta)&=\frac{\bra{B}D(\beta)^\dagger\ket{B}}{\langle B|B\rangle}\bra{A}D(\beta)\ket{A}, \\
    G_\sigma^{A,B}(\gamma)&=\frac{\int d^{2n}\beta\ e^{-|\beta-\gamma|^2/(2\sigma^2)}G^{A,B}(\beta)}{\int d^{2n}\beta\ e^{-|\beta|^2/(2\sigma^2)}G^{A,B}(\beta)}.
\end{align}
We then take the following lemma which controls a problem-specific information quantity as given. While Q$\Psi$ does not simplify the proof of this lemma, we will show here that it simplifies the rest of the proof substantially.
\begin{lemma}[Lemma S1 of Ref.~\cite{Oh_2024}]
\label{lemma:oh_random_direction_overlap}
If
\begin{equation}
    \sigma^2\leq\max\left\{\frac{1}{2}-2\sigma_\gamma^2,\sigma_\gamma^2\left(\sqrt{1+\frac{1}{4\sigma_\gamma^4}}-1\right)\right\},
\end{equation}
then for every normalized $\ket{A}$ and nonzero $\ket{B}$,
\begin{equation}
    \mathbb E_\gamma\left[\left|G_\sigma^{A,B}(\gamma)\right|^2\right]\leq\left(\frac{1+2\sigma^2}{1+2\sigma^2+4\sigma_\gamma^2}\right)^n.
\end{equation}
\end{lemma}

\begin{graydefinitionbox}
\begin{theorem}[Entanglement-free lower bound of Ref.~\cite{Oh_2024}]
\label{thm:qes_reproof_entanglement_free_displacement}
Let $n\geq 8$, $\kappa>0$, and $0<\epsilon\leq 0.24$. Suppose an adaptive entanglement-free protocol makes $N$ queries to an arbitrary $n$-mode displacement channel, then receives any $\beta$ satisfying $|\beta|^2\leq\kappa n$, and estimates the channel characteristic function at $\beta$ to error $\epsilon$ with success probability at least $2/3$. Then
\begin{equation}
    N\geq 0.01\epsilon^{-2}(1+1.98\kappa)^n.
\end{equation}
\end{theorem}
\end{graydefinitionbox}

\begin{proof}
Set $2\sigma_\gamma^2=0.99\kappa$, and fix $\sigma>0$ small enough that Lemma \ref{lemma:oh_random_direction_overlap} applies and $\sigma^2\leq\sigma_\gamma^2$. In Q$\Psi$ language, consider the Gaussian reference background and Fourier perturbation
\begin{equation}
    P_0(d\alpha)=\left(\frac{2\sigma^2}{\pi}\right)^n e^{-2\sigma^2|\alpha|^2}d^{2n}\alpha, \qquad h_\gamma(\alpha)=\sin\left(2\Omega(\gamma,\alpha)\right).
\end{equation}
Let $\epsilon_0=\epsilon/0.98$ and $\eta=4\epsilon_0<1$. Since $h_\gamma$ is centered under $P_0$ and $\|h_\gamma\|_\infty\leq 1$, the laws
\begin{equation}
    P_{s,\gamma}(d\alpha)=\left(1+s\eta h_\gamma(\alpha)\right)P_0(d\alpha), \qquad s\in\{\pm1\},
\end{equation}
are valid probability distributions. Their characteristic functions are
\begin{align}
    \lambda_0(\beta)&=e^{-|\beta|^2/(2\sigma^2)}, \\
    \lambda_{s,\gamma}(\beta)&=\lambda_0(\beta)+2is\epsilon_0e^{-|\beta-\gamma|^2/(2\sigma^2)}-2is\epsilon_0e^{-|\beta+\gamma|^2/(2\sigma^2)},
\end{align}
illustrating that the Q$\Psi$ hypothesis-testing formulation yields the same characteristic function distinguishing task considered in \cite{Oh_2024}. Now consider one entanglement-free query to the hypothesis channel. Revealing a pure-state decomposition of the input and refining the final POVM to rank-one outcomes can only increase the available information \cite{chen2021exponential}, so it suffices to use a pure probe $\ket{A}$ and an outcome proportional to a projector $\ketbra{B_y}{B_y}$. Substituting the characteristic functions above into the channel Fourier representation gives
\begin{equation}
    Q_{s,\gamma}(dy)=\left(1+s\eta R_{M,\gamma}(y)\right)Q_0(dy), \qquad R_{M,\gamma}(y)=-\operatorname{Im}G_\sigma^{A,B_y}(\gamma),
\end{equation}
where $Q_0$ is the outcome law under $P_0$. Thus $R_{M,\gamma}$ is the Q$\Psi$ score for the direction $h_\gamma$, and
\begin{align}
    \mathbb E_\gamma\left[\mathsf A_M(h_\gamma;P_0)\right]&=\int \mathbb E_\gamma\left[|R_{M,\gamma}(y)|^2\right]dQ_0(y) \\
    &\leq\Lambda_n, \qquad \Lambda_n\coloneqq\left(\frac{1+2\sigma^2}{1+2\sigma^2+4\sigma_\gamma^2}\right)^n,
\end{align}
where the inequality is Lemma \ref{lemma:oh_random_direction_overlap}. We now use Lemma \ref{lemma:semiclassical_quantum_measurement} and obtain that any entanglement-free protocol which achieves constant success bias for this hypothesis test requires
\begin{equation}
    N \geq \Omega(1/\epsilon_0^2\Lambda_n) \ .
\end{equation}
It remains to reduce characteristic-function learning to this test. Sample $\gamma\sim q_{\sigma_\gamma}$ and a uniform sign $s$, give the learner either $\mathcal{E}_{P_0}$ or $\mathcal{E}_{P_{s,\gamma}}$ with equal prior probability, and reveal $\gamma$ only after all channel queries are complete. On the event
\begin{equation}
    \mathcal G=\left\{2\sigma^2<|\gamma|^2\leq\kappa n\right\},
\end{equation}
querying the learned characteristic function at $\beta=\gamma$ separates the null value from either signed alternative because
\begin{equation}
    \left|\lambda_0(\gamma)-\lambda_{s,\gamma}(\gamma)\right|=2\epsilon_0\left(1-e^{-2|\gamma|^2/\sigma^2}\right)>2\epsilon.
\end{equation}
The Gaussian tail estimate of Ref.~\cite{Oh_2024} gives $\Pr(\mathcal G)\geq 0.49987$ for $n\geq 8$ and $\sigma^2\leq\sigma_\gamma^2$. The binary testing identity therefore gives
\begin{equation}
    \mathbb E_\gamma d_{\rm TV}\left(Q_0^{(N)},Q_{{\rm mix},\gamma}^{(N)}\right)\geq\frac{1}{3}\Pr(\mathcal G)\geq 0.1666.
\end{equation}
Combining the two total-variation bounds and using $\epsilon=0.98\epsilon_0$ gives
\begin{equation}
    N\geq 0.01\epsilon^{-2}\left(1+\frac{4\sigma_\gamma^2}{1+2\sigma^2}\right)^n=0.01\epsilon^{-2}\left(1+\frac{1.98\kappa}{1+2\sigma^2}\right)^n.
\end{equation}
Taking $\sigma\to0$ proves the theorem.
\end{proof}
This gives a greatly simplified proof relative to the original result in S3 of \cite{Oh_2024}.

\subsection{A hierarchy of exponential sensing lower bounds with increased control depth} \label{sec:depth_lower_bounds}

The separations above show that adding a single controllable qubit to an otherwise conventional bosonic sensor can dramatically increase the Fourier bandwidth of accessible sensing tasks. A natural next question is whether increasing the coherent processing capability of that qubit leads to a genuine hierarchy of sensing power. 

Proving such a separation presents an important technical challenge. Previously, most query-complexity lower bounds in quantum learning and sensing have operated in highly unconstrained settings such as allowing arbitrary POVMs; in some cases, additional constraints such as quantum memory \cite{chen2021exponential} or noise \cite{cotler2026noisy} have been accounted for with substantially more complex proofs. Circuit depth lower bounds on learning have been similarly limited beyond studies of full state tomography \cite{Zhao_2024}. This is because unconstrained lower bounds only require the use of simple properties of the measurement class, such as positivity and normalization of arbitrary POVMs. When additional constraints are present, it often becomes intractable to control their effect on quantities like one-sided likelihood ratios or statistical divergence bounds, which form the core of known information-theoretic lower bound techniques.

In this section, we prove precisely this kind of hierarchy lower bound.

\begin{graydefinitionbox}
    \begin{theorem}[Hardness of learning with slightly shallower depths] \label{thm:interleaved_control_lb_formal}
    For any $k\in\mathbb{Z}^+$ and $\gamma < \log_3 2 \approx 0.63$, there exist functions $h_k(\alpha)$ such that any protocol of control depth $\leq \gamma k$ (as in Definition \ref{def:depth_r_interleaved_measurement}) requires $\exp(\Omega(k))$ queries to estimate $\mathbb{E}[h_{k}(\alpha)]$ for any signal to constant accuracy.
    \end{theorem}
\end{graydefinitionbox}

Our proof elucidates the utility of the Q$\Psi$ formalism for characterizing the power of constrained quantum measurements. This first part of this section is devoted to instantiating the appropriate family of polynomials $h_k$ that will appear in the Lemma \ref{lemma:semiclassical_quantum_measurement} hypothesis test. This will require introducing harmonic-analytic techniques that connect the control-depth hierarchy to the Fourier support picture. Then, we use these results to obtain bounds on the \textit{support size} of depth-$r$ response kernels so that we may use Corollary \ref{cor:sparse_semiclassical_quantum_measurement} to obtain the main result of this section.

\subsubsection{Fourier response of depth-$r$ measurements and hypothesis testing}
\label{sec:hypothesis_testing_golay_rudin_shapiro}

Our aim is to use the semiclassical measurement lemma to control the hardness of hypothesis testing with bounded-depth circuits. To do so, we must first characterize the Fourier decomposition of response functions in this class. To instantiate a control depth-dependent bound, we choose a concrete circuit model. As described in Section \ref{sec:sensing-hierarchy}, we work with circuits that interleave qubit rotations with echoed conditional displacement (ECD) gates, because these are precisely the class of circuits we use in our experiments. These circuits are universal and can efficiently implement powerful primitives such as quantum signal processing, as realized in Ref.~\cite{prabhu2026quantumcomputationaldisplacementsensing}. We further work with binary measurements, which is without loss of generality when classical postprocessing is allowed \cite{chen2021exponential}. These choices are not limiting: the Q$\Psi$ proof technique enables lower bounds for other circuit models as long as one can characterize the Fourier support of their response kernels.

\begin{definition}[Depth-$r$ measurement class]
\label{def:depth_r_interleaved_measurement}
Consider a joint system of a bosonic mode and a qubit supported on $\{\ket{g},\ket{e}\}$. A signal-interleaved depth-$r$ measurement is any binary measurement on this system implemented as follows. The protocol starts from a fixed oscillator-qubit state $\rho_0$, for instance $\ketbra{0,g}{0,g}$ for concreteness. It then applies a circuit of the form
\begin{equation}
    V_r(\alpha)=S_r(\alpha)G_rS_{r-1}(\alpha)G_{r-1}\cdots S_1(\alpha)G_1S_0(\alpha),
\end{equation}
where each signal window is
\begin{equation}
    S_\ell(\alpha)=D(t_\ell\alpha)\otimes I_q
\end{equation}
for a known $t_\ell\in\{0,1\}$, and each control layer is
\begin{equation}
    G_j=\mathrm{ECD}(\beta_j)(I\otimes R_j).
\end{equation}
Here $R_j$ is an arbitrary single-qubit $\textnormal{\textsf{SU}}(2)$ rotation and
\begin{equation}
    \mathrm{ECD}(\beta_j)=D(\beta_j/2)\otimes \ketbra{e}{g}+D(-\beta_j/2)\otimes \ketbra{g}{e}.
\end{equation}
After the final signal or control window, a binary POVM is performed on the qubit. We denote by $\mathcal M_r$ the family of all binary measurements obtainable in this way, allowing arbitrary ECD displacement and qubit rotation parameters at every layer and arbitrary choices of the known interleaving variables $t_0,\ldots,t_r$.
\end{definition}

This is a physically-motivated model in which a signal imparts a displacement $\alpha$ sampled from its underlying displacement during each sensing window, and we can apply fast interleaved control to our sensor using the operations available on our circuit-QED platform (Fig. \ref{fig:ecd_pulse}. Setting a particular $t_\ell$ to $0$ simply denotes applying multiple consecutive controls. The displacement parameter $\alpha$ from the signal is, of course, refreshed between measurements, as we take each sensing experiment to be an independent query of the signal. We remark that all subsequent arguments go through identically for the two-parameter gate $\textnormal{JC}(\beta_1,\beta_2)$ from Equation \eqref{eq:gate_defns}; in that case the role of $\beta_j$ below is played by the branch difference $\beta_{j,2}-\beta_{j,1}$. We work with the symmetric ECD gate for clarity. The following lemmas connect bounded control depth to support size in the Weyl basis.

\begin{lemma}[Weyl expansion of depth-$r$ sensing circuits]
\label{lemma:finite_weyl_expansion_control_depth}
Let $V_r(\alpha)$ be a signal-interleaved depth-$r$ circuit as in Definition \ref{def:depth_r_interleaved_measurement}. Then there are branch labels indexed by $s=(s_1,\ldots,s_r)\in\{\pm1\}^r$, qubit operators $A_s$, signal-independent phases $\theta_s$, and phase-space labels $\Gamma_s,\Lambda_s\in\mathbb R^2$ such that
\begin{equation}
    V_r(\alpha)=\sum_{s\in\{\pm1\}^r}e^{i\theta_s}e^{i\Omega(\Lambda_s,\alpha)}D(T\alpha+\Gamma_s)\otimes A_s,
\end{equation}
where $T=\sum_{\ell=0}^r t_\ell$, $\Gamma_s=\frac{1}{2}\sum_{j=1}^r s_j\beta_j$, and
\begin{equation}
    \Lambda_s=\frac{1}{4}\sum_{j=1}^r s_j\left(T_{\ge j}-T_{<j}\right)\beta_j.
\end{equation}
Here
\begin{equation}
    T_{<j}=\sum_{\ell=0}^{j-1}t_\ell\,,
    \qquad
    T_{\ge j}=\sum_{\ell=j}^{r}t_\ell\,.
\end{equation}
In particular, the signal-interleaved circuit has at most $2^r$ coherent displacement branches.
\end{lemma}

\begin{proof}
Recall that an ECD gate can be written as
\begin{equation}
    \mathrm{ECD}(\beta)=\sum_{s\in\{\pm1\}}D(s\beta/2)\otimes F_s\,,
\end{equation}
where $F_+=\ketbra{e}{g}$ and $F_-=\ketbra{g}{e}$. A single-qubit rotation has the form $I\otimes R$, so it changes only the qubit-side operator. Expanding each ECD gate in the circuit gives a sum over branch strings $s=(s_1,\ldots,s_r)\in\{\pm1\}^r$. On the branch $s$, the oscillator displacement sequence appearing in $V_r(\alpha)$ is
\begin{equation}
    D(t_r\alpha)D(s_r\beta_r/2)D(t_{r-1}\alpha)\cdots D(t_1\alpha)D(s_1\beta_1/2)D(t_0\alpha).
\end{equation}
The corresponding qubit-side operator is a product of the $F_{s_j}$'s and the single-qubit rotations. We denote this operator by $A_s$.

We now compute the oscillator product along a fixed branch. We repeatedly use the Weyl relation
\begin{equation}
    D(a)D(b)=e^{-i\Omega(a,b)/2}D(a+b).
\end{equation}
For any ordered sequence of displacement labels $a_0,\ldots,a_m$, this relation implies
\begin{equation}
    D(a_m)\cdots D(a_0)=\exp\left(-\frac{i}{2}\sum_{u>v}\Omega(a_u,a_v)\right)D\left(\sum_{u=0}^m a_u\right).
\end{equation}
In our branch, the total displacement label is
\begin{equation}
    \sum_{\ell=0}^r t_\ell\alpha+\frac{1}{2}\sum_{j=1}^r s_j\beta_j=T\alpha+\Gamma_s.
\end{equation}
The Weyl phase in the product has three types of terms. The signal-signal terms vanish because $\Omega(\alpha,\alpha)=0$. The control-control terms are independent of $\alpha$ and can be collected into a signal-independent phase $e^{i\theta_s}$. The only signal-dependent terms come from pairs containing one signal displacement $t_\ell\alpha$ and one branch displacement $s_j\beta_j/2$.

Fix $j$. If the signal window $t_\ell\alpha$ occurs after the $j$-th ECD layer, meaning $\ell\ge j$, then the pair contributes
\begin{equation}
    -\frac{i}{2}\,\Omega(t_\ell\alpha,s_j\beta_j/2)=\frac{i}{4}s_jt_\ell\,\Omega(\beta_j,\alpha).
\end{equation}
If the signal window occurs before the $j$-th ECD layer, meaning $\ell<j$, then the pair contributes
\begin{equation}
    -\frac{i}{2}\,\Omega(s_j\beta_j/2,t_\ell\alpha)=-\frac{i}{4}s_jt_\ell\,\Omega(\beta_j,\alpha).
\end{equation}
Summing over all signal windows, the total signal-dependent phase associated with the $j$-th control displacement is
\begin{equation}
    \frac{i}{4}s_j\left(T_{\ge j}-T_{<j}\right)\Omega(\beta_j,\alpha).
\end{equation}
Therefore the full signal-dependent phase on branch $s$ is
\begin{equation}
    \exp\!\left(i\Omega\left(\frac{1}{4}\sum_{j=1}^r s_j\left(T_{\ge j}-T_{<j}\right)\beta_j,\alpha\right)\right).
\end{equation}
This is exactly $e^{i\Omega(\Lambda_s,\alpha)}$. Hence the branch contribution has the form
\begin{equation}
    e^{i\theta_s}e^{i\Omega(\Lambda_s,\alpha)}D(T\alpha+\Gamma_s)\otimes A_s.
\end{equation}
Summing over all $2^r$ branch strings proves the claim.
\end{proof}

\begin{lemma}[Weyl support of depth-$r$ measurements]
\label{lemma:depth_r_qcds_weyl_support}
Let $M\in\mathcal M_r$ be any signal-interleaved depth-$r$ measurement. Then there exists a set $S_M\subset\mathbb R^2$ with $|S_M|\le 3^r$ such that
\begin{equation}
    f_M(\alpha)=a_0+\sum_{\zeta\in S_M}a_\zeta\chi_\zeta(\alpha).
\end{equation}
Consequently, for any reference distribution $P_0$, the centered response satisfies
\begin{equation}
    f_M-\int f_M\ dP_0\in \operatorname{span}\{\widetilde\chi_\zeta:\zeta\in S_M\},
\end{equation}
where $\{\widetilde\chi_\zeta\}_{\zeta\in\mathbb R^2}$ is the centered Weyl family.
\end{lemma}

\begin{proof}
Let $E_q$ denote the accepting element of the final binary qubit POVM, so that $0\le E_q\le I_q$. By Lemma \ref{lemma:finite_weyl_expansion_control_depth}, the signal-dependent circuit has the expansion
\begin{equation}
    V_r(\alpha)=\sum_{s\in\{\pm1\}^r}e^{i\theta_s}e^{i\Omega(\Lambda_s,\alpha)}D(T\alpha+\Gamma_s)\otimes A_s.
\end{equation}
The response function is
\begin{equation}
    f_M(\alpha)=\Tr\!\left[(I\otimes E_q)V_r(\alpha)\rho_0V_r(\alpha)^\dagger\right].
\end{equation}
Equivalently, by cyclicity of trace,
\begin{equation}
    f_M(\alpha)=\Tr\!\left[\rho_0V_r(\alpha)^\dagger(I\otimes E_q)V_r(\alpha)\right].
\end{equation}
Substituting the expansion of $V_r(\alpha)$ gives
\begin{align}
    f_M(\alpha)
    &=
    \sum_{s,s'\in\{\pm1\}^r}
    e^{-i\theta_s}e^{i\theta_{s'}}
    e^{-i\Omega(\Lambda_s,\alpha)}e^{i\Omega(\Lambda_{s'},\alpha)} \\
    &\qquad \qquad \times
    \Tr\!\left[
    \rho_0
    \left(
    D(T\alpha+\Gamma_s)^\dagger D(T\alpha+\Gamma_{s'})
    \otimes
    A_s^\dagger E_q A_{s'}
    \right)
    \right].
\end{align}
We now compute the remaining oscillator displacement product. Since $D(x)^\dagger=D(-x)$,
\begin{equation}
    D(T\alpha+\Gamma_s)^\dagger D(T\alpha+\Gamma_{s'})
    =
    D(-T\alpha-\Gamma_s)D(T\alpha+\Gamma_{s'}).
\end{equation}
Using the Weyl relation,
\begin{align}
    D(-T\alpha-\Gamma_s)D(T\alpha+\Gamma_{s'})
    &=
    e^{-i\Omega(-T\alpha-\Gamma_s,T\alpha+\Gamma_{s'})/2}D(\Gamma_{s'}-\Gamma_s).
\end{align}
The exponent simplifies as follows:
\begin{align}
    \Omega(-T\alpha-\Gamma_s,T\alpha+\Gamma_{s'})
    &=
    \Omega(-T\alpha,T\alpha)+\Omega(-T\alpha,\Gamma_{s'})+\Omega(-\Gamma_s,T\alpha)+\Omega(-\Gamma_s,\Gamma_{s'}) \\
    &=
    -T\Omega(\alpha,\Gamma_{s'})-T\Omega(\Gamma_s,\alpha)-\Omega(\Gamma_s,\Gamma_{s'}) \\
    &=
    T\Omega(\Gamma_{s'},\alpha)-T\Omega(\Gamma_s,\alpha)-\Omega(\Gamma_s,\Gamma_{s'}) \\
    &=
    T\Omega(\Gamma_{s'}-\Gamma_s,\alpha)-\Omega(\Gamma_s,\Gamma_{s'}).
\end{align}
Therefore
\begin{equation}
    D(T\alpha+\Gamma_s)^\dagger D(T\alpha+\Gamma_{s'})
    =
    e^{-iT\Omega(\Gamma_{s'}-\Gamma_s,\alpha)/2}
    e^{i\Omega(\Gamma_s,\Gamma_{s'})/2}
    D(\Gamma_{s'}-\Gamma_s).
\end{equation}
Equivalently, the $\alpha$-dependent part of this product is $e^{i\Omega(T(\Gamma_s-\Gamma_{s'})/2,\alpha)}$.

Now combine all $\alpha$-dependent phases in the $(s,s')$ summand. The branch phases contribute $e^{i\Omega(\Lambda_{s'}-\Lambda_s,\alpha)}$, and the displacement product contributes $e^{i\Omega(T(\Gamma_s-\Gamma_{s'})/2,\alpha)}$. Thus the total $\alpha$-dependent phase is $e^{i\Omega(\zeta_{s,s'},\alpha)}$, where
\begin{equation}
    \zeta_{s,s'}=\Lambda_{s'}-\Lambda_s+\frac{T}{2}(\Gamma_s-\Gamma_{s'}).
\end{equation}
All remaining factors in the $(s,s')$ summand are independent of $\alpha$.

We now show that the possible labels $\zeta_{s,s'}$ form a set of size at most $3^r$. Using
\begin{equation}
    \Gamma_s=\frac{1}{2}\sum_{j=1}^r s_j\beta_j
\end{equation}
and
\begin{equation}
    \Lambda_s=\frac{1}{4}\sum_{j=1}^r s_j\left(T_{\ge j}-T_{<j}\right)\beta_j,
\end{equation}
we obtain
\begin{align}
    \zeta_{s,s'}
    &=
    \frac{1}{4}\sum_{j=1}^r(s'_j-s_j)(T_{\ge j}-T_{<j})\beta_j+\frac{T}{4}\sum_{j=1}^r(s_j-s'_j)\beta_j \\
    &=
    \frac{1}{4}\sum_{j=1}^r(s'_j-s_j)(T_{\ge j}-T_{<j}-T)\beta_j.
\end{align}
Since $T=T_{\ge j}+T_{<j}$, this becomes
\begin{equation}
    \zeta_{s,s'}=-\frac{1}{2}\sum_{j=1}^r(s'_j-s_j)T_{<j}\beta_j.
\end{equation}
Equivalently,
\begin{equation}
    \zeta_{s,s'}=\sum_{j=1}^r m_j T_{<j}\beta_j,
\end{equation}
where $m_j=(s_j-s'_j)/2$ lies in $\{-1,0,1\}$. Therefore every response frequency lies in the set
\begin{equation}
    \left\{\sum_{j=1}^r m_jT_{<j}\beta_j:m_j\in\{-1,0,1\}\right\}.
\end{equation}
This set has cardinality at most $3^r$. Grouping all summands with the same value of $\zeta_{s,s'}$ gives
\begin{equation}
    f_M(\alpha)=\sum_{\zeta\in S_M\cup\{0\}}a_\zeta\chi_\zeta(\alpha),
\end{equation}
where $|S_M|\le 3^r$ after absorbing the $\zeta=0$ contribution into the constant term $a_0$. Hence
\begin{equation}
    f_M(\alpha)=a_0+\sum_{\zeta\in S_M}a_\zeta\chi_\zeta(\alpha)
\end{equation}
with $|S_M|\le 3^r$. It remains only to prove the centered statement. Let $P_0$ be any reference distribution and define
\begin{equation}
    \mu_\zeta=\int \chi_\zeta(\alpha)dP_0(\alpha),
    \qquad
    \widetilde\chi_\zeta=\chi_\zeta-\mu_\zeta.
\end{equation}
Taking the $P_0$-mean of the expansion gives
\begin{equation}
    \int f_MdP_0=a_0+\sum_{\zeta\in S_M}a_\zeta\mu_\zeta.
\end{equation}
Therefore
\begin{align}
    f_M-\int f_MdP_0
    &=
    a_0+\sum_{\zeta\in S_M}a_\zeta\chi_\zeta-a_0-\sum_{\zeta\in S_M}a_\zeta\mu_\zeta \\
    &=
    \sum_{\zeta\in S_M}a_\zeta(\chi_\zeta-\mu_\zeta) \\
    &=
    \sum_{\zeta\in S_M}a_\zeta\widetilde\chi_\zeta.
\end{align}
Thus the same support set $S_M$ works after centering.
\end{proof}

The second ingredient in our semiclassical measurement lemma is a perturbation $h$ which has a sparse Fourier support in the same Weyl basis. We wish to construct a family of such functions $h_k$, indexed by positive integers $k\in\mathbb{Z}^+$, such that the following requirements are satisfied. Let $P_0$ denote the centered Gaussian distribution on $\mathbb{R}^2$ with covariance $\sigma^2 I$, where $\sigma$ is a positive constant.
\begin{enumerate}
    \item (Centering and normalization) For all $k$, we have $\int h(\alpha) \ dP_0(\alpha) = 0$ and $\|h_k\|_\infty \leq 1$
    \item (Flatness and sparsity) They have $\exp(k)$ many nonzero Fourier coefficients, all of order $\exp(-k)$ magnitude 
    \item (Exact realizability) A depth-$k$ measurement can distinguish the hypothesis signals $P_\pm = P_0(1\pm \eta h_k)$ with constant success probability, when $\eta < 1$ is a constant.
\end{enumerate}

To construct such a perturbation, we modify a sequence of polynomials known as Golay-Rudin-Shapiro polynomials in harmonic analysis. They were introduced as explicit examples of sign polynomials with unusually flat magnitude on the unit circle. This flatness is exactly the feature we need: no single Fourier coefficient of the resulting response should carry too much of the signal. Because of this property, we can fall back to simply counting the number of terms in the Fourier support, rather than having to bound their amplitudes.

We use the following definitions. Let $e\in\mathbb R^2$ be any fixed unit vector (for concreteness, one can take e.g. $(0, 1)$) and a constant spacing parameter $B>0$. Define
\begin{equation}
    \beta_j=B3^{j-1}e,
    \qquad
    j=1,\ldots,k.
\end{equation}
For $n=(n_1,\ldots,n_k)\in\{-1,0,1\}^k$, write
\begin{equation}
    \xi_n=\sum_{j=1}^k n_j\beta_j.
\end{equation}
It follows that if $n\neq m$, then $\xi_n\neq \xi_m$.

\begin{lemma}[Golay-Rudin-Shapiro flatness]
\label{lemma:golay_rudin_shapiro_flatness}
For every integer $k>0$, consider the functions $P_k^{\rm GRS}$ and $Q_k^{\rm GRS}$ generated by the recurrence
\begin{align}
    P_0^{\rm GRS}&=Q_0^{\rm GRS}=\frac{1}{\sqrt2}, \\
    P_j^{\rm GRS}&=\frac{P_{j-1}^{\rm GRS}+z_jQ_{j-1}^{\rm GRS}}{\sqrt2},
    \qquad
    Q_j^{\rm GRS}=\frac{P_{j-1}^{\rm GRS}-z_jQ_{j-1}^{\rm GRS}}{\sqrt2},
\end{align}
where
\begin{equation}
    z_j(\alpha)=\chi_{\beta_j}(\alpha)=e^{i\Omega(\beta_j,\alpha)}.
\end{equation}
Then the function
\begin{equation}
    h_k(\alpha)=|P_k^{\rm GRS}(\alpha)|^2-|Q_k^{\rm GRS}(\alpha)|^2
\end{equation}
satisfies $|h_k(\alpha)|\le 1$ for every $\alpha$, and has a Weyl expansion
\begin{equation}
    h_k(\alpha)=\sum_{n\in\{-1,0,1\}^k\setminus\{0\}}c_n\chi_{\xi_n}(\alpha)
\end{equation}
with coefficient flatness $|c_n|^2\le C2^{-k}$ for a universal constant $C$.
\end{lemma}

\begin{proof}
First, note that at each level of the recurrence, the next level is obtained by a unitary transformation of the current polynomials. Therefore,
\begin{equation}
    |P_j^{\rm GRS}(\alpha)|^2+|Q_j^{\rm GRS}(\alpha)|^2=|P_{j-1}^{\rm GRS}(\alpha)|^2+|Q_{j-1}^{\rm GRS}(\alpha)|^2
\end{equation}
for every $j$ and every $\alpha$. Since the initial sum is one, we have
\begin{equation}
    |P_k^{\rm GRS}(\alpha)|^2+|Q_k^{\rm GRS}(\alpha)|^2=1,
\end{equation}
which implies $|h_k(\alpha)|\le 1$.

Next we prove the coefficient bound. Define $G_j=P_j^{\rm GRS}\overline{Q_j^{\rm GRS}}$.
A direct expansion of the recurrence gives
\begin{equation}
    h_j=z_j\overline{G_{j-1}}+z_j^{-1}G_{j-1}
\end{equation}
and
\begin{equation}
    G_j=\frac{1}{2}\left(h_{j-1}+z_j\overline{G_{j-1}}-z_j^{-1}G_{j-1}\right).
\end{equation}
For any finite Weyl polynomial $F$, write
\begin{equation}
    F(\alpha)=\sum_{\xi}\widehat F(\xi)\chi_\xi(\alpha)
\end{equation}
and define $\operatorname{supp}(F)=\{\xi:\widehat F(\xi)\neq0\}$. It will be useful to track supports explicitly. Let
\begin{equation}
    T_j=\left\{\sum_{\ell=1}^j n_\ell\beta_\ell:n_\ell\in\{-1,0,1\}\right\}.
\end{equation}
We claim inductively that
\begin{equation}
    \operatorname{supp}(h_j)\subseteq T_j,
    \qquad
    \operatorname{supp}(G_j)\subseteq T_j.
\end{equation}
This is immediate for $j=0$, since $h_0=0$ and $G_0=P_0^{\rm GRS}\overline{Q_0^{\rm GRS}}=1/2$ has support contained in $T_0=\{0\}$. If the claim holds at step $j-1$, then multiplication by $z_j=\chi_{\beta_j}$ shifts Weyl labels by $+\beta_j$, multiplication by $z_j^{-1}$ shifts them by $-\beta_j$, and complex conjugation maps a label $\xi$ to $-\xi$. Since the set of points $T_{j-1}$ is preserved under sign flip $\xi\mapsto-\xi$, the recurrence relations for $h_j, G_j$ imply that the supports at step $j$ are contained in $T_j$.

Since all $\beta_j$ are scalar multiples of the fixed unit vector $e$, every label in $T_j$ lies on the line spanned by $e$. Write the scalar coordinate of a label $\xi$ on this line as $t(\xi)$, so that $\xi=t(\xi)e$. For labels in $T_{j-1}$,
\begin{equation}
    |t(\xi)|\le B\sum_{\ell=1}^{j-1}3^{\ell-1}=\frac{B(3^{j-1}-1)}{2}.
\end{equation}
Let$R_{j-1}=B(3^{j-1}-1)/2$. Then $T_{j-1}$ is contained in the interval $[-R_{j-1},R_{j-1}]$ along the $e$ direction. The shifted set $\beta_j+T_{j-1}$ is contained in the interval
\begin{equation}
    [B3^{j-1}-R_{j-1},B3^{j-1}+R_{j-1}],
\end{equation}
whose left endpoint is
\begin{equation}
    B3^{j-1}-R_{j-1}=\frac{B(3^{j-1}+1)}{2}>R_{j-1}.
\end{equation}
Thus $\beta_j+T_{j-1}$ is disjoint from $T_{j-1}$. Similarly, $-\beta_j+T_{j-1}$ is contained in an interval whose right endpoint is
\begin{equation}
    -B3^{j-1}+R_{j-1}=-\frac{B(3^{j-1}+1)}{2}<-R_{j-1},
\end{equation}
so $-\beta_j+T_{j-1}$ is also disjoint from $T_{j-1}$. Finally, $\beta_j+T_{j-1}$ and $-\beta_j+T_{j-1}$ are disjoint because the first lies entirely to the right of $R_{j-1}$ and the second lies entirely to the left of $-R_{j-1}$. Therefore the three sets
\begin{equation}
    T_{j-1},
    \qquad
    \beta_j+T_{j-1},
    \qquad
    -\beta_j+T_{j-1}
\end{equation}
are pairwise disjoint. This disjointness lets us read off coefficient magnitudes. Define
\begin{equation}
    a_j=\max_\xi |\widehat h_j(\xi)|\,,
    \qquad
    b_j=\max_\xi |\widehat G_j(\xi)|\,.
\end{equation}
Multiplication by phases $z_j$ or $z_j^{-1}$ only shifts the support, not the magnitude, and complex conjugation only changes $\widehat G_{j-1}(\xi)$ to the conjugate of a coefficient at the label $-\xi$. Thus these operations preserve absolute values of coefficients. Since the two supports in
\begin{equation}
    h_j=z_j\overline{G_{j-1}}+z_j^{-1}G_{j-1}
\end{equation}
are disjoint, no two coefficients add at the same Weyl label. Hence $a_j = b_{j-1}$. Likewise, the three supports in
\begin{equation}
    G_j=\frac{1}{2}\left(h_{j-1}+z_j\overline{G_{j-1}}-z_j^{-1}G_{j-1}\right)
\end{equation}
are disjoint, so every coefficient of $G_j$ is exactly one half of a coefficient coming either from $h_{j-1}$, from $G_{j-1}$, or from $\overline{G_{j-1}}$. Therefore
\begin{equation} \label{eq:bj_mag_recurrence}
    b_j=\frac{1}{2}\max(a_{j-1},b_{j-1}).
\end{equation}
The initial values are $ a_0=0, b_0=\frac12$. We now solve the recurrence on the coefficient magnitudes. We prove by induction that for all $j\ge0$,
\begin{equation}
    a_j\le 2^{-\lfloor j/2\rfloor-1},
    \qquad
    b_j\le 2^{-\lceil j/2\rceil-1},
\end{equation}
where the bound on $a_0$ is trivial. Suppose the bounds hold at step $j-1$. Then
\begin{equation}
    a_j=b_{j-1}\le 2^{-\lceil (j-1)/2\rceil-1}=2^{-\lfloor j/2\rfloor-1}.
\end{equation}
For $b_j$, we use Equation \eqref{eq:bj_mag_recurrence}. If $j=2t$ is even, then
\begin{equation}
    a_{j-1}\le 2^{-t},
    \qquad
    b_{j-1}\le 2^{-t-1},
\end{equation}
and hence $b_j\le 2^{-t-1}=2^{-\lceil j/2\rceil-1}$. If $j=2t+1$, then
\begin{equation}
    a_{j-1}\le 2^{-t-1},
    \qquad
    b_{j-1}\le 2^{-t-1},
\end{equation}
and hence $b_j\le 2^{-t-2}=2^{-\lceil j/2\rceil-1}$. This completes the induction.

Therefore every Weyl coefficient $c_n$ of $h_k$ satisfies $|c_n|\le a_k\le 2^{-\lfloor k/2\rfloor-1}$. Squaring gives
\begin{equation}
    |c_n|^2\le 2^{-2\lfloor k/2\rfloor-2}\le 2^{-k}.
\end{equation}
This proves the desired flatness bound.
\end{proof}

Finally, we record that the function just constructed is exactly realized by a depth-$k$ measurement circuit. This will be used later in Section \ref{sec:control_hierarchy_upper_bound} to give an algorithm for the hypothesis testing problem, completing the exponential depth-based separation.

\begin{lemma}[Depth-$k$ realization of the Golay-Rudin-Shapiro kernel]
\label{lemma:depth_k_realizes_grs_kernel}
There is a depth-$k$ circuit whose final binary qubit measurement produces an outcome $Y\in\{-1,1\}$ satisfying
\begin{equation}
    \mathbb E[Y\mid \alpha]=h_k(\alpha).
\end{equation}
\end{lemma}

\begin{proof}
    We now show that the same recurrence is exactly realized by a depth-$k$ phase-kick circuit built from controlled displacements and single-qubit rotations. We first analyze one controlled displacement, built from ECD gates compatible with our hardware (Fig. \ref{fig:ecd_pulse}), carefully. Let
\begin{equation}
    C(\beta)=D(\beta/2)\otimes\ketbra{g}{g}+D(-\beta/2)\otimes\ketbra{e}{e}.
\end{equation}
With our ECD convention,
\begin{equation}
    C(\beta)=\mathrm{ECD}(\beta)(I\otimes X_q),
\end{equation}
where $X_q=\ketbra{g}{e}+\ketbra{e}{g}$. Thus $C(\beta)$ is obtained from one ECD gate and a fixed qubit bit flip, and we count it as one control layer.

Fix an arbitrary oscillator state $\ket{\psi}$ and an arbitrary qubit state $P\ket g+Q\ket e$. Before the signal displacement, applying $C(\beta)$ gives
\begin{equation}
    C(\beta)\ket{\psi}(P\ket g+Q\ket e)
    =
    P D(\beta/2)\ket{\psi}\ket g
    +
    Q D(-\beta/2)\ket{\psi}\ket e.
\end{equation}
After the signal displacement $D(\alpha)$ acts on the oscillator, the state becomes
\begin{equation}
    P D(\alpha)D(\beta/2)\ket{\psi}\ket g
    +
    Q D(\alpha)D(-\beta/2)\ket{\psi}\ket e.
\end{equation}
Finally we uncompute the controlled displacement by applying $C(\beta)^\dagger$. Since
\begin{equation}
    C(\beta)^\dagger=D(-\beta/2)\otimes\ketbra{g}{g}+D(\beta/2)\otimes\ketbra{e}{e},
\end{equation}
the final state is
\begin{equation}
    P D(-\beta/2)D(\alpha)D(\beta/2)\ket{\psi}\ket g
    +
    Q D(\beta/2)D(\alpha)D(-\beta/2)\ket{\psi}\ket e.
\end{equation}
Using the Weyl relation $D(a)D(b)=e^{-i\Omega(a,b)/2}D(a+b)$, we have
\begin{align}
    D(-\beta/2)D(\alpha)D(\beta/2)=e^{i\Omega(\beta,\alpha)/2}D(\alpha) ,\\
        D(\beta/2)D(\alpha)D(-\beta/2)=e^{-i\Omega(\beta,\alpha)/2}D(\alpha).
\end{align}
Therefore the final state is
\begin{equation}
    D(\alpha)\ket{\psi}\left(P e^{i\Omega(\beta,\alpha)/2}\ket g+Q e^{-i\Omega(\beta,\alpha)/2}\ket e\right).
\end{equation}
After factoring out the global phase $e^{i\Omega(\beta,\alpha)/2}$, which has no effect on measurement probabilities, the qubit state is
\begin{equation}
    P\ket g+Q e^{-i\Omega(\beta,\alpha)}\ket e.
\end{equation}
Thus the echoed controlled displacement $C(\beta)^\dagger D(\alpha)C(\beta)$ implements a qubit phase kick with relative phase $e^{-i\Omega(\beta,\alpha)}$, while leaving only a common oscillator displacement $D(\alpha)$. Equivalently, using $C(-\beta)$ implements the phase kick with relative phase $e^{i\Omega(\beta,\alpha)}$. In the recurrence below we use $C(-\beta_j)$ so that the phase variable is
\begin{equation}
    z_j(\alpha)=e^{i\Omega(\beta_j,\alpha)}.
\end{equation}

We now iterate this primitive. Let $\ket{\eta_j(\alpha)}$ denote the oscillator state after the first $j$ echoed phase-kick primitives. Its exact form is irrelevant, because at every step the oscillator factor is common to both qubit branches and therefore does not affect the final qubit probabilities. We prove by induction that after $j$ rounds the joint state has the product form
\begin{equation}
    \ket{\eta_j(\alpha)}\left(P_j^{\rm GRS}(\alpha)\ket g+Q_j^{\rm GRS}(\alpha)\ket e\right),
\end{equation}
where $P_j^{\rm GRS}$ and $Q_j^{\rm GRS}$ obey the Golay-Rudin-Shapiro recurrence.

For $j=0$, take $\ket{\eta_0}=\ket{\psi}$ and prepare the qubit state
\begin{equation}
    \frac{1}{\sqrt2}\ket g+\frac{1}{\sqrt2}\ket e.
\end{equation}
Thus $P_0^{\rm GRS}=Q_0^{\rm GRS}=\frac{1}{\sqrt2}$. Assume the claim holds after $j-1$ rounds, so the state is
\begin{equation}
    \ket{\eta_{j-1}(\alpha)}\left(P_{j-1}^{\rm GRS}(\alpha)\ket g+Q_{j-1}^{\rm GRS}(\alpha)\ket e\right).
\end{equation}
Apply the echoed controlled displacement using $C(-\beta_j)$. By the single-layer calculation above, this maps the state to
\begin{equation}
    \ket{\eta'_{j}(\alpha)}\left(P_{j-1}^{\rm GRS}(\alpha)\ket g+z_j(\alpha)Q_{j-1}^{\rm GRS}(\alpha)\ket e\right),
\end{equation}
where $\ket{\eta'_j(\alpha)}$ is a common oscillator state and $z_j(\alpha)=e^{i\Omega(\beta_j,\alpha)}$. Now apply the Hadamard rotation
\begin{equation}
    H=\frac{1}{\sqrt2}
    \begin{pmatrix}
        1 & 1 \\
        1 & -1
    \end{pmatrix}
\end{equation}
in the ordered basis $(\ket g,\ket e)$. Since $H\ket g=(\ket g+\ket e)/\sqrt2$ and $H\ket e=(\ket g-\ket e)/\sqrt2$, the state becomes
\begin{equation}
    \ket{\eta_j(\alpha)}\left(\frac{P_{j-1}^{\rm GRS}(\alpha)+z_j(\alpha)Q_{j-1}^{\rm GRS}(\alpha)}{\sqrt2}\ket g+\frac{P_{j-1}^{\rm GRS}(\alpha)-z_j(\alpha)Q_{j-1}^{\rm GRS}(\alpha)}{\sqrt2}\ket e\right).
\end{equation}
Thus the new qubit amplitudes satisfy
\begin{equation}
    P_j^{\rm GRS}=\frac{P_{j-1}^{\rm GRS}+z_jQ_{j-1}^{\rm GRS}}{\sqrt2},
    \qquad
    Q_j^{\rm GRS}=\frac{P_{j-1}^{\rm GRS}-z_jQ_{j-1}^{\rm GRS}}{\sqrt2}.
\end{equation}
This is exactly the Golay-Rudin-Shapiro recurrence. The induction is complete. After $k$ rounds, the state has the form
\begin{equation}
    \ket{\eta_k(\alpha)}\left(P_k^{\rm GRS}(\alpha)\ket g+Q_k^{\rm GRS}(\alpha)\ket e\right).
\end{equation}
Measure the qubit in the computational basis and define $Y=1$ for outcome $\ket g$ and $Y=-1$ for outcome $\ket e$. Since the oscillator factor is common to both branches, tracing it out does not change the qubit probabilities. Hence
\begin{equation}
    \Pr[Y=1\mid \alpha]=|P_k^{\rm GRS}(\alpha)|^2
\end{equation}
and
\begin{equation}
    \Pr[Y=-1\mid \alpha]=|Q_k^{\rm GRS}(\alpha)|^2.
\end{equation}
Therefore
\begin{align}
    \mathbb E[Y\mid \alpha]
    &=
    \Pr[Y=1\mid \alpha]-\Pr[Y=-1\mid \alpha] \\
    &=
    |P_k^{\rm GRS}(\alpha)|^2-|Q_k^{\rm GRS}(\alpha)|^2 \\
    &=
    h_k(\alpha).
\end{align}
Thus the depth-$k$ phase-kick circuit realizes exactly the desired measurement response.
\end{proof}

\subsubsection{Proof of Theorem \ref{thm:interleaved_control_lb_formal}}
\label{sec:proof_of_control_depth_theorem}

Now we have all the ingredients to prove the control-depth lower bound.
\begin{proof}
    As usual, we reduce the learning task to hypothesis testing. Any algorithm for estimation of $\mathbb E[f(\alpha)]$ for $f$ in a fixed class of functions $\mathcal{F}$ and for general signals can distinguish two signals $P_0, P_1$ for which for some $f\in \mathcal{F}$, $|\mathbb{E}_{P_0}[f(\alpha)] - \mathbb{E}_{P_1}[f(\alpha)]| = \Theta(1)$. By demonstrating functions which no depth $<\gamma k$ algorithm can distinguish with subexponential signal queries, we establish a learning lower bound. 

    Let $P_0$ be the centered Gaussian distribution on $\mathbb R^2$ with covariance $\sigma^2 I$, and let $\mathcal W$ be the centered Weyl dictionary associated to $P_0$. Fix a constant $0<\eta\le 1/4$ and let $h_k$ be the Golay-Rudin-Shapiro kernel given in Lemma \ref{lemma:golay_rudin_shapiro_flatness}. Define
    \begin{equation}
        g_k(\alpha)=h_k(\alpha)-\int h_k(\alpha)dP_0(\alpha).
    \end{equation}
    Then define two signals induced by the distributions
    \begin{equation} \label{eq:control_depth_hypothesis_family}
        P_\pm^{(k)}(\alpha)=\left(1\pm \eta g_k(\alpha)\right)P_0(\alpha).
    \end{equation}
    These two hypotheses are valid probability distributions. By Lemma \ref{lemma:golay_rudin_shapiro_flatness}, $|h_k|\le 1$, and so $\|g_k\|_\infty\le 2$. Because $\eta\le 1/4$,
    \begin{equation}
        1\pm \eta g_k(\alpha)\ge \frac12
    \end{equation}
    for every $\alpha$. Also, by definition of $g_k$,
    \begin{equation}
        \int g_k(\alpha)dP_0(\alpha)=0.
    \end{equation}
    Therefore $P_\pm^{(k)}$ are valid probability distributions. Next, for $r\in\mathbb{Z}^+$, let $M$ be any single-query measurement in $\mathcal M_r$, and let $f_M$ be its acceptance response. By Lemma \ref{lemma:depth_r_qcds_weyl_support}, the centered response satisfies
    \begin{equation}
        f_M-\int f_MdP_0\in\operatorname{span}\{\widetilde\chi_\zeta:\zeta\in S_M\}
    \end{equation}
    for some set $S_M$ with $|S_M|\le 3^r$. Thus every measurement in $\mathcal M_r$ is $3^r$-sparse in the centered Weyl dictionary $\mathcal W$. Moreover, from Lemma \ref{lemma:golay_rudin_shapiro_flatness} we have that each Fourier coefficient $c_n$ of $g_k$ in the Weyl dictionary satisfies $|c_n|^2 \leq C2^{-k}$ for some universal constant $C > 0$. Therefore we have that
    \begin{equation}
         \mathsf{C}_{3^r}(g_k;\mathcal W,P_0)\le C_0 3^r2^{-k}
    \end{equation}
    By the semiclassical support-counting Corollary \ref{cor:sparse_semiclassical_quantum_measurement} with
    \begin{equation}
        h=g_k,
        \qquad
        L=3^r,
        \qquad
        \mathcal D=\mathcal W.
    \end{equation}
    we obtain that any adaptive protocol using measurements from $\mathcal M_r$ requires
    \begin{equation}
        N\ge \Omega\left(\eta^{-2}\left(C 3^r2^{-k}\right)^{-1}\right) = \Omega(2^k/3^r).
    \end{equation}
    For $r < k\log_3 2$, this is $\exp(\Omega(k))$.
\end{proof}

\newpage
\section{Exponential advantage in learning signals with Quantum Feature Sensing}\label{sec:advantage_upper_bounds}

Here we present the quantum-enhanced protocols and their sample complexity analyses that complete our exponential separations. We discuss our results through the Quantum Phase-Space Inference framework from Section \ref{sec:qes}, illustrating how each sensing protocol operationally maximizes the relevant Fourier overlap quantity under constraints, and how the upper bounds emerge from the same considerations as their respective lower bounds.

Our results imply an important operational message. Modern continuous-variable quantum sensors are extremely mature technologies that operate at the forefront of modern engineering, and can already achieve measurement precision orders of magnitude below the classical floor, making further improvements ever more challenging and increasingly incremental. Each of our separations, however, considers taking a conventional sensor and adding an \textit{auxiliary} resource in the form of quantum control. That is, our exponential sensing gains are enabled entirely by quantum information processing technology already present today, without requiring any precision improvement to the sensor itself. As such, these developments suggest a new paradigm in which development of sensing technologies can be accelerated by focusing on architectures which couple to rudimentary quantum processors.

\subsection{Learning with quantum probes without quantum control}
\label{sec:quantum_probe_upper_bounds}

We begin with the first separation in the hierarchy, where the learner may use nonclassical probe states but performs no coherent quantum control during the sensing interaction. This puts us in the class of conventional quantum sensors which can prepare Gaussian probe states. We will first utilize the estimator from Theorem \ref{thm:qes_matched_score_guarantee} and a lower bound of the AFI (Definition \ref{def:AFI}) to immediately establish an exponential advantage for the simpler hypothesis testing task. Then, we will give an estimator for the broader angular Fourier learning task and bound its sample complexity. 

\subsubsection{Fixed-amplitude learning with squeezed probes}
\label{sec:squeezed_fixed_amplitude_upper_bound}

First we recall the Q$\Psi$ hypothesis testing problem used to establish an exponential lower bound against coherent probe protocols in Section \ref{sec:coherent_probe_lb}. Consider a displacement distribution $P$ supported on the circle $\{Ae^{i\theta}:\theta\in[0,2\pi)\}$, for fixed $A = \Theta(\sqrt{k})$. We let $ \Lambda_k(P)=\mathbb{E}_{\theta\sim F}\left[e^{ik\theta}\right]$ denote the $k$-th angular Fourier coefficient. Then the two hypothesis distributions used in the coherent-probe lower bound,
\begin{equation}
    T_\pm^{(k)}(\theta)=\frac{1}{2\pi}\left(1\pm 0.9\cos(k\theta)\right),
\end{equation}
we have $\Lambda_k(P_\pm^{(k)})=\pm 0.45$, where $P_{\pm}$ is the pushforward of $T_{\pm}$ under the map $\theta \rightarrow Ae^{i\theta}$.

The class of response functions at hand now corresponds to the set of Gaussian sensing protocols. To utilize Theorem \ref{thm:qes_matched_score_guarantee}, we first compute the Gaussian protocol response kernels. For simplicity, consider the response upon performing standard squeezed homodyne detection. Let $r$ be the squeezing parameter and set $\nu=e^{-2r}/2$. If the signal applies the displacement $D(Ae^{i\theta})$ and we perform a squeezed homodyne measurement along angle $\varphi$, the outcome has the form $Y_\varphi=R\cos(\theta-\varphi)+G$, where $G\sim\mathcal N(0,\nu)$. Let $g_\nu(y)=\frac{1}{\sqrt{2\pi\nu}}\exp\!\left(-\frac{y^2}{2\nu}\right)$; under the reference distribution $P_0(d\theta) = d\theta/2\pi$ and perturbation $h_k(\theta) = 0.9 \cos(k\theta)$, we have
\begin{align}
    &q_{0,k}(y) = \frac{1}{2\pi}\int_{0}^{2\pi} g_\nu(y - A\sqrt{2}\cos\theta)d\theta \\ 
    &a_{h,k}(y) = \frac{0.9}{2\pi}\int_{0}^{2\pi} \cos(k\theta)g_\nu(y - A\sqrt{2}\cos\theta)d\theta \\ 
    &R_{h, k}(y) = \frac{a_{h,k}(y)}{q_{0,k}(y)} \ .
\end{align}
These are the quantities used in the AFI-optimal estimator for Theorem \ref{thm:qes_matched_score_guarantee}. Now we show that the AFI of this protocol saturates the achievable AFI for the given hypothesis test among any conventional Gaussian protocol, up to constants. 
\begin{lemma}
    Let $M_{\nu, \phi}$ denote the squeezed homodyne measurement strategy with squeezing strength $\nu$ and homodyne angle $\phi$. For the task of distinguishing displacement distributions $P_{\pm}$ above with $A = C\sqrt{k}$ and $C$ a constant, there exist constants $C_1, C_2$ such that
    \begin{equation}
        \mathsf A_{M_{\nu, 0}}(h_k; P_0) \geq C_1
    \end{equation}
    for any $\nu \leq C_2/k$. This is optimal up to constants for any measurement strategy.
\end{lemma}
\begin{proof}
By definition of AFI for the measurement $M_{\nu,0}$,
\begin{equation}
    \mathsf A_{M_{\nu,0}}(h_k;P_0)=\int_{\mathbb R}\frac{a_{h,k}(y)^2}{q_{0,k}(y)}dy.
\end{equation}
Let $R =A\sqrt{2}$, let $\Theta$ be uniformly distributed on $[0,2\pi)$, let $G\sim\mathcal N(0,\nu)$ be independent of $\Theta$, and set
\begin{equation}
    X=R\cos\Theta,\qquad Y=X+G.
\end{equation}
Then $q_{0,k}$ is the density of $Y$, and Bayes' rule gives
\begin{equation}
    \frac{a_k(y)}{q_{0,k}(y)}=\mathbb E[\cos(k\Theta)|Y=y].
\end{equation}
Consequently,
\begin{align}
    \int_{\mathbb R}\frac{a_k(y)^2}{q_{0,k}(y)}dy
    &=\mathbb E\left[\mathbb E[\cos(k\Theta)|Y]^2\right].
\end{align}
By Cauchy-Schwarz, for any $f:\mathbb{R}\rightarrow \mathbb{R}$ with bounded variance,
\begin{equation}
    \mathbb E\left[\mathbb E[\cos(k\Theta)|Y]^2\right]\ge \frac{\left|\mathbb E[f(Y)\cos(k\Theta)]\right|^2}{\mathbb E[f(Y)^2]}. \label{eq:coh_probe_variational_ub_step}
\end{equation}
We now choose a test function which reconstructs $\cos(k\Theta)$ from the noiseless homodyne outcome. Let $T_k$ be the $k$-th Chebyshev polynomial, and define
\begin{equation}
    \tau_k(y)=T_k\left(\max\left\{-1,\min\left\{1,\frac{y}{R}\right\}\right\}\right).
\end{equation}
Since $|T_k(x)|\le 1$ for $x\in[-1,1]$, we have $|\tau_k(y)|\le 1$ for all $y$. Moreover, because $X=R\cos\Theta$,
\begin{equation}
    \tau_k(X)=T_k(\cos\Theta)=\cos(k\Theta).
\end{equation}
Therefore,
\begin{equation}
    \mathbb E[\tau_k(X)\cos(k\Theta)]=\mathbb E[\cos^2(k\Theta)]=\frac{1}{2}.
\end{equation}
This is not yet what we want, as we have to show that upon replacing $X$ with $Y$ by adding Gaussian noise, this expected value remains large. Fix a small universal constant $\Delta_0\in(0,1/10)$, to be chosen below. Define the two bad events
\begin{equation}
    \mathcal B=\{|\sin\Theta|\le \Delta_0\},\qquad \mathcal C=\{|G|>R\Delta_0^2/4\}.
\end{equation}
The event $\mathcal B$ is the region where the map $\theta\mapsto\cos\theta$ approaches a turning point, while $\mathcal C$ is the event where the homodyne noise is large. For appropriately chosen $\Delta_0$, these bad events contain the cases in $\tau_k(Y)\cos(k\Theta)$ becomes too small compared to $\tau_k(X)\cos(k\Theta)$. Since $\Theta$ is uniform and $\sin(x) \leq x$, $\Pr(\mathcal B)\le 2\Delta_0$. Also, by a standard Gaussian tail bound,
\begin{equation}
    \Pr(\mathcal C)\le 2\exp\left(-\frac{R^2\Delta_0^4}{32\nu}\right).
\end{equation}
We next control the Lipschitz constant of $\tau_k$ on the complement of these events. On $\mathcal B^c$, note that $|\sin \Theta| > \Delta_0$ implies $|\cos\Theta| \leq \sqrt{1-\Delta_0^2}$. As such on $\mathcal B^c$ we have
\begin{equation}
    \left|\frac{X}{R}\right|=|\cos\Theta|\le \sqrt{1-\Delta_0^2}\le 1-\frac{\Delta_0^2}{2}.
\end{equation}
Moreover on $\mathcal C^c$, $|G|/R\leq \Delta_0^2/4$, so we also have $|Y/R-X/R|\le \Delta_0^2/4$. Hence every point $z$ between $X/R$ and $Y/R$ on this complement region satisfies $|z|\le 1-\Delta_0^2/4$. Within the complement region, due to our choice of normalization, $\tau_k = T_k$. Moreover, the Chebyshev derivative satisfies $T_k'(z)=kU_{k-1}(z)$, where $U_{k-1}$ is the Chebyshev polynomial of the second kind. Writing $z = \cos v$, it holds that $U_{k-1} = \sin(kv)/\sin(v)$, so we have the bound
\begin{equation}
    |U_{k-1}(z)|\le \frac{1}{\sqrt{1-z^2}},\qquad z\in(-1,1).
\end{equation}
This then gives us $|T_k'(z)|\le \frac{k}{\sqrt{1-z^2}}$. Since $|z|\le 1-\Delta_0^2/4$, we have
\begin{align}
    1-z^2
    &\ge 1-\left(1-\frac{\Delta_0^2}{4}\right)^2 \\
    &=\frac{\Delta_0^2}{2}-\frac{\Delta_0^4}{16} \\
    &\ge \frac{\Delta_0^2}{4},
\end{align}
where the last inequality uses $\Delta_0\le 1$. Therefore
\begin{equation}
    |T_k'(z)|\le \frac{2k}{\Delta_0}.
\end{equation}
Combining this with $|Y/R-X/R|=|G|/R$, we find that on $(\mathcal B\cup\mathcal C)^c$,
\begin{equation}
    |\tau_k(Y)-\tau_k(X)|\le \frac{2k}{R\Delta_0}|G| ,
\end{equation}
so the function $\tau_k$ is $2k/R\Delta_0$-Lipschitz on the complement region. On the bad event $\mathcal B\cup\mathcal C$, we use the trivial bound $|\tau_k(Y)-\tau_k(X)|\le 2$. Therefore,
\begin{align}
    \mathbb E[|\tau_k(Y)-\tau_k(X)|]
    &\le 2\Pr(\mathcal B)+2\Pr(\mathcal C)+\frac{2k}{R\Delta_0}\mathbb E[|G|] \\
    &\le 4\Delta_0+4\exp\left(-\frac{R^2\Delta_0^4}{32\nu}\right)+\frac{2k\sqrt{\nu}}{R\Delta_0}.
\end{align}
Now choose $\Delta_0>0$ small enough so that the first term is at most $1/12$. Then choose a universal constant $c_0>0$ small enough so that whenever
\begin{equation}
    \nu\le c_0\frac{R^2}{k^2},
\end{equation}
the last two terms together are at most $1/6$. With this choice, we have $\mathbb E[|\tau_k(Y)-\tau_k(X)|]\le 1/4$. It follows that
\begin{align}
    \left|\mathbb E[\tau_k(Y)\cos(k\Theta)]\right|
    &\ge \left|\mathbb E[\tau_k(X)\cos(k\Theta)]\right|-\mathbb E[|\tau_k(Y)-\tau_k(X)|] \\
    &\ge \frac{1}{2}-\frac{1}{4} =\frac{1}{4}.
\end{align}
Since $\mathbb E[\tau_k(Y)^2]\le 1$, plugging this into Equation \ref{eq:coh_probe_variational_ub_step} gives us
\begin{equation}
    \int_{\mathbb R}\frac{a_k(y)^2}{q_{0,k}(y)}dy\ge \frac{1}{16}.
\end{equation}
With constants included from $h_k(\theta)=0.9\cos(k\theta)$, we obtain
\begin{equation}
    \mathsf A_{M_{\nu,0}}(h_k;P_0)\ge \frac{0.9^2}{16}.
\end{equation}
Thus we may take $C_1=0.9^2/16$. Finally, since $A=C\sqrt{k}$ and $R=A\sqrt2$, the condition $\nu\le c_0R^2/k^2$ becomes
\begin{equation}
    \nu\le \frac{2c_0C^2}{k}.
\end{equation}
Taking $C_2=2c_0C^2$ proves the claimed statement. We also note that this constant lower bound is optimal up to universal factors, since AFI satisfies
\begin{equation}
    \mathsf A_M(h_k;P_0)\le \|h_k\|_{L^2(P_0)}^2=\frac{0.9^2}{2}
\end{equation}
for any measurement $M$. 
\end{proof}
Through the Q$\Psi$ framework, this result gives us our first exponential separation. 
\begin{graydefinitionbox}
\begin{corollary}[Exponential advantage of conventional quantum sensing over coherent classical probes]
\label{cor:squeezed_over_coherent}
There is a squeezed-probe protocol with mean photon number $O(k)$ per query which distinguishes $P_+^{(k)}$ from $P_-^{(k)}$ with $O(1)$ queries and constant success probability. In contrast, any coherent-probe protocol, regardless of energy, requires $\exp(\Omega(k))$ queries to distinguish the same two hypotheses with constant success probability.
\end{corollary}
\end{graydefinitionbox}

\begin{proof}
The squeezed homodyne protocol $M_{\nu, 0}$, using energy $O(k)$ (due to $\nu \leq O(1/k)$), has AFI $\Omega(1)$. By applying the estimator from Theorem \ref{thm:qes_matched_score_guarantee}, the Q$\Psi$ guarantee than gives a constant success probability using $O(1)$ samples. The coherent-probe lower bound is Theorem \ref{thm:classical_hardness_formal}.
\end{proof}

This result already enables a rigorous exponential separation. However, the given protocol is specialized to the stated hypotheses testing problem. More broadly, it is practically useful to be able to estimate general $k$-th angular Fourier coefficients. We now give an optimal Gaussian protocol for this more general task. 

In Section \ref{sec:qes} we introduced the Fourier gain and Theorem \ref{thm:qes_fourier_gain_global_observable} to handle the setting of global learning; we now switch from the local AFI language to this framework. On the fixed-amplitude circle, the relevant Fourier dictionary is
\begin{equation}
    \chi_\ell(\theta)=e^{i\ell\theta},\qquad \ell\in\mathbb Z.
\end{equation}
Let $\mathcal A_{\rm sq}(A,\nu)$ denote the class of randomized squeezed-homodyne protocols at fixed amplitude $A$ and homodyne noise variance $\nu$, together with arbitrary classical postprocessing. Its Fourier gain (introduced before Theorem \ref{thm:qes_fourier_gain_global_observable}) at angular frequency $\ell$ is $G^{A,\nu}(\ell)=G_{\mathcal A_{\rm sq}(A,\nu)}(\ell)$.
The goal is to show that $G_{\rm sq}^{A,\nu}(k)$ is constant once the squeezed quadrature has enough resolution to see one period of the $k$-th angular fringe.

For the remainder of this subsection, we work with the angular response 
\begin{equation}
    a_\ell(y)=\frac{1}{2\pi}\int_0^{2\pi}e^{-i\ell u}g_\nu(y-R\cos u)\,du\,,
\end{equation}
defined for each frequency $\ell\ge1$. Note that the replacement of the cosine with an exponential does not change the response because the sine term vanishes by the symmetry $u\mapsto -u$. Finally define
\begin{equation}
    I(\ell;A,\nu)=\int_{\mathbb R}\frac{a_\ell(y)^2}{q_0(y)}\,dy\,.
\end{equation}
If we treated $e^{-i\ell u}$ as a perturbation on top of a uniform background and asked about hypothesis testing, this would be exactly the AFI. Here we use a different symbol to conceptually differentiate from hypothesis testing, as we are concerned with global learning of angular Fourier coefficients. 

\begin{lemma}[Squeezed homodyne certifies angular Fourier gain]
\label{lemma:squeezed_homodyne_fourier_gain}
For every $\ell\ge1$,
\begin{equation}
    G^{A,\nu}(\ell)\ge I(\ell;A,\nu).
\end{equation}
Moreover, there exist universal constants $c_0,c_1>0$ such that $I(\ell;A,\nu)\ge c_1$ whenever $\nu\le c_0\frac{A^2}{\ell^2}$.
\end{lemma}

\begin{proof}
To show a Fourier gain lower bound in the broader Gaussian sensing class, we explicitly provide a protocol with Fourier gain exceeding this bound. The protocol draws $\varphi$ uniformly from $[0,2\pi)$, performs squeezed homodyne along angle $\varphi$, obtains $Y=R\cos(\theta-\varphi)+G$, and outputs the unnormalized score
\begin{equation}
    W_\ell=e^{i\ell\varphi}\frac{a_\ell(Y)}{q_0(Y)}.
\end{equation}
We now compute its response function and show that it is peaked at the Fourier character labeled by $\ell$. Conditioning on a deterministic angle $\theta$,
\begin{align}
    \mathbb E[W_\ell|\theta]
    &=
    \int_0^{2\pi}\frac{d\varphi}{2\pi}\int_{\mathbb R}dy \ e^{i\ell\varphi}\frac{a_\ell(y)}{q_0(y)}g_\nu(y-R\cos(\theta-\varphi)).
\end{align}
Make the change of variables $u=\theta-\varphi$. Then $e^{i\ell\varphi}=e^{i\ell\theta}e^{-i\ell u}$, so
\begin{align}
    \mathbb E[W_\ell|\theta]
    &=
    e^{i\ell\theta}
    \int_{\mathbb R}dy \frac{a_\ell(y)}{q_0(y)}
    \frac{1}{2\pi}\int_0^{2\pi}du \ e^{-i\ell u}g_\nu(y-R\cos u) \\
    &=
    e^{i\ell\theta}
    \int_{\mathbb R}\frac{a_\ell(y)^2}{q_0(y)}dy \\
    &=
    I(\ell;A,\nu)\,e^{i\ell\theta}.
\end{align}
Thus the response function of this one-shot protocol is
\begin{equation}
    f_{W_\ell}(\theta)=I(\ell;A,\nu)\chi_\ell(\theta),
\end{equation}
so its $\ell$-th Fourier coefficient is precisely $\widehat f_{W_\ell}(\ell)=I(\ell;A,\nu)$. In words, the Fourier support of the measurement class on the desired coefficient is precisely the AFI quantity $I(\ell;A,\nu)$. To obtain the Fourier gain and use Theorem \ref{thm:qes_fourier_gain_global_observable}, we require one more piece: the variance of the protocol's readout. This is essential because the Fourier gain is defined as the Fourier mass \textit{per unit readout variance}. Again conditioning on $\theta$,
\begin{align}
    \mathbb E[|W_\ell|^2|\theta]
    &=
    \int_0^{2\pi}\frac{d\varphi}{2\pi}\int_{\mathbb R}dy \ \frac{a_\ell(y)^2}{q_0(y)^2}\,g_\nu(y-R\cos(\theta-\varphi)).
\end{align}
Averaging the conditional density over the uniformly random homodyne angle gives $q_0(y)$:
\begin{equation}
    \int_0^{2\pi}\frac{d\varphi}{2\pi}\,g_\nu(y-R\cos(\theta-\varphi))=q_0(y).
\end{equation}
Therefore
\begin{align}
    \mathbb E[|W_\ell|^2|\theta]
    &=
    \int_{\mathbb R}\frac{a_\ell(y)^2}{q_0(y)}dy \\
    &=
    I(\ell;A,\nu).
\end{align}
Hence $V_{W_\ell}=\gamma_{\rm sq}(\ell;A,\nu)$, and the definition of Fourier gain gives
\begin{equation}
    G^{A,\nu}(\ell)
    \ge
    \frac{|\widehat f_{W_\ell}(\ell)|^2}{V_{W_\ell}}
    =
    I(\ell;A,\nu).
\end{equation}
It remains only to lower bound this certified gain. This is exactly the Chebyshev Lipschitz-continuity in the preceding lemma, with the factor $0.9$ removed and with $k$ replaced by $\ell$. Drawing on that result, we have
\begin{equation}
    \int_{\mathbb R}\frac{a_\ell(y)^2}{q_0(y)}\,dy\ge c_1
\end{equation}
whenever $\nu\le c_0\frac{R^2}{\ell^2}$.
\end{proof}

By Lemma \ref{lemma:squeezed_homodyne_fourier_gain}, constant Fourier gain at frequency $\ell$ is achieved whenever
\begin{equation}
    r\ge \log\left(\frac{\ell}{A}\right)+O(1) \ ,
\end{equation}
which gives us that a Gaussian protocol with energy $O(\ell^2/A^2)$ suffices to learn the frequency-$\ell$ angular Fourier coefficient of a distribution with amplitude $A$. To formalize this and conclude, we state the full algorithm and rigorously combine Theorem \ref{thm:qes_fourier_gain_global_observable} and Lemma \ref{lemma:squeezed_homodyne_fourier_gain}. 

\begin{algorithm}[h!]
\caption{Randomized squeezed-homodyne estimator for fixed-amplitude angular Fourier learning}
\label{alg:randomized_squeezed_homodyne_fixed_amplitude_gain}
\begin{algorithmic}[1]
\Require Frequency $k\ge1$, fixed amplitude $A>0$, squeezing parameter $r$, accuracy parameters $\epsilon,\delta$, and query access to $\mathcal E_P$.
\Ensure An estimate $\widehat{\Lambda}_k$ of $\Lambda_k(P)=\mathbb E_{\theta\sim F}[e^{ik\theta}]$.
\State Set $\nu=e^{-2r}/2$ and $R=A\sqrt2$.
\State Compute
\begin{equation}
    q_0(y)=\frac{1}{2\pi}\int_0^{2\pi}g_\nu(y-R\cos u)du,
\end{equation}
\begin{equation}
    a_k(y)=\frac{1}{2\pi}\int_0^{2\pi}\cos(ku)g_\nu(y-R\cos u)du,
\end{equation}
and
\begin{equation}
    \gamma_{\rm sq}(k;A,\nu)=\int_{\mathbb R}\frac{a_k(y)^2}{q_0(y)}dy.
\end{equation}
\State Take
\begin{equation}
    N=C\frac{1}{\gamma_{\rm sq}(k;A,\nu)\epsilon^2}\log\left(\frac{4}{\delta}\right)
\end{equation}
shots for a sufficiently large universal constant $C$.
\For{$j=1,\ldots,N$}
    \State Draw $\varphi_j$ uniformly from $[0,2\pi)$.
    \State Prepare a squeezed vacuum whose squeezed quadrature is the homodyne quadrature at angle $\varphi_j$.
    \State Send the probe through $\mathcal E_P$ and homodyne the same quadrature, obtaining $Y_j$.
    \State Set
    \begin{equation}
        Z_j=e^{ik\varphi_j}\frac{a_k(Y_j)}{q_0(Y_j)\gamma_{\rm sq}(k;A,\nu)}.
    \end{equation}
\EndFor
\State Output
\begin{equation}
    \widehat{\Lambda}_k=\frac{1}{N}\sum_{j=1}^N Z_j.
\end{equation}
\end{algorithmic}
\end{algorithm}

\begin{graydefinitionbox}
\begin{theorem}[Fixed-amplitude angular Fourier learning from squeezed Fourier gain]
\label{thm:fixed_amplitude_squeezed_fourier_gain_learning}
Let $P$ be any displacement distribution with amplitude $A$. Then Algorithm \ref{alg:randomized_squeezed_homodyne_fixed_amplitude_gain}, with $r\ge \Omega(\log\left(k/A\right))$, outputs an estimate $\widehat{\Lambda}_k$ satisfying $|\widehat{\Lambda}_k-\Lambda_k(P)| \leq \epsilon$ with probability at least $1-\delta$ given 
\begin{equation}
    N=
    O\!\left(
        \epsilon^{-2}\log\!\left(\frac{1}{\delta}\right)
    \right).
\end{equation}
samples. This requires total energy or mean photon number $O(k^2/A^2)$. 
\end{theorem}
\end{graydefinitionbox}
\begin{proof}
For the single-coefficient learning problem, take the property kernel $\psi(\theta)=\chi_k(\theta)=e^{ik\theta}$. Its Fourier support is the singleton $S=\{k\}$ and $\widehat{\psi}(k)=1$. The Fourier-overlap guarantee in Theorem \ref{thm:qes_fourier_gain_global_observable} therefore gives a protocol with sample complexity
\begin{equation}
    N=
    O\left(
        \frac{1}{G^{A,\nu}(k)\epsilon^2}
        \log\left(\frac{1}{\delta}\right)
    \right).
\end{equation}
Algorithm \ref{alg:randomized_squeezed_homodyne_fixed_amplitude_gain} is exactly this protocol, specialized to the randomized-homodyne measurement. By Lemma \ref{lemma:squeezed_homodyne_fourier_gain}, $G^{A,\nu}(k)\ge I(k;A,\nu)$, and if $\nu\le c_0A^2/k^2$, then $I(k;A,\nu)\ge c_1$ for a universal constant $c_1>0$. Since $\nu=e^{-2r}/2$, this condition is equivalent to choosing $r\ge \log\left(\frac{k}{A}\right)+O(1)$. Substituting the constant gain lower bound into the Fourier-overlap sample complexity gives
\begin{equation}
    N=
    O\!\left(
        \epsilon^{-2}\log\!\left(\frac{1}{\delta}\right)
    \right).
\end{equation}
Finally, choosing $r=\log(k/A)+O(1)$ gives mean photon number $\sinh^2 r=O(e^{2r})=O(k^2/A^2)$.
\end{proof}

\subsubsection{Experimental proxy protocol} \label{sec:exp_proxy}
Our experimental results in Figure \ref{fig:coherent_and_memory_experiments}(b) aim to illustrate the Corollary \ref{cor:squeezed_over_coherent} separation in Fourier amplitude learning. However due to the incompatibility of squeezed homodyne measurements with our microwave cavity superconducting circuit apparatus, we implement a different protocol than the one above. Here we describe this protocol and rigorously demonstrate that it is strictly weaker than the squeezed homodyne strategy, so that any separation present in our experiment is no larger than the one that would be observed upon performing squeezed homodyne sensing with similar fidelity. 

We use the following fact about Bessel functions of the first kind:

\begin{fact}
    Let $J_k(z)$ denote the $k$-th Bessel function of the first kind. There exist universal constants $c_B,C_B>0$ such that, for every $k\ge1$,
    \begin{equation}
        c_Bk^{-1/3}\le \sup_{z\in\mathbb R}|J_k(z)|\le C_Bk^{-1/3}.
    \end{equation}
\end{fact}
\noindent In the subsequent discussion we let $z_k$ be a point satisfying $|J_k(z_k)|\ge c_Bk^{-1/3}$ and $|z_k|=O(k)$.

\begin{algorithm}[h!]
\caption{Experimental proxy estimator for angular Fourier learning}
\label{alg:experimental_proxy_fixed_amplitude}
\begin{algorithmic}[1]
\Require Frequency $k\ge1$, fixed amplitude $A>0$, accuracy parameters $\epsilon,\delta$, and query access to $\mathcal E_P$.
\Ensure An estimate $\widehat{\Lambda}_k$ of $\Lambda_k(P)=\mathbb E[e^{ik\theta}]$.
\State Set $R=A\sqrt2$ and $t=z_k/R$.
\State Take
\begin{equation}
    N=Ck^{2/3}\epsilon^{-2}\log\left(\frac{4}{\delta}\right)
\end{equation}
shots for a sufficiently large universal constant $C$.
\For{$j=1,\ldots,N$}
    \State Draw $\varphi_j$ uniformly from $[0,2\pi)$.
    \State Set
    \begin{equation}
        \beta_{t,\varphi_j}=-\frac{it}{\sqrt2}e^{i\varphi_j}.
    \end{equation}
    \State Prepare
    \begin{equation}
        \ket{\Psi_{t,\varphi_j}}=\frac{1}{\sqrt2}\left(\ket{g}\ket{0}+\ket{e}D(\beta_{t,\varphi_j})\ket{0}\right).
    \end{equation}
    \State Send the sensing mode through $\mathcal E_P$.
    \State Apply $D(-\beta_{t,\varphi_j})$ on the $\ket{e}$ arm.
    \State With probability $1/2$, measure $X$ on the qubit, obtain $S_j\in\{\pm1\}$, and set $W_j=2S_j$.
    \State With probability $1/2$, measure $Y$ on the qubit, obtain $S_j\in\{\pm1\}$, and set $W_j=2iS_j$.
    \State Set
    \begin{equation}
        Z_j=\frac{i^{-k}e^{ik\varphi_j}}{J_k(z_k)}W_j.
    \end{equation}
\EndFor
\State Output
\begin{equation}
    \widehat{\Lambda}_k=\frac{1}{N}\sum_{j=1}^NZ_j.
\end{equation}
\end{algorithmic}
\end{algorithm}

\begin{lemma}[Sample complexity of the experimental proxy]
\label{thm:experimental_proxy_fixed_amplitude}
Algorithm \ref{alg:experimental_proxy_fixed_amplitude} estimates $\Lambda_k(P)$ to additive error $\epsilon$ with probability at least $1-\delta$ for every fixed-amplitude distribution $P$, using
\begin{equation}
    N= O\!\left(k^{2/3}\epsilon^{-2}\log\!\left(\frac{1}{\delta}\right)\right)
\end{equation}
queries. The mean photon number in the sensing mode before the signal is $O(k^2/A^2)$, and hence is $O(k)$ when $A=\Theta(\sqrt{k})$.
\end{lemma}

\begin{proof}
Fix a deterministic signal displacement $D(Ae^{i\theta})$. The Weyl relation gives
\begin{equation}
    D(-\beta_{t,\varphi})D(Ae^{i\theta})D(\beta_{t,\varphi})
    =
    \exp\left(2i\operatorname{Im}(Ae^{i\theta}\overline{\beta}_{t,\varphi})\right)D(Ae^{i\theta}).
\end{equation}
By the choice of $\beta_{t,\varphi}$,
\begin{equation}
    2\operatorname{Im}(Ae^{i\theta}\overline{\beta}_{t,\varphi})=tR\cos(\theta-\varphi).
\end{equation}
Thus after the inverse conditional displacement $D(-\beta_{t,\varphi_j})$, the sensing mode factors out and the qubit carries the relative phase $e^{itR\cos(\theta-\varphi)}$. The $X/Y$ then gives
\begin{equation}
    \mathbb E[W_j|\theta,\varphi]=e^{itR\cos(\theta-\varphi)}.
\end{equation}
Since $tR=z_k$, the Jacobi-Anger expansion gives
\begin{equation}
    e^{iz_k\cos(\theta-\varphi)}
    =
    \sum_{m\in\mathbb Z}i^mJ_m(z_k)e^{im(\theta-\varphi)}.
\end{equation}
Averaging over the random angle $\varphi$,
\begin{align}
    \mathbb E[Z_j|\theta]
    &=
    \frac{i^{-k}}{J_k(z_k)}
    \int_0^{2\pi}\frac{d\varphi}{2\pi}e^{ik\varphi}e^{iz_k\cos(\theta-\varphi)} \\
    &=
    e^{ik\theta}.
\end{align}
Averaging over $\theta$ proves $\mathbb E[Z_j]=\Lambda_k(P)$. Moreover, $|W_j|=2$ and $|J_k(z_k)|\ge c_Bk^{-1/3}$, so $|Z_j|\le \frac{2}{c_B}k^{1/3}$. Hoeffding's inequality applied to the real and imaginary parts gives
\begin{equation}
    \Pr\left[\left|\frac{1}{N}\sum_{j=1}^NZ_j-\Lambda_k(P)\right|>\epsilon\right]\le 4\exp\left(-c\frac{N\epsilon^2}{k^{2/3}}\right)
\end{equation}
for a universal constant $c>0$, which proves the stated sample complexity.

The sensing mode is in vacuum on one arm and in the coherent state $D(\beta_{t,\varphi})\ket{0}$ on the other arm, so its mean photon number is at most a universal constant times
\begin{equation}
    |\beta_{t,\varphi}|^2=\frac{t^2}{2}=\frac{z_k^2}{2R^2}=O\left(\frac{k^2}{A^2}\right).
\end{equation}
For $A=\Theta(\sqrt{k})$, this is $O(k)$.
\end{proof}

For the hypothesis testing distributions $P_\pm^{(k)}$, the proxy therefore distinguishes the two cases with $O(k^{2/3})$ samples and $O(k)$ energy when $A=\Theta(\sqrt{k})$. This is polynomially worse than the squeezed homodyne protocol, which achieves $O(1)$ samples at the same energy scaling. Thus the experimental proxy is strictly less informative for this task than the squeezed homodyne strategy. It is nevertheless still exponentially more sample-efficient than coherent-probe sensing by Theorem \ref{thm:classical_hardness_formal}, so observing the separation with this weaker protocol gives a conservative experimental demonstration of the Fourier analysis advantage in Corollary \ref{cor:squeezed_over_coherent}.

\subsection{Learning with quantum control}
\label{sec:quantum_control_upper_bounds}

We next move one level higher in the sensing hierarchy. The lower bound in Section \ref{sec:conventional_gaussian_lb} shows that conventional Gaussian sensing protocols with bounded energy have exponentially small overlap with high-frequency Fourier modes. Here, we show that a single control qubit and one controlled displacement convert this exponentially small overlap into an inverse-polynomial one, enabling efficient Quantum Feature Sensing. As in the previous subsection, we begin with the hypothesis testing task from Section \ref{sec:conventional_gaussian_lb}, and use the Q$\Psi$ hypothesis testing guarantee based on an AFI lower bound to demonstrate the exponential advantage. We then use the global Fourier-gain theorem (Theorem \ref{thm:qes_fourier_gain_global_observable}) to give a broader learning algorithm for efficiently learning high-frequency Fourier coefficients.

First we recall the hypothesis testing task. We work in quadrature coordinates $z=(x,p)\in\mathbb R^2$, and let
\begin{equation}
    g(x,p)=\frac{1}{2\pi}\exp\left(-\frac{x^2+p^2}{2}\right)
\end{equation}
be the standard Gaussian reference distribution. For $k\geq 1$, recall that the hypothesis families were given by
\begin{align} \label{eq:one_qubit_depth_one_defns}
    &c_k=\mathbb E_g[\cos(kX)]=e^{-k^2/2}\\
    &h_k(x,p)=\frac{\cos(kx)-c_k}{1+c_k} \\
    &P_{\pm}^{k}(x,p)=g(x,p)\left(1\pm h_k(x,p)\right).
\end{align}
These signals differ in the observable
\begin{equation}
    \Psi_R(P)=\mathbb E_{(X,P')\sim P}[\cos(kX)].
\end{equation}
by a constant margin. Naturally, estimating the Fourier coefficient $\mathbb{E}[e^{ikX}]$ and taking the real part would reveal this quantity and solve the hypothesis testing problem. Previously we demonstrated that any adaptive protocol using Gaussian measurements and energy budget $\leq ck$, for any constant $c$ independent of $k$ requires $\exp(\Omega(k))$ queries to the signal to solve this task with constant bias.

Now consider the following protocol for the hypothesis testing task. For a constant $c>0$, fix an energy budget $E$ and assume $k^2\ge 8E$. Set
\begin{equation}
    p_E=\frac{4E}{k^2},\qquad \beta_k=-\frac{ik}{2}.
\end{equation}
Prepare the state
\begin{equation}
   \sqrt{1-p_E}\ket{g}\ket{0}+\sqrt{p_E}\ket{e}D(\beta_R)\ket{0}
\end{equation}
by initializing an ancilla qubit in the state $\sqrt{1-p_E}|0\rangle + \sqrt{p_E}|1\rangle$ and applying $\textnormal{ECD}(\beta_k)$, send the sensing mode through the displacement channel, apply $\textnormal{ECD}(\beta_k)^\dagger$ on the $\ket{e}$ arm, and then measure $X$ on the qubit. This straightforwardly defines a Ramsey sequence using single-qubit ancilla control that requires only a single controlled gate to prepare the probe state and a single inverse gate to perform the readout. We denote the resulting binary measurement by $M^{E,k}$ and its outcome by $W\in\{\pm1\}$. Importantly, at no point during the protocol does the sensor state have a mean photon number $> E$.

\begin{lemma}[AFI of the depth-one controlled sensing protocol]
\label{lemma:depth_one_control_AFI}
For the hypothesis testing direction $h_R$ above,
\begin{equation}
    \mathsf A_{M^{E,k}}(h_k;g)\ge C\frac{E}{k^2}
\end{equation}
for a universal constant $C>0$, whenever $k\ge1$ and $k^2\ge8E$. Furthermore, this lower bound is tight up to constants for any depth-one sensing protocol with energy $\leq E$.
\end{lemma}

\begin{proof}
We first compute the response of the protocol. Fix a deterministic signal displacement $D(x+ip')$. The Weyl relation gives
\begin{equation}
    D(-\beta_k)D(x+ip')D(\beta_k)=\exp\left(2i\operatorname{Im}((x+ip')\overline{\beta_k})\right)D(x+ip').
\end{equation}
Since $\beta_k=-ik/2$, $2\operatorname{Im}((x+ip')\overline{\beta_k})=kx$. Thus after the inverse controlled displacement, the sensing mode factors out and the qubit is in the state
\begin{equation}
    \sqrt{1-p_E}\ket{g}+\sqrt{p_E}e^{ikx}\ket{e}.
\end{equation}
Measuring $X$ gives
\begin{equation}
    \mathbb E[W|x,p']=2\sqrt{p_E(1-p_E)}\cos(kx).
\end{equation}
Let $s_E=2\sqrt{p_E(1-p_E)}$. We now compute the AFI, noting that because we are now measuring the qubit, the response functions will simply be two-outcome discrete probabilities; nevertheless, the Q$\Psi$ formalism applies directly. The response under reference distribution $g$ is 
\begin{equation}
    Q_0(W=w)=\frac{1}{2}\left(1+ws_Ec_k\right),\qquad w\in\{\pm1\}.
\end{equation}
The response function under the perturbed hypothesis is
\begin{align}
    A_{h_k}(W=w)
    &=
    \int h_k(x,p')\frac{1}{2}\left(1+ws_E\cos(kx)\right)g(x,p')dxdp' \\
    &=
    \frac{ws_E}{2}\mathbb E_g[h_k(X,P')\cos(kX)] \\
    &=
    \frac{ws_E}{2}\frac{v_k}{1+c_k} \ ,
\end{align}
where $v_k$ is given by
\begin{equation}
    v_k=\mathbb E_g[\cos^2(RX)]-\mathbb E_g[\cos(RX)]^2 \ .
\end{equation}
Therefore
\begin{align}
    \mathsf A_{M^{E,R}}(h_R;g)
    &=
    \sum_{w=\pm1}\frac{A_{h_R}(W=w)^2}{Q_0(W=w)} \\
    &=
    \frac{s_E^2v_R^2}{(1+c_R)^2(1-s_E^2c_R^2)}.
\end{align}
Since $k^2\ge8E$, we have $p_E\le1/2$, and hence $s_E^2=4p_E(1-p_E)\ge2p_E$. Also, for $R\ge1$,
\begin{equation}
    v_k=\frac{1+e^{-2k^2}}{2}-e^{-k^2}\ge \frac{1}{2}-\frac{1}{e}>\frac{1}{8},
\end{equation} 
so $v_k>1/8$ and $1+c_k\le2$. Thus
\begin{equation}
    \mathsf A_{M^{E,R}}(h_R;g)\ge C p_E=C\frac{E}{k^2}
\end{equation}
for a universal constant $C>0$. Next we show that this is optimal for depth-one protocols. Given a single qubit and a single controlled displacement, the state of the sensor must always be of the form

\begin{equation}
    \ket{\Psi_0}
    =
    \sqrt a\ket{0}\ket{\beta_1}
    +
    e^{i\chi}\sqrt{1-a}\ket{1}\ket{\beta_2}
\end{equation}
for some $a\in[0,1], \chi \in \mathbb R$, and $\beta_1,\beta_2\in\mathbb C$. The mean photon number in the sensing mode is
\begin{align}
    E_{\Psi}
    &=
    a|\beta_1|^2+(1-a)|\beta_2|^2 \\
    &=
    \left|a\beta_1+(1-a)\beta_2\right|^2+a(1-a)|\beta_2-\beta_1|^2.
\end{align}
Thus, if $\Delta=\beta_2-\beta_1$, the energy constraint implies
\begin{equation}
    a(1-a)|\Delta|^2\le E.
\end{equation}
The first term in the energy is a common displacement of both branches; as is evident from the rest of this work, the more refined resource is not ordinary energy but \textit{nonclassical} energy, so energy allocated towards the first term will not asymptotically change performance. It can only contribute the ordinary coherent-probe response, whereas the controlled part of the protocol is governed by the branch separation $\Delta$. This is made concrete by the following argument. 

Let $z=(x,p)$, let $\sigma_z=D(z)\ketbra{0}{0}D(z)^\dagger$, and write
\begin{equation}
    \xi_\Delta\cdot z=2\operatorname{Im}(z\overline{\Delta}),\qquad \Delta=\beta_2-\beta_1.
\end{equation}
Conditioning on a displacement $D(z)$, the sensor state after the inverse controlled displacement becomes 
\begin{equation}
    \ket{\Psi_z}
    = \left(
        \sqrt a\ket{0}+
        e^{i\chi}\sqrt{1-a}e^{i\xi_\Delta\cdot z}\ket{1}
    \right)D(z)\ket{0},
\end{equation}
Averaging over the signal distribution $g(z)(1+\eta h_k(z))$, the pre-measurement state is
\begin{equation}
    \rho_\eta=\int dz \ g(z)(1+\eta h_k(z))\ketbra{\Psi_z}{\Psi_z}.
\end{equation}
We introduced $\eta$ so that we can let $\dot\rho=\left.\frac{d}{d\eta}\rho_\eta\right|_{\eta=0}$. This will give us a clean way to write the AFI, in a QFI-reminiscent manner. The diagonal blocks of $\dot\rho$ are proportional to $\int dz \ g(z)h_k(z)\sigma_z$, independent of the relative separation in the initial state and have exponentially small operator norm in $k$ by the conventional Gaussian lower bound. The off-diagonal block is
\begin{equation}
    \dot\rho_{\rm off}
    =
    \sqrt{a(1-a)}
    \left(
        e^{-i\chi}\ketbra{0}{1}\otimes T_\Delta
        +
        e^{i\chi}\ketbra{1}{0}\otimes T_\Delta^\dagger
    \right),
\end{equation}
where
\begin{equation}
    T_\Delta=\int dz \ g(z)h_R(z)e^{i\xi_\Delta\cdot z}\sigma_z.
\end{equation}
Now fix an arbitrary POVM $\{M_s\}_s$ applied to the final qubit-sensor state. If $p_s(\eta)=\Tr(M_s\rho_\eta)$, then the classical AFI of this POVM in the direction $\dot\rho_{\rm off}$ is
\begin{equation}
    \sum_s\frac{\dot p_s^2}{p_s(0)},\qquad
    \dot p_s=\Tr(M_s\dot\rho_{\rm off}).
\end{equation}
Using the block form above and Cauchy-Schwarz in the Hilbert-Schmidt inner product, one obtains
\begin{equation}
    \sum_s\frac{\dot p_s^2}{p_s(0)}
    \le
    C a(1-a)\Tr\left(T_\Delta^\dagger\sigma_0^{-1}T_\Delta\right),
    \label{eq:controlled_block_chi_square_bound}
\end{equation}
where $\sigma_0=\int dz \ g(z)\sigma_z$ and $C>0$ is a universal constant. 
Now we control the operator norm in \eqref{eq:controlled_block_chi_square_bound}. Since
\begin{equation}
    h_k(z)=\frac{\cos(Rx)-c_k}{1+c_k},
\end{equation}
the function $h_k(z)e^{i\xi_\Delta\cdot z}$ is a linear combination of Gaussian Fourier modes at frequencies $\xi_\Delta+(R,0)$, $\xi_\Delta-(R,0)$, and $\xi_\Delta$. The Gaussian coherent-state channel damps a Fourier mode of phase-space frequency $\kappa$ by a factor exponentially small in $|\kappa|^2$. Omitting the Gaussian integration algebra, we find
\begin{equation}
    \Tr\left(T_\Delta^\dagger\sigma_0^{-1}T_\Delta\right)
    \le
    C\left(
        e^{-c|\xi_\Delta-(k,0)|^2}
        +
        e^{-c|\xi_\Delta+(k,0)|^2}
        +
        e^{-ck^2}e^{-c|\xi_\Delta|^2}
    \right)
\end{equation}
for universal constants $c,C>0$. We have that if e.g. $|\xi_\Delta|<k/2$, the controlled contribution to the AFI is exponentially small in $k^2$. Any response that is significant has $|\xi_\Delta|\ge k/2$ (and must indeed be $\sim k$, but the cruder bound will suffice). Since $|\xi_\Delta|=2|\Delta|$, this implies $|\Delta|\ge k/4$. Combining this with the energy constraint $a(1-a)|\Delta|^2\le E$ gives
\begin{equation}
    a(1-a)\le \frac{16E}{k^2}.
\end{equation}
Substituting into \eqref{eq:controlled_block_chi_square_bound}, the contribution to the AFI is at most $O(E/k^2)$, up to terms exponentially small in $k^2$. The diagonal block contributes only exponentially small AFI as discussed earlier. This proves that no final POVM can extract more than $O(E/k^2)$ AFI for this hypothesis testing problem once an energy-$E$ probe state is created using a single controlled displacement.
\end{proof}

This Lemma gives us the Q$\Psi$ hypothesis testing separation.

\begin{graydefinitionbox}
\begin{corollary}[Exponential advantage from one ancilla qubit and constant control depth]
\label{cor:one_qubit_control_hypothesis_advantage}
There is an optimal depth-one single-qubit control protocol using energy $k$ per query which distinguishes $P_+^{k}$ from $P_-^{k}$ in Equation \eqref{eq:one_qubit_depth_one_defns} with constant bias using $O(k)$ queries. In contrast, any conventional Gaussian sensing protocol with the same energy budget requires $\exp(\Omega(k))$ queries to do so.
\end{corollary}
\end{graydefinitionbox}
\begin{proof}
Lemma \ref{lemma:depth_one_control_AFI} and Theorem \ref{thm:qes_matched_score_guarantee} give sample complexity
\begin{equation}
    O\!\left(\frac{1}{\mathsf A_{M^{E=k,k}}(h_{k};g)}\log\!\left(\frac{1}{\delta}\right)\right)
    =
    O\left(k\right),
\end{equation}
since $\delta$ is a fixed constant. The protocol uses energy $E=k$ by construction. The lower bound is Theorem \ref{thm:conventional_hardness_formal}. This is exactly the statement that conventional Gaussian sensing with energy $k$ has exponentially small access to the frequency-$k$ phase-space Fourier mode, while the one-qubit controlled-displacement experiment has AFI $\Omega(1/k)$ and therefore succeeds with only $O(k)$ queries.
\end{proof}
We now turn to the genuinely global version of this learning problem. The previous hypothesis test fixed the Fourier fringe along the $x$ axis. In practice, however, the relevant Fourier direction may not be known when the data are collected, and will then need to be ascertained from the collected data. For this we consider a directional analog of the characteristic function,
\begin{equation}
    \lambda_P(k,\theta)=\mathbb E_{z\sim P}\left[e^{ik u_\theta\cdot z}\right],\qquad u_\theta=(\cos\theta,\sin\theta).
\end{equation}
The controlled-displacement protocol can learn this object post-hoc on an angular grid by randomizing the controlled-displacement angle and storing the angle used in each shot. Let $\Theta_m=\{2\pi r/m:r=0,\ldots,m-1\}$. For $\theta\in[0,2\pi)$, let $\Pi_m(\theta)$ denote a nearest point in $\Theta_m$.

\begin{algorithm}[h!]
\caption{Post-hoc directional Fourier learning with one control qubit}
\label{alg:posthoc_directional_fourier_control}
\begin{algorithmic}[1]
\Require Frequency $k\ge1$, energy $E$ with $k^2\ge8E$, angular net size $m$, query access to $\mathcal E_P$, and number of shots $N$.
\Ensure A classical transcript which can estimate $\lambda_P(k,\theta)$ for post-hoc directions $\theta$.
\State Set $p_E=4E/k^2$ and $s_E=2\sqrt{p_E(1-p_E)}$.
\For{$j=1,\ldots,N$}
    \State Draw $\varphi_j$ uniformly from $\Theta_m$ and set $\beta_j=-ik e^{i\varphi_j}/2$.
    \State Prepare $\sqrt{1-p_E}\ket{g}\ket{0}+\sqrt{p_E}\ket{e}D(\beta_j)\ket{0}$.
    \State Send the sensing mode through $\mathcal E_P$ and apply $D(-\beta_j)$ on the $\ket{e}$ arm.
    \State Measure $X$ with probability $1/2$ and set $W_j=2S_j$, or measure $Y$ with probability $1/2$ and set $W_j=2iS_j$.
    \State Store $(\varphi_j,W_j)$.
\EndFor
\State After receiving a query $\theta$, set $\theta^*=\Pi_m(\theta)$ and output
\begin{equation}
    \widehat{\lambda}_P(k,\theta)=\frac{m}{Ns_E}\sum_{j=1}^N \mathds{1}\{\varphi_j=\theta^*\}W_j.
\end{equation}
\end{algorithmic}
\end{algorithm}

\begin{graydefinitionbox}
\begin{theorem}[Post-hoc directional Fourier learning]
\label{thm:posthoc_directional_fourier_control}
For every displacement distribution $P$ and every $\theta\in\Theta_m$, Algorithm \ref{alg:posthoc_directional_fourier_control} estimates $\lambda_P(k,\theta)$ to additive error $\epsilon$ with probability at least $1-\delta$ using
\begin{equation}
    N = O\!\left(\frac{mk^2}{E\epsilon^2}\log\left(\frac{m}{\delta}\right)\right)
\end{equation}
queries, simultaneously for all $\theta\in\Theta_m$. If, moreover, $P$ satisfies $\mathbb E_{z\sim P}[\|z\|]\le B$, then choosing $m=O(kB/\epsilon)$ gives a post-hoc estimator for every $\theta\in[0,2\pi)$ with sample complexity
\begin{equation}
    N = O\!\left(\frac{k^3B}{E\epsilon^3}\log\left(\frac{kB}{\epsilon\delta}\right)\right).
\end{equation}
\end{theorem}
\end{graydefinitionbox}

\begin{proof}
Fix $z=(x,p)$ and a chosen angle $\varphi$. By the Weyl relation and the choice $\beta=-ik e^{i\varphi}/2$, the echo imprints the relative phase $e^{ik u_\varphi\cdot z}$ on the qubit. Hence the randomized $X/Y$ readout satisfies
\begin{equation}
    \mathbb E[W_j|z,\varphi_j]=s_Ee^{ik u_{\varphi_j}\cdot z}.
\end{equation}
For a fixed net point $\theta\in\Theta_m$, the postprocessed one-shot variable
\begin{equation}
    Z_{\theta,j}=\frac{m}{s_E}\mathds{1}\{\varphi_j=\theta\}W_j
\end{equation}
therefore satisfies $\mathbb E[Z_{\theta,j}|z]=e^{ik u_\theta\cdot z}$. Averaging over $z\sim P$ gives $\mathbb E[Z_{\theta,j}]=\lambda_P(k,\theta)$.

The second moment obeys $\mathbb E[|Z_{\theta,j}|^2]\le 4m/s_E^2$. Since $k^2\ge8E$, we have $p_E\le1/2$, so $s_E^2=4p_E(1-p_E)=\Omega(E/k^2)$. Bernstein's inequality, applied to real and imaginary parts and union bounded over the $m$ net points, gives
\begin{equation}
    N=O\left(\frac{mk^2}{E\epsilon^2}\log\left(\frac{m}{\delta}\right)\right).
\end{equation}
This is exactly the Fourier-overlap theorem with gain $\Omega(E/(mk^2))$ per net direction, because the randomized protocol spends a $1/m$ fraction of its shots on each Fourier response.

For the continuum statement, let $\theta^\sharp=\Pi_m(\theta)$. Since $|e^{ia}-e^{ib}|\le |a-b|$,
\begin{align}
    |\lambda_P(k,\theta)-\lambda_P(k,\theta^\sharp)|
    &\le k\mathbb E_{z\sim P}\left[|(u_\theta-u_{\theta^\sharp})\cdot z|\right] \\
    &\le kB|\theta-\theta^\sharp| \\
    &\le \frac{2\pi kB}{m}.
\end{align}
Choosing $m=O(kB/\epsilon)$ makes this discretization error at most $\epsilon/2$, and applying the net guarantee with accuracy $\epsilon/2$ proves the claim.
\end{proof}

This is the global Fourier-overlap picture for one-qubit control. A known direction requires gain $\Omega(E/k^2)$ and hence $O(k^2/(E\epsilon^2))$ samples. If the direction is not known when data are collected, randomized cat angles spread this gain over the angular net, producing gain $\Omega(E/(mk^2))$ per post-hoc direction. In the main-text scaling $E=k$, this remains polynomial in $k$, while the conventional Gaussian lower bound is exponential for the same high-frequency features.

\subsubsection{Alternative approach using $\textnormal{\textsf{SU}}(1, 1)$ interferometry or Gaussian boson sampling} \label{sec:gbs_alternate}

Gaussian boson samplers are most often studied in settings where sampling from the circuit output distribution is itself the objective. However, here we demonstrate that the same hardware can instead be used to learn a property of an external signal more efficiently than any conventional Gaussian sensor. We consider the same angular Fourier coefficient task above and show that the echo implemented by a canonical $\mathrm{SU}(1,1)$ interferometry protocol \cite{yurke_su2_1986}, followed by photon-number parity measurement, estimates $\Lambda_k(P)$ with polynomially many signal queries and mean photon number $O(k)$.

Let $S(r,\phi)$ denote a single-mode squeezing unitary oriented so that
\begin{equation}
S(r,\phi)^\dagger D(Ae^{i\theta})S(r,\phi)=D\left(e^{i\phi}A\left(e^{-r}\cos(\theta-\phi)+ie^r\sin(\theta-\phi)\right)\right).
\end{equation}
We take $A=\sqrt{k}$ and $r=\log A$. We also define the following normalization constant:
\begin{equation}
\gamma_k=\frac{1}{2\pi}\int_0^{2\pi}du \exp\left(-2(1+\cos u)^2-2k^2\sin^2u\right)e^{-iku}.
\end{equation}
This coefficient depends only on $k$ and is used in the following estimator.

\begin{algorithm} \label{alg:su_11_estimator}
\caption{Randomized $\textsf{SU}(1, 1)$ interferometry for angular Fourier learning}
\begin{algorithmic}[1]
\Require Frequency $k\in\mathbb{Z}^+$, accuracy parameters $\epsilon,\delta$, and query access to $\mathcal{E}_P$, where $P$ has fixed amplitude $A=\sqrt{k}$.
\Ensure An estimate $\widehat{\Lambda}_k$ of $\Lambda_k(P)=\mathbb{E}_{\theta\sim F}[e^{ik\theta}]$.
\State Set $r=\log A$ and
\begin{equation}
N=\left\lceil \frac{C}{|\gamma_k|^2\epsilon^2}\log\left(\frac{4}{\delta}\right)\right\rceil
\end{equation}
for a sufficiently large universal constant $C$.
\For{$j=1,\ldots,N$}
\State Draw $\phi_j$ uniformly from $[0,2\pi)$.
\State Prepare $D(e^{i\phi_j})\ket{0}$ and apply $S(r,\phi_j)$.
\State Send the mode through $\mathcal{E}_P$ and apply $S(r,\phi_j)^\dagger$.
\State Measure $\hat{\Pi}=(-1)^{\hat{n}}$, obtaining $Y_j\in\{-1,1\}$.
\State Set $Z_j=e^{ik\phi_j}Y_j/\gamma_k$.
\EndFor
\State Output
\begin{equation}
\widehat{\Lambda}_k=\frac{1}{N}\sum_{j=1}^N Z_j.
\end{equation}
\end{algorithmic}
\end{algorithm}

\begin{theorem}[Angular Fourier learning with $\textsf{SU}(1, 1)$ interferometry]
Let $P$ be any displacement distribution supported on $\{Ae^{i\theta}:\theta\in[0,2\pi)\}$ with $A=\sqrt{k}$. Algorithm \ref{alg:su_11_estimator} outputs $\widehat{\Lambda}_k$ satisfying
\begin{equation}
\left|\widehat{\Lambda}_k-\Lambda_k(P)\right|\leq\epsilon
\end{equation}
with probability at least $1-\delta$ using
\begin{equation}
N = O\!\left(\frac{k^2}{\epsilon^2}\log\!\left(\frac{1}{\delta}\right)\right)
\end{equation}
signal queries. The mean photon number used is $O(k)$.
\end{theorem}

\begin{proof}
Fix a deterministic displacement $Ae^{i\theta}$ and write $u=\theta-\phi$. After applying the inverse squeezer, the output is, up to a global phase, the coherent state
\begin{equation}
\ket{\psi_{\theta,\phi}}=\ket{e^{i\phi}\left(1+\cos u+ik\sin u\right)}
\end{equation}
which follows from the commutation relation between squeezing and displacement operators. Since $\langle\alpha|\hat{\Pi}|\alpha\rangle=e^{-2|\alpha|^2}$, the measurement response is given by
\begin{equation}
\mathbb{E}[Y|\theta,\phi]=q_k(\theta-\phi),
\end{equation}
where
\begin{equation}
q_k(u)=\exp\left(-2(1+\cos u)^2-2k^2\sin^2u\right).
\end{equation}
Averaging over the random squeezing angle and applying the normalization factor gives
\begin{align}
\mathbb{E}[Z|\theta]&=\frac{1}{\gamma_k}\frac{1}{2\pi}\int_0^{2\pi}d\phi e^{ik\phi}q_k(\theta-\phi)\\
&=e^{ik\theta}.
\end{align}
It follows that $\mathbb{E}[Z]=\Lambda_k(P)$ for every angular distribution $F$. Next we control the sample complexity. The response $q_k$ is concentrated in neighborhoods of width $\Theta(1/k)$ around $u=0$ and $u=\pi$. Rescaling these neighborhoods by $u=s/k$ and $u=\pi+s/k$ gives
\begin{equation}
\eta_k^{\mathrm{par}}=\frac{e^{-1/8}}{2\sqrt{2\pi}k}\left(e^{-8}+(-1)^k\right)+o(k^{-1}).
\end{equation}
Consequently, there are universal constants $c,C>0$ such that
\begin{equation}
\frac{c}{k}\leq|\eta_k^{\mathrm{par}}|\leq\frac{C}{k}.
\end{equation}
The normalization is nonzero for every $k$: indeed,
\begin{equation}
q_k(u)=e^{-(k^2+3)}e^{-4\cos u}e^{(k^2-1)\cos(2u)},
\end{equation}
and its modified-Bessel expansion shows that the sign of its $k$-th Fourier coefficient is $(-1)^k$. Since $|Y|=1$, each postprocessed outcome satisfies $|Z|=O(k)$. Hoeffding's inequality applied to the real and imaginary parts therefore gives
\begin{equation}
\Pr\left[\left|\widehat{\Lambda}_k-\Lambda_k(P)\right|>\epsilon\right]\leq4\exp\left(-c\frac{N\epsilon^2}{k^2}\right),
\end{equation}
which proves the sample-complexity claim.

Finally, the state incident on the signal is $S(r,\phi)D(e^{i\phi})\ket{0}$. Its mean photon number is bounded by
\begin{equation}
\overline{n}_{\mathrm{probe}}\leq e^{2r}+\sinh^2r=O(k).
\end{equation}
The signal displacement has amplitude $A=\sqrt{k}$, so the occupation remains $O(k)$ throughout the sensing window. Moreover, squeezing preserves photon-number parity, so $[S(r,\phi),\hat{\Pi}]=0$ and the inverse squeezer can equivalently be absorbed into the parity readout. Thus it does not require a larger probe-energy budget.
\end{proof}

The non-Gaussian resource in this protocol is the photon-counting readout itself. At the level of non-Gaussian resource counting, the bosonic phase supplied by the ECD echo may be represented by a Kerr unitary of the form
\begin{equation}
K_{\pi/2}=\exp\left(\frac{i\pi}{2}\hat{n}^2\right),
\end{equation}
where $\hat{n}=\hat{a}^\dagger\hat{a}$ and $\hat{n}^2$ is fourth order in the ladder operators. At this special phase,
\begin{equation}
K_{\pi/2}=\frac{1+i}{2}I+\frac{1-i}{2}\hat{\Pi}\,,
\end{equation}
because its eigenvalue is $1$ on even photon numbers and $i$ on odd photon numbers. A single photon-number measurement therefore supplies exactly the spectral information needed to evaluate this quartic phase. Under this accounting, the squeezed-parity protocol uses the same one layer of non-Gaussian resource as the ECD protocol.

\subsection{Learning with quantum control and memory}
\label{sec:quantum_memory_upper_bounds}

Next, we illustrate the power of a single qubit of long-lived quantum memory for sensing. In Section \ref{sec:memoryless_lower_bounds} we established that energy-constrained protocols without long-lived degrees of freedom, even with arbitrary ancillas, modes, and measurements, incur an exponential cost to estimate temporal Fourier coefficients. Here we show that a single qubit of memory coherent over the lifetime of a time-varying signal enables efficient estimation of these temporal correlations, even if the sensor itself decoheres quickly.

First let us understand how temporal Fourier coefficients capture properties of time-varying signals. 
Let $P$ denote the probability law of the time-dependent displacement signal whose induced displacements at different points in time are described by a stochastic process $\{Z_t\}_{t\in\mathcal T}$, where $Z_t=(X_t,P_t)\in\mathbb R^2$. For a finite list of times $\mathbf t=(t_1,\ldots,t_m)$, the resulting marginal distribution is $P_{\mathbf t}$ on $(\mathbb R^2)^m$ defined by
\begin{equation}
    P_{\mathbf t}(A_1\times\cdots\times A_m)=P[Z_{t_1}\in A_1,\ldots,Z_{t_m}\in A_m] .
\end{equation}
In Quantum Signal Learning language, physical observables of such time-varying signals are defined by a kernel function $F:(\mathbb R^2)^m\to\mathbb C$, such that a property is given by
\begin{equation}
 \mathbb E_{P_{\mathbf t}}[F]=\int_{(\mathbb R^2)^m}F(z_1,\ldots,z_m)\,p_{\mathbf t}(z_1,\ldots,z_m)\prod_{j=1}^m d^2z_j .
\end{equation}
where $P_{\mathbf t}$ has a density $p_{\mathbf t}$. The temporal Fourier coefficient is the special case in which the kernel is a product of Weyl characters:
\begin{equation}
    \widehat P(\mathbf t,\boldsymbol{\zeta})=\int_{(\mathbb R^2)^m}\prod_{j=1}^m e^{i\Omega(\zeta_j,z_j)}P_{\mathbf t}(dz_1\cdots dz_m).
    \label{eq:temporal_characteristic_integral_form}
\end{equation}
This is the characteristic function of the joint distribution $P_{\mathbf t}$. Note, importantly, that $P_{\mathbf t}$ is generally not the product $P_{t_1}\otimes\cdots\otimes P_{t_m}$. The temporal correlations are exactly the information contained in the joint distribution that is not captured by single-time marginals.

This picture shows us that other temporal observables can be recovered from measuring temporal Fourier coefficients, because these objects can be used to identify the characteristic function of the signal in an informationally-complete manner. Let $\psi:(\mathbb R^2)^m\to\mathbb C$ be any property kernel with symplectic Fourier transform
\begin{equation}
    \widehat\psi(\boldsymbol{\zeta})=\frac{1}{(2\pi)^{2m}}\int_{(\mathbb R^2)^m}\psi(\mathbf z)\exp\left(-i\sum_{j=1}^m\Omega(\zeta_j,z_j)\right)\prod_{j=1}^m d^2z_j ,
\end{equation}
and assume $\widehat\psi\in L^1$. The inverse symplectic Fourier transform gives
\begin{equation}
    \psi(z_1,\ldots,z_m)=\int_{(\mathbb R^2)^m}\widehat\psi(\boldsymbol{\zeta})\exp\left(i\sum_{j=1}^m\Omega(\zeta_j,z_j)\right)\prod_{j=1}^m d^2\zeta_j .
\end{equation}
Consequently, the corresponding temporal observable $\Psi_\psi(P_{\mathbf t})=\mathbb E_{P_{\mathbf t}}[\psi(Z_{t_1},\ldots,Z_{t_m})]$ can be written as
\begin{equation}
    \Psi_\psi(P_{\mathbf t})=\int_{(\mathbb R^2)^m}\widehat\psi(\boldsymbol{\zeta})\widehat P(\mathbf t,\boldsymbol{\zeta})\prod_{j=1}^m d^2\zeta_j .
    \label{eq:temporal_observable_from_characteristic_function}
\end{equation}
We therefore see that estimation of $\widehat P(\mathbf t,\boldsymbol{\zeta})$, the appropriate temporal Fourier coefficients, enables downstream estimation of $\Psi_\psi(P_{\mathbf{t}})$ using classical postprocessing (with a corresponding conditioning-factor overhead depending on the complexity of $\psi$). For many observables with controlled Fourier support the additional measurement overhead required for this postprocessing can be modest. A practical example of physical observables that can be extracted from temporal Fourier coefficients are ordinary temporal moments, which emerge from taking derivatives at the origin. For a quadrature direction $u\in\mathbb R^2$, let $\zeta(u)$ denote the phase-space vector satisfying $\Omega(\zeta(u),z)=u^Tz$ for all $z\in\mathbb R^2$. If the moment exists and the characteristic function is differentiable at the origin, then
\begin{equation}
    \mathbb E_{P_{\mathbf t}}\left[\prod_{j=1}^m u_j^TZ_{t_j}\right]=(-i)^m\left.\frac{\partial^m}{\partial\lambda_1\cdots\partial\lambda_m}\widehat P\left(\mathbf t,(\lambda_1\zeta(u_1),\ldots,\lambda_m\zeta(u_m))\right)\right|_{\lambda_1=\cdots=\lambda_m=0}.
    \label{eq:temporal_moments_from_characteristic_function}
\end{equation}
Thus we have seen that the temporal Fourier coefficients are a general and highly expressive object for characterizing time-varying signals. Now we give a quantum feature sensing protocol using a long-lived memory qubit that estimates these coefficients directly.

\begin{algorithm}[h!]
\caption{Temporal Fourier coefficient estimation with one memory qubit}
\label{alg:temporal_weyl_memory_estimator}
\begin{algorithmic}[1]
\Require Times $\mathbf t=(t_1,\ldots,t_m)$, phase-space frequencies $\boldsymbol{\zeta}=(\zeta_1,\ldots,\zeta_m)$, number of trajectories $N$, and query access to independent realizations of the time-dependent displacement process.
\Ensure An estimate $\widehat G$ of $\widehat P(\mathbf t,\boldsymbol{\zeta})$.
\For{$\ell=1,\ldots,N$}
    \State Prepare the memory qubit in $\ket{+}=(\ket g+\ket e)/\sqrt2$.
    \For{$j=1,\ldots,m$}
        \State Prepare or retain the sensor mode in any state independent of the qubit.
        \State Apply $C_{\zeta_j}$.
        \State Let the signal act on the sensor mode at time $t_j$, applying the unknown displacement $D(Z_{t_j})$ for the current trajectory.
        \State Apply $C_{\zeta_j}^{\dagger}$.
        \State Discard, reset, or leave the sensor mode idle before the next time point.
    \EndFor
    \State With probability $1/2$, measure $X_q$ on the memory qubit, obtain $S_\ell\in\{\pm1\}$, and set $W_\ell=2S_\ell$.
    \State With probability $1/2$, measure $Y_q$ on the memory qubit, obtain $S_\ell\in\{\pm1\}$, and set $W_\ell=2iS_\ell$.
\EndFor
\State Output
\begin{equation}
    \widehat G=\frac{1}{N}\sum_{\ell=1}^N W_\ell .
\end{equation}
\end{algorithmic}
\end{algorithm}

\begin{graydefinitionbox}
\begin{theorem}[Temporal Fourier learning with one memory qubit]
\label{thm:temporal_fourier_memory_upper}
For any time-dependent displacement process $P$, Algorithm \ref{alg:temporal_weyl_memory_estimator} estimates $\widehat P(\mathbf t,\boldsymbol{\zeta})$ to additive error $\epsilon$ with probability at least $1-\delta$ using
\begin{equation}
    N = O\!\left(\epsilon^{-2}\log\!\left(\frac{1}{\delta}\right)\right)
\end{equation}
independent signal trajectories. Each trajectory uses $m$ controlled-displacement echoes and one qubit that remains coherent over the time interval containing $\mathbf t$. The sensor mode need not remain coherent between time points after each echo is closed.
\end{theorem}
\end{graydefinitionbox}

\begin{proof}
Fix a deterministic trajectory $z_1,\ldots,z_m$, where $z_j$ is the displacement at time $t_j$. We first compute one echo. By the Weyl relation,
\begin{equation}
    C_\zeta^\dagger(D(z)\otimes I_q)C_\zeta=D(z)\otimes\left(e^{-i\Omega(\zeta,z)/2}\ketbra{g}{g}+e^{i\Omega(\zeta,z)/2}\ketbra{e}{e}\right).
    \label{eq:single_temporal_echo_identity}
\end{equation}
The oscillator displacement $D(z)$ is common to both qubit branches. Therefore, up to an irrelevant global phase, the qubit transformation is
\begin{equation}
    \ket g+\xi\ket e\mapsto \ket g+\xi e^{i\Omega(\zeta,z)}\ket e .
\end{equation}
Applying this identity at the $m$ requested time points gives, for the fixed trajectory,
\begin{equation}
    \ket{+}\mapsto \frac{1}{\sqrt2}\left(\ket g+\Phi_{\boldsymbol{\zeta}}(z_1,\ldots,z_m)\ket e\right),
\end{equation}
where the accumulated phase is
\begin{equation}
    \Phi_{\boldsymbol{\zeta}}(z_1,\ldots,z_m)=\prod_{j=1}^m e^{i\Omega(\zeta_j,z_j)}=\exp\left(i\sum_{j=1}^m\Omega(\zeta_j,z_j)\right).
\end{equation}
Since the oscillator factor is common to the two qubit branches after every echo, tracing out or resetting the oscillator after a time point does not change the qubit coherence.

Averaging over the trajectory law $P_{\mathbf t}$, the final qubit satisfies
\begin{equation}
    \mathbb E[X_q]=\operatorname{Re}\widehat P(\mathbf t,\boldsymbol{\zeta}),\qquad \mathbb E[Y_q]=\operatorname{Im}\widehat P(\mathbf t,\boldsymbol{\zeta}).
\end{equation}
The randomized $X_q/Y_q$ readout used in the algorithm therefore has expectation
\begin{equation}
    \mathbb E[W_\ell]=\widehat P(\mathbf t,\boldsymbol{\zeta}).
\end{equation}
Moreover $|W_\ell|=2$ for every shot. Hoeffding's inequality applied to the real and imaginary parts gives
\begin{equation}
    \Pr\left[\left|\frac{1}{N}\sum_{\ell=1}^N W_\ell-\widehat P(\mathbf t,\boldsymbol{\zeta})\right|>\epsilon\right]\le 4\exp\left(-cN\epsilon^2\right)
\end{equation}
for a universal constant $c>0$. Taking $N=O(\epsilon^{-2}\log(1/\delta))$ proves the stated query complexity.
\end{proof}

Finally, the parity instance from Theorem \ref{thm:constant_frequency_parity_memory_lower} can be directly solved with this protocol. In that construction, the signal alphabet at time $t_j$ is
\begin{equation}
    Z_{t_j}=B_j(x_\star,0),\qquad B_j\in\{0,1\}.
\end{equation}
Choose $\zeta_\star$ so that $\Omega(\zeta_\star,(x,p))=\lambda_\star x$ and $\lambda_\star x_\star=\pi$. Then the local Weyl character is
\begin{equation}
    e^{i\Omega(\zeta_\star,Z_{t_j})}=e^{i\pi B_j}=(-1)^{B_j}=1-\frac{2X_{t_j}}{x_\star}.
\end{equation}
Applying Algorithm \ref{alg:temporal_weyl_memory_estimator} with $\boldsymbol{\zeta}=(\zeta_\star,\ldots,\zeta_\star)$ therefore estimates
\begin{align}
    \widehat P((t_1,\ldots,t_m),(\zeta_\star,\ldots,\zeta_\star))
    &=
    \mathbb E\left[\prod_{j=1}^m(-1)^{B_j}\right] \\
    &=
    \mathbb E\left[\prod_{j=1}^m\left(1-\frac{2X_{t_j}}{x_\star}\right)\right].
\end{align}
Under the two distributions $P_\sigma$ used in the lower bound, this expectation is $\sigma\epsilon_{\rm par}$, giving us the desired separation.

\begin{graydefinitionbox}
\begin{corollary}
    Given a list of times $\mathbf{t} = {t_1, t_2,...,t_m}$ separated by at most $\Delta$, a single quantum oscillator that may be coherent for time $\ll \Delta$, and a single qubit coherent over the time span of $\mathbf{t}$, there is an algorithm that can learn any temporal Fourier coefficient indexed by $\mathbf{t}$ using $O_m(1)$ signal queries using an energy $E_0$ independent of $m$. Any strategy with energy $E_0$ including arbitrary ancilla qubits or modes, and any non-Gaussian control, but without resources coherent for time $> \Delta$ requires $\exp(\Omega(m))$ signal queries. 
\end{corollary}    
\end{graydefinitionbox}

\subsection{A quadratic advantage over infinite-energy conventional quantum sensing}\label{sec:Gaussian_testing_upper_bound}
In Section \ref{sec:gaussian_testing_lower_bound}, we demonstrated a quadratic lower bound for estimating the variance of a single-mode quantum state. We considered the physically-motivated task of learning the strength of a weak thermal background by bounding the sample complexity of discriminating the two single-mode states
\begin{equation}
\rho_0^{(k)} =\ketbra{0}{0}\,, \qquad \rho_1^{(k)} = \tau_{1/k}\,,
\end{equation}
where $\tau_{1/k}$ is the thermal state with mean occupation number $1/k$:
\begin{equation}
\tau_t = \sum_{n=0}^\infty \frac{t^n}{(1+t)^{n+1}}\ketbra{n}{n} \ ,
\end{equation}
and any conventional Gaussian sensing protocol, even with access to arbitrarily high energy, requires at least $\Omega(k^2)$ samples to estimate their variance to the required accuracy. Here we show that there is a simple non-Gaussian measurement that estimates this variance using $O(k)$ samples.

\begin{lemma}\label{lemma:gaussian_testing_upper_bound}
There is a non-Gaussian measurement which estimates the variance of the above states to the required accuracy using $O(k)$ samples.
\end{lemma}
\begin{proof}
Measure the photon number $\hat{n}$ on each copy. For an arbitrary single-mode state $\rho$, the photon-counting outcome $X$ satisfies
\begin{equation}
\mathbb{E}[X]=\Tr(\hat{n}\rho) \ ,
\end{equation}
so the empirical photon number directly estimates the phase-space variance $v(\rho)=1+2\Tr(\hat{n}\rho)$. For the thermal state $\tau_t$, the photon-counting distribution is geometric, with
\begin{equation}
\mathbb{E}[X]=t, \qquad \operatorname{Var}(X)=t(1+t) \ .
\end{equation}
Let $\widehat t$ be the empirical mean of $N$ photon-counting outcomes and define $\widehat v=1+2\widehat t$. For the states above, $t\in\{0,1/k\}$, and hence
\begin{equation}
\operatorname{Var}(X)\leq \frac{1}{k}\left(1+\frac{1}{k}\right) \ .
\end{equation}
Standard concentration for geometric random variables therefore gives, for a universal constant $c>0$,
\begin{equation}
\Pr\left(\left|\widehat t-t\right|\geq \frac{1}{4k}\right)\leq 2\exp\left(-c\frac{N}{k}\right) \ .
\end{equation}
On the complementary event,
\begin{equation}
\left|\widehat v-v(\rho)\right|=2\left|\widehat t-t\right|\leq \frac{1}{2k} \ .
\end{equation}
Taking $N \geq Ck\log(2/\delta)$ for a sufficiently large universal constant $C$ makes the failure probability at most $\delta$. Thus $O(k\log(1/\delta))$ samples suffice, and in particular $O(k)$ samples suffice for constant success probability.
\end{proof}

For solving only the two-hypothesis discrimination task, one can also obtain a quadratic speedup over Gaussian protocols by using a parity measurement with POVM $\{\Pi_{\rm even}, \Pi_{\rm odd}\}$. Let $\Pi_{\rm odd}$ be the projector onto the odd Fock subspace. For a thermal state $\tau_t$,
\begin{equation}
\textnormal{Pr(odd)} = \sum_{l\geq 0}\frac{t^{2l+1}}{(1+t)^{2l+2}} = \frac{t}{1+2t} \ .
\end{equation}
Thus
\begin{equation}
\textnormal{Pr(odd)}_{\rho_0}=0 \qquad \textnormal{Pr(odd)}_{\rho_1}=\frac{1}{k+2} \ .
\end{equation}
More generally, if $p_{\rm odd}=\textnormal{Pr(odd)}$ for a thermal state, then
\begin{equation}
t=\frac{p_{\rm odd}}{1-2\,p_{\rm odd}} \ .
\end{equation}
Estimating $p_{\rm odd}$ by the empirical odd-outcome frequency and applying the same concentration argument therefore estimates $v(\tau_t)=1+2t$ to accuracy $1/(2k)$ using $O(k\log(1/\delta))$ samples. This parity measurement can be realized using the Ramsey sequence
\begin{equation}
\ket{+} \longrightarrow \exp((-i\pi/2)\sigma_z \hat{n}) \longrightarrow R_y(-\pi/2) \longrightarrow Z\textnormal{-measurement} \ .
\end{equation}

Together, Lemma \ref{lemma:gaussian_testing_upper_bound} and Theorem \ref{thm:Gaussian_testing_lower_bound} illustrate a simple but illuminating result.
\begin{graydefinitionbox}
\begin{corollary} \label{thm:gaussian_testing_combined}
There exists a family of single-mode Gaussian states whose variance can be estimated by a non-Gaussian protocol in quadratically fewer samples than by any arbitrarily high-energy adaptive Gaussian protocol.
\end{corollary}
\end{graydefinitionbox}

This result illustrates that non-Gaussianity as is afforded by single-qubit control is a resource that cannot be replicated by \textit{any} high-energy Gaussian processing. While this is well-established in the case of computation~\cite{Lloyd_1999}, where non-Gaussian operations are analogous to the non-Clifford gates that enable universal quantum computation, the above result establishes that non-Gaussianity is \textit{also} an irreplaceable resource in learning and sensing. This is operationally crucial given that conventional cutting-edge sensing platforms such as optical cavities utilize high-performance \textit{Gaussian} modes that by design struggle to prepare non-Gaussian states and perform non-Gaussian measurements.

\subsection{Entanglement-free multimode advantage for learning characteristic functions with one qubit}
\label{sec:char_function_learning_gkp}

In Section \ref{sec:multimode_charfun_lower_bound}, we identified a gap in the entanglement-free lower bounds of Ref.~\cite{Oh_2024}. Here we show that the gap is operationally meaningful. When the displacement distribution is sufficiently concentrated in phase-space, single-qubit control of the bosonic sensor enables a non-Gaussian readout that approximately samples the actual displacement drawn by the channel. Once such samples are obtained, post-hoc characteristic-function learning is just classical empirical Fourier estimation. 

\subsubsection{Displacement sensing with quantum phase estimation and GKP states}
\label{sec:gkp-preliminaries}

We first provide a basic overview Gottesman-Kitaev-Preskill (GKP) states as they apply to the characteristic-function learning protocol. Here we do not review their more well-known role in bosonic quantum error correction. For a single bosonic mode let $\hat x,\hat p$ denote canonical quadratures satisfying $[\hat x,\hat p]=i$. For $(x,p)\in\mathbb{R}^2$, we write the standard displacement operator as $D(x,p)=\exp\left(i(p\hat x-x\hat p)\right)$. In these coordinates, the Weyl commutation relations, Equation \eqref{eq:weyl_relation}, imply
\begin{align}
    D(x,p)D(x',p') &= \exp\left(-\frac{i}{2}(xp'-px')\right)D(x+x',p+p') \ , \\
    D(x,p)D(x',p') &= \exp\left(-i(xp'-px')\right)D(x',p')D(x,p) \ .
\end{align}
For $n$ modes, we write $\alpha=((x_1,p_1),\ldots,(x_n,p_n))\in\mathbb{R}^{2n}$ and use
\begin{equation}
    D(\alpha)=\bigotimes_{m=1}^n D(x_m,p_m) \ .
\end{equation}
We also write the symplectic form as
\begin{equation}
    \Omega(\beta,\alpha)=\sum_{m=1}^n(x_m p_m'-p_m x_m')
\end{equation}
when $\beta=((x_1,p_1),\ldots,(x_n,p_n))$ and $\alpha=((x_1',p_1'),\ldots,(x_n',p_n'))$. With this convention, the characteristic function of a displacement distribution $P$ is
\begin{equation}
    \lambda_P(\beta)=\mathbb{E}_{\alpha\sim P}\left[e^{i\Omega(\beta,\alpha)}\right] \ .
\end{equation}
We now define the square GKP grid state. Let $L=\sqrt{2\pi}$ and consider the two displacement operators
\begin{equation}
    S_x=D(L,0),\qquad S_p=D(0,L) \ .
\end{equation}
These commute because the their commutator, from the Weyl relation, is the phase $\exp(-iL^2)=\exp(-i2\pi)=1$. The ideal square GKP grid state is the simultaneous $+1$ eigenstate of $S_x$ and $S_p$. In the $x$ basis, it may be written formally as the Dirac comb
\begin{equation}
    \ket{\mathrm{GKP}}
    \propto
    \sum_{s\in\mathbb{Z}}\ket{x=sL} \ .
\end{equation}
However, this is not a normalizable physical state, and we will shortly discuss a more physical realization; nevertheless, the Dirac comb representation is the simplest way to understand the properties of a GKP state. Importantly, it has peaks separated by $L$ in $x$, and because it is also a $+1$ eigenstate of $D(0,L)$, it has the corresponding periodic structure in $p$. Now suppose an unknown signal acts on a GKP state by applying a displacement $D(a)$, where $a=(x_a,p_a)$. Then
\begin{align}
    S_xD(a)\ket{\mathrm{GKP}} &= e^{-iLp_a}D(a)\ket{\mathrm{GKP}} \ , \\
    S_pD(a)\ket{\mathrm{GKP}} &= e^{iLx_a}D(a)\ket{\mathrm{GKP}} \ .
\end{align}
So after the displacement, the phases of the two stabilizers contain $p_a$ and $x_a$ modulo $L$ (more precisely, modulo $2\pi / L = L$). If one could then measure the GKP stabilizer phases, the resulting two quantities form $a \bmod L\mathbb{Z}^2$. Importantly, simultaneous measurements of these phases are possible precisely because $S_x$ and $S_p$ commute. This idea extends to the multimode setting. For $n$ modes, we consider the product grid state $\ket{\mathrm{GKP}}^{\otimes n}$. The corresponding measurement of all $2n$ phases, via the corresponding commuting stabilizers, returns $\alpha \bmod L\mathbb{Z}^{2n}$.  We will call the centered fundamental cell
\begin{equation}
    \mathcal{V}_L=\left[-\frac{L}{2},\frac{L}{2}\right)^{2n}
\end{equation}
the GKP decoding cell. From the above, we see that the true displacement $\alpha$ lies in $\mathcal{V}_L$, then the modular outcome can be used to recover $\alpha$ itself. However if $\alpha$ lies outside $\mathcal{V}_L$, then the measurement wraps it back into the cell and we will misidentify the true displacement. We refer to such an error as an \textit{aliasing} error. 

However, as we discussed, the ideal GKP state is not physically realizable, and it remains to explain how to extract the stabilizer phases. We will resolve these issues simultaneously by using Quantum Phase Estimation (QPE) to demonstrate a quantum circuit which uses ancilla qubit ECD control of bosonic sensing modes to prepare and read out displacement phases from GKP probe states. First, we recall the well-known QPE primitive. Let $M=2^r$ and let $\mathbb{Z}_M=\{0,\ldots,M-1\}$. The phase-estimation circuit takes a unitary $U$ and tries to learn an eigenphase $\theta\in[0,1)$, where $U\ket{\theta}=e^{2\pi i\theta}\ket{\theta}$. The procedure, described in Algorithm \ref{alg:qpe_unitary}, works by coherently applying many powers of $U$ and then Fourier transforming the accumulated phases.

\begin{algorithm}[h!]
\caption{$M$-bin quantum phase estimation for a unitary $U$}
\label{alg:qpe_unitary}
\begin{algorithmic}[1]
\Require A system state $\rho$, controlled access to the powers $U^{2^j}$ for $j=0,\ldots,r-1$, and $M=2^r$.
\Ensure A classical outcome $b\in\mathbb{Z}_M$.
\State Prepare an $r$-qubit register in $\ket{0}^{\otimes r}$.
\State Apply Hadamards to the register, obtaining $\frac{1}{\sqrt{M}}\sum_{t=0}^{M-1}\ket{t}$.
\For{$j=0,\ldots,r-1$}
    \State Apply controlled-$U^{2^j}$ from the $j$-th register qubit to the system.
\EndFor
\State The joint operation has mapped $\ket{t}\ket{\psi}$ to $\ket{t}U^t\ket{\psi}$. Apply the inverse Fourier transform $F_M^\dagger$ on the register, where $F_M^\dagger\ket{t}=\frac{1}{\sqrt{M}}\sum_{b=0}^{M-1}e^{-2\pi ibt/M}\ket{b}$.
\State Measure the register in the computational basis and output the resulting bitstring $b$.
\end{algorithmic}
\end{algorithm}

When the given state is an eigenstate of $U$, the algorithmic guarantee is as follows. If the system is in eigenstate $\ket{\theta}$, then after the controlled powers of $U$ the joint state is
\begin{equation}
    \frac{1}{\sqrt{M}}\sum_{t=0}^{M-1}e^{2\pi it\theta}\ket{t}\ket{\theta} \ .
\end{equation}
Thus the ancilla register contains a length-$M$ complex phase with frequency $\theta$. The inverse Fourier transform is exactly the operation that converts this phase into a computational-basis label. Assuming $\theta = b_0/M$ lies exactly on an equally spaced $M$-point grid of phases,

\begin{align}
    F_M^\dagger
    \left(
        \frac{1}{\sqrt M}
        \sum_{t=0}^{M-1}e^{2\pi i t b_0/M}\ket{t}
    \right)
    &=
    \frac{1}{M}
    \sum_{b=0}^{M-1}
    \sum_{t=0}^{M-1}
    e^{2\pi i t b_0/M}
    e^{-2\pi i bt/M}
    \ket{b} \\
    &=
    \sum_{b=0}^{M-1}
    \left(
        \frac{1}{M}
        \sum_{t=0}^{M-1}
        e^{2\pi i t(b_0-b)/M}
    \right)
    \ket{b} \\
    &=
    \ket{b_0} \ .
\end{align}
For a general continuous phase, a similar calculation shows that the output is distributed according to
\begin{equation}
    Q_M(b|\theta)
    =
    \frac{1}{M^2}
    \left|
        \sum_{t=0}^{M-1}e^{2\pi it(\theta-b/M)}
    \right|^2
    =
    \frac{1}{M^2}
    \frac{\sin^2(\pi M(\theta-b/M))}{\sin^2(\pi(\theta-b/M))} \ ,
\end{equation}
The second equality is defined away from any grid point $\theta = b/M$, but the exact distribution is continuous and well-defined everywhere. The distribution $Q_M(b|\theta)$ is sharply peaked near the closest grid point to $\theta$. For a general input state, Algorithm \ref{alg:qpe_unitary} defines a measurement with Kraus operators
\begin{equation}
    A_b^{(M)}(U)
    =
    \frac{1}{M}
    \sum_{t=0}^{M-1}e^{-2\pi ibt/M}U^t \ .
\end{equation}
These operators satisfy $\sum_{b=0}^{M-1}A_b^{(M)}(U)^\dagger A_b^{(M)}(U)=I$, which follows from the finite Fourier identity $\sum_{b=0}^{M-1}e^{2\pi ib(t-t')/M}=M\delta_{t,t'}$. QPE then measures the expected value of the phase of $U$, with finite resolution $1/M$.

We can now see that QPE gives us a natural way to extract displacement-induced phases from GKP states. To align notation, we define
\begin{equation}
    U_x=S_p=D(0,L),\qquad U_p=S_x^\dagger=D(-L,0) \ .
\end{equation}
This choice simply fixes a sign convention. The unitary $U_x$ is used to estimate an $x$ displacement, and $U_p$ is used to estimate a $p$ displacement. Indeed, if the signal applies $D(x_a,p_a)$, the Weyl relation gives
\begin{align}
    U_xD(x_a,p_a) &= e^{iLx_a}D(x_a,p_a)U_x \ , \\
    U_pD(x_a,p_a) &= e^{iLp_a}D(x_a,p_a)U_p \ .
\end{align}
Since $L^2=2\pi$, the phases $e^{iLx_a}$ and $e^{iLp_a}$ are equal to $e^{2\pi ix_a/L}$ and $e^{2\pi ip_a/L}$. Therefore, if we estimate the phases of $U_x$ and $U_p$ with respect to the probe state before and after the unknown displacement, their differences estimate $x_a/L$ and $p_a/L$ modulo $1$. Multiplying by $L$ gives the displacement modulo $L\mathbb{Z}^2$.  For $k\in\mathbb{Z}_M$, let $\operatorname{cent}_M(k)$ denote the unique integer representative of $k$ in $[-M/2,M/2)$. Thus $L\operatorname{cent}_M(k)/M$ lies in $[-L/2,L/2)$ and will encode the displacement coordinates which one obtains from their corresponding phase. We now give the actual finite-GKP sensing primitive in Algorithm \ref{alg:gkp_qpe_sensing}.

\begin{algorithm}[h!]
\caption{Finite-GKP displacement sensing by phase estimation}
\label{alg:gkp_qpe_sensing}
\begin{algorithmic}[1]
\Require One query to an $n$-mode displacement channel, $L=\sqrt{2\pi}$, and $M=2^r$.
\Ensure A decoded modular displacement estimate $\widehat{\alpha}\in\mathcal{V}_L$.
\State Prepare the $n$-mode vacuum state. For each mode $m$, draw $(\xi_{x,m},\xi_{p,m})$ uniformly from $[-L/2,L/2)^2$ and apply the known displacement $D(\xi_{x,m},\xi_{p,m})$.
\For{$m=1,\ldots,n$}
    \State Run Algorithm \ref{alg:qpe_unitary} on $U_{x,m}=D_m(0,L)$ and record $b_{x,m}^{(0)}$.
    \State Run Algorithm \ref{alg:qpe_unitary} on $U_{p,m}=D_m(-L,0)$ and record $b_{p,m}^{(0)}$.
\EndFor
\State Send the $n$ modes through the unknown displacement channel.
\For{$m=1,\ldots,n$}
    \State Run Algorithm \ref{alg:qpe_unitary} on $U_{x,m}=D_m(0,L)$ and record $b_{x,m}^{(1)}$.
    \State Run Algorithm \ref{alg:qpe_unitary} on $U_{p,m}=D_m(-L,0)$ and record $b_{p,m}^{(1)}$.
\EndFor
\For{$m=1,\ldots,n$}
    \State Set $k_{x,m}=b_{x,m}^{(1)}-b_{x,m}^{(0)} \bmod M$ and $k_{p,m}=b_{p,m}^{(1)}-b_{p,m}^{(0)} \bmod M$.
    \State Set $\widehat{x}_m=L\operatorname{cent}_M(k_{x,m})/M$ and $\widehat{p}_m=L\operatorname{cent}_M(k_{p,m})/M$.
\EndFor
\State Output $\widehat{\alpha}=((\widehat{x}_1,\widehat{p}_1),\ldots,(\widehat{x}_n,\widehat{p}_n))$.
\end{algorithmic}
\end{algorithm}

The first line of Algorithm \ref{alg:gkp_qpe_sensing} is a known randomization performed by the learner. Its role is to randomize the initial stabilizer phases, so that the difference between the post-channel and pre-channel phase estimates has an exact shift-covariant distribution. After state preparation, the structure of the algorithm is simple: perform QPE on the initial state, evolve by the unknown channel, and perform QPE once again. After the first two phase-estimation measurements on a mode, the state sent through the unknown channel is precisely the finite-resolution physical instantiation of a GKP probe state. For example, conditional on preparation outcomes $b_x^{(0)}$ and $b_p^{(0)}$, and on the known initial displacement $\xi$, the unnormalized one-mode probe is
\begin{equation}
    A_{b_p^{(0)}}^{(M)}(U_p)A_{b_x^{(0)}}^{(M)}(U_x)D(\xi)\ket{0}
    =
    \frac{1}{M^2}
    \sum_{s,t=0}^{M-1}
    e^{-2\pi i(sb_p^{(0)}+tb_x^{(0)})/M}
    D(-sL,tL)D(\xi)\ket{0} \ .
\end{equation}
The probe is a finite superposition of coherent states, and the parameter $M$ controls how sharply it approximates a stabilizer-phase eigenstate.

We will next demonstrate that Algorithm \ref{alg:gkp_qpe_sensing} is precisely the sensing primitive we want. Define the one-coordinate difference kernel
\begin{equation}
    H_M(k|s) = \sum_{b=0}^{M-1}\int_0^1 Q_M(b|\theta)Q_M(b+k|\theta+s)\,d\theta
\end{equation}
for $k\in\mathbb{Z}_M$ and $s\in[0,1)$, where $b+k$ is interpreted modulo $M$. Here $\theta$ is the initial stabilizer phase, $s$ is the phase shift caused by the displacement, $b$ is the pre-channel QPE outcome, and $b+k$ is the post-channel QPE outcome. Expanding the finite-QPE kernel in Fourier modes gives the equivalent expression
\begin{equation}
    H_M(k|s)
    =
    \frac{1}{M}
    \sum_{j=-(M-1)}^{M-1}
    \left(1-\frac{|j|}{M}\right)^2
    e^{2\pi ij(s-k/M)} \ .
\end{equation}
The definition as an average of products of probabilities makes clear that $H_M(k|s)\geq0$, and summing over $k\in\mathbb{Z}_M$ gives $1$. Once the difference bitstring $k$ has been obtained, the algorithm outputs the corresponding displacement phase by centering and normalizing $k$. For notational convenience, define the one-dimensional decoded grid
\begin{equation}
    \mathcal{G}_{M,L}
    =
    \left\{
        \frac{Lj}{M}:
        j\in\mathbb{Z},\ -\frac{M}{2}\leq j<\frac{M}{2}
    \right\}
    \subset
    \left[-\frac{L}{2},\frac{L}{2}\right) \ .
\end{equation}
For $y=Lj/M\in\mathcal{G}_{M,L}$, define the one-coordinate readout kernel
\begin{equation}
    \mathsf{H}_{M,L}(y|a)
    =
    H_M\left(j \bmod M \,\middle| \,\frac{a}{L}\right) \ .
\end{equation}
Thus $\mathsf{H}_{M,L}(y|a)$ is the probability that the finite-QPE readout of a single coordinate returns the decoded value $y$, when the true coordinate displacement is $a$. Mathematically these probabilities are one-to-one with $H_M(k|s)$, but here $y$ and $a$ correspond to estimates and actual phase-space coordinates of the sensed displacement respectively. With these definitions, we can exactly characterize the output distribution of Algorithm \ref{alg:gkp_qpe_sensing}.

\begin{lemma}[Algorithmic guarantee for GKP-based sensing]
\label{lemma:exact_finite_gkp_readout}
Suppose the channel applies a deterministic displacement $\alpha=(\alpha_1,\ldots,\alpha_{2n})\in\mathbb{R}^{2n}$, where the coordinates are ordered as $(x_1,p_1,\ldots,x_n,p_n)$. Then Algorithm \ref{alg:gkp_qpe_sensing} outputs a grid point $    \widehat{\alpha}\in\mathcal{G}_{M,L}^{2n}\subset\mathcal{V}_L
$ with probability law
\begin{equation}
    \Pr\left[\widehat{\alpha}=y\middle|\alpha\right]
    =
    \prod_{j=1}^{2n}\mathsf{H}_{M,L}(y_j|\alpha_j)
\end{equation}
for every $y=(y_1,\ldots,y_{2n})\in\mathcal{G}_{M,L}^{2n}$. Furthermore, for every $0<h\leq L/2$,
\begin{equation}
    \Pr\left[
        \max_{j\in\{1,\ldots,2n\}}
        \min_{q\in\mathbb{Z}}
        \left|\widehat{\alpha}_j-\alpha_j+qL\right|
        >h
        \,\middle|\,
        \alpha
    \right]
    \leq
    \frac{4nL}{Mh} \ .
\end{equation}
The minimax expression simply quantifies the error in the estimate after taking the true displacement $\alpha$ modulo the GKP period $L$ and centering.
\end{lemma}
\begin{proof}
We first prove the one-coordinate statement. Consider the $x$ coordinate of one mode, so the relevant unitary is $U_x=D(0,L)$. The known random displacement applied at the beginning shifts the initial eigenphase of $U_x$ by a uniform random element of $[0,1)$. Thus the initial phase $\theta$ may be treated as uniform. If the true displacement has $x$ coordinate $x_a$, then the phase of $U_x$ is shifted by $s=x_a/L$ modulo $1$. By Algorithm \ref{alg:qpe_unitary}, the first phase-estimation outcome is distributed as $Q_M(b_0|\theta)$, and the second is distributed as $Q_M(b_1|\theta+s)$. Therefore the difference $k=b_1-b_0\bmod M$ has probability
\begin{equation}
    \sum_{b=0}^{M-1}\int_0^1 Q_M(b|\theta)Q_M(b+k|\theta+s)d\theta
    =
    H_M(k|s) \ .
\end{equation}
The same argument applies to the $p$ coordinate using $U_p=D(-L,0)$, for which the phase shift is $p_a/L$. Since all the stabilizers used in the procedure commute, the calculation applies jointly to all $2n$ coordinates, giving the product formula. It remains to prove the concentration bound. For one phase-estimation measurement on an eigenphase $\theta$, the outcome kernel satisfies
\begin{equation}
    Q_M(b|\theta)
    =
    \frac{1}{M^2}
    \frac{\sin^2(\pi M(\theta-b/M))}{\sin^2(\pi(\theta-b/M))} \ .
\end{equation}
away from any exact grid points. Let $\|u\|_{\mathbb T}$ denote distance to the nearest integer. Since $|\sin(\pi u)|\geq 2\|u\|_{\mathbb T}$ for $\|u\|_{\mathbb T}\leq 1/2$, we have
\begin{equation}
    Q_M(b|\theta)
    \leq
    \frac{1}{4M^2\|\theta-b/M\|_{\mathbb T}^2}
\end{equation}
whenever $b/M\neq\theta$ modulo $1$. Summing this bound over all grid points with $\|\theta-b/M\|_{\mathbb T}>\tau$ gives
\begin{equation}
    \Pr\left[
        \left\|b/M-\theta\right\|_{\mathbb T}>\tau
    \right]
    \leq
    \frac{1}{2M\tau}
\end{equation}
for every $0<\tau\leq 1/2$. The decoded coordinate error is the difference of two phase-estimation errors, multiplied by $L$. If this error exceeds $h$ on the circle of circumference $L$, then at least one of the two phase-estimation errors must exceed $h/(2L)$. Therefore, for one coordinate,
\begin{equation}
    \Pr\left[
        \textnormal{coordinate error}>h
    \right]
    \leq
    \frac{2L}{Mh} \ .
\end{equation}
A union bound over the $2n$ coordinates gives $4nL/(Mh)$.
\end{proof}

Let us now understand the resource requirements of this sensing protocol. One $M$-bin QPE measurement uses $r=\log_2 M$ controlled powers of the relevant stabilizer. In our setting these powers are controlled displacements of sizes $L,2L,\ldots,2^{r-1}L$, so the largest controlled displacement has magnitude $ML/2$. Algorithm \ref{alg:gkp_qpe_sensing} performs two stabilizer phase estimations per mode before the channel and two after the channel, for a total of $4n\log_2 M$ controlled displacements per channel query. 

The probe sent through the unknown channel has finite energy. To see the scaling, consider one mode and average over the preparation outcomes. For a single phase-estimation measurement with Kraus operators
\begin{equation}
    A_b^{(M)}(U)=\frac{1}{M}\sum_{t=0}^{M-1}e^{-2\pi ibt/M}U^t \ ,
\end{equation}
summing over $b$ gives the map
\begin{equation}
    \sum_{b=0}^{M-1}A_b^{(M)}(U)\rho A_b^{(M)}(U)^\dagger
    =
    \frac{1}{M}\sum_{t=0}^{M-1}U^t\rho U^{-t} \ .
\end{equation}
Thus the two pre-channel stabilizer phase-estimation measurements act, after averaging over their outcomes, like a uniform mixture over the displacements $D(-sL,tL)$ with $s,t\in\{0,\ldots,M-1\}$. Starting from vacuum, the average photon number is therefore
\begin{align}
    \frac{1}{M^2}
    \sum_{s,t=0}^{M-1}
    \frac{L^2s^2+L^2t^2}{2}
    &=
    \frac{L^2(M-1)(2M-1)}{6} \ .
\end{align}
The initial random displacement inside the GKP cell contributes only $O(L^2)$ additional energy on average. Hence one mode has average energy $O(L^2M^2)$, and the $n$-mode product probe has average energy $O(nL^2M^2)=O(nM^2)$ for fixed $L=\sqrt{2\pi}$. To summarize, the total energy used in the protocol is $O(nM^2 + nM\log M)  = O(nM^2)$.

The presentation above used an $r$-qubit phase-estimation register because it is the cleanest way to see the measurement. However, the same measurement can be implemented with a single controllable qubit by semiclassical iterative phase estimation. For the $j$-th bit, one prepares the qubit in $\ket{+}$, applies the controlled stabilizer power $U^{2^j}$, applies a qubit phase correction determined by the previously measured bits, measures the qubit, and resets it. Repeating this for $j=0,\ldots,r-1$ produces exactly the same outcome distribution as Algorithm \ref{alg:qpe_unitary}. Each phase-estimation unitary is a standard controlled-displacement gate that can be implemented with a single ECD, qubit rotation, and mode displacement. As such, the GKP sensing protocol is implementable within the same single-qubit control model used throughout this work.

\subsubsection{Characteristic-function learning from phase estimation}
\label{sec:gkp-characteristic-function-learning}

The GKP sensing protocol turns each channel use into an approximate displacement sample from the underlying distribution, up to potential modular errors. We next show that this enables efficient learning of characteristic functions without entangled resources. Fix a query radius $R>0$. For $0<h<L/2$, define the shrunken GKP cell
\begin{equation}
    \mathcal{V}_{L,h}
    =
    \left[-\frac{L}{2}+h,\frac{L}{2}-h\right]^{2n}
\end{equation}
and the corresponding functional
\begin{equation}
    \mathfrak{a}_P(h)
    =
    \Pr_{\alpha\sim P}\left[\alpha\notin\mathcal{V}_{L,h}\right] \ .
\end{equation}
This quantity controls the likelihood of an aliasing error, and is small exactly when the random displacement drawn by the channel usually lies inside the GKP cell, and not too close to its boundary. The role of the boundary margin is only to ensure that a small modular readout error can be interpreted as an ordinary displacement error.

\begin{algorithm}[h!]
\caption{Post-hoc characteristic-function learning from GKP sensing}
\label{alg:gkp_characteristic_learning}
\begin{algorithmic}[1]
\Require Query access to an $n$-mode displacement channel $\mathcal{E}_P$, phase-estimation resolution $M=2^r$, and number of channel uses $N$.
\Ensure After a query $\beta$ is revealed, an estimate $\widehat{\lambda}(\beta)$ of $\lambda_P(\beta)$.
\For{$t=1,\ldots,N$}
    \State Run Algorithm \ref{alg:gkp_qpe_sensing} on one use of $\mathcal{E}_P$ and store the decoded outcome $\widehat{\alpha}^{(t)}\in\mathcal{V}_L$.
\EndFor
\State After receiving a query $\beta\in\mathbb{R}^{2n}$, output
\begin{equation}
    \widehat{\lambda}(\beta)
    =
    \frac{1}{N}\sum_{t=1}^N e^{i\Omega(\beta,\widehat{\alpha}^{(t)})} \ .
\end{equation}
\end{algorithmic}
\end{algorithm}

This algorithm is geared towards the difficult post-hoc characteristic function learning task, as the data $\widehat{\alpha}^{(1)},\ldots,\widehat{\alpha}^{(N)}$ are collected before a query $\beta$ is provided. The postprocessing is just empirical Fourier estimation as done in Refs.~\cite{Oh_2024, cotler2026quantumadvantagesensingproperties}. We next study the performance guarantees of this learning algorithm.

\begin{theorem}
\label{thm:gkp_instance_dependent_learning}
Let $P$ be any distribution on $\mathbb{R}^{2n}$, let $R>0$, and let $0<h<L/2$. Run Algorithm \ref{alg:gkp_characteristic_learning} with phase-estimation resolution $M$. Then for every post-hoc query $\beta$ satisfying $|\beta|\leq R$, with probability at least $1-\delta$,
\begin{equation}
    \left|
        \widehat{\lambda}(\beta)-\lambda_P(\beta)
    \right|
    \leq
    2\sqrt{\frac{\log(4/\delta)}{N}}
    +
    \Delta_P(R,M,h)
\end{equation}
where
\begin{equation}
    \Delta_P(R,M,h)
    =
    4R\sqrt{2n}\,\frac{L(1+\log M)}{M}
    +
    2\mathfrak{a}_P(h)
    +
    \frac{8nL}{Mh} \ .
\end{equation}
Consequently, if $M$ and $h$ are chosen so that $\Delta_P(R,M,h)\leq \epsilon/2$, then
\begin{equation}
    N
    \geq
    16\epsilon^{-2}\log\left(\frac{4}{\delta}\right)
\end{equation}
channel uses suffice to estimate $\lambda_P(\beta)$ to additive error $\epsilon$ with success probability at least $1-\delta$.
\end{theorem}

\begin{proof}
Let $\widehat{\alpha}$ denote the output of one run of Algorithm \ref{alg:gkp_qpe_sensing}, and let $\alpha\sim P$ denote the true displacement drawn by the channel in that run. We first bound the bias
\begin{equation}
    \left| \mathbb{E}\!\left[e^{i\Omega(\beta,\widehat{\alpha})}\right] - \mathbb{E}\!\left[e^{i\Omega(\beta,\alpha)}\right]
    \right| \ .
\end{equation}
Let $B_h$ be the bad event that either $\alpha\notin\mathcal{V}_{L,h}$ or the finite-QPE modular error exceeds $h$ in at least one coordinate. By Lemma \ref{lemma:exact_finite_gkp_readout},
\begin{equation}
    \Pr[B_h]
    \leq
    \mathfrak{a}_P(h)+\frac{4nL}{Mh} \ .
\end{equation}
On the complement of $B_h$, the modular readout is also an ordinary readout in $\mathbb{R}^{2n}$, so $\widehat{\alpha}-\alpha$ is an ordinary coordinate-wise error vector. We next bound the average size of this QPE error. From the proof of Lemma \ref{lemma:exact_finite_gkp_readout}, the error in one decoded coordinate has tail at most $2L/(Mu)$ at error scale $u$. Therefore
\begin{align}
    \mathbb{E}\left[|\widehat{\alpha}_j-\alpha_j|_{\mathrm{mod}}\right]
    &\leq
    \frac{2L}{M}
    +
    \int_{2L/M}^{L/2}\frac{2L}{Mu}du \\
    &\leq
    \frac{4L(1+\log M)}{M} \ .
\end{align}
Here $|\cdot|_{\mathrm{mod}}$ denotes the coordinate-wise error after reducing modulo $L$ and choosing the centered representative. On $B_h^c$ this is the same as the ordinary coordinate error. Using $|e^{iu}-e^{iv}|\leq |u-v|$, we have on $B_h^c$
\begin{equation}
    \left|
        e^{i\Omega(\beta,\widehat{\alpha})}
        -
        e^{i\Omega(\beta,\alpha)}
    \right|
    \leq
    |\Omega(\beta,\widehat{\alpha}-\alpha)| \ .
\end{equation}
The symplectic form is a signed sum of coordinate products, so
\begin{equation}
    |\Omega(\beta,\widehat{\alpha}-\alpha)|
    \leq
    \sum_{j=1}^{2n}|\beta_j||\widehat{\alpha}_j-\alpha_j| \ .
\end{equation}
Taking expectation and using $|\beta|_1\leq \sqrt{2n}|\beta|\leq R\sqrt{2n}$ gives
\begin{equation}
    \mathbb{E}\left[
        \left|
            e^{i\Omega(\beta,\widehat{\alpha})}
            -
            e^{i\Omega(\beta,\alpha)}
        \right|
        \mathbf{1}_{B_h^c}
    \right]
    \leq
    4R\sqrt{2n}\frac{L(1+\log M)}{M} \ .
\end{equation}
On the bad event, the two phases can differ by at most $2$. Hence
\begin{equation}
    \left|
        \mathbb{E}\left[e^{i\Omega(\beta,\widehat{\alpha})}\right]
        -
        \lambda_P(\beta)
    \right|
    \leq
    4R\sqrt{2n}\frac{L(1+\log M)}{M}
    +
    2\mathfrak{a}_P(h)
    +
    \frac{8nL}{Mh} \ .
\end{equation}
This is exactly $\Delta_P(R,M,h)$. It remains to bound sampling error. Each random variable $e^{i\Omega(\beta,\widehat{\alpha}^{(t)})}$ has magnitude $1$. Applying Hoeffding's inequality to the real and imaginary parts and taking a union bound gives
\begin{equation}
    \Pr\left[
        \left|
            \widehat{\lambda}(\beta)
            -
            \mathbb{E}\widehat{\lambda}(\beta)
        \right|
        >
        2\sqrt{\frac{\log(4/\delta)}{N}}
    \right]
    \leq
    \delta \ .
\end{equation}
Combining the sampling bound with the bias bound proves the theorem.
\end{proof}

The theorem has a useful interpretation. For any distribution $P$, the finite-GKP learner succeeds to the extent that $P$ is contained in the GKP dynamic range. If $P$ has substantial mass outside $\mathcal{V}_L$, then this particular product-GKP sensor sees only the displacement modulo the GKP lattice, and full characteristic-function learning is impossible without additional prior information. If, however, $P$ is concentrated well inside the cell, the learner produces genuine approximate displacement samples, and characteristic-function learning becomes ordinary empirical Fourier estimation. To understand how restrictive this condition actually is, we show that any family which (in displacement space) is contained within a Gaussian envelope of width no broader than $O(\log n)$ satisfies the constraint. In Fourier space, this corresponds to characteristic functions with width at least $\Omega(\log n)$. The condition itself is simply a bound on the tails of $P$, such that with high probability most of $P$ fits inside the GKP decoding cell. Let
\begin{equation}
    g_\sigma(\alpha)
    =
    \left(\frac{2\sigma^2}{\pi}\right)^n
    e^{-2\sigma^2|\alpha|^2}
\end{equation}
be the centered Gaussian density whose real phase-space coordinates have variance $1/(4\sigma^2)$. The following result formalizes the above.

\begin{corollary}[Efficient learning for Gaussian-enveloped displacement channels]
\label{cor:gkp_gaussian_envelope}
Suppose $P$ is dominated by a Gaussian envelope in the sense that $P(\alpha)\leq C_0g_\sigma(\alpha)$ for all $\alpha\in\mathbb{R}^{2n}$, where $C_0$ is a constant. Fix $R^2=\kappa n$ and $\epsilon,\delta\in(0,1)$. If
\begin{equation}
    \sigma^2
    \geq
    \frac{4}{\pi}
    \log\left(\frac{64C_0n}{\epsilon}\right) \ ,
\end{equation}
then Algorithm \ref{alg:gkp_characteristic_learning}, with
\begin{equation}
    M = O\!\left(
        \frac{(\sqrt{\kappa}+1)n}{\epsilon} \log\!\left(\frac{(\sqrt{\kappa}+1)n}{\epsilon}\right)
    \right) \ ,
\end{equation}
estimates $\lambda_P(\beta)$ for any post-hoc query $|\beta|^2\leq \kappa n$ to additive error $\epsilon$ with success probability at least $1-\delta$ using
\begin{equation}
    N  = O\!\left(\epsilon^{-2}\log\left(\frac{1}{\delta}\right)\right)
\end{equation}
channel queries.
\end{corollary}

\begin{proof}
Take $h=L/4$. Then $\mathcal{V}_{L,h}=[-L/4,L/4]^{2n}$. Since $P\leq C_0g_\sigma$, a union bound over the $2n$ coordinates and the Gaussian tail bound give
\begin{align}
    \mathfrak{a}_P(L/4)
    &\leq
    4C_0n\exp\left(-2\sigma^2(L/4)^2\right) \\
    &=
    4C_0n\exp\left(-\frac{\pi\sigma^2}{4}\right) \ .
\end{align}
The specified lower bound on $\sigma^2$ makes $2\mathfrak{a}_P(L/4)\leq \epsilon/8$. The two finite-QPE resolution terms in Theorem \ref{thm:gkp_instance_dependent_learning} are
\begin{equation}
    4\sqrt{2\kappa}n\frac{L(1+\log M)}{M}
    +
    \frac{32n}{M} \ ,
\end{equation}
where we used $R^2=\kappa n$ and $h=L/4$. The stated choice of $M$ makes these terms at most $3\epsilon/8$, up to increasing the universal constant. Thus $\Delta_P(R,M,h)\leq \epsilon/2$, and Theorem \ref{thm:gkp_instance_dependent_learning} gives the claimed channel-query complexity.
\end{proof}

Using this result, we can evaluate how well the entanglement-free GKP sensing protocol performs for learning the three-peak hard family introduced in Ref.~\cite{Oh_2024}. This three-peak displacement density can be written as
\begin{equation}
    P_{s,\gamma}^{\eta_0,\sigma}(\alpha)
    =
    \left(\frac{2\sigma^2}{\pi}\right)^n
    e^{-2\sigma^2|\alpha|^2}
    \left[
        1+4s\eta_0\sin\left(2\Omega(\gamma,\alpha)\right)
    \right] \ .
\end{equation}
In Ref.~\cite{Oh_2024}, this is precisely the family used to show an exponential lower bound against (separately) Gaussian and all entanglement-free strategies. However, for the latter, their proof required that $\sigma$ be upper bounded by a constant dependent on the radius parameter $\kappa$. While it may have seemed that this constraint is only an artifact of the proof strategy, and that an entanglement-enabled protocol was the only way to achieve a superpolynomial advantage, our next corollary shows that this is not the case: for modestly broad distributions, non-Gaussian sensing enabled by single-qubit control is already sufficient to realize a superpolynomial speedup over Gaussian strategies, and to match the query-complexity scaling of the entanglement-enabled Bell sampling strategy.

\begin{graydefinitionbox}
\begin{corollary}[Phase-estimation learning of the three-peak family, formal version of Theorem \ref{thm:displacement_channel_advantage_informal}]
\label{cor:gkp_three_peak_family}
For the three-peak family of Ref.~\cite{Oh_2024}, if $R^2=\kappa n$ and
\begin{equation}
    \sigma^2
    \geq
    C\log\left(\frac{n}{\epsilon}\right)
\end{equation}
for a sufficiently large universal constant $C$, then the finite-GKP learner estimates $\lambda_{s,\gamma}^{\eta_0,\sigma}(\beta)$ for any post-hoc query $|\beta|^2\leq \kappa n$ using
\begin{equation}
    N = O\!\left(\epsilon^{-2}\log\left(\frac{1}{\delta}\right)\right)
\end{equation}
channel queries. Any Gaussian strategy requires $\exp(\Omega(n/\log n))$ samples to do so.
\end{corollary}
\begin{proof}
For $\eta_0\leq 1/4$, the three-peak density obeys $P_{s,\gamma}^{\eta_0,\sigma}(\alpha)\leq 2g_\sigma(\alpha)$ uniformly in $s$ and $\gamma$. Therefore Corollary \ref{cor:gkp_gaussian_envelope} applies with $C_0=2$. The lower bound follows from Equation \eqref{eq:characteristic_function_lb_Gaussian}:
\begin{equation}
    N_{\rm Gauss}
    \geq
    0.01\epsilon^{-2}
    \min\left\{
        \left(1+\frac{0.99\kappa}{C\log n}\right)^{n/2},
        \left(1+\frac{1.98\kappa}{1+2C\log n}\right)^n
    \right\}
    =
    \epsilon^{-2}\exp\left(\Omega_\kappa\left(\frac{n}{\log n}\right)\right) \ .
\end{equation}
\end{proof}
\end{graydefinitionbox}

The resource scaling of this approach is modest. Each channel query uses $O(n\log M)$ controlled displacements, and the largest controlled displacement has size $O(ML)$. The average photon number in the $n$-mode probe is
\begin{equation}
    E_{\rm GKP}   =  O(nM^2) = O\!\left(
        \frac{\kappa n^3}{\epsilon^2}
        \log^2\left(\frac{(\sqrt{\kappa}+1)n}{\epsilon}\right)
    \right)
\end{equation}
for fixed $L=\sqrt{2\pi}$. It is useful to compare this to the entanglement-assisted Bell sampling protocol. A two-mode squeezed Bell measurement produces data of the form $Z=\alpha+G_{\rm Bell}$, where $G_{\rm Bell}$ is Gaussian noise with variance set by the inverse EPR squeezing. To keep characteristic-function deconvolution stable for all $|\beta|^2\leq \kappa n$, one needs Gaussian noise variance $O(1/(\kappa n))$. This corresponds to total two-mode-squeezed energy $E_{\rm Bell} = O(\kappa n^2)$. By using only an $O(n)$ larger amount of energy, entanglement-free GKP based sensing matches the query-complexity scaling of the Bell sampling algorithm while eliminating the need for $n$ ancillary memory modes. Moreover, our construction is neither energy-optimized nor tailored precisely for characteristic function estimation. We use product GKP probe states and vanilla phase-estimation, but it may be possible to fully close the gap in the parameter $\sigma$ to a constant by designing non-Gaussian lattice probe states with more nontrivial geometries that widen the ``good" section of the sensing hypercube. This would rigorously demonstrate that an entanglement-enabled advantage only persists when features of the displacement channel are simultaneously high-frequency and extremely sharp, and that single-qubit control is already sufficient to solve this difficult learning task when either requirement is relaxed.

\subsection{Learning with depth-$k$ quantum control} \label{sec:control_hierarchy_upper_bound}
Here we complete the exponential depth-based separation by proving that sensing protocols with control depth $k$ can efficiently distinguish the hypotheses instantiated in Section \ref{sec:depth_lower_bounds}.

\begin{lemma} \label{lemma:depth_upper_bound}
    There is a depth-$k$ protocol, as in Definition \ref{def:depth_r_interleaved_measurement}, that can distinguish the signals induced by the distributions in Equation \eqref{eq:control_depth_hypothesis_family} with constant bias using $O(1)$ sensing rounds. 
\end{lemma}
\begin{proof}
\noindent Consider the QSL property
\begin{equation}
    \Psi_k(P)=\int g_k(\alpha)P(d\alpha).
\end{equation}
Under the two hypotheses from Equation \eqref{eq:control_depth_hypothesis_family},
\begin{align}
    \Psi_k(P_+^{(k)})-\Psi_k(P_-^{(k)})
    &=
    2\eta\int g_k(\alpha)^2dP_0(\alpha) \\
    &=
    2\eta\|g_k\|_{L^2(P_0)}^2.
\end{align}
From the construction of $g_k$, we know that there exists a constant $c_0$ independent of $k$ such that $\|g_k\|_{L^2(P_0)}^2 \geq c_0$. As such,
\begin{equation}
    \Psi_k(P_+^{(k)})-\Psi_k(P_-^{(k)})\ge 2c_0\eta.
\end{equation}
By Lemma \ref{lemma:depth_k_realizes_grs_kernel}, there is a depth-$k$ QCDS circuit producing a random variable $Y\in\{-1,1\}$ with $\mathbb E[Y\mid \alpha]=h_k(\alpha)$. Let $m_k=\int h_k(\alpha)dP_0(\alpha)$. Under $P_\pm^{(k)}$,
\begin{align}
    \mathbb E_{P_\pm^{(k)}}[Y]
    &=
    \int h_k(\alpha)\left(1\pm \eta g_k(\alpha)\right)dP_0(\alpha) \\
    &=
    m_k\pm \eta\int h_k(\alpha)g_k(\alpha)dP_0(\alpha).
\end{align}
Since $g_k=h_k-m_k$,
\begin{align}
    \int h_k(\alpha)g_k(\alpha)dP_0(\alpha)
    &=
    \int (g_k(\alpha)+m_k)g_k(\alpha)dP_0(\alpha) \\
    &=
    \int g_k(\alpha)^2dP_0(\alpha)+m_k\int g_k(\alpha)dP_0(\alpha) \\
    &=
    \|g_k\|_{L^2(P_0)}^2.
\end{align}
Therefore
\begin{equation}
    \mathbb E_{P_\pm^{(k)}}[Y]
    =
    m_k\pm \eta\|g_k\|_{L^2(P_0)}^2.
\end{equation}
The two means are separated by at least $2c_0\eta$. Run the depth-$k$ circuit independently for $N$ queries and let
\begin{equation}
    \overline Y=\frac{1}{N}\sum_{j=1}^N Y_j.
\end{equation}
Then outputs $(+)$ if $\overline Y\ge m_k$ and output $(-)$ otherwise. Since each $Y_j\in[-1,1]$, Hoeffding's inequality gives
\begin{equation}
    \Pr[\textnormal{error}]
    \le
    \exp\left(-\frac{N\eta^2\|g_k\|_{L^2(P_0)}^4}{2}\right).
\end{equation}
Using $\|g_k\|_{L^2(P_0)}^2 \geq c_0$, it suffices to take $N=O(\eta^{-2}\log(1/\delta))$ to achieve failure probability at most $\delta$. Using constant $\eta, \delta$ completes the proof.
\end{proof}
Finally, by combining Lemma \ref{lemma:depth_upper_bound} with Theorem \ref{thm:interleaved_control_lb_formal}, we obtain the full exponential control-depth separation.

\begin{graydefinitionbox}
\begin{corollary} \label{cor:depth_separation_complete}
    There exist QSL properties with kernels that are polynomials of degree $k$ such that a protocol of control depth $k$ can estimate the property to constant precision $\epsilon$ using $O(1)$ queries to the signal, whereas for $\gamma < \log_3 2$, any protocol using control depth at most $\gamma k$ per measurement requires $\exp(\Omega(k))$ queries to do so.
\end{corollary}    
\end{graydefinitionbox}

\newpage

\section{Self-contained list of results} \label{sec:all_results}

Our appendices contain several contributions not detailed in the main text. For the reader's convenience, we provide a self-contained reference list of the main results in our work. 

\begin{center}
   \textit{Main separations}
\end{center}
We begin with informal statements of all main hierarchy separations in ascending order.

\begin{theorem}[Theorem \ref{thm:conventional_vs_classical_informal}, Exponential advantage in Fourier analysis with nonclassical Gaussian probes] 
    A conventional quantum protocol using squeezed homodyne measurements of energy $O(k)$ estimates the $k^{\rm th}$ angular Fourier coefficient of a signal using $O(1)$ queries. Any protocol exposing coherent-state probes to the signal, regardless of available energy in ancillas or measurements, requires $\exp(\Omega(k))$ queries to do so. A non-Gaussian quantum protocol with $O(k)$ energy also achieves an exponential advantage over all coherent-probe protocols.
\end{theorem}

\begin{theorem}[Theorem \ref{thm:one_qubit_advantage_informal}, Exponential advantage in Fourier analysis with a single qubit] 
    Using a sensor probe with energy $O(k)$, one ancilla qubit, and one control operation, any $k^{\rm th}$ directional Fourier coefficient of a signal can be estimated with $O(k)$ signal queries.  Any adaptive Gaussian protocol with the same energy scaling requires $\exp(\Omega(k))$ queries.
\end{theorem}

\begin{theorem}[Theorem \ref{thm:memory_informal}, Exponential advantage in learning temporal correlations with one memory qubit]
    Let $t_1 < t_2 < \cdots < t_m$ be a list of times separated by at most $\Delta$. A sensor that decoheres on a timescale much shorter than $\Delta$, when coupled to a single qubit that remains coherent up to time $t_m$, can estimate the corresponding $m$-point temporal correlator using $O(1)$ queries to the time-dependent signal. By contrast, any protocol with the same energy and no quantum degree of freedom coherent for times much longer than $\Delta$ requires $\exp(\Omega(m))$ queries, even if it has access to arbitrarily many short-lived ancilla qubits or modes.
\end{theorem}

\begin{theorem}[Theorem \ref{cor:depth_separation_complete}, Exponential advantage of increasing circuit depth for learning polynomial functionals of signals]
    There exist QSL properties with kernels that are polynomials of degree $k$ such that a protocol of control depth $k$ can estimate the property to constant precision $\epsilon$ using $O(1)$ queries to the signal, whereas for $\gamma < \log_3 2$, any protocol using control depth at most $\gamma k$ per measurement requires $\exp(\Omega(k))$ queries to do so.
\end{theorem}

These separations establish that each level of the sensing hierarchy in Figure \ref{fig:schematic_main} is exponentially separated from the last, and that adding a single quantum resource to an otherwise identical sensor can enable exponential resource savings. We also prove other separations that operate in the multimode sensing setting or provide speedups against infinite-energy Gaussian sensing.

\begin{theorem}[Theorem \ref{cor:gkp_three_peak_family}, Superpolynomial entanglement-free advantage in learning a displacement channel] \label{thm:displacement_channel_advantage_informal}
    Let $\kappa > 0$, $n \in \mathbb{Z}^+$. For any displacement channel with frequency-domain peaks of width $\Omega(\log n)$, an entanglement-free protocol using a single ancilla qubit, poly$(n)$ energy, and no ancilla modes can learn its characteristic function within a phase-space radius $\kappa n$ using a number of measurements independent of $n$ (as efficiently as the best-known entanglement-enabled strategy). Meanwhile, any Gaussian protocol with unconstrained energy requires $\exp(\Omega(n/\log n))$ measurements to do so. 
\end{theorem}
\noindent The Gaussian lower bound for this task is due to Ref.~\cite{Oh_2024}; surprisingly, we showed that a superpolynomial advantage can be achieved without entangled sensors, and by using only a single ancilla qubit. Our protocol used one qubit to implement polynomial-energy approximate GKP state quantum phase estimation. Therefore, the entanglement-enabled advantage of Ref.~\cite{Oh_2024}, achieved by performing two-mode squeezed vacuum Bell measurements, is only necessary in edge cases when the displacement channel has extremely fine peaks in unknown locations. 
\begin{theorem}[Theorem \ref{thm:gaussian_testing_combined}, Quadratic single-mode advantage over infinite-energy Gaussian sensing]
Any Gaussian protocol of arbitrary energy which can estimate the variance of a (Gaussian) thermal state to accuracy $1/k$ requires $\Omega(k^2)$ samples. There exist non-Gaussian measurements using $O(1)$ energy, including parity and photon-counting measurement, that can do so using $O(k)$ samples.
\end{theorem}
\noindent This theorem establishes the existence of single-mode Gaussian states which require non-Gaussian measurements to be distinguished optimally. Moreover, it provides a rigorous backing for why non-Gaussian photon-counting measurements provide a large practical speedup for stochastic sensing tasks which involve estimating small perturbations to white background power spectra. The measurements which enable this speedup are easily implemented in superconducting-circuit platforms with a transmon qubit coupled to a microwave cavity, as we use in this work.

\begin{center}
   \textit{Q$\Psi$ toolkit}
\end{center}
We also provide a high-level, intuitive recap of the Q$\Psi$ toolkit. Here, for clarity, we present Q$\Psi$ restricted to single-mode quantum sensing with bosonic oscillators; the most general formulation, which encapsulates much more general quantum-enhanced experiments, is given in Appendix \ref{sec:qes}. 

\vspace{2mm}
\noindent\textbf{Q$\Psi$ inputs.} We let $P_0$ denote a reference signal, whose representation on phase space is compact or approximately compact (in the sense of Definition \ref{def:character_gram} and Lemma \ref{lemma:compact_orthog}). An example of such a reference is a Gaussian signal. With the reference fixed, $Q\Psi$ requires three inputs: a family of weaker "conventional" experiments, $\mathcal{M}_{\rm conv}$, a family of quantum-enhanced experiments, $\mathcal{M}_{\rm QIP}$, and a physical observable specified by its phase-space representation as a function $h$ on $L^2(P_0)$. Mathematically, a family of experiments is represented by a collection of response functions which record classical probability distributions over potential outcomes from each allowed experiment. The function $h$ can represent an arbitrary physical property such as a Fourier amplitude, phase, moment, or temporal correlation. 

\vspace{2mm}
\noindent\textbf{AFI.} The \textit{accessible feature information} (AFI), denoted by $\mathsf A_M(h; P_0)$ for any experiment $M$ in $\mathcal M_{\rm conv}$ or $\mathcal{M}_{\rm QIP}$ is a real-valued function of these inputs (Definition \ref{def:AFI}). For the entire family of experiments $\mathcal{M} = \mathcal M_{\rm conv}$ or $\mathcal{M}_{\rm QIP}$, the AFI $\mathsf A_{\mathcal{M}}(h; P_0)$ is simply $\sup_{M\in \mathcal{M}}\mathsf A_M(h; P_0)$. 

\vspace{2mm}
\noindent\textbf{Semiclassical quantum measurement lemma and tightness certificate.} The primary lower-bound tool of Q$\Psi$ is the \textit{semiclassical quantum measurement lemma}:
\begin{theorem}[Informal statement of Theorem \ref{lemma:semiclassical_quantum_measurement}, Semiclassical Quantum Measurement Lemma]
    Consider an arbitrary experimental protocol for learning property $h$ to absolute precision $\epsilon$, in which each experiment is in the class $\mathcal{M}$ and arbitrary classical adaptivity is allowed between experiments. Any such protocol then requires $N\geq \Omega(\epsilon^{-2} \mathsf A_{\mathcal M}(h; P_0)^{-1})$ queries to the unknown signal (or more generally, an uncharacterized physical system). 
\end{theorem}
\noindent This theorem is obtained by combining information-theoretic lower bound techniques with harmonic analysis of phase-space response functions, and the learning lower bound is established by lifting discrimination of the signals $(1\pm \epsilon h)P_0$ to the harder task of learning $h$ to absolute precision $\epsilon$. The next element of the Q$\Psi$ toolkit establishes that the semiclassical measurement lemma is tight:
\begin{theorem}[Informal statement of Theorem \ref{thm:qes_matched_score_guarantee}, Q$\Psi$ saturability guarantee]
    There exists an adaptive experimental protocol, with each round in the class of experiments $\mathcal{M}$, which discriminates the signals $P_{\pm} = (1\pm \epsilon h)P_0$ with constant bias using $N = O(\epsilon^{-2} \mathsf A_{\mathcal M}(h; P_0)^{-1})$ queries.
\end{theorem}
\noindent This upper bound both certifies the existence of a lower bound-saturating protocol and also produces the protocol directly, as detailed in the proof of Theorem \ref{thm:qes_matched_score_guarantee}. Together, they produce the following implication:
\begin{corollary}[Certifiably optimal quantum advantage, informal]
    Let $h^{(k)}$ be a family of observables parameterized by values of $k> 0$. If  $ A_{\mathcal M_{\rm conv}}(h^{(k)}; P_0) \leq c(k)$ and $A_{\mathcal M_{\rm QIP}}(h^{(k)}; P_0) \geq q(k)$, then the quantum-enhanced family of experiments achieves an $\Omega(q(k)/c(k))$ quantum advantage for learning the properties $h^{(k)}$, and this advantage is the largest possible.
\end{corollary}
In the present paper, we have shown that several physically natural properties such as frequency-$k$ Fourier coefficients, order-$k$ temporal correlations, and degree-$k$ polynomial transformations, all force $c(k)\leq \exp(-\Omega(k))$ for the respective lower rungs of the hierarchy, while adding a single resource like squeezing, single-qubit control, memory, or circuit depth, enables $q(k) \geq \Omega(1)$. $Q\Psi$ then immediately outputs a certificate of quantum advantage. 

\vspace{2mm}
\noindent\textbf{Global learning.} Q$\Psi$ makes new quantum separations transparent by recasting the design of optimal quantum experiments as a classical problem of statistical overlap. In particular, the optimal physical experiment is the one whose phase-space response functions have the largest overlap with the phase-space representation of the property of interest. In the frequency domain, this leads to a direct optimization: compute the Fourier representations of the experimental responses and the target property, and choose the measurement with the largest response on the Fourier characters supporting the property. Importantly, this optimization is entirely classical and commutative, suggesting the automated design principles discussed in Section \ref{sec:open_questions}.

This phase-space overlap formulation also allows Q$\Psi$ to produce optimal global-learning algorithms beyond the hypothesis-testing setting. We define the \textit{Fourier gain} of an experimental family as follows. We choose a Fourier dictionary ${\chi_\omega}$ that forms a complete basis for the physically relevant signals. Given a protocol with response function $f_j(x)$, where $x$ denotes the unobserved signal realization, we write its Fourier expansion as $f_j(x) = \sum_\omega \widehat f_j(\omega)\chi_\omega(x)$. The Fourier gain of the family $\mathcal{M}$ at frequency $\omega$ is $G_{\mathcal{M}}(\omega) = \sup_{j\in \mathcal{M}} |\widehat f_j(\omega)|^2/V_j$. Here $V_j$ is a protocol-dependent normalization. Thus, $G_{\mathcal{M}}(\omega)$ quantifies the largest phase-space response that the experimental class can achieve at frequency $\omega$.

\begin{theorem}[Informal statement of Theorem \ref{thm:qes_fourier_gain_global_observable}, Q$\Psi$ global learning guarantee] Let $h$ be a property of interest, and define $\widehat h$ by the Fourier transform 
\begin{equation}
    h(x) = \sum_{\omega\in S} \widehat h(x)\chi_\omega(x) \ ,
\end{equation}
where $S$ is the Fourier-domain support of $h$. Then there is a learning algorithm in the class $\mathcal{M}$ which learns $h$ to absolute accuracy $\epsilon$ using 
\begin{equation}
    N\propto \frac{1}{\epsilon^2}\sum_{\omega \in S} \frac{|\widehat{h}(\omega)|^2}{G_{\mathcal{M}}(\omega)}
\end{equation}
experiments.
\end{theorem}
A more detailed sample complexity statement is given in Theorem \ref{thm:qes_fourier_gain_global_observable}. In short, the resource complexity is directly controlled by the ratio of the observable's Fourier support at each frequency and the best response synthesizable by any measurement in the class at that frequency. The algorithm which achieves this scaling allocates experiments based on their relative overlap with the desired observable and is given in Section \ref{sec:qes}. 

\vspace{2mm}
\noindent\textbf{Summary.} Q$\Psi$ is therefore a unifying framework of quantum learning, metrology, and estimation theory useful for understanding the power of quantum information processing in experimental science. It reduces a quantum learning problem into phase-space overlap of classical statistical response functions, produces provably optimal lower bounds against conventional experiments and upper bounds achieved by quantum-enhanced ones, and certifies a rigorous quantum advantage while designing an asymptotically optimal quantum experiment. The central quantities are the AFI and Fourier gain, which characterize the achievable measurement complexity and guide the design of optimal quantum-enhanced experiments.

\newpage

\section{Open directions} 
\label{sec:open_questions}

\subsection{Applications of Quantum Feature Sensing}
\label{sec:qfs_applications}

In this work, we have established that Quantum Feature Sensing, the family of quantum processing-enhanced bosonic sensing protocols that emerge from Q$\Psi$, can enable quantum advantages for basic sensing primitives. Here, we have covered estimation of Fourier coefficients, moments, temporal correlations, and polynomial functionals. Our point of connection between these physical tasks and the Q$\Psi$ framework was Quantum Signal Learning, which translates arbitrary physical observables of classical signals incident on a quantum oscillator into mathematical descriptions that can be utilized by Q$\Psi$. Importantly, this QSL formalism circumscribes many of the sensing tasks studied in e.g. fundamental particle searches, device calibration, classical noise spectroscopy, and receivers, making these broad disciplines amenable to QFS speedups that have thus far been difficult to systematically discover. Here, we provide context for these applications and present concrete open questions of when and how the techniques introduced here can enable speedups for a wide array of classical sensing applications.

\subsubsection{Fundamental particle searches}
\label{sec:qfs_particle_search}

Cavity haloscopes search for wave-like dark matter candidates, such as axions and dark photons, by monitoring a high-quality microwave cavity for a feeble drive at an unknown frequency set by the particle mass. The galactic velocity dispersion endows this drive with a narrow stochastic lineshape of quality factor $\sim 10^6$, so the hypothetical signal is, in our language, a displacement channel whose density $P$ is the thermal background perturbed by a small excess occupation $n_a \ll 1$ concentrated near one frequency in a multi-gigahertz band. Detection is a hypothesis test between background and background-plus-excess, and the figure of merit is the \textit{scan rate}: the bandwidth that can be excluded or discovered per unit integration time at fixed statistical significance.

The historical development of dark-matter sensing traces the first rungs of our sensing hierarchy. Initially, squeezed-vacuum receivers were shown to roughly double the scan rate \cite{malnou_squeezed_2018, Backes_2021}. Then, replacing quadrature readout with qubit-based photon counting was shown to improve search speed by up to three orders of magnitude. In practice, these measurements were implementable by a transmon dispersively coupled to a microwave cavity, which is precisely the architecture we have used in this work \cite{lamoreaux2013analysis, dixit2021searching}. Moreover, proposals of entanglement-assisted receivers are prominent \cite{brady_entangled_2022, Shi_2023}. This progression through quantum-enhanced architectures has been assembled empirically, once the constraints of each level of the hierarchy became clear. In light of the present work, Q$\Psi$ provides a unifying understanding of why each additional quantum resource enabled speedups for these tasks, as well as a systematic strategy to discover future quantum enhancements.

\begin{graydefinitionbox}
\noindent \textbf{Open Problem (Quantum information processing for fundamental particle search).} Can we formulate the detection and characterization of wave-like dark matter, and weak stochastic signals more broadly, as Quantum Signal Learning problems, and use Q$\Psi$ to chart which quantum information-processing resources provably accelerate them? In particular: do the higher rungs of the hierarchy in Figure~\ref{fig:schematic_main}(b) confer advantages that are qualitatively beyond single-photon detection, and how large can these advantages be?
\end{graydefinitionbox}

Our results supply both rigor for the enhancements this field has already found and reason to believe it has explored only the first rung of what quantum control offers. The vacuum-versus-thermal separation of Sections~\ref{sec:gaussian_testing_lower_bound} and \ref{sec:Gaussian_testing_upper_bound} proves that photon-counting-style readout achieves a sample-complexity scaling that no Gaussian receiver of \emph{any} energy can match --- a precise version of the folklore underlying Refs.~\cite{lamoreaux2013analysis, dixit2021searching}, and an explanation for why the leap from squeezing to counting was a change in kind rather than in degree. Moreover, section~\ref{sec:gbs_alternate} shows that the single echoed conditional displacement powering our exponential advantage is interchangeable with photon-counting readout as a non-Gaussian resource. In Figure \ref{fig:applications}(a) and Section \ref{sec:axion_stream} we give concrete evidence that a superconducting circuit platform with a single qubit may produce asymptotic improvements for important tasks in the characterization of axionic dark matter.  

Single-photon detection is therefore not an endpoint but the \emph{first} unit of quantum control, and each additional resource in our hierarchy, including even a single additional non-Gaussian control, corresponds to a capability that no photon counter possesses. A single memory qubit can coherently accumulate the dark-matter field's phase across many of its coherence windows (Theorem~\ref{thm:memory_informal}), suggesting a route to estimating the signal lineshape (equivalently, the local dark-matter velocity distribution \cite{foster2018revealing, Foster_2021}) as temporal Fourier data rather than by prolonged power integration; deeper coherent control synthesizes matched filters directly as polynomial functionals of the signal (Sections~\ref{sec:depth_lower_bounds} and \ref{sec:control_hierarchy_upper_bound}). 

\subsubsection{Stochastic and spatiotemporal noise learning}
\label{sec:qfs_noise_learning}

Characterizing environmental noise is essential for calibrating quantum platforms. A common technique is dynamical-decoupling noise spectroscopy, in which one applies pulse sequences whose filter functions concentrate the sensor's response near a chosen frequency and thereby reconstructs the noise spectral density $S(\omega)$ from measured decay rates \cite{Alvarez_2011, Bylander2011NoiseSpectroscopy, Degen_2017}. In our language, the fluctuating environment is a stochastic process $\{Z_t\}$, a pulse sequence followed by destructive readout is a memoryless response kernel, and $S(\omega)$ is two-point temporal Fourier data. Every such protocol inherits a structural limitation: a sensor with coherence time $T_2$ produces response kernels of spectral linewidth $\sim 1/T_2$, and cannot resolve finer structure no matter how many measurements are taken. Experiments have breached this limit by storing phase in a long-lived nuclear-spin memory \cite{rosskopf_quantum_2017}, but a general account of which resolution--measurement tradeoffs are possible with bounded quantum resources is not present.

\begin{graydefinitionbox}
\noindent \textbf{Open Problem (Quantum information processing for noise spectroscopy).} Can we use Q$\Psi$ to chart which quantum information-processing resources provably extend the reach of noise spectroscopy?  What resolution and measurement-complexity tradeoffs become achievable with quantum memory, deeper coherent control, or entanglement across sensors, and which of these are provably impossible for existing filter-function protocols?
\end{graydefinitionbox}

Our results suggest a number of applications to measurement of spectral densities, higher-resolution spectral features, and their spatiotemporal counterparts across sensor arrays. Theorem~\ref{thm:memory_informal} is a proof that the mechanism observed in memory-assisted spectrometers \cite{rosskopf_quantum_2017} constitutes an \emph{exponential} quantum advantage; Section~\ref{sec:memoryless_lower_bounds} shows the linewidth limit of memoryless sensing is an information-theoretic barrier rather than a practical inconvenience, while a single long-lived qubit crosses it. Sub-linewidth estimation of a feature of width $\delta\omega \ll 1/T_2$ spans $m \sim (\delta\omega\, T_2)^{-1}$ sensor coherence windows, and certifying an exponential-versus-polynomial separation there, under realistic constraints on pulse bandwidth and total wall-clock time, would elucidate the role of memory as a resource for noise spectroscopy. In the spatiotemporal setting, arrays of sensors probing a correlated noise field target cross-spectra and spatial Fourier modes, and in the spirit of Section~\ref{sec:char_function_learning_gkp} one may ask which correlation structures single-qubit control resolves without any entanglement, and where entanglement across sensors is provably necessary. In the bosonic noise spectroscopy setting, QSL already provides the relevant language, whereas the spin-sensor paradigm requires a qubit-native formalism. The multiqubit correlation-spectroscopy protocols of Refs.~\cite{pazsilva2017multiqubit, szankowski2017qubits, von_lupke_two-qubit_2020} provide conventional baselines for this direction.

\subsubsection{Estimating polyspectra}
\label{sec:qfs_polyspectra}

A Gaussian stochastic process is fully characterized by its spectral density which is a function of a single frequency, but non-Gaussian noise is not: its complete frequency-domain description requires \textit{polyspectra}, the Fourier transforms of higher-order cumulants, beginning with the bispectrum $S_2(\omega_1,\omega_2)$. Polyspectra are standard objects in classical signal processing, where they detect phase coupling and nonlinearity invisible to the power spectrum, and they have entered the quantum domain through dynamical-decoupling protocols for reconstructing the dephasing bispectrum \cite{Norris_2016}, realized experimentally on a superconducting qubit \cite{sung_non-gaussian_2019}. A classical estimator of an order-$m$ spectrum averages products of $m$ noisy samples, so its variance compounds multiplicatively with the order, precisely the mechanism depicted in Figure~\ref{fig:coherent_and_memory_experiments}(a), often rendering orders beyond the bispectrum inaccessible. This higher-order data has broad utility: polyspectra discriminate
targets from non-Gaussian clutter in radar and sonar \cite{nikias1993signal},
constrain primordial non-Gaussianity in cosmology \cite{planck2019nongaussianity},
and identify the microscopic origin of a qubit's noise
environment. 

\begin{graydefinitionbox}
\noindent \textbf{Open Problem (Quantum information processing for higher-order spectra).} Formulate the estimation of polyspectra of non-Gaussian noise as Quantum Signal Learning, and determine through Q$\Psi$ whether quantum memory and coherent control enable provably efficient access to higher-order polyspectra relevant to the above applications. How do the achievable advantages scale jointly in the frequency and the order of the targeted coefficient?
\end{graydefinitionbox}

Theorem~\ref{thm:memory_informal} and Algorithm~\ref{alg:temporal_weyl_memory_estimator} show that \emph{full} $m$-point temporal Fourier coefficients, the moments of the process, accumulate on a single qubit coherence with a trajectory count independent of $m$, while any memoryless protocol provably pays the exponential compounding cost. Polyspectra are formed from the full order-$m$ coefficient minus all products of lower-order coefficients. As such, one must determine whether varying the control parameters of each gate in a memory-enabled algorithm can isolate the polyspectrum coefficient within the qubit coherence itself, and whether a memoryless lower bound holds against a strengthened adversary granted all correlations of order $<m$ for free. More broadly, the remark concluding Section~\ref{sec:memoryless_lower_bounds} suggests that a valuable direction is to investigate separations scaling both with the polyspectrum frequency and temporal order. A numerical starting point for this direction is to benchmark the frequency-dependent and order-dependent advantages of memory-enabled QFS against the canonical comb-based reconstruction of Ref.~\cite{Norris_2016}.

\subsection{Q$\Psi$ and future quantum advantages in experimental science}

\subsubsection{Multiparameter $Q\Psi$ and optimization}
Multiparameter estimation is a rich subject in canonical quantum metrology, and central to many realistic sensing tasks. The usual approach promotes the quantum Fisher information to a matrix and the quantum Cramer-Rao bound to a matrix-valued inequality. Upon doing so, the definition of an optimal protocol is dependent on a choice of cost function which encodes which parameters are more important to the estimation task at hand. Due to incompatibility of different measurement bases for estimating noncommuting observables, the matrix-valued Cramer-Rao bound is generally unattainable, so the central goal moves towards designing algorithms that output measurement configurations that approach the lower bound.

Importantly, the multiparameter extension of the QFI/QCRB picture inherits many of the challenges associated with its single-parameter counterpart. In particular, the model still assumes unbiased estimation of all parameters in a differentiable statistical model; moreover, entries of the QFI matrix are often difficult to estimate due to the noncommutative structure of the SLD and observables at hand. 

Given this context, a natural and high-value direction is to extend the Q$\Psi$ formalism to the multiparameter setting. The basic extensions are immediate, and similar to the QFI formalism. Let $P_0$ be a reference law on the signal space and let $h=(h_1,\ldots,h_d)$ be a vector of features, with $h_a\in L^2(P_0)$. For an experiment $M$ with response kernel $K_M(dy|x)$, define
\begin{equation}
    Q_{M,0}(dy)=\int K_M(dy|x)\,dP_0(x),
    \qquad
    A_{M,a}(dy)=\int h_a(x)K_M(dy|x)\,dP_0(x).
\end{equation}
The corresponding score vector is
\begin{equation}
    R_M(y)=\frac{dA_M}{dQ_{M,0}}(y)
    =
    \mathbb E[h(X)\mid Y=y],
\end{equation}
where $X\sim P_0$ and $Y$ is the measurement outcome. The natural Q$\Psi$ signal family is
\begin{equation}
    P_\theta(dx) = \Big(1 + \textstyle\sum_{a=1}^d \theta_a h_a(x)\Big)\,P_0(dx), \qquad \theta \in \mathbb{R}^d .
\end{equation}
For a measurement $M$ with scores $R_{M,h_a}(y) = \mathbb{E}[h_a(X)\,|\,Y=y]$ as in Definition~\ref{def:AFI}, we can define the \textit{matrix-valued AFI}
\begin{equation}
    \big[\mathsf{A}_M(\mathbf{h};P_0)\big]_{ab}
    =
    \int R_{M,h_a}(y)\,R_{M,h_b}(y)\;dQ_{M,0}(y) .
    \label{eq:matrix_afi}
\end{equation}
Upon fixing a particular linear combination of the features via a vector $\theta$, so that we obtain the local statistical model $P_\theta(dx)=(1+\theta^T h(x))P_0(dx)$, we find that $\mathsf A_M({\bf h};P_0)$ is exactly the classical Fisher information matrix of the outcome distribution induced by $M$ at $\theta=0$. Thus the matrix AFI is the feature-learning analogue of classical Fisher information after the restrictions of the sensing architecture have been imposed.

Then, for a matrix of weights $W \succeq 0$ encoding the cost function, the optimal weighted error of $N$ adaptive queries from $\mathcal{M}$ should satisfy
\begin{equation}
\inf_{\widehat{\theta}}\;\mathbb{E}\big[(\widehat{\theta}-\theta)^T W (\widehat{\theta}-\theta)\big]
    \;=\;
    \Theta\!\left(\frac{1}{N}\,\mathcal{E}_W(\mathcal{M})\right),
    \qquad
    \mathcal{E}_W(\mathcal{M}) = \inf_{F \in \mathcal{K}(\mathcal{M})} \mathrm{Tr}\big[W F^{-1}\big] ,
    \label{eq:expected_multiparameter_bound}
\end{equation}
with achievability by allocating shots among finitely many measurements and applying the matched-score estimator of Theorem~\ref{thm:qes_matched_score_guarantee} componentwise.

\begin{graydefinitionbox}
\noindent \textbf{Open Problem (Multiparameter Q$\Psi$).} Prove the matrix-valued semiclassical measurement lemma establishing Equation~\eqref{eq:expected_multiparameter_bound}, including full classical adaptivity, and protocols with persistent quantum memory? When is the information-region bound $\mathcal{E}_W$ tight, and which multiparameter sensing problems exhibit large gaps between the rungs of the sensing hierarchy under simultaneous estimation?
\end{graydefinitionbox}

The proof of the multiparameter bound does not directly emerge from the transcript factorization underlying Lemma~\ref{lemma:semiclassical_quantum_measurement}, because Le Cam's two-point method does not address incompatibility between different features. Natural approaches include Assouad's hypercube method, or first developing a Bayesian analog of AFI theory along the lines of van Trees' inequality and applying it to the proof. 

This direction would further clarify the advantages of an AFI approach to metrology. We  expect that many problems in multiparameter quantum metrology may become more numerically tractable due to the phase-space reduction of noncommutative quantum measurements to scalar statistical response functions, enabling efficient classical search. This also suggests a natural direction for convex optimization theory and numerical search.

\subsubsection{Automating the discovery of quantum advantages}
By a Schur complement, the quantity $\mathcal{E}_W$ of Equation~\eqref{eq:expected_multiparameter_bound} is the value of the convex program
\begin{equation}
    \mathcal{E}_W(\mathcal{M}) = \min_{V,\,F}\;\;\mathrm{Tr}[W V]
    \qquad
    \text{s.t.}
    \qquad
    \begin{pmatrix} V & I \\ I & F \end{pmatrix} \succeq 0\,,
    \qquad
    F \in \mathcal{K}(\mathcal{M}) .
    \label{eq:qpsi_sdp_outer}
\end{equation}
Here, $K(\mathcal M):=\overline{\operatorname{conv}}\left\{\mathsf A_M(h;P_0):M\in\mathcal M\right\}\subseteq \mathbb S_+^d$ is the convex hull of AFI matrices obtained by randomizing over experiments.
Moreover, the phase-space structure of Q$\Psi$ suggests that efficient descriptions of the information region $\mathcal{K}(\mathcal{M})$ exist. An immediate finite-dimensional optimization can be performed by fixing a candidate family of experimental protocols $M_1,\ldots,M_r\in\mathcal M$, with AFI matrices $A_i=A_{M_i}(h;P_0)$. Given this family, the problem reduces to optimizing over convex combinations $F=\sum_i p_iA_i$, where $p_i\geq0$ and $\sum_i p_i=1$. However, a more interesting question is how to parameterize the entire architecture class $\mathcal{K}(\mathcal{M})$, and whether such a description is amenable to numerical optimization. We expect that it is in many cases, because positivity and energy constraints often admit convex descriptions, while the achievable information region $K(\mathcal M)$ is convex by construction under classical randomization. Constraints such as Gaussianity or bounded circuit depth are generally nonconvex at the level of the underlying protocols, and may instead require restricted parameterizations, semidefinite relaxations, or heuristic optimization.

These observations suggest a broad and fruitful intersection between quantum experiments, convex optimization, and deep learning.

\begin{graydefinitionbox}
\noindent \textbf{Open Problem (Q$\Psi$-guided automated discovery of quantum advantage).}
Can we build a broad, closed-form formulation of the multiparameter Q$\Psi$ design problem \eqref{eq:qpsi_sdp_outer}, and thereby enable numerical design of optimal quantum experiments via:
\begin{enumerate}
    \item Convex optimization, when experimental constraints are convex, or
    \item Q$\Psi$-guided machine learning, more generally?
\end{enumerate}
Can this lead to automated discovery of quantum advantages, where quantum experimental problems and realistic constraints are inserted into classical machine learning, and optimal experimental designs emerge alongside rigorous certificates of quantum advantage?
\end{graydefinitionbox}
{
\let\oldaddcontentsline\addcontentsline
\renewcommand{\addcontentsline}[3]{}
\putbib
\let\addcontentsline\oldaddcontentsline
}
\end{bibunit}

\endgroup

\end{document}